\documentclass[a4paper,11pt] {book}
\usepackage[latin1]{inputenc}
\usepackage[T1]{fontenc}
\usepackage[english,francais]{babel}
\usepackage{amsmath} 
\usepackage{amssymb} 
\usepackage{amsfonts}
\usepackage[vcentermath]{youngtab}
\usepackage{slashed}
\usepackage[top=2.5cm,bottom=2.5cm,left=1.95cm,right=1.35cm,asymmetric,a4paper]{geometry}
 \usepackage{feynmp}
 \usepackage[pdftex]{graphicx}
\newcommand{\rmd}{{\rm d}}
\newcommand{\Tr}{\text{Tr}}
\newcommand{\tr}{\text{tr}}

\begin{document}
\selectlanguage{english}
\begin{fmffile}{fgraphs}
\pagestyle{empty}
\begin{titlepage}
\begin{center}

\begin{figure}
\center
\begin{minipage}{0.2\textwidth}
\center
\includegraphics[width=0.6\textwidth]{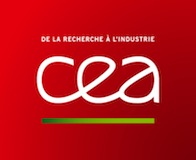}
\end{minipage}
\begin{minipage}{0.2\textwidth}
\center
\includegraphics[width=0.6\textwidth]{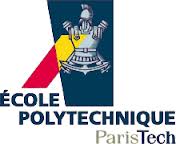}
\end{minipage}
\begin{minipage}{0.2\textwidth}
\center
\includegraphics[width=0.6\textwidth]{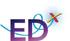}
\end{minipage}
\begin{minipage}{0.2\textwidth}
\center
\includegraphics[width=0.6\textwidth]{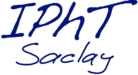}
\end{minipage}
\end{figure}

\textsc{\large \'ECOLE POLYTECHNIQUE \& INSTITUT DE PHYSIQUE TH\'EORIQUE - CEA/SACLAY}\\[1.5cm]

\textsc{\large \bf Th\`ese de doctorat}\\[1cm]

\text{Sp\'ecialit\'e}\\[1cm]

\textsc{\large \bf Physique th\'eorique}\\[1cm]

\text{pr\'esent\'ee par}\\[1cm]

\textsc{\large \bf Julien Laidet}\\[1cm]

\text{pour obtenir le grade de}\\[1cm]

\textsc{\large \bf Docteur de L'\'Ecole Polytechnique}\\[1cm]

\hrulefill \\[.5cm]
{\Large \bfseries High\,Energy\,Collisions of Dense\,Hadrons in Quantum\,Chromodynamics\,:\\[.3cm] LHC Phenomenology and Universality of Parton Distributions}\\[0.5cm]
\hrulefill

\vfill
Soutenue le 11 septembre 2013 devant le jury compos\'e de :\\

  \begin{tabular}{p{0.3\linewidth}p{0.3\linewidth}}
    \multicolumn{1}{c}{~} & \multicolumn{1}{c}{~} \\
Yuri DOKSHITZER & Examinateur\\
Emilian DUDAS & Pr\'esident du jury\\
Fran\c{c}ois G\'ELIS & Directeur de th\`ese\\
Edmond IANCU & Directeur de th\`ese\\
Al MUELLER & Examinateur\\
St\'ephane MUNIER & Examinateur\\
Lech SZYMANOWSKI & Rapporteur\\
Samuel WALLON & Rapporteur \\
  \end{tabular}

\end{center}
\end{titlepage}

\cleardoublepage

\begin{flushright}
\`a mes parents\\
\`a mes fr\`eres

\end{flushright}

\cleardoublepage

\begin{center}
\begin{normalsize}{\bf Abstract :}\end{normalsize}
\end{center}
\begin{quotation}
\noindent  As the value of the longitudinal momentum carried by partons in a ultra-relativistic hadron becomes small, one observes a growth of their density. When the parton density becomes close to a value of order $1/\alpha_s$, it does not grow any longer, it saturates. These high density effects seem to be well described by the Color Glass Condensate effective field theory. On the experimental side, the LHC provides the best tool ever for reaching the saturated phase of hadronic matter. For this reason saturation physics is a very active branch of QCD during these past and coming years since saturation theories and experimental data can be compared. I first deal with the phenomenology of the proton-lead collisions performed in winter 2013 at the LHC and whose data are about to be available. I compute the di-gluon production cross-section which provides the simplest observable for finding quantitative evidences of saturation in the kinematic range of the LHC. I also discuss the limit of the strongly correlated final state at large transverse momenta and by the way, generalize parton distribution to dense regime. The second main topic is the quantum evolution of the quark and gluon spectra in nucleus-nucleus collisions having in mind the proof of its universal character. This result is already known for gluons and here I detail the calculation carefully. For quarks universality has not been proved yet but I derive an intermediate leading order to next-to leading order recursion relation which is a crucial step for extracting the quantum evolution. Finally I briefly present an independent work in group theory. I detail a method I used for computing traces involving an arbitrary number of group generators, a situation often encountered in QCD calculations.
\end{quotation}~\\[.5cm]
\selectlanguage{francais}
\begin{center}
\begin{normalsize}{\large \bf R\'esum\'e :}\end{normalsize}
\end{center}
\begin{quotation}
\noindent  Lorsque l'impulsion longitudinale des partons contenus dans un hadron ultra-relativiste diminue, on observe un accroissement de leur densit\'e. Quand la densit\'e approche une valeur d'ordre $1/\alpha_s$, elle n'augmente plus, elle sature. Ces effets de haute densit\'e semblent \^etre correctement d\'ecrits par la th\'eorie effective du "Color Glass Condensate". Du point de vue exp\'erimental, le LHC est le meilleur outil jamais disponible pour atteindre la phase satur\'ee de la mati\`ere hadronique. Pour cette raison, la physique de la saturation est une branche tr\`es active de la QCD dans les ann\'ees pass\'ees et \`a venir car la th\'eorie et les exp\'eriences peuvent \^etre compar\'ees. En premier lieu, je discute de la ph\'enom\'enologie des collisions proton-plomb qui ont eu lieu \`a l'hiver 2013 et dont les donn\'ees sont sur le point d'\^etre disponibles. Je calcule la section efficace pour la production de deux gluons qui est l'observable la plus simple pour trouver des preuves quantitatives de la saturation dans le r\'egime cin\'ematique du LHC. Je traite \'egalement la limite des \'etats finaux fortement corr\'el\'es \`a grandes impulsions transverses et, par la m\^eme occasion, g\'en\'eralise la distribution de partons au r\'egime dense. Le second sujet principal est l'\'evolution quantique des spectres de gluons et de quarks dans les collisions noyau-noyau, ayant \`a l'esprit son caract\`ere universel. Ce r\'esultat est d\'ej\`a connu pour les gluons et je d\'etaille ici le calcul avec attention. Pour les quarks, l'universalit\'e n'a toujours pas \'et\'e prouv\'ee mais je d\'erive une formule de r\'ecursion interm\'ediaire entre l'ordre dominant et l'ordre sous-dominant qui constitue une \'etape cruciale dans l'extraction de l'\'evolution quantique. Enfin, je pr\'esente brievement un travail ind\'ependant de th\'eorie des groupes. Je d\'etaille une m\'ethode personnelle permettant de calculer des traces impliquant un nombre arbritraire de g\'en\'erateurs, une situation souvent rencontr\'ee dans les calculs de QCD. 
\end{quotation}

\cleardoublepage

~\\[1cm]
\begin{normalsize}{\bf Remerciements :}\end{normalsize}
~\\[1cm]

J'aimerais commencer la liste des remerciements par mes directeurs de th\`ese :  Edmond Iancu et Fran\c{c}ois G\'elis. En effet, merci \`a vous deux de m'avoir permis de travailler dans de bonnes condition, une ambiance sereine et stimulante et d'\^etre l\`a quand j'avais besoin de discuter des questions d'ordre scientifiques. Je vous remercie de votre disponibilit\'e. Merci aussi de m'avoir beaucoup appris pour, \`a l'issue de ces trois ans, dominer \`a peu pr\`es mon sujet et avoir publi\'e des r\'esultats nouveaux et dans l'\`ere du temps.\\

Merci aux membres du groupe de QCD de Saclay qui ont toujours \'et\'e tr\`es disponibles pour les diverses questions scientifiques que j'ai pu avoir \`a leur poser pendant ma th\`ese. Je remercie Jean-Paul Blaizot, Robi Peschanski, Jean-Yves Ollitrault, Yacine Mehtar-Tani, Fabio Dominguez et Gr\'egory Soyez. Je remercie \'egalement ce groupe pour ne jamais avoir entrav\'e ma mobilit\'e pour assister \`a des conf\'erences scientifiques. J'en ai \`a chaque fois tir\'e une tr\`es bonne exp\'erience. Je garde en particulier en t\^ete deux voyages qui m'ont beaucoup marqu\'e : ma collaboration avec Al Mueller \`a l'Universit\'e de Columbia \`a New York en janvier 2013 et la conf\'erence Low-X en Israel en juin 2013. Je reviendrai dans un instant sur la port\'ee scientifique de ces exp\'eriences mais d'un point de vue personnel, elles resteront inoubliables.\\

Lors de ces conf\'erences, ou encore les s\'eminaires plus locaux que j'ai eu l'occasion de donner, j'ai pu rencontrer foule de personnes avec qui j'ai eu des discussions scientifiques (ou non) tr\`es \'eclairantes. Tout d'abord je tenais \`a exprimer ma gratitude \`a Al Mueller pour sa gentillesse et sa patience ajout\'ees \`a l'\'eminence du scientifique. On apprend \'enorm\'ement en parlant, m\^eme bri\`evement, avec lui. Son sens physique et sa facult\'e \`a faire des calculs de th\'eorie des champs avec des formules ne prenant pas plus d'une ligne m'ont beaucoup impressionn\'e. Je voulais remercier St\'ephane Munier qui m'expliquait les choses simplement lorsque je ne les comprenais pas et aussi d'avoir accept\'e d'\^etre dans mon jury. Je remercie \'egalement Lech Szymanowski et Samuel Wallon. Je les remercie tous deux d'avoir accept\'e d'\^etre mes rapporteurs. Petite parenth\`ese concernant Samuel ; sache que j'ai beaucoup appr\'eci\'e t'avoir fait d\'ecouvrir les joies de la plong\'ee en apn\'ee apr\`es la conf\'erence \`a Eilat. Je remercie \'egalement Christophe Royon avec qui j'ai pass\'e la plupart du temps libre \`a la conf\'erence de Cracovie en 2011 puis celle d'Israel en 2013 ainsi que la petite (et magnifique) excursion en Jordanie qui en a suivi. Je remercie aussi Cyrille Marquet pour sa sympathie. Nous nous sommes crois\'es \`a de multiples reprises et ce fut \`a chaque fois des moments conviviaux. Je tenais \`a remercier ceux avec qui j'ai fait des rencontres plus br\`eves mais illuminantes scientifiquement : Alex Kovner, Yuri Kovchegov ou Kevin Dusling pour ne citer qu'eux. Il y a aussi ceux que j'ai pu rencontrer \`a des conf\'erences avec qui j'ai eu l'occasion de sympathiser : Guillaume Boeuf, Benoit Roland et j'en oublie s\^urement \'egalement. Enfin, je remercie ceux de mon jury que je n'ai pas cit\'es jusqu'ici pour avoir accept\'e d'en faire partie : Yuri Dokshitzer et Emilian Dudas.\\

Passons maintenant au laboratoire. Je tiens \`a remercier les directeurs successifs Henri Orland puis Michel Bauer pour m'avoir accueilli ici. Mention sp\'eciale \`a l'\'equipe administrative : Anne Capdepon, Catherine Cataldi, Sylvie Zaffanella, Laure Sauboy, Emilie Qu\'er\'e, Lo\"ic Bervas et, depuis peu, Morgane Moulin ; de l'informatique Pascale Beurtet, Patrick Berthelot, Philippe Caresmel et Laurent Sengmanivanh ainsi qu'Emmanuelle De Laborderie de la documentation. Toujours disponibles, efficaces et avenants, je vous f\'elicite pour ce que vous faites au sein du laboratoire. Je me souviens avoir bien ri avec certains d'entre eux (ou gr\^ace \`a certains d'entre eux quand mes oreilles tra\^inaient) qui se reconna\^itront. Merci aussi aux "grands fr\`eres" des th\'esards : Olivier Golinelli puis \`a sa succession St\'ephane Nonnenmacher. Des oreilles attentives, des conseils pr\'ecieux, une facult\'e \`a rassurer, je ressortais toujours des petits entretiens annuels (quoique pas si petits que \c{c}a en fait) avec le sourire et l'esprit plus clair quant \`a mon sombre avenir de jeune travailleur dans un monde qui part en eau de sucette. J'ai trouv\'e qu'il r\'egnait globalement une tr\`es bonne ambiance ici \`a l'IPhT offrant ainsi un lieu o\`u les \'echanges, scientifiques ou non, \'etaient ais\'es avec les coll\`egues.\\

Je tenais \`a remercier Chantal Rieffel, mon professeur de physique de terminale, sans qui je ne serais peut \^etre pas all\'e dans cette voie. Elle a su me donner gout \`a cette science \`a laquelle je ne me destinais pas forc\'ement (ainsi qu'\`a la chimie mais cette lubie m'est vite pass\'ee une fois \`a la fac o\`u j'ai rapidement pr\'ef\'er\'e les math\'ematiques). Chantal, vous n'\^etes pas \'etrang\`ere \`a mon cheminement, voil\`a pourquoi vous m\'eritiez largement de figurer ici \'egalement.\\

Sur le plan plus personnel, je veux remercier en tout premier lieu ma maman et mon papa. Je vous pr\'esente mes excuses pour cet incident regrettable du 17 septembre 1987 et esp\`ere que vous ne vous en mordez pas trop les doigts. Blague \`a part, merci d'\^etre l\`a quand j'ai besoin, merci de m'avoir soutenu dans ce que je faisais, merci pour mille autres choses encore dont la liste pourrait faire l'objet de l'appendice G, sans vous je ne sais pas si je serais l\`a aujourd'hui et c'est sans l'ombre d'une h\'esitation que je vous d\'edie cette th\`ese (vous avez int\'er\^et \`a la lire !). Ces remerciements sont l'occasion pour moi de vous dire ce qui a tant de mal \`a sortir mais que je pense profond\'ement : je vous aime. Enfin, je suis tr\`es content de faire avec vous ce voyage de fin d'\'etude dans le Far West am\'ericain \`a la place de commencer un nouveau travail ! Au passage, j'ai \'egalement une pens\'ee pour mes fr\`eres, Pascal et St\'ephane, qui comptent beaucoup pour moi ainsi qu'\`a toute leur famille.\\

Apr\`es mes parents je veux remercier les copains. Tout d'abord je voulais remercier de loin les deux meilleurs amis que j'ai : Geoffroy, engag\'e dans la Marine et Guillaume, qui n'est pas dans la marine mais qui rame. Merci \`a vous deux pour tous ces bons moments qu'on passe ensemble depuis maintenant un paquet d'ann\'ees qu'il s'agisse des soir\'ees, de la cueillette des champignons, de la bonne ripaille, du bon vin et j'en passe. R\'eunis par une passion commune : la p\^eche, je ne compte plus les bi\`eres et les blagues salaces partag\'ees avec vous. Merci pour cette franche camaraderie. Merci \`a Marine, pour ses succulents d\^iners o\`u on est toujours re\c{c}u comme de rois. Marine, tu sais que je t'estime beaucoup, j'en profite d'ailleurs pour te dire que j'attendrai le temps qu'il faudra que tu te lib\`eres enfin de cet homme qui ne te m\'erite pas. Je pense aussi aux copains du lyc\'ee que je perds un peu de vue au fil des ann\'ees : Goli, Isa, Mickey et Marc. Je remercie les copains que je me suis faits pendant les \'etudes et notamment ceux du M2 de Physique Th\'eorique que je continue de voir plus ou moins : Melody, Fabinou, Pilou, Ahmad, Axel et Adrien notamment. Melody, ma camarade du M2, nous travaillions ensemble et nous motivions l'un et l'autre, j'en garde un tr\`es bon souvenir. Mais je me rappelle aussi et surtout de nos moments de d\'econnade o\`u sous tes airs de petite fleur fragile et immacul\'ee se cachait quelqu'un de tr\`es bon public et ouvert \`a toutes formes d'humour. En \'ecrivant ces lignes, je me rappelle de la vilaine gu\^epe qui t'emb\^etait et qui a fini en tache de jus sur mon poly de cours. Les rares fois o\`u nous nous sommes revus ensuite ont toujours \'et\'e un plaisir pour moi. Fabinou, tout ce que j'ai \`a dire c'est, dommage que la Nature t'aie faite homme. Un plaisir de te recroiser \'egalement. Pilou, la petite fleur bleue, j'ai toujours bien aim\'e ton c\^ot\'e un peu \`a c\^ot\'e de tes pompes. Tu me fais penser \`a Droopy avec la t\^ete de Tintin. Avec Ahmad, nous \'etions les deux seuls \`a comprendre mutuellement nos blagues un peu trop \'elabor\'ees pour les petites gens. Le destin nous a r\'eunis en th\`ese lors de ma visite au CERN o\`u l'on m'a plac\'e \`a ton bureau. Enfin Axel et Adrien, je repense avec nostalgie \`a ces journ\'ees \`a travailler ensemble les DM de fin d'ann\'ee dans une ambiance \`a la fois studieuse et relax\'ee. Adrien, c'\'etait tr\`es sympa de faire la conf\'erence Low-X avec toi entre la plong\'ee, la Jordanie, la d\'egustation de testicules de dindons, le road trip o\`u on a pass\'e la derni\`ere nuit dans la voiture sur un terrain vague et bien d'autres souvenirs encore. Enfin, les derniers mais pas les moindres : les copains du labo. Ceux qui font qu'on se l\`eve le matin en se disant "aujourd'hui je vais au labo, je vais travailler bien sur, mais je vais aussi croiser les copains". Je distinguerai trois g\'en\'erations. Tout d'abord les "ante", Jean-Marie, H\'el\`ene, Emeline, Nicolas, Roberto et surtout Bruno avec qui j'ai partag\'e mon bureau dans une parfaite cohabitation. Ensuite il y a ceux de ma g\'en\'eration : Pitou et Romain, camarades de M2 volontairement omis plus haut pour figurer ici, Alexandre (le pauvre, il \'etait le bouc-\'emissaire de Pitou) et tous les autres. Pitou, despote auto-proclam\'e chef des th\'esards, Romain, qui m'a attribu\'e le titre d'organisateur du s\'eminaire des th\'esards et Alexandre qui avait toujours la petite blague apr\`es laquelle on entendait r\'esonner une corne de brume. Enfin, il y a les "post" : Katya, Eric, Thiago, Alexander, Beno\^it, Hannah, Antoine, R\'emi, J\'erome, H\'el\`ene et d'autres encore. Ces derniers constituent une belle rel\`eve, les plaisanteries grivoises et de mauvais go\^ut ont encore de beaux jours devant elles le midi \`a la cantine. A tous, merci d'avoir contribu\'e \`a la bonne ambiance - bien que platonique - au sein des th\'esards.

\selectlanguage{english}

\tableofcontents
\addtocontents{toc}{\protect\thispagestyle{empty}} 

\chapter*{Introduction}
\setcounter{page}{1}
\pagestyle{headings}
\addcontentsline{toc}{chapter}{Introduction}
\chaptermark{Introduction}
\indent

So far we know four interactions in Nature : gravity, electromagnetism, strong and weak interactions. All of them still have their mysteries and open problems. Gravity, is described at the classical level by general relativity \cite{Einstein:1914bt,Einstein:1914bw,Einstein:1915ca}, a gauge theory invariant under the group of diffeomorphism or an $SO(3,1)$ gauge theory in its vierbeins formulation. General relativity has predicted plenty of very accurate results which agree strikingly with experiments. The most impressive one is the period decrease of the binary pulsar PSR B1913+16 by gravitational radiation measured by Hulse and Taylor in 1974 \cite{Hulse:1974eb} which shows a 1\% agreement with the post-newtonian developments of general relativity and, by the way, provides an indirect evidence of gravitational waves. Although general relativity describes with a great accuracy astrophysical and cosmological observational phenomena, it has a singular short distance behavior. Theoretical troubles arise for instance in the $r=0$ limit of the Schwarzschild solution for black holes or in the $t=0$ limit of cosmological solutions. These singularities motivate a quantum description of gravity for understanding them. However, due to its inherent geometrical interpretation of space-time, gravity must be distinguished from others interactions. Problems arise when one tries to quantize gravity : one first faces conceptual problems when defining the Hilbert space since, the main difference with the others interactions is that gravity is not the theory of particles moving in a given background but the dynamics of the background itself. Due to this particular nature of gravity it is probable that it cannot be described at the quantum level in the same way as the other interactions. The quantum theory of gravity may even not be a field theory. Some alternative approaches have been proposed like loop quantum gravity or string theory. However we are quickly lost in the complexity of these theories and the predicted phenomenology, allowing to check whether or not they seem to be correct, so far lies beyond the scope of accessible experiments. If one tries however to apply quantum field theory techniques to gravity one faces to a technicality making the calculations quickly very cumbersome : gravitational interactions are non renormalizable and one has to consider the infinite serie of interactions allowed by diffeomorfic invariance. Since high order interactions do not play role at low loop level and/or in the computation of Green function with a small number of legs, one can however proceed step by step for renormalizing the couplings one by one (see for instance \cite{'tHooft:2002xp}). Anyway, either a crucial point has been missed with gravity or we have the right theories but in which the quantum description of gravity is still not clear. Concerning electromagnetism, the situation is better. It is described at the quantum level by the $U(1)$ abelian gauge theory known as quantum electrodynamics (QED) \cite{Tomonaga:1946zz,Schwinger:1948iu,Schwinger:1948yk,Feynman:1949zx,Feynman:1949hz,Feynman:1950ir,Dyson:1949bp,Dyson:1949ha}. The low energy sector of QED is nowadays under control. The radiative corrections to the fine structure constant are now known up to five loops \cite{Aoyama:2012wj} and the computed value agrees with experiment with an accuracy of order $10^{-9}$, for sure, one of the best successes of theoretical physics. The QED beta function is positive and its Landau pole is reached $10^{286}$ eV. Of course this energy is much beyond accessible experiments and it is probable that QED is replaced by an unknown new physics long before reaching this scale. In everyday life experiments, QED is perturbative and is nowadays well understood. QED has even been unified with the theory of weak interactions in the $SU(2)\times U(1)$ gauge theory known as the Glashow-Weinberg-Salam (GWZ) model \cite{Glashow:1961tr,Goldstone:1962es,Salam:1964ry,Weinberg:1967tq,Salam:1968rm}. This very elegant and simple model provides, through the Higgs mechanism \cite{Englert:1964et,Higgs:1964pj,Guralnik:1964eu}, explanations to puzzling experimental phenomena like maximal parity violation (there does not exist right handed neutrinos in Nature) and electrically charged gauge bosons $W^{\pm}$. The GWS model predicts new features which have been checked experimentally : in addition of being charged, the gauge bosons have to be massive and, in addition, the GWS model predicts the existence of a new neutral, massive gauge boson, the $Z^0$. Both the massive character and the $Z_0$ have been observed at CERN in 1974 \cite{Denegri:2004tv}. The corner stone of the GWS model is the Higgs boson which was the missing piece of the standard model until it has been finally observed at the LHC in 2012 \cite{CMS:2012wwa,ATLAS:2012eoa}. The coupling of weak interactions have a nice behavior : it has no Landau poles. At low energy - i.e. energies lower than the $W$ and $Z$ bosons masses - the beta function is positive as in QED. At higher energy, gauge bosons balance the growth of the coupling which then decreases at high energy. Everything would be perfect with weak interactions if they do not show up $CP$ violations first discovered in 1964 in kaon decay \cite{Christenson:1964fg}. $CP$ violations still remain obscure for theoreticians.\\

\begin{figure}[h]
\begin{center}
\includegraphics[width=0.30\paperwidth]{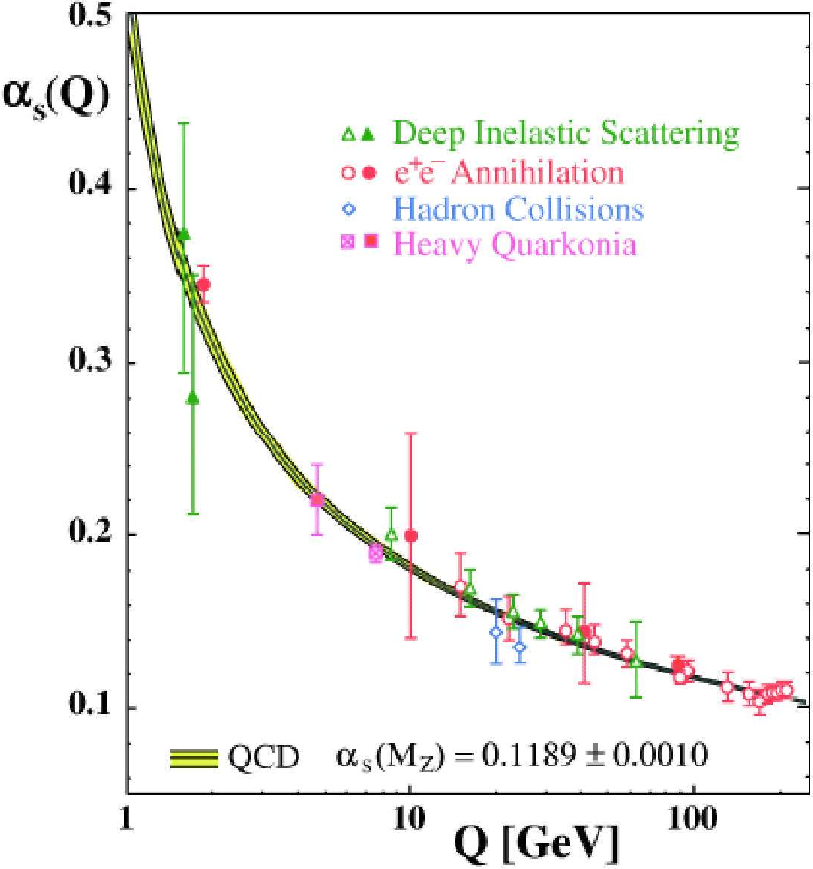}
\caption{Experimental evidence of the running coupling and asymptotic freedom. The coupling diverges at $\Lambda_{QCD}\sim 200$ MeV.\label{runcoupling}}
\end{center}
\end{figure}

The last type of interaction is the strong interaction, the one considered in this thesis. In its modern formulation, the theory of strong interactions is known as quantum chromodynamics (QCD), the theory of quarks and gluons. It turns out that matter made of quarks and gluons represents the main percentage of identified matter in our universe\footnote{Cosmological observations have shown that what we call "matter" actually represent 5\% of the whole cosmological cocktail. 68\% of the content of our universe is dark energy and 27\%, dark matter.}. This is one reason - not the only one - why QCD is so important since it governs the dynamics of most of the known matter. The quarks form so-called \emph{color} multiplets and furnish a fundamental representation of an $SU(3)$ gauge symmetry whose vector bosons are the gluons, lying in the adjoint representation of the gauge group. Just as weak interactions, the gauge group is non abelian and gluons, represented by a vector field $A^A_{\mu}$ ($A,B,C...$ denoting the adjoint representation color index), are described by the Yang-Mills lagrangian :
\begin{equation*}
\mathcal{L}_{YM}=-\frac{1}{4}F^A_{\mu\nu}F^{A\mu\nu}
\end{equation*}
where $F^A_{\mu\nu}$ is the field strength tensor
\begin{equation*}
F^A_{\mu\nu}=\partial_{\mu}A^A_{\nu}-\partial_{\nu}A^A_{\mu}+gf^{ABC}A^B_{\mu}A^C_{\nu}.
\end{equation*}
$g$ is the coupling constant, $f^{ABC}$ are the gauge group structure constant and we adopt the convention of an implicit summation over repeated indices. The Yang-Mills lagrangian alone corresponds to a theory containing only gluons. Quarks enter as Dirac fields $\psi^a$ ($a,b,c...$ denoting the fundamental color indices) minimally coupled to the gauge field via covariant derivatives\footnote{In many cases fundamental color indices can be understood to alleviate notations. Moreover we shall often write $A_{\mu}=A^A_{\mu}T^A$.} : $D^{ab}_{\mu}=\delta^{ab}\partial_{\mu}-igA^A_{\mu}(T^A)_{ab}$, where the $T^A$'s are the gauge group generators in the fundamental representation. Since there can be several copies of quark multiplets they will be denoted, in general with a \emph{flavor} label $f$. Thus the lagrangian corresponding to quarks reads (with all indices, except spinor ones, explicitly written, soon dropped out) :
\begin{equation*}
\mathcal{L}_{\rm quark}=\sum\limits_f \bar{\psi}^a_f\left(i\slashed{D}^{ab}-\delta^{ab}m\right)\psi^b_f.
\end{equation*}
The total lagrangian $\mathcal{L}_{YM}+\mathcal{L}_{\rm quark}$ is the QCD lagrangian. Non abelian gauge invariance requires that all fields couple with the same coupling constant\footnote{This is a fundamental difference between abelian and non abelian gauge theories. In abelian theories the gauge parameter can be arbitrarily rescaled for every fields and the coupling constant can be assigned any value, whereas in non abelian theories it cannot and the coupling is quantized. It is still an open question why the electric charges in Nature are all integral multiples of $e/3$, where $e$ is the electron's charge. QED alone do not predict such quantization condition. It can be explained for instance by assuming the existence of magnetic monopoles \cite{Dirac:1931kp}.} $g$. Up to now, the formalism applies for any simple gauge group and we shall deal with $SU(N_c)$ instead of $SU(3)$, $N_c$ being the number of colors. The first interesting physical consequence of the QCD lagrangian is the behavior of the running coupling. At one loop level, the beta function reads :
\begin{equation*}
\beta(g)=-\frac{g^3}{3(4\pi)^2}\left(11N_c-2n_f\right),
\end{equation*}
where $n_f$ is the number of flavors. So far we know 6 types of quark flavors which make the beta function negative in QCD (that is with $N_c=3$). Such a beta function predicts the following behavior of the fine structure constant of strong interactions $\alpha_s=g^2/4\pi$ with the energy $\mu$ :
\begin{equation*}
\alpha_s(\mu)=\frac{2\pi}{(11-\frac{2}{3}n_f)\ln(\mu/\Lambda_{QCD})}.
\end{equation*}
Thus $\alpha_s$ has a Landau pole at $\Lambda_{QCD}$, experimentally measured to be approximately $200$ MeV. On the one hand, at energies larger than $\Lambda_{QCD}$, the coupling decreases and tends to zero. This property is known as the \emph{asymptotic freedom} \cite{Gross:1973id,Politzer:1973fx} (see plot \ref{runcoupling}). At high energy, QCD is weakly coupled and perturbation theory is allowed. On the other hand, at energies less than $\Lambda_{QCD}$, we are in the non perturbative regime and perturbation theory, which is the only available tool for analytical calculations, breaks down. The main non perturbative property of QCD is the \emph{confinement}. It turns out that the range of strong interactions is very short $\sim 10^{-15}$ m and, in addition, the physical spectrum only contains color singlet bound states like mesons and baryons, called hadrons. Why gluons are confined, that is they cannot be observed directly. Why quarks cannot be observed individually but combined in color singlets. How these bound states follow from the QCD lagrangian. These are closely related and still unanswered questions awarded with a one million dollars price for the one who will show these properties theoretically. These properties indeed seem to follow from the QCD lagrangian according to lattice calculations.\\

The most powerful tool ever for exploring matter at very short scales, unreached so far, has been switched on since 2008. This tool is the Large Hadron Collider (LHC) at the CERN in Geneva. Accessible energies will be of order $10$ TeV in the center of mass frame. Concerning its applications to QCD, energies available at the LHC lie far beyond $\Lambda_{QCD}$, that is, in the perturbative regime. Hence, having in mind applications to LHC experiments, the calculations performed in this thesis use perturbative QCD. They will also apply to physics occurring in the others large accelerators : the Hadron-Electron Ring Accelerator (HERA), located at DESY in Hamburg and the Relativistic Heavy Ion Collider (RHIC), located at the Brookhaven National Laboratory, where the center of mass energy is of order $100$ GeV. In all these accelerators, the QCD coupling constant is larger than the electro-weak ones. Thus the strong interactions dominate all the processes involving hadronic matter, another reason for putting QCD on a pedestal.\\

Let us enter in more detail into the high energy behavior of hadronic matter. Historically, the first attempts for building a theory of strong interactions was a theory whose fundamental particles were the hadrons (mesons and baryons). In the early sixties experimental phenomena such as the Bjorken scaling showed the composite nature of hadrons interpreted as bound states involving \emph{partons}, that is, valence quarks interacting via gluons. Connecting these results with the works of Yang and Mills have led to the modern formulation of QCD. The asymptotic freedom property justifies, at high energy, the parton model proposed by Feynman \cite{Feynman:1969ej} based free partons - quarks and gluons - within the hadron interacting weakly and not coherently \cite{Bjorken:1969ja}. We shall see that, in this context, parton distributions naturally merge and count the number of partons carrying a given momentum value. Of course low energy partons enter into the unknown non perturbative part of the wave function but given a low energy configuration up to some scale, it is possible to look at the variation of the distribution under a small change of the scale. This leads to evolution equations and they are of two types : one is with the virtuality of the partons and leads to the DGLAP renormalization group equation \cite{Altarelli:1977zs,Dokshitzer:1977sg,Gribov:1972ri} and the other one is at given parton energy but evolving the energy transferred by the parent hadron and leads at first to the BFKL evolution \cite{Lipatov:1976zz,Kuraev:1977fs,Balitsky:1978ic}. The DGLAP evolution knew a great success since it explained many phenomena like the Bjorken scaling deviations while BFKL was ignored. The BFKL equation revived in the nineties when HERA data for deep inelastic electron-proton scattering showed for the first time the structure of a high energy hadron. The most striking phenomena was the rapid rise of the distribution with decreasing transferred energy. This is the first evidence of high parton density, the main purpose of this report. Nowadays, saturation is better understood. The raise of parton density must be finite otherwise there would be troubles with unitarity. At high density, recombinations between partons balance the growth of density and their number \emph{saturates} to a fixed value of order $1/\alpha_s$ per phase space element. In the high energy saturated regime, a large number of partons form a sort of "soup" with many interactions among them. Although the coupling is weak the large number of partons involves collective phenomena. Due to the very large number of particles, strongly entangled with each others, usual Feynman diagram techniques become vain. The saturated regime seems to be well described by the Color Glass Condensate (CGC) effective field theory whose validity is confirmed by its predictions for proton-proton and deuteron-gold collisions at RHIC. A natural continuation is the prediction of CGC at the LHC, where, for the first time, saturated hadronic matter is fully expected. This topic is treated in chapter \ref{pAchapter}. The other main interesting question risen up in this thesis is the universal character of parton distributions :  they are intrinsic properties of the hadrons, independent of the reaction or observable considered. This will be checked in nucleus-nucleus collisions in chapter \ref{AAcollisions}. This last case shows a particular complexity due to the presence of two dense media.

~\\[1cm]
\begin{normalsize}{\bf How to read this thesis ?}\end{normalsize}
~\\[1cm]

How to handle this report and what is contained inside ? The main body includes the following chapters :\\

Chapter \ref{LC-QFT} sets the normalization conventions in light-cone quantized field theories. First it can be helpful for the reader who is not familiar with light-cone quantum field theory and related miscellaneous like the light-cone gauges. The derivation of quantum field theory in light-cone coordinates presents very few differences with the quantization in Minkowski coordinates and we just point out some of the small differences. For the reader used to this formalism this chapter can be skipped. It is also useful for the reader interested in following precisely the calculations. Indeed all the normalization conventions of states, fields, creation and annihilation operators... are set here and are the ones used in the whole report.\\

Chapter \ref{QCDsat} is an introduction to small $x$ and saturation physics that will be the basic physical ground of the following chapters. Of course it would have been too long to detail it in an exhaustive way and useless since this topic is widely covered by textbooks. I rather tried to emphasize the physical insight and the key ideas that lead to the concept of dense QCD matter. For clarity I preferred to not discuss notions that will not be used in the following - like the dipole model for instance - although they are the cornerstone for a more rigorous derivation. The small $x$ evolution equations are motivated and explained from the intuitive point of view but once again I preferred to state the results avoiding long derivations which can be found in the already existing literature. To summarize, I tried to introduce tools and ideas necessary for the following but not more than that. This section will use important results from appendices \ref{nonabelianBF} and \ref{eikonalprop} whose long and technical derivation obliged me to put them outside of the main line for clarity.\\
 
Chapter \ref{pAchapter} is the first one dealing with new results. Especially the di-gluon production cross-section is the main result of our paper \cite{Iancu:2013dta} with Edmond Iancu. Many physical ideas introduced in the previous chapter are used in this one. I spent some pages to motivate why the di-gluon production has a special importance in p-A collisions at the LHC for finding quantitative evidences of saturation physics. To get the formal background, I first deal with the simplest case : the single quark production. Indeed it shows the emergence of color operators and playing with it enables us to introduce almost everything we will need for the di-gluon production. Then when I discuss the di-gluon case, I will not have to set plenty of definitions all along the discussion. Appendices \ref{pAcrosssections} and \ref{shockwaveFeynrules} will be used in this chapter for cross-sections and Feynman rules respectively. These appendices are the derivation of results which are intuitive in the sense that the structure of the cross-sections and the Feynman rules can be more or less guessed by a familiar reader who is more interested by an overall understanding rather than a careful check of prefactors.\\

Chapter \ref{AAcollisions} is more theoretical. It deals with the small $x$ evolution in nucleus-nucleus collisions that shows a universal character encoded into the factorization property. I fully detail calculations leading to the LO to NLO recursion relations both for gluons and quarks. The former is the new result, mentioned in our paper with Francois Gelis \cite{Gelis:2012ct}. The starting point of this chapter will make use of the Schwinger-Keldysh formalism treated in appendix \ref{Keldyshappendix}. There are also gauge fixing questions whose deep existence are justified in appendix \ref{BRSTspectrum}.\\

Chapter \ref{invtensors} deals with Lie algebra and is independent. I motivate the chapter with an example of calculation following from chapter \ref{pAchapter} where it can find applications. However results exposed here are not used anywhere else in the thesis. I deal with a method for computing traces containing an arbitrary number of group generators, a situation sometimes encountered in QCD computations. This chapter is formal and closer to mathematics but remains in the spirit of mathematics for physicists.\\

Appendices contain either too long or too technical calculations that would have broken the continuity of the discussions in the main body or elements of unusual formalisms with which even experts in the domain may not be used to.\\

Appendix \ref{nonabelianBF} is the derivation and the justification of the external field approximation in non abelian gauge theories. Although it may be intuitive to consider classical Yang-Mills fields in some situations, it is a highly non trivial result assuming strong hypothesis. This approximation is well known in QED but breaks down if transposed to the non abelian case. I will discuss the physical conditions allowing such an approximation in the non abelian case and make the calculation explicitly.\\

Appendix \ref{eikonalprop} is the natural continuation of the previous one. In \ref{nonabelianBF}, I justified to which extend a field radiated by sources can be treated at the classical level, here I deal with the interaction of quantum fields evolving in this classical background field in the eikonal approximation - justified by the way. I show how the color structure of the S-matrix in encoded into Wilson lines corresponding to particles colliding the background field in the eikonal approximation.\\

Appendix \ref{pAcrosssections} is a setup about the structure of cross-sections in p-A collisions. The presence of a background field associated to the target makes the ordinary relations between amplitude and cross-sections break down. Here I write the corresponding relation in this specific case. Moreover I will discuss collinear-factorization which leads to a simple contribution of the proton to the total cross-section.\\

Appendix \ref{shockwaveFeynrules} is a derivation of Feynman rules widely used in \ref{pAchapter}. Once one has understood the role played by the Wilson lines in the eikonal approximation, the corresponding Feynman rules are rather intuitive. However, here I make the derivation carefully with various phases and prefactors included.\\

Appendix \ref{Keldyshappendix} details the Schwinger-Keldysh formalism, well known in condensed matter physics but more rare in quantum field theory. It would have been too long to detail it in \ref{AAcollisions} but its results may seem non obvious to the unfamiliar reader. I thus detail how it comes out in quantum field theory for computing inclusive observables. The Schwinger-Keldysh formalism leads to generalizations of path integrals and Feynman rules, also detailed in this appendix.\\ 

Appendix \ref{BRSTspectrum} is the determination of the physical spectrum in light-cone gauge using the BRST symmetry. I discuss how the physical spectrum is given by the BRST cohomology. I show the well known result that ghosts and anti-ghosts are absent from the physical spectrum and that the physical gauge field's degrees of freedom are given by the transverse components only. I thought it was worthwhile to detail this proof which uses beautiful mathematics which are unusual in QCD. \\

\chapter{Light-cone quantum field theory\label{LC-QFT}}

\setcounter{equation}{0}
\indent

The physical issues discussed in this thesis are related to high energy collisions of hadrons in accelerators. In the lab frame, the two hadrons have opposite velocities close to the speed of light\footnote{Note that the lab frame and the center of mass frame are distinct but similar in the sense that in both of them the speeds of the two hadrons are comparable to the speed of light in the nowadays available accelerators.}. In such ultra-relativistic collisions it is easier to work with the so called light-cone coordinates rather than the Minkowski ones. Moreover, the velocities of involved particles are so large that one can neglect their mass and treat them as light-like representations of the Lorentz group. Although light-cone quantization breaks Lorentz invariance, it provides, in most of cases, a very convenient choice for making practical calculations applied to di-hadron collisions.\\

This preliminary section introduces the basic tools and conventions. It can be skipped by the reader who is not interested in following calculations in details in the next chapters. The aim of this section is to set a precise catalog of normalization conventions used in all this thesis since there are as many conventions as there are authors. The reader can refer to this section at any time to check the prefactors in various formulas. The first section introduces the light-cone coordinates system and some properties of the four-momentum in this system. The second section is a catalog of the normalization conventions used for states, creation and annihilation operators and fields consistent with light-cone quantization (we mimic the conventions used in \cite{PeskinSchroeder}). The third section is devoted to the axial gauge which will be the gauge used in most of the following when we deal with gluons.

\section{Light-cone kinematics}
\indent

As long as we are working in a frame in which particles travel with velocities close to the speed of light, it is useful to work in light-cone coordinates rather than in the Minkowski ones. Let $a$ be a 4-vector whose Minkowski coordinates read $(a^0,\mathbf{a})$, we define the light-cone coordinates as :
\begin{equation}
a^{\pm}=\frac{a^0\pm a^3}{\sqrt{2}}~~~~~~~\mathbf{a}_{\perp}=(a^1,a^1).
\end{equation}
In these coordinates, we conventionally order the components as follow : $a=(a^+,a^-,\mathbf{a}_{\perp})$. The scalar product of two vectors $a$ and $b$ reads $a\cdot b=a^+b^-+a^-b^+-\mathbf{a}_{\perp}\cdot\mathbf{b}_{\perp}$. The light-cone components do not require to make a distinction with the Minkowski ones, the are labeled with Greek indices $\mu, \nu...$. The transverse components are denoted with a Latin index $i, j...$ running over the values $1$ and $2$. Since we never use explicitly spatial Minkowski coordinates there will be no confusion possible and $i, j...$ indices will always refer to transverse components in light-cone coordinates (in chapter \ref{AAcollisions} we shall go back to the Minkowski coordinate system but we will not have to use these labels explicitly for denoting spatial components).\\

Let us focus on the four-momentum vector in light-cone coordinates. We consider a single free particle of mass $m$. In light-cone coordinates, its momentum reads $p=(p^+,p^-,\mathbf{p}_{\perp})$, the mass-shell condition $p^2=m^2$ constrains $p^{\pm}$ to be $p^{\pm}=(\mathbf{p}_{\perp}^2+m^2)/2p^{\mp}$. Moreover  $p^+$ and $p^-$ have to be both positive. Indeed, the mass-shell condition tells us that $p^+p^-$ must be positive (or possibly zero in the massless case) and therefore $p^+$ and $p^-$ must have the same sign, but physically consistent Lorentz group representations must have $p^0=(p^++p^-)/\sqrt{2} >0$ and then both $p^+$ and $p^-$ are positive. The Lorentz invariant measure becomes in light cone coordinates :
\begin{equation}
\label{LorentzInvMeasure}
\int\rmd p^+\rmd p^-\rmd^2p_{\perp}\theta(p^++p^-)\delta(p^2-m^2)f(p)=\int_{p^{\pm}>0}\left.\frac{\rmd p^{\pm}\rmd^2p_{\perp}}{2p^{\pm}}f(p)\right|_{p^{\mp}=\frac{\mathbf{p}_{\perp}^2+m^2}{2p^{\pm}}}.
\end{equation}
As long as we deal with right-moving (resp. left-moving) particles, the $x^+$ (resp. $x^-$) direction in space-time plays the role of time. Therefore it is natural to refer $p^-$ (resp. $p^+$) as the "energy" and $\vec{p}=(p^+,\mathbf{p}_{\perp})$ (resp. $\vec{p}=(p^-,\mathbf{p}_{\perp})$) as the "spatial" components of the momentum and one chooses the upper (resp. lower) sign in the Lorentz invariant measure \eqref{LorentzInvMeasure}. Although this interpretation is convenient for dealing with ultra-relativistic reactions, Lorentz invariance is broken since the range of accessible frames consistent with these conventions is restricted to the ones that conserve the right-moving (resp. left-moving) character of considered particles. Actually for the purposes considered here it is not really a problem since the frame will be fixed once for all as we shall see. Now let us investigate the underlying quantum theory in the light-cone coordinates language.

\section{One-particle states and quantum fields}
\indent

In this section we are going to deal with right-moving fields, the transposition to left-moving ones will be obvious by just changing the plus components into minus ones and conversely. Of course this section will not be a far-reached and complete rederivation of quantum field theory in light-cone coordinates which does not show up particular difficulties and is rather straightforward. The consistency of conventions can be checked by the reader following the procedure detailed in \cite{WeinbergI} of but with conventions of \cite{PeskinSchroeder}. We rather give a catalog of various conventions for the normalization of states and fields that will be used in all the following. The natural way is to describe a one-particle state by the quantum number $\vec{p}$ and some possible discrete quantum numbers like spin, colors... generically denoted $\sigma$. These states are conventionally normalized in the restricted Lorentz invariant manner explained above so that :
\begin{equation}
\label{statesnormalization}
\left\langle \left. \vec{q}, \bar{\sigma}\right|\vec{p}, \sigma\right\rangle=(2\pi)^32p^+\delta_{\sigma\bar{\sigma}}\delta^{(3)}(\vec{p}-\vec{q}).
\end{equation}
The state $\left.\left.\right|\vec{p}, \sigma\right\rangle$ is created from the vacuum (normalized to unity) by a creation operator $a^{\dagger}_{\vec{p}, \sigma}$ satisfying the commutation (minus sign), if they are bosons, or anti-commutation (plus sign) relations if they are fermions :
\begin{equation}
\begin{split}
\left[a_{\vec{p}, \sigma};a^{\dagger}_{\vec{q}, \bar{\sigma}}\right]_{\mp}&=(2\pi)^3\delta_{\sigma\bar{\sigma}}\delta^{(3)}(\vec{p}-\vec{q})\\
\left[a_{\vec{p}, \sigma};a_{\vec{q}, \bar{\sigma}}\right]_{\mp}&=\left[a^{\dagger}_{\vec{p}, \sigma};a^{\dagger}_{\vec{q}, \bar{\sigma}}\right]_{\mp}=0.
\end{split}
\end{equation}
The normalization of one-particle states and (anti-)commutation relations uniquely fix (up to an irrelevant phase set to one) the action of a creation operator on the vacuum as :
\begin{equation}
\left. \left.\right|\vec{p}, \sigma\right\rangle=\sqrt{2p^+}a^{\dagger}_{\vec{p}, \sigma}\left. \left.\right|0\right\rangle.
\end{equation}
The construction of multi-particle states from tensor product of one-particle states is straightforward. Let us just mention a sign ambiguity for fermionic multi-particle states. The multi-particle state built from the tensor product of $n$ single-particle states has a phase fixed as follow :\begin{equation}
\left. \left.\right|\vec{p}_1, \sigma_1; ...; \vec{p}_n, \sigma_n\right\rangle\equiv\sqrt{2p^+_1...2p^+_n}a^{\dagger}_{\vec{p}_1, \sigma_1}...a^{\dagger}_{\vec{p}_n, \sigma_n}\left. \left.\right|0\right\rangle.
\end{equation}
For creation and annihilation operators associated to fermions, their order matters. Our convention is so that a creation operator acting on a state creates the particle labeled on the leftmost side of the ket.\\

The completeness relation with the correct normalization factors reads : 
\begin{equation}
\label{completenessrelation}
1=\left|0\left\rangle\right\langle0\right|+\displaystyle{\sum_{n=1}^{\infty}}\frac{1}{n!}\displaystyle{\sum_{\sigma_1...\sigma_n}}\int\frac{\rmd^3p_1}{(2\pi)^32p^+_1}...\frac{\rmd^3p_n}{(2\pi)^32p^+_n}\left|\vec{p}_1,\sigma_1;...;\vec{p}_n,\sigma_n\left\rangle\right\langle\vec{p}_1,\sigma_1;...;\vec{p}_n,\sigma_n\right|,
\end{equation}
where $\rmd^3p$ stands for $\rmd p^{+}\rmd^2p_{\perp}$. Written in the form \eqref{completenessrelation}, the completeness relation concerns only one particle species. If the theory contains several kinds of particle, the full completeness relation is merely given by the tensor product of completeness relations for each type of particles.\\

The natural question is now how to build field operators from the creation and annihilation operators in order to get a consistent S-matrix theory fulfilling the very first physical requirements such as micro-causality, cluster decomposition principle and (at least restricted) Lorentz invariance. There is actually a very few differences with the procedure in Minkowski coordinates. Taking $\sigma$ to be the spin $s$, and omitting possible other quantum numbers like color charge to alleviate notations (such quantum numbers are carried by the field operators and cration and annihilation operators), a general field operator is defined as :
\begin{equation}
\label{genericfield}
\phi_{l}(x)=\displaystyle{\sum_s}\int \frac{\rmd^3p}{(2\pi)^3\sqrt{2p^+}}\left[a_{\vec{p},s}u^s_l(p)e^{-ip\cdot x}+b^{\dagger}_{\vec{p},s}v^s_l(p)e^{ip\cdot x}\right].
\end{equation}
Note that just the integration measure is changed with respect to the Minkowski case. $b^{\dagger}_{\vec{p},s}$ is the creation operator for the antiparticle. $u^s_l(p)$ and $v^s_l(p)$ are the coefficient functions for respectively the particle and the anti-particle and furnish representations of the Lorentz group (not necessarily irreducible).

\section{The axial gauge \label{axialgauge}}
\indent

When dealing with ultra-relativistic collisions in the framework of gauge theories, there is an often convenient gauge known as the \emph{axial gauge}. Since this is the gauge that will be used in almost all the following it is not useless to discuss it in the very beginning for the reader not used to it. We first define the axial gauge condition in a generalized sense. We shall derive the equations of motion and the propagator in this gauge. Then we focus on the subset of axial gauges of interest : the \emph{light-cone gauge}. On the one hand, in light-cone gauge most of the formulas from the general formulation simplify a lot but on the other hand, this special case may cause trouble with singularities as the gauge-fixing parameter $\xi$ goes to zero. This section shows how to handle light-cone gauge in a rigorous way when such singularities occur. At the end we sketch the proof of a very nice property of light-cone gauges : the ghosts decouple from the gauge field.

\subsection{Definition and Green functions}
\indent

In general an axial gauge is a constrain on some linear combination of $A_{\mu}$ components that formally reads :
\begin{equation}
\label{generalizedaxialgauge}
n\cdot A(x)=\rho(x)~;
\end{equation}
with $n$ a constant vector and $\rho$ a function of the coordinates. Such gauge fixing requires the following additional term in the Yang-Mills lagrangian\footnote{To be precise, this additional term in the lagrangian holds for a gaussian-distributed set of gauge conditions of the form \eqref{generalizedaxialgauge} so that $|\int\rho|\lesssim\xi$ (see for instance \cite{PeskinSchroeder} where se procedure is mimicked for the Lorenz gauge).} $-\frac{1}{4}F^2$ :
\begin{equation}
\label{Lgf}
\mathcal{L}_{\rm gf}=\frac{1}{2\xi}(n\cdot A^A(x))^2~;
\end{equation}
with $\xi$ a real gauge parameter. The free Green\footnote{Its color structure being trivially $\delta^{AB}$ it is omitted by the replacement $\Delta^{\mu\nu}_{0~AB}\rightarrow \delta^{AB}\Delta^{\mu\nu}_0$.} function $\Delta^{\mu\nu}_0(x-y)$ is given by the inverse of the quadratic piece of the Yang-Mills lagrangian $-\frac{1}{4}(\partial_{\mu}A_{\nu}-\partial_{\nu}A_{\mu})(\partial^{\mu}A^{\nu}-\partial^{\nu}A^{\mu})$ plus the gauge fixing term $\mathcal{L}_{\rm gf}$. That is, it satisfies the equation :
\begin{equation}
\label{axialGF}
\begin{split}
&(g_{\mu\sigma}\partial^2_x-\partial_{x\mu}\partial_{x\sigma}+\frac{n_{\mu}n_{\sigma}}{\xi})\Delta^{\sigma\nu}_0(x-y)= i\delta^{\nu}_{\mu}\delta^{(4)}(x-y).
\end{split}
\end{equation}
Depending on the prescription, $\Delta^{\mu\nu}_0$ stands for the Feynman, retarded, advanced or anti-Feynman propagator as well. Let us forget about the prescription which does not matter for present discussion, writing the Fourier representation of the Green function as :
\begin{equation}
\Delta^{\mu\nu}_0(k)=\frac{i\Pi^{\mu\nu}(k)}{k^2},
\end{equation}
where 
\begin{equation}
\Pi^{\mu\nu}(k)=\sum\limits_{\text{pol.}}\epsilon^{\mu}(k)\epsilon^{\nu *}(k).
\end{equation}
Equation \eqref{axialGF} is satisfied for $\Pi^{\mu\nu}$ given by the following expression :
\begin{equation}
\Pi^{\mu\nu}(k)=-g^{\mu\nu}+\frac{n^{\mu}k^{\nu}+n^{\nu}k^{\mu}}{n.k}+\frac{\xi k^2-n^2}{(n.k)^2}k^{\mu}k^{\nu}.
\end{equation}
This is the general case but one can go a bit further since we shall work only in particular axial gauges satisfying the two further requirements :
\begin{itemize}
\item $\xi=0 \rightarrow$ the gauge is fixed so that $n\cdot A(x)=0$. Moreover the theory is then ghost free as it is shown below.
\item a light-like $n$ vector $\rightarrow n^2=0$.
\end{itemize}
Such specific axial gauge is called the \emph{light-cone gauge}. In light-cone gauge, the $\Pi_{\mu\nu}$ tensor becomes simpler :
\begin{equation}
\label{piaxial}
\Pi_{\mu\nu}(k)=-g_{\mu\nu}+\frac{n_{\mu}k_{\nu}+n_{\nu}k_{\mu}}{n.k},
\end{equation}
and satisfies the properties :
\begin{equation}
\label{gaugeid}
n^{\mu}\Pi_{\mu\nu}(k)=0~;
\end{equation}
and for on shell $k$ :
\begin{equation}
\label{oswardid}
k^{\mu}\Pi_{\mu\nu}(k)=0.
\end{equation}
However there is at this point a small problem that needs to be mentioned. The propagator numerator \eqref{piaxial} is a projector and is no longer invertible. There seems to be an incompatibility with \eqref{axialGF} which is ill-defined as $\xi\rightarrow 0$. Using \eqref{piaxial} for the propagator numerator, one has :
\begin{equation}
\label{axialGF1}
\begin{split}
&(g_{\mu\sigma}\partial^2_x-\partial_{x\mu}\partial_{x\sigma})\Delta^{\sigma\nu}_0(x-y)= i\delta^{\nu}_{\mu}\delta^{(4)}(x-y)-i\int\frac{\rmd^4k}{(2\pi)^4}\frac{n^{\mu}k^{\nu}}{n.k}e^{-ik\cdot(x-y)}.
\end{split}
\end{equation}
As we will check explicitly in our applications, the extra term of the r.h.s actually plays no role in Lorentz invariant quantities. The rigorous way to work in light-cone gauge is to work with a finite $\xi$ when it causes trouble and to send it to zero at the end of the calculation hoping there will not be singularities anymore - we always get rid of them in all the problems studied bellow.

\subsection{Ghosts}
\indent

Here we emphasize a nice property of axial gauges $n\cdot A=0$ : they are \emph{ghost free}. The gauge condition enters naturally in the path integral following the De Witt - Faddeev - Popov method whose detailed calculation can be found for instance in \cite{WeinbergII}. The gauge condition is rewritten as an integral over two independent Grassmann fields $\omega$ and $\bar{\omega}$ known as \emph{ghosts} and \emph{anti-ghosts} respectively\footnote{Ghosts and anti-ghosts are different fields, unrelated by any complex conjugation or charge conjugation operation.}. Ghosts are Lorentz scalar fields in the adjoint representation of the gauge group. The lagrangian density corresponding to ghosts reads :
\begin{equation}
\label{Lghost}
\mathcal{L}_{\rm ghosts}=\bar{\omega}^{A}n^{\mu}(\delta^{AB}\partial_{\mu}-gf^{ABC}A^C_{\mu})\omega^B.
\end{equation}
The interaction term between ghosts and gauge fields is proportional to $n\cdot A$ which is zero by the gauge condition. For the axial gauge $n\cdot A=0$ the ghosts decouple from the theory and are completely absent from calculations.

\chapter{Perturbative QCD phase diagram and saturation physics\label{QCDsat}}
\indent

In this thesis, the main purpose is the study of phenomena that have to do with saturation effects. The saturated state of hadronic matter is a very active branch of QCD since it is right now accessible to experiments occurring in accelerators. These last past years, the RHIC has shown evidences of this phase of QCD matter while the LHC is about to explore it deeper. Saturation is a consequence of the raise of parton density as the emitted partons are soft with respect to the parent ones.\\

The available tools for an experimental investigation of QCD matter in accelerators are two-body collisions. These can be either protons or nuclei. For definiteness, one of the two colliding hadrons is chosen to travel along the positive $z$ axis and is referred to as the \emph{projectile} while the other one, traveling in the negative $z$ direction is referred to as the \emph{target}. Due to the (very) large number of particles produced in such high energy collisions, we shall consider only \emph{inclusive} observables : the final state is summed over all possible configurations of unobserved particles.\\

The goal of this chapter is to introduce all the needed framework. We shall motivate the saturation phenomena from QCD and then detail the appropriate formalism to deal with it. First we will study the particle content of a fast hadron, that is an uncolored QCD bound state composed of valence quarks, such as a proton or a nucleus. We will see that the virtual fluctuations (called partons together with the valence quarks) occurring in the hadron are described in terms of parton distribution functions which count the number of partons present in some phase space region. Then we shall see that the probability of emission of a parton diverges as the longitudinal momentum carried by the parton becomes small. The physical consequence is that the parton density increases for small longitudinal momenta. When the number of partons becomes very large one has to consider also recombination effects so that the number of partons does not grow indefinitely and converges to a fixed value to be precised. This is known as the \emph{saturation} phenomenon. Along the way we shall see the emergence of a new intrinsic energy scale : the \emph{saturation scale}. Once these things are understood we shall propose an alternative formalism that well describes the saturated regime : the Color Glass Condensate (CGC), an effective theory following from QCD. We shall study the physical motivations and write the associated evolution equations.\\

From now, in this chapter and in the following, all the masses will be neglected since the energy scales considered are much larger than the masses of the colliding particles.

\section{The parton picture}

\subsection{The hadronic content and deep inelastic scattering\label{DIS}}
\indent

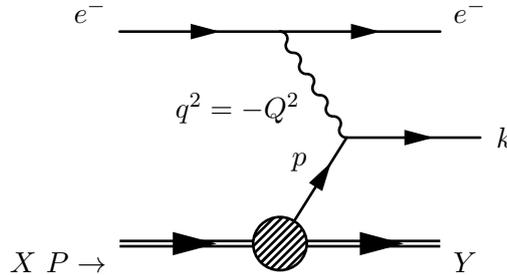
\begin{figure}[h]
\centering
%(along, up)
\begin{fmfgraph*}(150,80)
    \fmfleft{i1,i2,i3}
    \fmfright{o1,o2,o3}
      \fmf{fermion}{i3,v1,o3}
      \fmf{dbl_plain_arrow}{i1,v2,o1}
    \fmffreeze
      \fmf{fermion,label=$p$}{v2,v3}
      \fmf{fermion}{v3,o2}
      \fmf{photon,label=$q^2=-Q^2$}{v1,v3}
      \fmflabel{$X~P\rightarrow$}{i1}
      \fmflabel{$Y$}{o1}
      \fmflabel{$e^-$}{o3}
      \fmflabel{$e^-$}{i3}
      \fmflabel{$k$}{o2}
      \fmfblob{20.}{v2}
\end{fmfgraph*}
\vspace{0.5cm}
\caption{Feynman diagram for the deep inelastic scattering process. The photon interacts with one of the quarks within the proton.\label{DISpicture}}
\end{figure}

Here we shall see the physical picture of a hadron and its content. For brevity, the hadron shall refer to a proton in this section but the considerations are valid for any other hadrons, and even for nuclei. The full description of a proton lies in the scope of non perturbative QCD. A proton is a bound state of QCD composed of three valence quarks including radiative corrections to all order in powers of the interaction. Neglecting electromagnetic and weak interaction effects, the fluctuations are either gluons or quark-antiquark pairs. In the rest frame of the proton, the typical life time of quantum fluctuations is of the order of $1/\Lambda_{QCD}$ and thus enter into the strong coupling regime. However the situation changes when one chooses a frame in which the proton has a velocity close to the speed of light called the \emph{infinite momentum frame}, conventionally taken along the positive $z$ axis. The boost affects hadronic fluctuations which live much longer by Lorentz time dilatation and their energies are increased by a boost factor large enough to lie in the perturbative regime of QCD. At energies available in current accelerators, both the projectile and the target acquire sufficiently large velocities to be seen in an infinite momentum frame from the lab. Thus in the lab frame hadronic fluctuations have a typical lifetime that is very long with respect to the duration of the scattering process. Of course this assumption holds only for partons that have energies much smaller than the total energy of the projectile-target system, denoted $\sqrt{s}$. When we will discuss the partonic content of a hadron we shall see that most of the partons are very soft with respect to $\sqrt{s}$ but obviously partons' energies cannot be larger than $\sqrt{s}$ by mere kinematic considerations. To probe the parton content of a proton, the academic process considered is the \emph{deep inelastic scattering} (DIS) represented on figure \ref{DISpicture}. In the DIS, the proton content is probed with the exchange of a virtual, space-like photon of momentum $q$ and virtuality $Q^2=-q^2$ between the hadron and an electron, say. Any other process involving other kind of particles exchanged would not bring new qualitative phenomena for present considerations.
The momentum $P$ of the proton is chosen so that $P=(P^+,0,\mathbf{0})$, with $P^+$ very large with respect to the proton mass. The observed quark carries, before it scatters off the photon, a momentum $p$ with longitudinal component parametrized as $p^+=xP^+$. $x$ is called the \emph{Bjorken variable} and is the fraction of the longitudinal momentum carried by the quark. Provided $p^+$ is much larger than the transverse momentum of the quark\footnote{Theoretically, this can always be fulfilled by an appropriate choice of frame boosted enough. Practically, even though the speed of colliding particles in accelerators is close to the speed of light, the boost is not arbitrary large and this assumption holds only if the transverse momenta are not too large with respect to the longitudinal ones. We shall see in the next sections that the transverse momentum of partons present in a hadron is bounded by the saturation scale. This assumption is for instance fulfilled at the LHC where $\sqrt{s}\sim$ TeV while $p_{\perp}\sim$ GeV.}, the observed quark is initially almost on-shell and it makes sense to consider it as an asymptotic initial state. The total cross-section for the $Xe^-\rightarrow Ye^-$ process is hence easily written in terms of the cross-section corresponding to the sub-process $qe^-\rightarrow qe^-$, where $q$ denotes the quark :
\begin{equation}
\label{DISfactorization}
\sigma(Xe^-\rightarrow Ye^-)=\int_0^1 \rmd x ~q(x,Q^2)\sigma(q(p=xP)e^-\rightarrow qe^-).
\end{equation}
$q(x,Q^2)$ is called the \emph{integrated quark distribution} and represents the average number of quarks with a momentum fraction $x$ within the proton. Such factorization of the cross-section is known as \emph{collinear factorization}. The integrated quark distribution depends on the virtuality scale $Q^2$ since the virtuality of the exchanged photon determines the probing resolution. In the next section we shall introduce analogously the \emph{gluon distribution}. The example of DIS enabled us to see the emergence of parton distribution functions. The parton distribution functions play a central role since they encode the distribution of partons within the proton. Especially, saturation is reached when the parton distribution takes a large value to be precised.

\subsection{Parton distribution functions}
\indent

Through the DIS process one has introduced the concept of quark distribution function. Similarly, one can extend the concept of distribution functions to anti-quarks and to gluons as well. For instance one can consider that the probed quark is actually a sea quark merging from a gluon splitting into a $q\bar{q}$ pair : a \emph{color dipole}. We will not deal with details about the dipole scattering in the context of DIS to avoid technical complications, spurious for the present purpose. The interested reader in the so-called \emph{dipole factorization} can find more details in \cite{ForshawRoss}.  Considerations made for quarks intuitively motivate as well the concept of integrated gluon distribution function, denoted $G(x,Q^2)$, whose physical interpretation is the average number of gluons in the hadron that carry a longitudinal momentum fraction $x$ and a transverse momentum bounded by the energy scale $\sqrt{Q^2}\ll p^+$. Thus its definition in terms of the number of gluons per phase space volume element $\rmd N/\rmd x\rmd^2k_{\perp}$ is straightforward :
\begin{equation}
G(x,Q^2)=\int_{\mathbf{k}_{\perp}^2<Q^2}\rmd^2k_{\perp}\frac{\rmd N_g}{\rmd x\rmd^2k_{\perp}}.
\end{equation}
Instead of dealing with the $x$ variables one sometimes rather uses the \emph{rapidity} parametrization defined as $Y=-\ln x$. Generalizing the integrated distribution function one introduces the \emph{unintegrated distribution function} $f_Y(\mathbf{k}_{\perp})$ which is, up to a conventional $1/\pi\mathbf{k}_{\perp}^2$ prefactor, the average number of gluons per unit of phase space volume defined as :
\begin{equation}
\label{UGPDF}
xG(x,Q^2)=\int_{\mathbf{k}_{\perp}^2<Q^2}\rmd^2k_{\perp}\frac{\rmd N_g}{\rmd Y\rmd^2k_{\perp}}=\int_{\mathbf{k}_{\perp}^2<Q^2}\frac{\rmd^2k_{\perp}}{\pi\mathbf{k}_{\perp}^2}f_Y(\mathbf{k}_{\perp}).
\end{equation}
By construction, the parton distribution functions take a single exchange into account. That is the probe interacts with the hadron only via a single quark or gluon. For multiple scatterings involving possible interactions between the exchanged particles it is possible in some cases to generalize the concept of unintegrated parton distribution functions. We shall see such examples in sections \ref{diluteqq} and \ref{backtoback}. The physical meaning of the parton distribution is promoted to an \emph{effective} parton distribution that is the probability that some given total momentum is transferred between the probe and the hadron.\\

Note that quark and gluon distributions are not independent. Of course, the quark distribution contains the valence quarks. The other quarks can only come into $q\bar{q}$ pairs from virtual gluons. They are called \emph{sea quarks}. If the density of gluons becomes large - and we shall see this actually happens as $x$ becomes small enough, sea quarks dominate the quark distribution. The precise relation between quark and gluon distribution does not matter for further discussion here but it exists and easily follows from the previous consideration. Furthermore, the creation of a  $q\bar{q}$ pair from a parent gluon requires an additional vertex since a gluon can be directly emitted by a valence quark whereas a  $q\bar{q}$ pair requires at least an intermediate gluon. Hence the quark distribution is suppressed by an additional power of $\alpha_s$ with respect to the gluon distribution.\\

Since parton distributions encode the parton content of hadrons, their computation would give direct information about this. Of course the form of parton distributions \emph{a priori} also differs depending on the nature of the hadron concerned. Unfortunately parton distribution functions also encode non perturbative physics. For instance, equation \eqref{DISfactorization}, splits the total DIS process into a hard (high energy process) computable thanks to perturbation theory and soft phenomena including the hadron wave function that lies in the scope of non perturbative QCD. Collinear factorization, is physically motivated by the separation of time scales : soft processes are frozen during the characteristic time scales of the hard processes. Thus, concerning the form of distribution functions, the best we can do is getting them from \emph{evolution equations}, that is their behavior by varying the value of $Y$ and/or $Q^2$. This is the aim of the following.

\section{The raise of parton density at small $x$\label{Partondistribution}}
\indent

In this section we investigate the behavior of parton distribution functions with kinematics and especially their variation with the rapidity $Y$. First we roughly extract the physical behavior of parton distributions from very simple considerations. For this purpose we first focus on the Bremsstrahlung process : the emission of a gluon by a quark. It turns out that the probability for this elementary process shows up a logarithmic divergence at small $x$. Hence, even though the coupling constant is small, the $qqg$ vertex comes together with a large logarithm in the Bjorken variable. Perturbation theory breaks down and one has to resum all these large contributions. We shall see that the leading log contribution to soft gluon emission cascades easily sums. On the physical side this divergence is interpreted as a growth of the gluon distribution as $x$ becomes small. Of course this growth cannot be infinite otherwise the unitarity bound would be violated. At large density the recombination effect also becomes important and tames the growth. This naturally leads to the concept of saturation. Then we shall make a more quantitative treatment of saturation. However, in order to avoid the introduction of new notions and to follow the main line, we instead use a toy model and make correspondence with actual results. We shall think about the creation and recombination of gluons in terms of a reaction-diffusion process. Although the toy model does not govern the right physical quantities, precisely we deal with gluon occupation number whereas the physical quantity to consider is the dipole amplitude, it is easier to understand and contains relevant the physics. The reason is because the number of gluons is \emph{not} and observable and it has to be \emph{defined} in terms of existing observables. We shall see that in the dilute regime the number of gluons has an unambiguous interpretation as being proportional to the unintegrated gluon distribution but it becomes less clear in the saturated regime. The main result will be the emergence of the saturation scale. From our toy model we will even be able to sketch a crude analytical expression valid at high energy for the saturation scale.

\subsection{Soft Bremsstrahlung\label{Brems}}
\indent

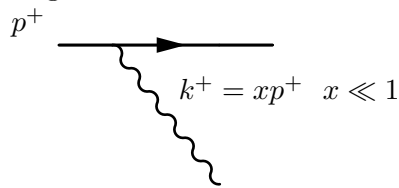
\begin{figure}[h]
\centering
%(along, up)
\begin{fmfgraph*}(80,60)
    \fmftop{t1,t5}
    \fmfbottom{b1,b2,b3,b4,b5}
      \fmf{plain}{t1,t2,t3,t4,t5}
      \fmf{fermion}{t1,t5}
    \fmffreeze
      \fmf{photon,label=$k^+=xp^+~~x\ll 1$}{t2,b4}
      \fmflabel{$p^+$}{t1}
\end{fmfgraph*}
\caption{Soft Bremsstrahlung of a gluon by a parent quark.\label{singlegbremsstrahlung}}
\end{figure}

An easy first step for studying how the partonic content evolves with kinematics is to consider the elementary process $q \rightarrow qg$ represented on figure \ref{singlegbremsstrahlung}\footnote{Present considerations would have led to the same conclusion if the parent parton were a gluon.}.
The emission of a real gluon by a quark is called the \emph{Bremsstrahlung}. We focus on the soft part of phase space, that is the energy of the emitted gluon is small compared to the energy of the parent quark, $x\ll 1$. A straightforward calculation (see \cite{PeskinSchroeder} for instance) shows that the differential probability for emitting such a gluon with momentum $k$ behaves like :
\begin{equation}
\label{Pbrem}
\rmd\mathcal{P}_{\rm brem}(k)\simeq\frac{\alpha_s(\mathbf{k}_{\perp}^2)C_R}{\pi^2}\frac{\mathrm{d}^2k_{\perp}}{\mathbf{k}_{\perp}^2}\frac{\mathrm{d}x}{x},
\end{equation}
where $C_F=(N_c^2-1)/2N_c$ is the Casimir of the fundamental representation of the gauge group\footnote{If the parent parton were a gluon, $C_R$ will be replaced by the adjoint representation Casimir $C_A=N_c$.} and $\alpha_s=g^2/4\pi$. This last expression \eqref{Pbrem} shows up two kinds of logarithmic divergences for the total probability :
\begin{itemize}
\item a \emph{collinear} divergence as $\mathbf{k}_{\perp}$ goes to zero,\\
\item a \emph{soft} divergence as the longitudinal momentum or $x$ goes to zero.
\end{itemize}
A naive perturbative expansion in powers of $\alpha_s$ breaks down if one of these logarithms becomes large since such a vertex contributes as $\sim\alpha_s\ln(\mathbf{k}_{\perp}^2/m^2)\ln (1/x)$ which is not necessarily small with respect to 1 even though $\alpha_s$ is small. The resummation of the collinear divergences is not considered here. The careful procedure is well known since the 70's and the underlying evolution equation is the Dokshitzer-Gribov-Lipatov-Altarelli-Parisi (DGLAP) equation \cite{Altarelli:1977zs,Dokshitzer:1977sg,Gribov:1972ri}. The DGLAP equation is the Callan-Symanzik equation for QCD, that is, it governs, at least in the leading log approximation, the behavior of the distribution functions with the energy. Let us rather focus on the small $x$ divergence to be seen in the next section.

\subsection{Gluon cascades : the BFKL evolution}
\indent

\begin{figure}[h]
\centering
%(along, up)
\begin{fmfgraph*}(80,80)
    \fmftop{t1,t5}
    \fmfbottom{b1,b2,b3,b4,b5}
    \fmfright{o}
      \fmf{plain}{t1,t2,t3,t4,t5}
      \fmf{fermion}{t1,t5}
    \fmffreeze
      \fmf{photon}{t2,v,b4}
      \fmf{photon,tension=0}{v,o}
      \fmflabel{$p^+$}{t1}
      \fmflabel{$k_1^+=x_1p^+$}{o}
      \fmflabel{$k_2^+=x_2k_1^+~~x_2\ll x_1$}{b4}
\end{fmfgraph*}
\vspace{.5cm}
\caption{Successive $x$-ordered emission of two gluons.\label{doublegbremsstrahlung}}
\end{figure}
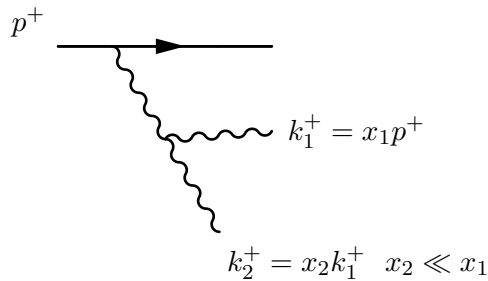

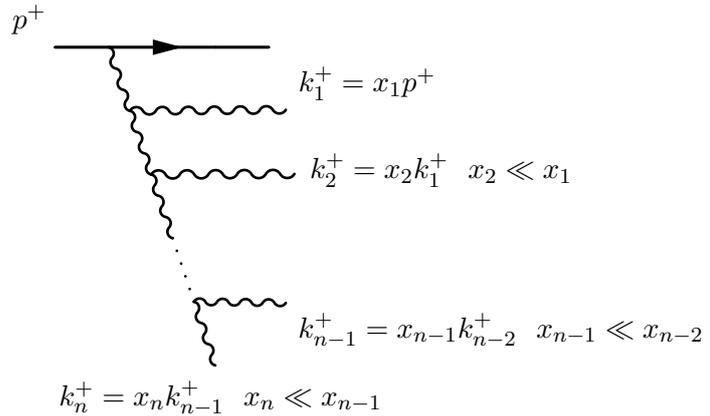
\begin{figure}[h]
\vspace{.5cm}
\centering
%(along, up)
\begin{fmfgraph*}(100,120)
    \fmfleft{i1,t1}
    \fmfright{b5,o4,o3,o2,o1,t5}
      \fmf{phantom}{i1,b2,b3,b4,b5}
      \fmf{plain}{t1,t2,t3,t4,t5}
      \fmf{fermion}{t1,t5}
    \fmffreeze
      \fmf{phantom}{t2,v1,v2,v3,v4,b4}
      \fmf{dots}{v3,v4}
      \fmf{photon}{t2,v1,v2,v3}
      \fmf{photon}{v4,b4}
      \fmf{photon,tension=0}{v1,o1}
      \fmf{photon,tension=0}{v2,o2}
      \fmf{photon,tension=0}{v4,o4}
      \fmflabel{$p^+$}{t1}
      \fmflabel{$k_1^+=x_1p^+$}{o1}
      \fmflabel{$k_2^+=x_2k_1^+~~x_2\ll x_1$}{o2}
      \fmflabel{$k_{n-1}^+=x_{n-1}k_{n-2}^+~~x_{n-1}\ll x_{n-2}$}{o4}
      \fmflabel{$k_n^+=x_nk_{n-1}^+~~x_n\ll x_{n-1}$}{b4}
\end{fmfgraph*}
\vspace{.5cm}
\caption{Successive $x$-ordered emission of $n$ gluons.\label{BFKLladder}}
\end{figure}

\begin{figure}[h]
\begin{center}
\includegraphics[width=0.45\paperwidth]{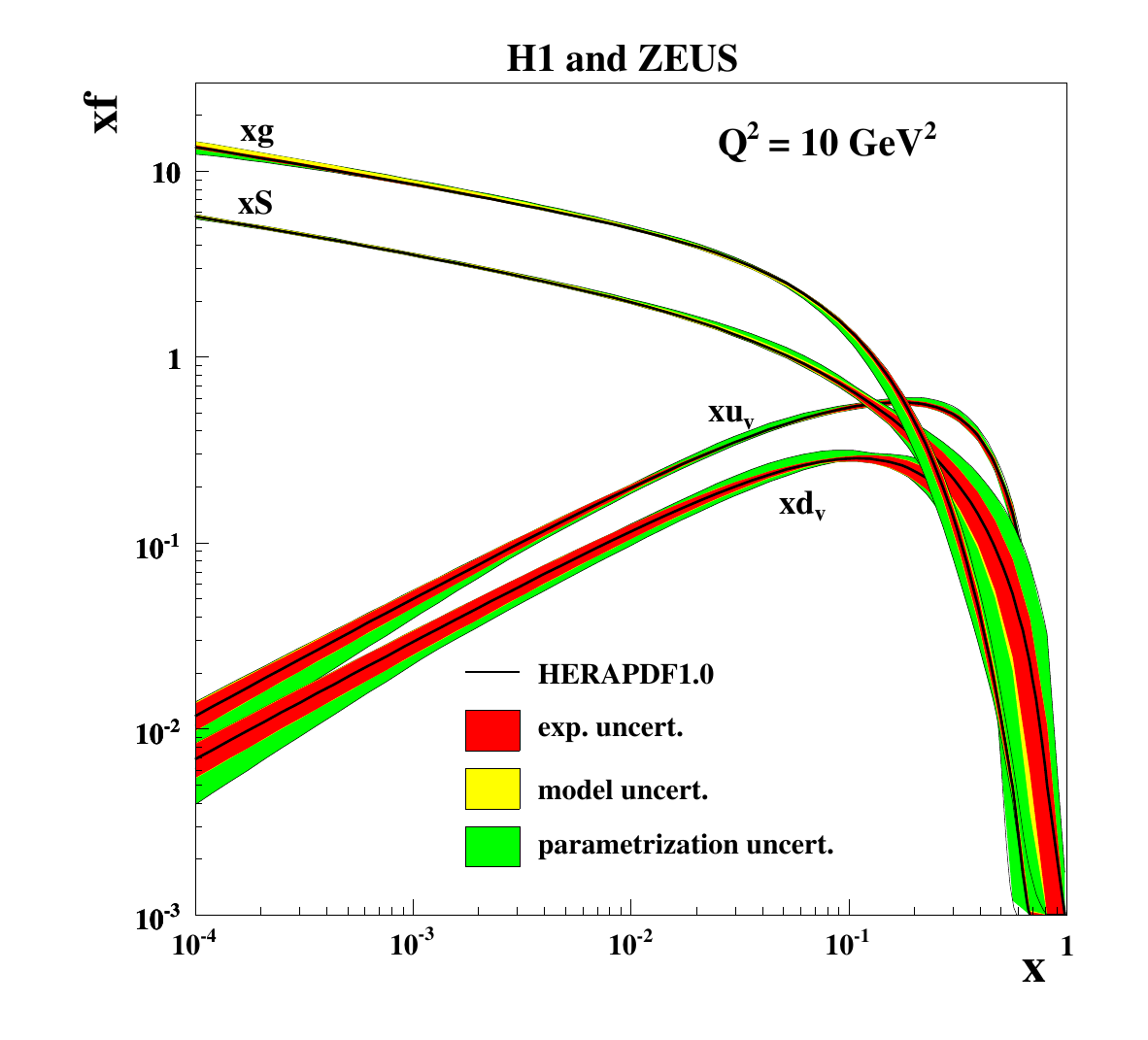}
\caption{Parton distributions (which are matched to denstity) as functions of the Bjorken variable $x$ for a proton. $u_v$ and $d_v$ denote the valence quarks whose distribution is peaked in the vicinity of $x=1/3$ since there are three valence quarks that typically carry $1/3$ of the proton momentum. $g$ denotes the gluon distribution that shows up the expected raise at small $x$. $S$ represent the sea quark density. Since sea quarks cannot merge directly from a valence quark but come into $q\bar{q}$ pairs from emitted gluons, their density is suppressed by an additional power of $\alpha_s$ compared to the gluons.\label{qandgsmallx}}
\end{center}
\end{figure}

The small $x$ divergence suggests the necessity for summing all possible emissions since each successive emission brings a factor $\alpha_s\ln (1/x)$ that is not necessarily small even though $\alpha_s$ is. Let us consider the case in which the radiated gluon emits in turn another gluon as shown on figure \ref{doublegbremsstrahlung}.
For small $x$, the largest contribution comes from the region of phase space where the momentum fraction of the parent quark carried by the second gluon is smaller than the one carried by the first one. We say that the gluons are \emph{strongly ordered} in the longitudinal direction. The total probability goes like :
\begin{equation}
\alpha_s^2\int_x^1 \frac{\rmd x_1}{x_1}\int_{x_1}^1 \frac{\rmd x_2}{x_2}=\frac{\alpha_s}{2}(\ln\frac{1}{x})^2.
\end{equation}
The other regions of phase space bring sub-leading contributions like $\ln\ln(1/x)$, this is why the strong ordering assumption is know as the \emph{leading log approximation}. The process of figure \ref{doublegbremsstrahlung} contributes as much as the single emission \ref{singlegbremsstrahlung}, for $\alpha_s\ln (1/x)\sim 1$. Repeating the calculation for the successive emission of $n$ gluons ordered in the $x$ variable contributes as $\frac{1}{n!}(\alpha_s\ln(1/x))^n$. This cascade, represented on diagram \ref{BFKLladder}, is known as a Balitsky-Fadin-Kuraev-Lipatov (BFKL) ladder.
For small $x$, so that $\alpha_s\ln (1/x)\sim 1$ there is no perturbative expansion in the number of final gluons. The sum of all the ladders must be taken into account for small $x$ and they exponentiate. The number of gluons carrying the fraction $x$ is easily obtained from the total probability and reads :
\begin{equation}
\label{raiseofG}
x\frac{\rmd N_g}{\rmd x}\sim \frac{1}{x^{\omega\alpha_s}},
\end{equation}
with $\omega$ some positive constant of order one. From these very simple considerations, \eqref{raiseofG} shows up a fast raise of gluon density as $x$ becomes small. A more careful analysis within the dipole framework which includes also sea quarks\footnote{In the dipole framework $q\bar{q}$ pairs and gluons are on the same pedestal since in the large $N_s$ limit of an $SU(N_c)$ gauge theory, a single gluon, which is a particle in the adjoint representation and a $q\bar{q}$ pair, which is composed of two particles in the fundamental representation are equivalent. This directly follows from the fundamental and adjoint representation properties of $SU(N_c)$. We use similar properties in section \ref{largeNc} when we write adjoint representation matrices in terms of the fundamental ones.} and transverse momentum sharing along the successive splittings, leads to an evolution equation in the $x$ - or more conveniently the rapidity $Y$ - variable for the unintegrated gluon distribution \eqref{UGPDF}. This evolution equation is known as the BFKL equation \cite{Lipatov:1976zz,Kuraev:1977fs,Balitsky:1978ic} and reads :
\begin{equation}
\label{BFKLeq}
\frac{\partial f_Y(\mathbf{k}_{\perp})}{\partial Y}=\alpha_s N_c\int\frac{\rmd^2p_{\perp}}{\pi^2}\frac{\mathbf{k}_{\perp}^2}{\mathbf{p}^2_{\perp}(\mathbf{k}_{\perp}-\mathbf{p}_{\perp})^2}\left[f_Y(\mathbf{p}_{\perp})-\frac{1}{2}f_Y(\mathbf{k}_{\perp})\right].
\end{equation}
The evolution with $Y$ is \emph{not} an evolution with the virtuality that is provided by the evolution is the transverse momentum. Indeed the evolution can be understood as an evolution with the energy \emph{difference} between the produced partons and the parent ones but at a fixed energy of the produced one. In other words the relevant parameter is the energy fraction rather than the energy itself. The BFKL equation is linear since the ladders do not interact with each others. Solutions to BFKL equation confirm the power growth of the gluon distribution at small $x$ or equivalently its exponential growth at large $Y$. This suggests that the gluon density (and also the sea quark density) grows indefinitely. It agrees with the sketchy consideration \eqref{raiseofG} which shows up a power raise of the gluon number as $x$ becomes small but also with experimental data \ref{qandgsmallx}.
If one considers only the BFKL ladders, one ends up with an infinite growth of the gluon density as $x$ decreases. This is unsatisfactory from the physical side since such a growth would cause trouble with the unitarity requirement. In the next section we shall see that considering only BFKL branching processes is not enough as the density becomes high. Unitarity is restored by taking into account additional processes that are negligible as long as the system is dilute but become important at large density. To see how these effects can be added to the evolution equation we shall first mimic the BFKL resummation in a naive but intuitive and faithful way that will allow us to see what happens at large density.\\

For motivating the BFKL equation, let us take a point of view inspired by the reaction-diffusion techniques of statistical physics. For this purpose it is convenient to introduce the occupation number $n$ which is the number of gluon per phase space volume element. We shall see why when we shall deal with saturation. The gluonic phase space volume element is $\rmd k^+\rmd^2k_{\perp}/(2\pi)^32k^+=\rmd Y\rmd^2k_{\perp}/2(2\pi)^3$. Furthermore, one has to consider two degrees of freedom for the helicity and $N_c^2-1$ for the color. More over there is an impact parameter degree of freedom $\mathbf{b}_{\perp}$ : one has to consider the number of gluons per phase space element in a given region of transverse space. If one assumes that the hadron is homogeneous, the density is the same everywhere and one has to merely divide by $S_{\perp}$ the transverse surface of the hadron\footnote{It is reasonable to further assume axial symmetry so that $S_{\perp}=\pi R^2$, with $R$, the radius of the hadron. This further refinement will be useless and we shall only deal with $S_{\perp}$.}. Therefore the occupation number reads\footnote{Since the impact parameter dependence is assumed to be trivial it has been omitted.} :
\begin{equation}
\label{occupationnumberdef0}
n_Y(\mathbf{k}_{\perp})\equiv \frac{2(2\pi)^3}{2(N_c^2-1)S_{\perp}}\frac{\rmd N_g}{\rmd Y\rmd^2k_{\perp}}.
\end{equation}
Using \eqref{UGPDF} the gluon occupation number also have a clear interpretation in term of the unintegrated gluon distribution which reads :
\begin{equation}
\label{occupationnumberdef}
n_Y(\mathbf{k}_{\perp})\equiv \frac{8\pi^2}{(N_c^2-1)S_{\perp}}\frac{f_Y(\mathbf{k}_{\perp})}{\mathbf{k}_{\perp}^2}.
\end{equation}
Let us consider the occupation number at some given rapidity $Y$ and perform a step in rapidity from $Y$ to $Y+\rmd Y$ (recall that increasing rapidity decreases $x$). What can happens in the rapidity slice is the splitting of the last gluon into two gluons. This occurs with a probability proportional to $n$ itself at rapidity $Y$. However the transverse momentum of the parent gluon is shared between the two produced ones. For this reason, the evolution equation is \emph{non-local} in transverse momenta. Thus the evolution equation must take the form :
\begin{equation}
\label{BFKLn}
\frac{\partial n_Y(\mathbf{k}_{\perp})}{\partial Y}=\alpha_s \int\frac{\rmd^2p_{\perp}}{\pi^2}K(\mathbf{k}_{\perp},\mathbf{p}_{\perp}) n_Y(\mathbf{p}_{\perp}),
\end{equation}
where $K$ is a positive definite kernel and the factor $\alpha_s$ has been kept explicit since the splitting probability is proportional to $\alpha_s$. The BFKL equation \eqref{BFKLeq} governs the evolution of the unintegrated gluon distribution which is, according to \eqref{occupationnumberdef} proportional to the gluon occupation number. Thus the BFKL equation indeed governs the occupation number and we can identify :
\begin{equation}
K(\mathbf{k}_{\perp},\mathbf{p}_{\perp})=N_c\left[\frac{1}{(\mathbf{k}_{\perp}-\mathbf{p}_{\perp})^2}-\frac{1}{2}\delta^{(2)}(\mathbf{k}_{\perp}-\mathbf{p}_{\perp})\int\rmd^2q_{\perp}\frac{\mathbf{k}_{\perp}^2}{\mathbf{q}^2_{\perp}(\mathbf{k}_{\perp}-\mathbf{q}_{\perp})^2}\right].
\end{equation}
We will justify soon that the validity range of the BFKL equation assumes the occupation number to be small, i.e. it is valid in dilute hadrons. In the dense regime, the unintegrated gluon distribution loses its physical interpretation as the gluon occupation number. A natural generalization is to define it as the Fourier transform of the dipole amplitude (see \ref{diluteqq} for instance) which is the physical observable governed by the evolution equations. This is why considerations on occupation number become sketchy at high density since it is more a matter of definition rather than a real physical picture. The number of gluons does not really make sense in the dense regime, the only requirement is to recover its canonical definition \eqref{occupationnumberdef} in the dilute regime.\\

We expand the r.h.s of \eqref{BFKLn} in powers of $\mathbf{k_{\perp}}$. It has been proved that at high energy, i.e. for transverse momenta that are large with respect to the non perturbative scale $\Lambda_{QCD}$, the BFKL equation is well approximated by its second order expansion in $\mathbf{k_{\perp}}$. Moreover the first order vanishes. This leads to a diffusion equation which is the high energy limit of \eqref{BFKLn} :
\begin{equation}
\label{BFKLndif}
\frac{\partial n_Y(\mathbf{k}_{\perp})}{\partial Y}\simeq a\alpha_s n_Y(\mathbf{k}_{\perp})+b\alpha_s\frac{\partial^2 n_Y(\mathbf{k}_{\perp})}{\partial (\ln\mathbf{k_{\perp}}^2)^2},
\end{equation}
where $a$ and $b$ are positive number of order unity. Written in this form it will be easy to motivate the additional terms that restore unitarity and to see the emergence of saturation.

\subsection{Toward saturation : the BK equation\label{BKequation}}
\indent

\begin{figure}[h]
\vspace{.5cm}
\centering
%(along, up)
\begin{fmfgraph*}(120,120)
    \fmfleft{i1,i4,i3,i2,i1,t1}
    \fmfright{b5,o4,o3,o2,o1,t5}
      \fmf{phantom}{i1,b2,b3,b4,b5}
      \fmf{plain}{t1,t2,t3,t4,t5}
      \fmf{fermion}{t1,t5}
    \fmffreeze
      \fmf{phantom}{t3,v1,v2,v3,v4,b5}
      \fmf{photon}{t4,vv1,vv2,vv3,vv4,b5}
    \fmffreeze
      \fmf{photon}{t2,vvv1,vvv2,vv3}
      \fmf{photon,tension=0}{vvv1,v1}
      \fmf{photon,tension=0}{vv1,o1}
      \fmf{photon,tension=0}{vv2,o2}
      \fmf{photon,tension=0}{vv4,o4}
\end{fmfgraph*}
\vspace{.5cm}
\caption{Typical recombination of BFKL ladders leading to non-linear evolution.\label{recombination}}
\end{figure}
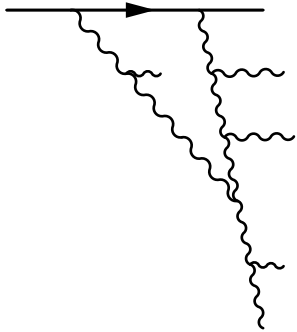

Fortunately all possible mechanisms were not taken into account when we ressummed BFKL ladders neglecting possible interactions among them. Indeed if the number of gluons becomes large it is possible that some of them recombine together \cite{Gribov:1984tu} as represented on figure \ref{recombination}.
If two gluons are very separated in frequency, they have a very low probability to recombine, they are transparent to each other. It means that for recombination to become important, occupation numbers of gluons in neighboring phase space volume must be at least of order one. One sees the advantage of dealing with the occupation number : they provide a quantitative criterion for the transition to saturation. We also understand that the linear BFKL equation, which neglects the recombination effects, is valid if and only if the occupation number is small. Obviously, the probability for two gluons to recombine is proportional to $n^2$ since the recombination process requires two initial gluons. Since these two gluons must be close in phase space, this contribution is roughly local in $n^2$. Concerning the $\alpha_s$ counting one would naively say that it is also proportional to a single power of $\alpha_s$ coming from the vertex. However our toy model breaks down here : in the dipole framework, it turns out that the recombination of two dipoles is only possible via a a double-gluon exchange process at leading order which brings an $\alpha_s^2$ contribution to the recombination rate (an early attempt of BK equation has been provided by a more intuitive approach by Gribov, Levin and Riskin known as the GLR equation \cite{Gribov:1984tu,Mueller:1985wy}). We conclude that the recombination effects enters as a negative term (since it lowers the occupation numbers) to r.h.s of the diffusion equation \eqref{BFKLndif} which becomes :
\begin{equation}
\label{BKndif}
\frac{\partial n_Y(\mathbf{k}_{\perp})}{\partial Y}=a\alpha_s n_Y(\mathbf{k}_{\perp})+b\alpha_s \frac{\partial^2 n_Y(\mathbf{k}_{\perp})}{\partial (\ln\mathbf{k_{\perp}}^2)^2}-c\alpha_s^2n_Y^2(\mathbf{k}_{\perp}),
\end{equation}
where $c$ is again a constant of order unity. The above equation is a toy model of the evolution equation known as the Balitsky-Kovchegov (BK) equation \cite{Balitsky:1995ub,Kovchegov:1999yj,Kovchegov:1999ua}, derived in the dipole framework. Before discussing its physical content, let us emphasize its toy model character. We argued in the previous section that, even though the BFKL governs the evolution of the dipole amplitude, it holds as well for the occupation number. This breaks down for the BK equation since in the dense regime where non-linear effects are important, the dipole amplitude cannot be expressed as the Fourier transform of the unintegrated gluon distribution and by the way, the occupation number. Thus equation \eqref{BKndif} does not govern the evolution of the right physical quantity. However the physics we shall extract from this equation is - at least qualitatively - the same as the physics contained in the BK equation. Thus we shall continue to close our eyes on this subtlety.\\

\begin{figure}[h]
\begin{center}
\includegraphics[width=0.45\paperwidth]{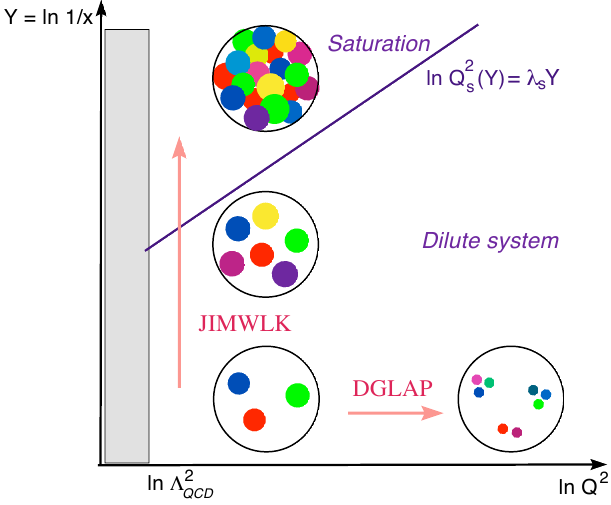}
\caption{QCD phase diagram. From left to right, the evolution is the ordinary evolution with the energy that is governed by the DGLAP equation and increases the resolution of the theory at short lengths. In the vertical direction is shown the evolution with $Y$ and the growth of parton density. In the dilute regime it is governed by the BFKL linear equation but requires corrective terms as one approach large densities that is contained in the BK equation. Above the saturation scale the system is saturated. This is described by the CGC effective theory whose fundamental equation is the JIMWLK equation to be seen later. $Q_s$ is the frontier scale between the dense and dilute regimes, to be discussed in more details.\label{figQCDPS}}
\end{center}
\end{figure}

Let us first make a sketchy analysis of equation \eqref{BKndif} to see the form of the solutions and the behavior of the gluon density with rapidity. At small $Y$ (or $x$ close to $1$) we have seen that the perturbative expansion in powers of $\alpha_s$ alone holds and the leading contribution to BFKL ladders is trivial : nothing is emitted. This gives an initial condition for both the toy version of BFKL and BK equation : $n_{Y=0}=0$, which is a fixed point of both \eqref{BFKLndif} and \eqref{BKndif}. Of course if the parton density is strictly zero, it remains zero. It actually acquire a small value thanks to the higher $\alpha_s$ orders. The BFKL equation predicts an exponential growth of the parton density in the rapidity variable at fixed transverse momentum. As long as $n$ is small the toy BK equation \eqref{BKndif} reduces to \eqref{BFKLndif} since the $n^2$ term in the r.h.s is negligible. The point is that \eqref{BKndif} have another fixed point for $n=a/c\alpha_s\sim 1/\alpha_s$. This point is fixed at large rapidity when the density is large and recombination balances exactly the splitting processes. The parton density \emph{saturates} at a value of $n\sim 1/\alpha_s$. Therefore the toy BK equation \eqref{BKndif} has solutions that interpolate between $n=0$ and $n\sim 1/\alpha_s$ as one evolves with the rapidity variable at fixed transverse momentum. We already see the emergence of a saturation scale : at fixed transverse momentum, there is a value of rapidity at which the system becomes dense and saturation effects become important. This is illustrated on figure \ref{figQCDPS}.

\subsection{The saturation momentum\label{QCDPS}}
\indent

\begin{figure}[h]
\begin{center}
\includegraphics[width=0.4\paperwidth]{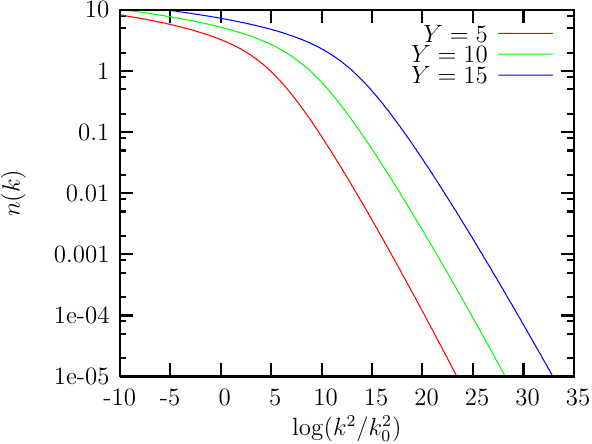}
\caption{Shape of the traveling wave solution to the toy BK equation \eqref{BKndif}. At fixed rapidity, the occupation number varies from $\sim 1/\alpha_s$ to $0$ over a short range interval of $\ln\mathbf{k}_{\perp}^2$. This wave goes forward along the  $\ln\mathbf{k}_{\perp}^2$ axis as $Y$ increases without deformation. \label{travelingwave}}
\end{center}
\end{figure}

\begin{figure}[h]
\begin{center}
\includegraphics[width=0.4\paperwidth]{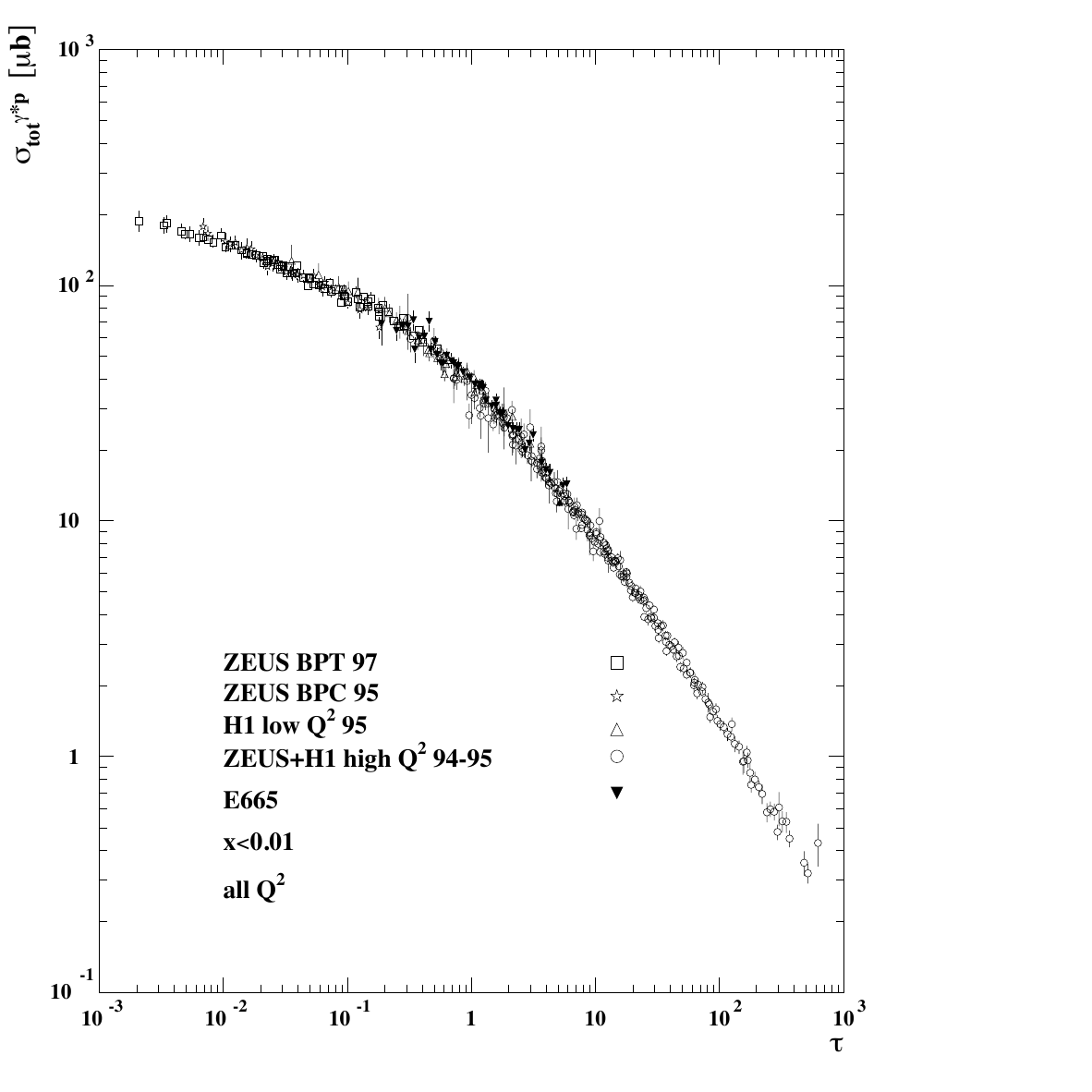}
\caption{Experimental evidence of the geometric scaling. Experimental data for the DIS $\gamma^*p$ cross-section which is a function of the dipole amplitude fits along a straight line in the $(Y,\tau\equiv\ln\mathbf{k}_{\perp}^2)$ plane. Recall the dipole amplitude is the physical quantity governed by the BK equation, the approach in terms of the occupation number is only a toy model.\label{geomscaling}}
\end{center}
\end{figure}

A more quantitative description of saturation is provided by solving the BK equation. Once again let us consider our toy version \eqref{BKndif}. It turns out that this is an already well known equation in reaction-diffusion theory and fluid mechanics. The reaction-diffusion BK equation is equivalent to the so called Fisher-Kolmogorov-Petrovsky-Piscounov (FKPP) equation. It is known that the FKPP asymptotic solutions are \emph{traveling waves} represented on figure \ref{travelingwave} \cite{Munier:2003vc}. The waves progress as rapidity increases without deformation, which means that the solution $n_Y(\mathbf{k}_{\perp})$ actually depend on the single variable $\ln\mathbf{k}_{\perp}^2-\lambda Y$, a non trivial property known as the \emph{geometric scaling}, confirmed experimentally \cite{Stasto:2000er,Gelis:2006bs} as shown on figure \ref{geomscaling}. $\lambda$ is a positive constant interpreted as the "speed" of the wave front (in correspondence with reaction-diffusion processes, $Y$ plays the role of time and $\ln\mathbf{k}_{\perp}^2$, of a spatial coordinate).
Here enters a very far-reached concept of QCD : the geometric scaling property shows the emergence of a dynamically generated intrinsic scale in QCD other than $\Lambda_{QCD}$. Indeed, the single variable dependence of the occupation numbers - or the dipole amplitude in the accurate approach - is denoted $\mathbf{k}_{\perp}^2/Q_s^2(Y)$ instead of $\ln\mathbf{k}_{\perp}^2-\lambda Y$. By identification we have :
\begin{equation}
\label{Qs1}
Q^2_s(Y)=Q^2_0e^{\lambda Y}.
\end{equation}
$Q_s$ is called the \emph{saturation momentum}. A more accurate analysis \cite{Mueller:2002zm,Munier:2003sj} of the BK equation actually shows deviations to geometric scaling taking the running coupling into account and equation \eqref{Qs1} is in fact an approximate expression valid for high energies, where the variation of the QCD coupling constant is slow. Such analysis also leads to $\lambda\simeq 4.9\alpha_s$. We omit another parameter that affects the value of $Q_s$ : $Q_s$ may depend on the nature of the hadron. Just from first QCD principles, gauge invariance does not distinguish between protons and neutrons, thus it must depend only on the total number of nucleons $A$ in the considered nucleus. The number of gluons scales like $A$ and so the occupation number, that is the number of gluons per unit of transverse surface which scales roughly like $A^{2/3}$ for a large nucleus, scales like $A^{1/3}$ and so does $Q_s$. Therefore taking the nuclear size into account, equation \eqref{Qs1} is modified according to :
\begin{equation}
\label{Qs2}
Q^2_s(Y,A)=Q^2_0(A)e^{\lambda Y}\sim A^{1/3}e^{\lambda Y}.
\end{equation}\\

\begin{figure}[ht]
\begin{minipage}[b]{0.45\linewidth}
\centering
\includegraphics[width=\textwidth]{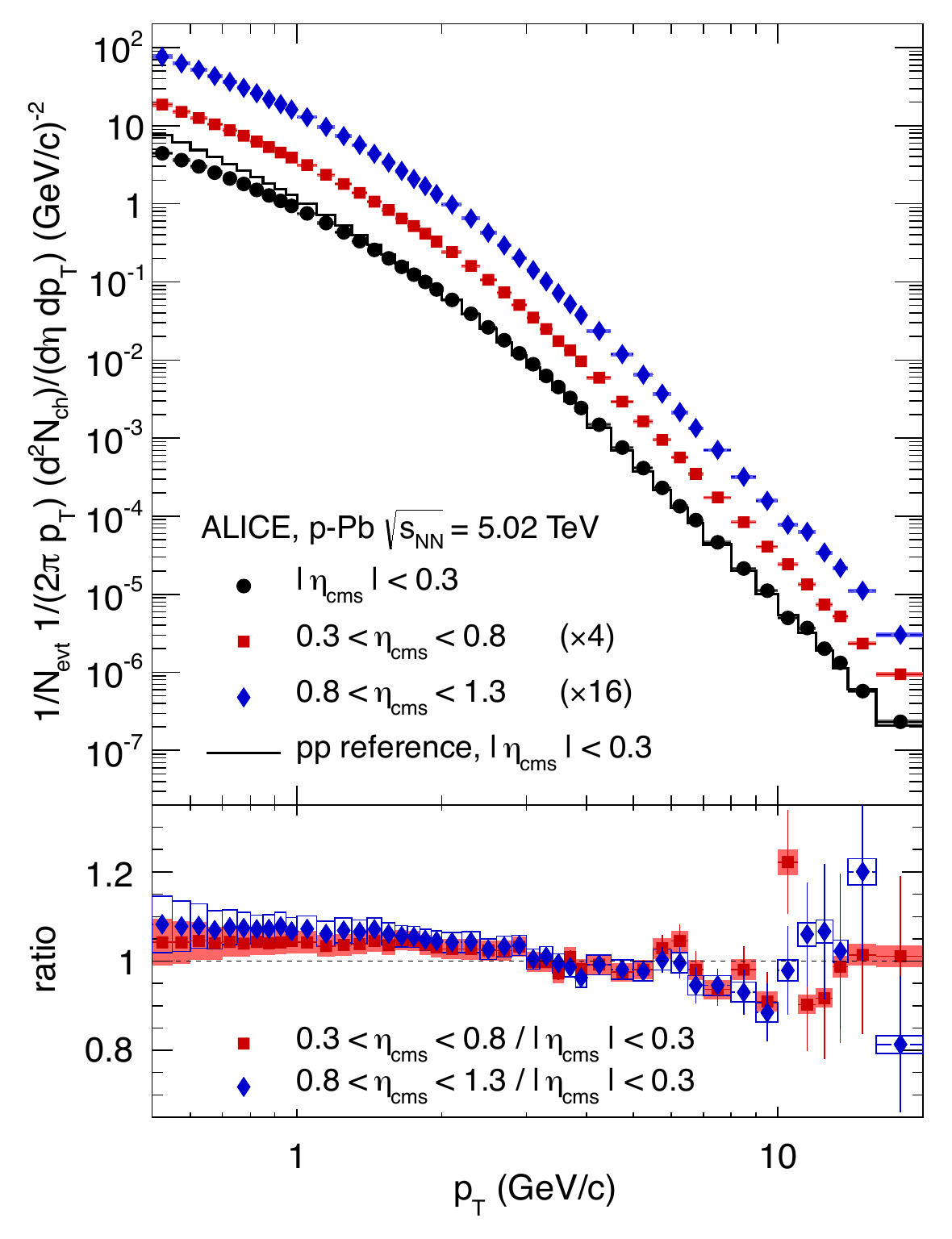}
\end{minipage}
\hspace{0.5cm}
\begin{minipage}[b]{0.45\linewidth}
\centering
\includegraphics[width=\textwidth]{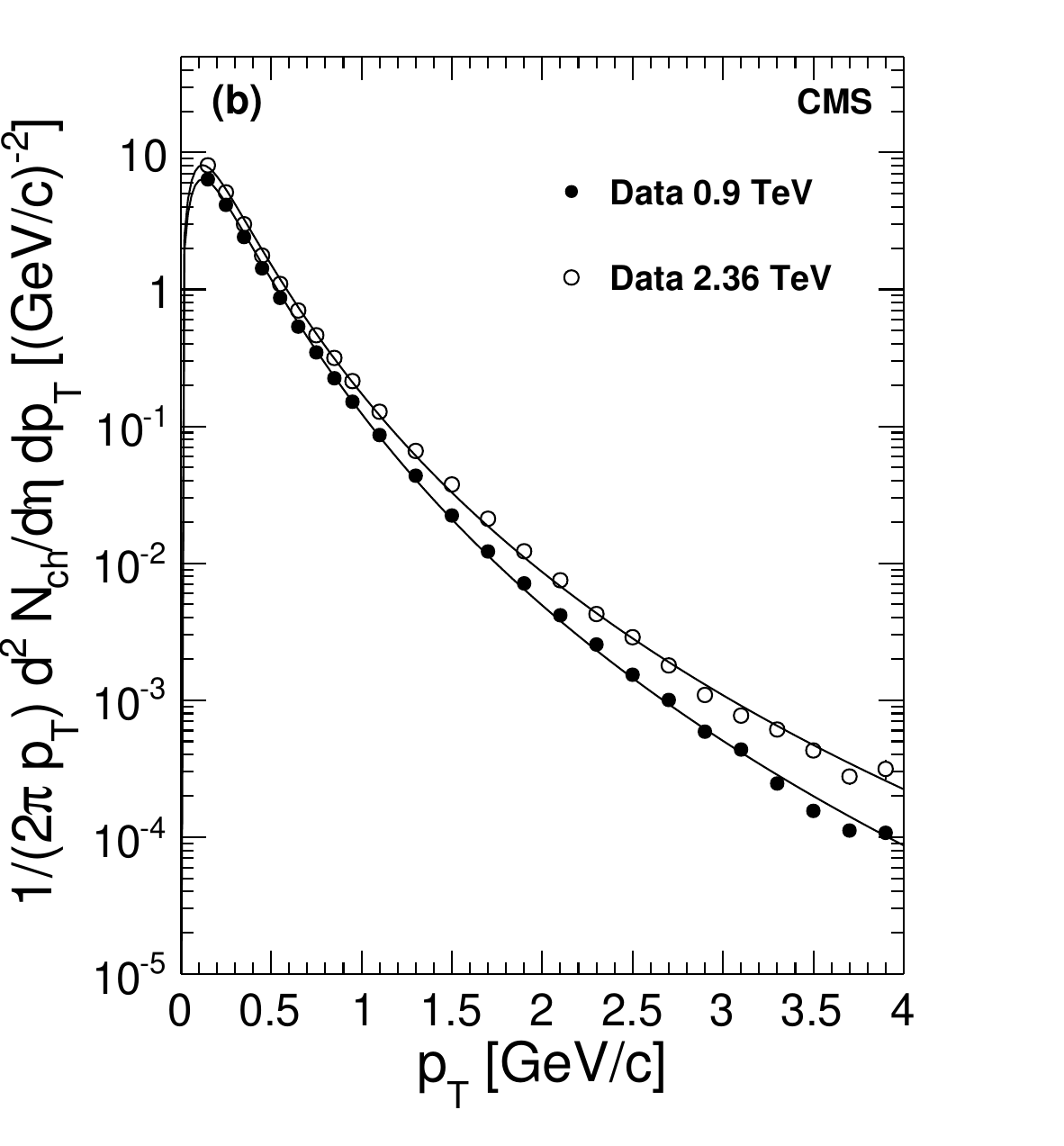}
\end{minipage}
\caption{Experimental evidence of the intrinsic scale in both p-A (ALICE) and p-p (CMS) experiments. The density of partons falls off above values of transverse momenta of the order of the GeV.\label{ALICE-CMS-Qs}}
\end{figure}

Traveling waves solutions enable us to determine the transverse momentum distribution of partons within the hadron. At constant, large enough rapidity\footnote{"Large enough" means that saturation is rerached, if the rapidity is small, the system is always dilute - at least in the perturbative region.} the occupation number interpolates between $n_Y(\mathbf{k}_{\perp}^2)\sim 1/\alpha_s$ for $k_{\perp}\lesssim Q_s$ and $n_Y(\mathbf{k}_{\perp}^2)\rightarrow 0$ for $k_{\perp}\gtrsim Q_s$ with a sharp fall off around the saturation scale. It means that almost all the partons have $k_{\perp}\lesssim Q_s$ in the hadron. The saturation scale turns out to be the typical transverse momentum scale. While in the dilute regime partons may have all possible values of transverse momentum with small occupation numbers, in the dense regime, the transverse momentum is bounded by the saturation scale. This agrees with experiments as shown on figure \ref{ALICE-CMS-Qs}.

\section{Dense media and color glass condensate\label{nucleardescription}}
\indent

The BFKL equation well describes the growth of parton density in a dilute hadron but breaks down once recombination becomes important. The BK equation takes recombination into account and describes the transition to saturation. Both BFKL and BK equations govern the evolution of a color dipole which is a simplified limit - to be discussed - of an infinite hierarchy of equations that couples higher rank correlation functions known as the Balitsky-JIMWLK hierarchy. An alternative description of dense QCD matter is provided by the \emph{Color Glass Condensate} (CGC) effective field theory to be discussed in this section. The aim is to give an intuitive motivation to the topic, to briefly set the framework and to state some known results, especially concerning the evolution at small $x$. More exhaustive approaches can be found in \cite{Gelis:2010nm, Iancu:2012xa}. First we shall see how a dense medium (or at least some of its degrees of freedom) is described by a classical field. These considerations will motivate the CGC formalism and the underlying renormalization group approach. We will discuss the evolution equations of the CGC and the possible simplifying assumptions. In this section the projectile/target description of high energy collisions plays a central role. Indeed, although the CGC provides intrinsically the description of a dense medium, saturation is measured with a probe which gives access to physical observables.

\subsection{The hardness hierarchy and separation of scales}
\indent

\begin{figure}[h]
\begin{center}
\includegraphics[width=0.75\paperwidth]{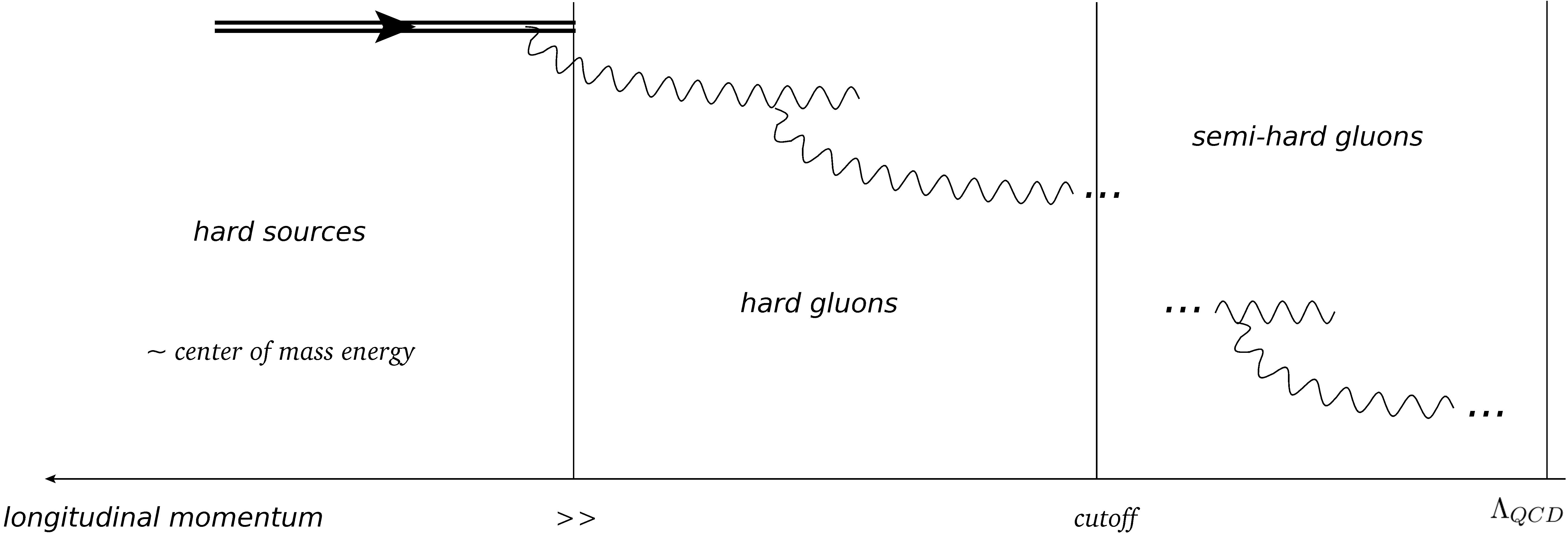}
\caption{Illustration of the terminology. Hard gluons are soft compared with the sources but hard with respect to the cutoff $\Lambda$. Classical sources are used to describe them. Those softer than $\Lambda$ are called semi-hard since they are harder than the non perturbative scale $\Lambda_{QCD}$.\label{scalesep}}
\end{center}
\end{figure}

As seen in the previous sections the softer are the hadronic modes and the more numerous they are. In a frame where the hadron has a high energy, the valence quarks within the hadron are hard and radiate gluons that are mainly soft with respect to them according to the soft divergence \eqref{Pbrem} of the gluon radiation probability and thus, from section \ref{BKequation}, have large occupation numbers of order $1/\alpha_s$. They do in turn radiate mostly softer and softer gluons according to the leading log contribution \eqref{raiseofG}. This hierarchy in the cascade process is a very important feature for motivating the CGC effective theory. A large density of hard partons radiating softer ones is properly described at the classical level (see appendix \ref{nonabelianBF} for the proof of this assertion - it has been proved for quarks emitting gluons but holds for gluons emitting gluons as well). From this consideration, the hardest partons, are described by a classical color source $\mathcal{J}^{A\mu}(x)$. The form of the current will be discussed in the next section and does not matter for the present considerations. The point is that hard particles described by a classical field do have longitudinal momenta greater than some arbitrary scale $\Lambda$. That is one considers successive emissions and recombinations at the classical level up to this scale and the modes that are below this scale are ordinary quantum fields. However they are assumed to carry an energy greater than the non perturbative scale $\Lambda_{QCD}$ to allow perturbation theory. For this reason they are called \emph{semi-hard} rather than soft. This hierarchy is summarized on figure \ref{scalesep}.

\subsection{Background field associated to the target\label{BFtarget}}
\indent

Here we shall discuss more specific properties of the current associated to the nuclear target. The nucleus is conventionally taken to be a left-mover so that the associated classical current is single component $\mathcal{J}^{A-}(x)$. Moreover the calculation of appendix \ref{nonabelianBF} shows that this current does not depend on the variable $x^-$ \footnote{An alternative way is to use covariant conservation. In the gauge $A^+=0$, covariant conservation reads $D\cdot\mathcal{J}=\partial^+\mathcal{J}^-=0$ which completes the proof.}. From the physical side this is rather easy to understand : in the lab frame the longitudinal momentum of quantum fluctuations $p^-$ is small with respect to the total center of mass energy $\sqrt{s}$. Hence their life-time is long with respect to the duration of the collision between the two hadrons : they can be considered as \emph{static}, i.e. $x^-$-independent, at least over times scales that are of the order of the collision process between the projectile and the target. In high energy collision, the lab and center of mass frame are more or less the same : both the projectile and the target carry very large, opposite longitudinal momenta. It means that there is a strong Lorentz length contraction in this frame especially concerning the target. Thus, the source distribution is sharply peaked around $x^+=0$. This allows to refer to the target as a \emph{shockwave}. These considerations enable us to parametrize the source as follows :
\begin{equation}
\mathcal{J}^{A-}(x^+,\mathbf{x}_{\perp}) = g\delta_{\epsilon}(x^+)\rho^A(\mathbf{x}_{\perp}),
\end{equation}
where $\delta_{\epsilon}$ is a representation of the delta function. In actual calculations we send $\epsilon$ to zero as long as this limit is well defined and then $\delta_{\epsilon}$ is merely $\delta$. $\rho^A$ is some arbitrary function of transverse coordinates only. A reasonable physical assumption is that its support is compact : the target extension is finite in the transverse plane. An equivalent description of classical source is the associated background field that satisfies the classical equation of motion in presence of these classical source (see \ref{BFaction} for details). In the present case of a left-moving, static source, the corresponding background field\footnote{The gauge used for the proof is the Landau gauge $\partial\cdot\mathcal{A}=0$. At the classical level the Landau gauge turns out to be the same as the axial gauge $A^+=0$.} is the single component field $\mathcal{A}^{A-}(x^+,\mathbf{x}_{\perp})$ as shown by equation \eqref{classicalA} :
\begin{equation}
\label{classicalA-}
\mathcal{A}^{A-}(x^+,\mathbf{x}_{\perp})= g\delta_{\epsilon}(x^+)\int\frac{\rmd^2y_{\perp}}{2\pi}\ln\left|\mathbf{x}_{\perp}-\mathbf{y}_{\perp}\right|\rho^A(\mathbf{y}_{\perp}).
\end{equation}
The non trivial coordinate dependence of the background field makes the system spatially inhomogeneous. It is therefore inevitable to keep a trace of the coordinate dependence in the Feynman rules. In the dense regime, the occupation number being of order $1/\alpha_s$, the sources and, by the way, the associated classical field are of order $1/g$. Therefore the insertion of one background field in some interaction vertex exactly cancels one power of $g$ coming from the vertex (a more accurate analysis will be performed in section \ref{powercounting}). Thus, $n$ background field legs plugged on a vertex of order $g^n$ is of order one and perturbation theory as an expansion in the number of vertices breaks down if the background field is attached to these vertices. One has then to resum all the diagrams with an arbitrary number of insertions of the classical field. Fortunately the narrowness in the $x^+$ space-time direction of the background field in the axial gauge makes interactions of fast particles with the shockwave very simple. Particles coming from the projectile are not necessarily described by a CGC even though they are hard. Indeed, the CGC gives a description of hard modes contained in the dense medium, i.e. the target only (chapter \ref{AAcollisions} is devoted to the case where both the projectile and the target are dense and thus described by a CGC). For typical momentum scales of particles from the projectile that are assumed to be large with respect to the ones exchanged with the target, one can use the \emph{eikonal} approximation. The justification of this approximation is carefully performed in appendix \ref{eikonalprop}. From the physical point of view this means that the path of a hard particle is almost unaffected by the soft gluons exchanged with the shockwave. For a shockwave in the eikonal approximation, it turns out that the dependence of an observable in the background field only comes through Wilson lines (see appendix \ref{eikonalprop} for a proof) :
\begin{equation}
\label{WilsonLine}
\Omega_{ab}(\mathbf{x}_{\perp})=\mathcal{P}\exp\left[ig\int\rmd x^+\mathcal{A}^{-A}(x^+,\mathbf{x}_{\perp})(T^A)_{ab}\right]
\end{equation}
where $\mathcal{P}$ denotes the path ordering operator along the $x^+$ direction. The Wilson line describes the interaction of particles with the target and encodes their color precession. As we will see through examples and as shown in general in appendix \ref{eikonalprop}, there is one Wilson line per particle traveling through the shockwave in the amplitude corresponding to a given process and each of them are in the gauge group representation of the corresponding field.\\

Let us say a little bit more about the hard modes described by the classical source. Their $x^-$-independence is a consequence of Lorentz time dilatation that \emph{freezes} these modes over large time scales - larger than the duration of the scattering process. Indeed the life time of quantum fluctuations is typically $\tau_{\rm fluc.}\sim 1/p^+\sim 1/x\sqrt{s}$ which is much greater, for small $x$ than the collision time which lasts $\tau_{\rm col.}\sim 1/\sqrt{s}$. The former mechanism of emission-recombination that generates a given source configuration $\mathcal{J}^{A-}$ at the collision time fundamentally obeys the laws of quantum mechanics and is thus \emph{random}. This is a very important feature. It is impossible experimentally to constrain the source configuration in a collision. Two occurrences of the same collision, even with the same initial conditions will be performed \emph{a priori} with different realizations of source's configuration. A brute force computation of an observable, which is in general a functional of the source distribution via the associated classical field $\mathcal{A}^-$, in a given configuration is therefore meaningless. To give a physical meaning to observables, one has to average over the range of all possible source configurations. Obviously some configurations are more likely realizable than others and do obey a probability distribution denoted $\mathcal{W}$ known as the \emph{CGC weight function}. This probability also depends on the cutoff $\Lambda$ that separates the classical modes from the quantum ones. Instead of dealing with a longitudinal momentum cutoff $\Lambda$, one rather introduces the corresponding value of $x_{\rm cut}$ and the associated rapidity $Y=-\ln x_{\rm cut}$. One also prefers to work in the background field $\mathcal{A}^{A-}$ representation of sources instead of the - equivalent - source representation $\mathcal{J}^{A-}$. The probability distribution is denoted $\mathcal{W}_Y[\mathcal{A}^{A-}]$. Observables are averaged thanks to this distribution according to :
\begin{equation}
\label{fundamentalCGC}
\left\langle\mathcal{O}\right\rangle_{Y}=\int \mathcal{D}\mathcal{A}^{A-} W_{Y}[\mathcal{A}^{A-}]\mathcal{O}[\mathcal{A}^{A-}].
\end{equation}
This last expression is one of the fundamental relations of the CGC framework. As long as fields are described by a CGC, one has to perform averages systematically. In order for the theory to be self-consistent one has to be able to work out explicitly $\mathcal{W}_Y[\mathcal{A}^{A-}]$ or, at least, to derive an equation whose solution is $\mathcal{W}_Y[\mathcal{A}^{A-}]$. This can be performed thanks to a renormalization group approach to be discussed in the next section.

\subsection{The renormalization group approach}
\indent

The aim of this section is not the rederivation of the evolution equation that will be stated. The interested reader can find further details in the references quoted in the header \ref{nucleardescription}. The idea of how to find the evolution equation is the same as the classic Wilsonian approach for deriving the Callan-Symanzik equations although the evolution in rapidity $Y$ is \emph{not} an evolution with the energy scale of the theory. The cutoff $Y$ that appears for instance in \eqref{fundamentalCGC} is physically the separation between "hard" and "semi-hard" modes, that is the scale so that for rapidities larger than $Y$ the fields are classical and bellow they are quantum. But this cutoff is arbitrary\footnote{Arbitrary within a range where the external field approximation is justified and perturbation theory holds.}. In mathematical terms, in order to avoid double counting, the loop integrals arising from the computation of observables in perturbation theory must be cut at longitudinal momenta corresponding to the cutoff $Y$ in the ultraviolet. The result is that these observables explicitly contain logarithms of the longitudinal momentum cutoff that can be reabsorbed into a redefinition of the CGC weight function. The $Y$ independence of observables is guaranteed provided the CGC weight function satisfies a renormalization group equation know as the Jalilian Marian-Iancu-McLerran-Weigert-Leonidov-Kovner (JIMWLK) equation \cite{JalilianMarian:1997gr,JalilianMarian:1997jx,JalilianMarian:1997dw,Iancu:2000hn,Ferreiro:2001qy,Iancu:2001ad} which reads as follow :
\begin{equation}
\label{JIMWLKeq}
\frac{\partial\mathcal{W}_Y}{\partial Y}[\mathcal{A}^{-}]=\mathcal{H}\left[\mathcal{A}^{-},\frac{\delta}{\delta\mathcal{A}^{-}}\right]\mathcal{W}_Y[\mathcal{A}^{-}].
\end{equation}
$\mathcal{H}$ is a hermitian, second order functional differential operator known as the JIMWLK hamiltonian. Its explicit form is :
\begin{equation}
\label{JIMWLKham}
\mathcal{H}\left[\mathcal{A}^{-},\frac{\delta}{\delta\mathcal{A}^{-}}\right]=\frac{1}{2}\int \rmd^2x_{\perp} \rmd^2y_{\perp}\frac{\delta}{\delta\mathcal{A}^{A-}(0,\mathbf{x}_{\perp})}\eta^{AB}(\mathbf{x}_{\perp},\mathbf{y}_{\perp})\frac{\delta}{\delta\mathcal{A}^{B-}(0,\mathbf{y}_{\perp})}.
\end{equation}
The integral kernel $\eta^{AB}$ is a functional of the classical field through adjoint representation Wilson lines $\tilde{\Omega}$ which reads :
\begin{equation}
\begin{split}
\eta^{AB}(\mathbf{x}_{\perp},\mathbf{y}_{\perp})=\frac{1}{4\pi^3}\int \rmd^2z_{\perp}\frac{(\mathbf{x}_{\perp}-\mathbf{z}_{\perp})\cdot (\mathbf{y}_{\perp}-\mathbf{z}_{\perp})}{(\mathbf{x}_{\perp}-\mathbf{z}_{\perp})^2(\mathbf{y}_{\perp}-\mathbf{z}_{\perp})^2}\left[\tilde{\Omega}^{\dagger}(\mathbf{x}_{\perp})-\tilde{\Omega}^{\dagger}(\mathbf{z}_{\perp})\right]^{AC}\left[\tilde{\Omega}(\mathbf{y}_{\perp})-\tilde{\Omega}(\mathbf{z}_{\perp})\right]^{CB}.
\end{split}
\end{equation}\\

The background field averaged observables \eqref{fundamentalCGC} also satisfy an evolution equation. The equation that governs the evolution of observables follows from plugging \eqref{JIMWLKeq} into \eqref{fundamentalCGC} and integrating the functional integral by part using the hermiticity property of the JIMWLK hamiltonian. This leads, for an observable $\mathcal{O}$, to :
\begin{equation}
\frac{\partial}{\partial Y}\left\langle\mathcal{O}\right\rangle_{Y}=-\left\langle\mathcal{H}\mathcal{O}\right\rangle_{Y}.
\end{equation}
This equation is - a bit improperly - called the JIMWLK equation as well and will be the one that is understood when we deal with the evolution of observables. The JIMWLK equation holds only for gauge invariant observables.

\subsection{Multipoles and B-JIMWLK hierarchy\label{B-JIMWK}}
\indent

Such gauge invariant quantities are provided by squared amplitudes and cross-sections summed over the initial and final colors. This always leads to a dependence of observables on the background field through terms of the form :
\begin{equation}
\label{Coloroperator}
\begin{split}
&\tr\left[\Omega_{R_1}(\mathbf{x}_{1,\perp})\Omega^{\dagger}_{R_1}(\mathbf{x}_{2,\perp})\Omega_{R_1}(\mathbf{x}_{3,\perp})...\Omega_{R_1}^{\dagger}(\mathbf{x}_{2n,\perp})\right]\\
&~~~~~\times\tr\left[\Omega_{R_2}(\mathbf{y}_{1,\perp})\Omega^{\dagger}_{R_2}(\mathbf{y}_{2,\perp})\Omega_{R_2}(\mathbf{y}_{3,\perp})...\Omega_{R_2}^{\dagger}(\mathbf{y}_{2m,\perp})\right]\times...
\end{split}
\end{equation}
where $R_i$ is some representation of the gauge group depending on the nature of the fields involved in the reaction. A single trace operator of $2n$ Wilson lines is called a $2n$-rank \emph{multipole}. The CGC requires the average of color operators of the form \eqref{Coloroperator} over the classical field configurations. This leads to an infinite hierarchy of coupled differential equations known as the Balitsky-JIMWLK hierarchy. The simplest color operator is the dipole amplitude corresponding to a $q\bar{q}$ pair. From the CGC framework it is straightforward to show that the $q\bar{q}$ pair amplitude scattering a dense medium is proportional to :
\begin{equation}
S(\mathbf{x}_{\perp},\mathbf{y}_{\perp})=\frac{1}{N_c}\tr\left[\Omega(\mathbf{x}_{\perp})\Omega^{\dagger}(\mathbf{y}_{\perp})\right].
\end{equation}
where the Wilson lines are in the fundamental representation. The evolution equation for the averaged dipole is given by :
\begin{equation}
\label{firstB-JIMWLK}
\begin{split}
\frac{\partial}{\partial Y}\left\langle S(\mathbf{x}_{\perp},\mathbf{y}_{\perp})\right\rangle_Y=-\frac{\alpha_s N_c}{2\pi^2}\int\rmd^2z_{\perp}&\frac{(\mathbf{x}_{\perp}-\mathbf{y}_{\perp})^2}{(\mathbf{x}_{\perp}-\mathbf{z}_{\perp})^2(\mathbf{z}_{\perp}-\mathbf{y}_{\perp})^2}\\
&\left[\left\langle S(\mathbf{x}_{\perp},\mathbf{y}_{\perp})\right\rangle_Y-\left\langle S(\mathbf{x}_{\perp},\mathbf{y}_{\perp})S(\mathbf{z}_{\perp},\mathbf{y}_{\perp})\right\rangle_Y\right].
\end{split}
\end{equation}
The point is that the $<SS>$ term itself obeys another evolution equation whose r.h.s contains quadrupoles, i.e. operators made of trace of four Wilson lines and so on... this dependence of evolution equations on each other is the B-JIMWLK hierarchy. However one can simplify the equation in the limit of a large dense target which contains a lot of partons, $n$ for definiteness. If, for instance, two dipoles from the projectile interact with two different partons from the target they do not talk to each other. The mathematical consequence is that the average of the product factorizes into the product of the averages due to the independence of the scatterings. The number of ways to plug the two dipoles on the same parton in the target goes like $n$ and on two different partons, it goes like $n(n-1)/2\sim n^2$ for large $n$. Thus for a large target, that is as $n$ becomes large, one can neglect the entangled scatterings with respect to the independent ones (similar arguments arise in another context in the end of appendix \ref{nonabelianBF}). Thus for a large target, one has :
\begin{equation}
\left\langle S(\mathbf{x}_{\perp},\mathbf{y}_{\perp})S(\mathbf{z}_{\perp},\mathbf{y}_{\perp})\right\rangle_Y~~~\rightarrow~~~\left\langle S(\mathbf{x}_{\perp},\mathbf{y}_{\perp})\right\rangle_Y\left\langle S(\mathbf{z}_{\perp},\mathbf{y}_{\perp})\right\rangle_Y.
\end{equation}
Plugging this property into \eqref{firstB-JIMWLK} makes that equation \eqref{firstB-JIMWLK} becomes closed. The large target limit of \eqref{firstB-JIMWLK} is nothing but the BK equation discussed in section \ref{BKequation}.

\subsection{Alternative simplified approaches for computing CGC averages}
\indent

The brute force computation of the background field expectation value of multipoles is practically impossible, even numerically due to the infinite B-JIMWLK hierarchy. We here mention the available methods and approximations used for actual computation.

\subsubsection{The gaussian approximation\label{gaussianJIMWLK}}
\indent

To guess possible simplifications, let us focus on what is taken into account in the JIMWLK evolution. The CGC provides the leading small $x$ logs contribution of gluon cascades in the saturated regime. That is, it is a non-linear generalization of the BFKL evolution in which the BFKL ladders can recombine together. In the BFKL evolution the ladders freely propagate and do not interact with each other. In that case, the evolution is said to be linear. The non-linearity of the JIMWLK evolution arises from the possible fusion between ladders. Let us close our eyes for a moment on non-linearities considering only the free evolution of ladders in the CGC framework. In that case, the distribution $\mathcal{W}_Y[\mathcal{A}^{-}]$ is \emph{gaussian}, that is :
\begin{equation}
\left\langle \mathcal{A}^{A_1-}(x^+_1,\mathbf{x}_{1,\perp})...\mathcal{A}^{A_n-}(x^+_n,\mathbf{x}_{n,\perp})\right\rangle_Y = \left\{
    \begin{array}{ll}
        &0  \mbox{~~if~}n\mbox{~is odd}\\
	&\\
        &\left\langle \mathcal{A}^{A_1-}(x^+_1,\mathbf{x}_{1,\perp})\mathcal{A}^{A_2-}(x^+_2,\mathbf{x}_{2,\perp})\right\rangle_Y...\\
&...\times\left\langle \mathcal{A}^{A_{n-1}-}(x^+_{n-1},\mathbf{x}_{n-1,\perp})\mathcal{A}^{A_n-}(x^+_n,\mathbf{x}_{n,\perp})\right\rangle_Y\\
&~~~~~~~~~ + \mbox{perm.} \mbox{~~if~}n\mbox{~is even}.
    \end{array}
\right.
\end{equation}
In the gaussian limit, the distribution is encoded into the 2-point function $\left\langle \mathcal{A}^{A-}(x^+,\mathbf{x}_{\perp})\mathcal{A}^{B-}(y^+,\mathbf{y}_{\perp})\right\rangle_Y$ only. Its structure is furthermore constrained by physical requirements : gauge invariance requires the color structure to be trivial, i.e. proportional to $\delta^{AB}$, causality requires locality in the $x^+$ variable, i.e. the 2-point function is proportional to $\delta(x^+-y^+)$ and homogeneity of the nucleus in the transverse direction requires a dependence in the transverse coordinates only through $\mathbf{x}_{\perp}-\mathbf{y}_{\perp}$. Thus, without any loss of generality, the 2-point function can be parametrized as follows :
\begin{equation}
\label{AAtwopfunction}
\left\langle \mathcal{A}^{A-}(x^+,\mathbf{x}_{\perp})\mathcal{A}^{B-}(y^+,\mathbf{y}_{\perp})\right\rangle_Y=\delta^{AB}\delta(x^+-y^+)\gamma_Y(x^+,\mathbf{x}_{\perp}-\mathbf{y}_{\perp}).
\end{equation}
Although these considerations assume the evolution of ladders to be free, it has been shown \cite{Iancu:2011nj,Iancu:2011ns} that it remains valid even in the presence of non-linear effects. A mean field approximation can be performed identifying a ladder resulting from the recombination of two ladders as an effective single ladder. Under this mean field approximation, the JIMWLK evolution remains gaussian even if one takes non-linear effects into account. This only affects the form of the function $\gamma_Y$ in \eqref{AAtwopfunction} compared to the BFKL gaussian evolution. The gaussian approximation to JIMWLK has already shown a great success for the comparison of theory vs. experiments \cite{Kovchegov:2008mk,Marquet:2010cf}. Moreover it is much easier to implement \cite{Rummukainen:2003ns,Lappi:2011ju,Dumitru:2011vk} and also leads to accurate analytical results.

\subsubsection{Initial conditions, dilute regime and McLerran-Venugopalan model\label{MVmodel}}
\indent

From previous considerations the gaussian approximation is obviously valid in the dilute medium limit in which non-linear effects can be omitted. In the dilute limit the background field is weak and one can expand perturbatively the Wilson lines in powers of the background field in correlators. The first non trivial order is the second order in powers of the background field where the 2-point function \eqref{AAtwopfunction} can be identified. This low energy limit is known as the McLerran-Venugopalan (MV) model \cite{McLerran:1994vd}. Furthermore, the low energy limit can be taken as an initial condition for solving the JIMWLK equation since it is easy to work out correlators at large $x$ - or small $Y$ - thanks to the MV model. From the physical point of view the dilute limit takes only the dominant single scattering contribution into account. In this context the concept of parton distribution makes sense and it is actually possible to relate the unintegrated gluon distribution $f_Y(\mathbf{p}_{\perp})$, defined in \eqref{UGPDF}, to the 2-point function \eqref{AAtwopfunction} according to
\begin{equation}
\label{MVUGD}
\frac{f_Y(\mathbf{p}_{\perp})}{\mathbf{p}_{\perp}^2}=\frac{(N_c^2-1)S_{\perp}}{(2\pi)^2}\mathbf{p}_{\perp}^2\int \rmd x^+\rmd^2x_{\perp}\gamma_Y(x^+,\mathbf{x}_{\perp}-\mathbf{y}_{\perp})e^{-i\mathbf{p}_{\perp}\cdot(\mathbf{x}_{\perp}-\mathbf{y}_{\perp})}.
\end{equation}
This result will be proved in section \ref{diluteqq}.

\chapter{Di-hadron production in proton-nucleus collisions at the LHC \label{pAchapter}}
\indent

\setcounter{equation}{0}

For reasons to be explained bellow, an accurate, quantitative exploration of saturation is provided by proton-nucleus collisions, i.e. a dilute projectile, colliding a dense target. The first data for particle production in p-Pb collisions at the LHC have just become available \cite{:2012xs, :2012mj} and more data will be taken during the run scheduled for 2013. A main feature of the LHC for our present purposes is to provide experimental data at values of the Bjorken variable $x$ smaller than ever. This enables the exploration of the saturated regime of QCD whose natural framework is the CGC effective field theory. The precise boundary between the dilute and saturated regimes is not yet firmly established neither on the experimental nor on the theory side. However, the phenomenological success of CGC-based predictions for various observables measured at RHIC \cite{RHIC2012} suggest that the saturated regime has been - at least marginally - reached at RHIC. This makes it encouraging to conjecture that this regime will be fully reached and well explored by the p-Pb collisions at the LHC. In experiments, one of the easiest quantitative evidence of saturation is the measurement of di-hadron correlations. Working out the di-gluon production inclusive cross-section will be the main aim of this chapter having in mind the study of saturation at the LHC.\\

First we shall discuss what is expected from saturation and especially the decoherence effect between produced particles. Next we shall emphasize the di-hadron kinematics and by the way, why di-hadron production cross-sections in p-A collisions provide the simplest observable for a quantitative evidence of saturation. We will also motivate the choice of di-gluon production rather than any other process at the LHC. Then, as an appetizer we shall compute the single quark production cross-section. This is the simplest calculation in the spirit of what we are doing. This will be a good exercise to see how things work. The last section will be devoted to the computation of the di-gluon cross section and will be our main result that is closely related to current LHC experiments. This will be computed in the general case analytically. Then we shall push further analytic calculations by looking at the hard scattering limit where the transverse momenta of final gluons are large with respect to $Q_s$. This limit shows up a strongly correlated character and enables us to define new non-linear generalizations of gluon distributions. This chapter is based on our work \cite{Iancu:2013dta}.

\section{Final state decoherence}
\indent

For a saturated target, a parton coming from the projectile will receive multiple scatterings when traveling through the dense medium. The situation really differs from a single scattering where the projectile interacts with the target via a single particle exchange that is computable by perturbation theory. In the dense regime the multiple exchanges are very complicated and cannot be computed by Feynman diagrams techniques. For instance CGC faces this problem by assuming that it is well described by a random classical field whose multiple interactions lead - in the eikonal approximation - to the emergence of Wilson lines that fully encode the scattering process. From the physical point of view, our lack of knowledge about multiple scatterings within the saturated medium makes impossible to claim the existence of momentum conservation law. In other words, information about the initial momentum of an incoming parton is lost as it scatters the target off. This is broadly what we call the \emph{decoherence} effect, the main evidence of high density regime.\\

\begin{figure}[ht]
\begin{minipage}[b]{0.45\linewidth}
\centering
\includegraphics[width=\textwidth]{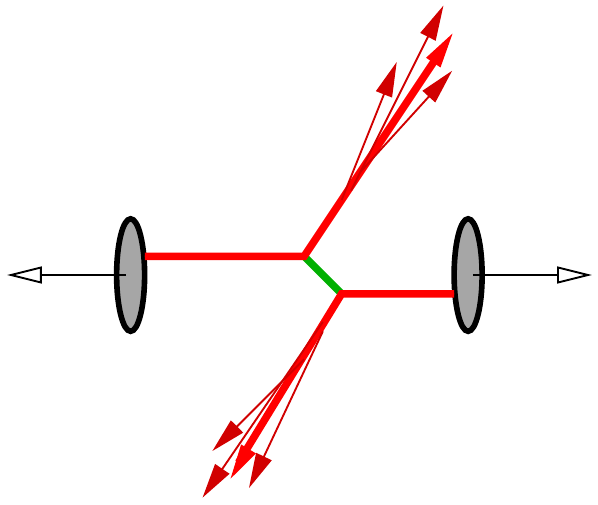}
\caption{Typical process between two dilute hadrons which produces two back-to-back jets.\label{dilutedilute}}
\end{minipage}
\hspace{0.5cm}
\begin{minipage}[b]{0.45\linewidth}
\centering
\includegraphics[width=\textwidth]{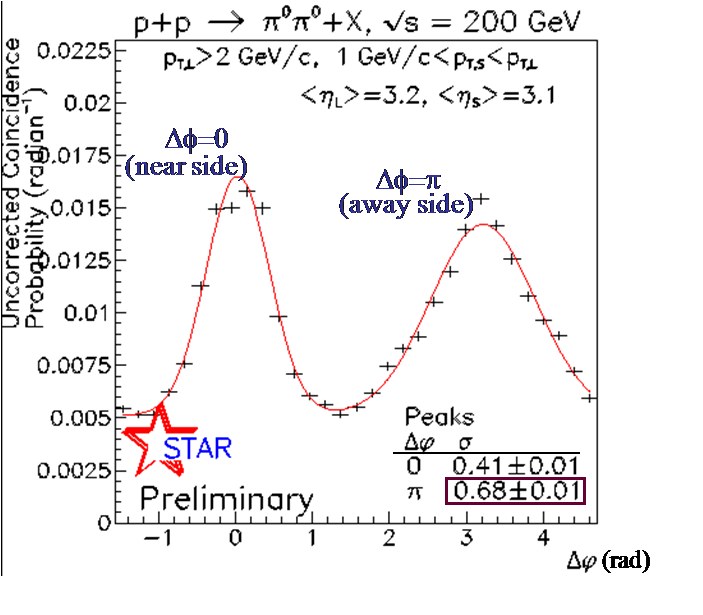}
\caption{Plot of the angular distribution - in the transverse plane - between two final particles. It shows up two sharp peaks at $\Delta\phi=0$ and $\Delta\phi=\pi$. This result comes from p-p experiments performed at RHIC\cite{Albacete:2010rh}.\label{dilutedilute0}}
\end{minipage}
\end{figure}

In our precise context, the term "decoherence" generally refers to the decorrelation between particles in the final state. To understand what is meant let us consider some arbitrary high energy collision between two dilute hadrons as represented on figure \ref{dilutedilute}.
Such a reaction involves the exchange of a single soft particle between the two hadrons. The produced particles are arranged in two beams, called \emph{jets}, that have opposite directions in the transverse plane and for this reason are called \emph{back-to-back}. At high energy, particles in each jet are almost collinear. This merely follows from the collinear divergence \ref{Pbrem}. Therefore if one picks randomly any two particles in the final state, either they belong to the same jet and have a very small angular separation or they belong to two different jets and they have quasi opposite directions. Then if one looks at the angular separation between final particles one finds a distribution sharply peaked around angular separations equal to $0$ or $\pi$, as show on plot \ref{dilutedilute0}. In that dilute-dilute case we say that particles are strongly correlated : given the momentum of one of the particles in the final state, there is a large probability that the other ones have been emitted in the same direction or in the opposite one.\\

\begin{figure}[ht]
\centering
\includegraphics[width=0.4\linewidth]{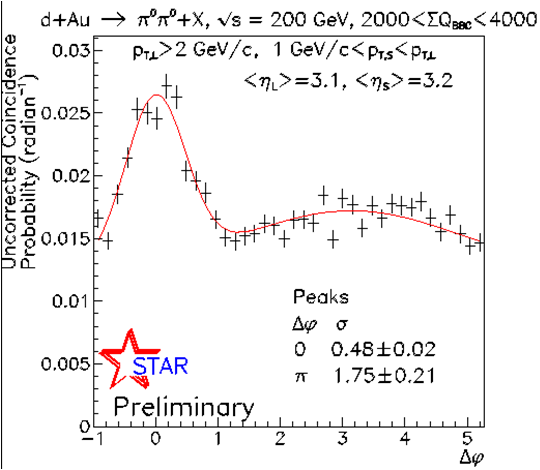}
\caption{Angular distribution between two final particles in d-Au collisions performed at RHIC \cite{Albacete:2010rh}. Correlations, especially around $\Delta\phi=\pi$ are widely suppressed compared to the p-p case. The $\Delta\phi=0$ peak remains in such an asymmetric collision since only one of the peaks is broadened. There is still a large probability that two produced particles are close to each other.\label{dAuRHIC}}
\end{figure}

The situation is very different if at least one of the two colliding hadrons is dense. Indeed, there is no longer a single particle exchanged but many of them. Each of them can couple to various particles within the jets : the angular distribution of particles in the jet is broadened. The transverse momentum broadening is the main observational evidence of saturation as shown on figure \ref{dAuRHIC}.
The disappearance of correlations between particles is what is referred to \emph{decoherence} or \emph{decorrelation} in this context. From now let us focus on correlations between two hadrons.

\section{Di-hadron kinematics\label{CGCdihadronpheno}}
\indent

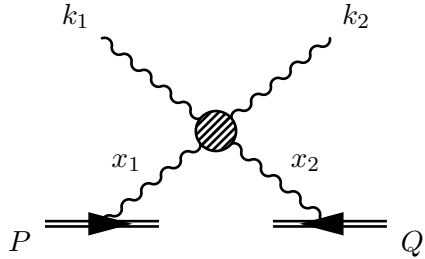
\begin{figure}[h]
\centering
%(along, up)
\vspace{.5cm}
\begin{fmfgraph*}(160,70)
    \fmfleft{b1,t1}
    \fmfright{b7,t7}
    \fmf{phantom}{t1,t2,t3,t4,t5,t6,t7}
    \fmf{phantom}{b1,b2,b3,b4,b5,b6,b7}
    \fmffreeze
      \fmf{dbl_plain_arrow}{b1,b3}
      \fmf{dbl_plain_arrow}{b7,b5}
    \fmffreeze
      \fmf{phantom}{b2,v,t6}
      \fmf{phantom}{b6,v,t2}
    \fmffreeze
      \fmf{photon,label=$x_1$}{v,b2}
      \fmf{photon,label=$x_2$}{b6,v}
      \fmf{photon}{t2,v,t6}
      \fmflabel{$P$}{b1}
      \fmflabel{$Q$}{b7}
      \fmflabel{$k_2$}{t6}
      \fmflabel{$k_1$}{t2}
      \fmfblob{15.}{v}
\end{fmfgraph*}
\vspace{.5cm}
\caption{A single scattering exclusive di-hadron production process.\label{DiHadronkinematics}}
\end{figure}

Inclusive di-hadron production cross-sections provide the simplest quantitative tool for studying saturation effects. Before making kinematic considerations for justifying the choice of this observable, let us motivate first the suitable experimental conditions provided by dilute-dense collisions for probing saturated QCD matter. The question, why does one prefer p-A collisions arises from the following consideration : the most accurate way to measure the physical effects of a saturated medium - the target - is to get rid of possible saturation effects coming from the probe, i.e. the projectile. A dense projectile would bring its intrinsic noise that would perturb the measurement. We come to the conclusion that the saturated target must be probed with a dilute projectile. A naive guess for motivating the p-A collisions rather than p-p or A-A ones follows from the intuitive picture that one has from hadrons : there are more valence quarks in a nucleus than in a proton and therefore it is denser. In section \ref{QCDPS}, we have seen that the saturation scale behaves like $Q_s\sim A^{1/3}$. At some given small value of $x$, saturation is easier to reach as the hadron considered has a larger number of valence quarks. However a large number of valence quarks increases the \emph{double parton scattering} (DPS) effect \cite{Lappi:2012nh}. From the physical point of view, the DPS effect is due to pair of partons created in the remote past. It can be shown that they must be included into the initial proton wave function and are indistinguishable from a double emission process by two independent sources. Obviously this effect is more important as the number of valence quarks increases. DPS effect also brings an independent contribution to momentum broadening that has nothing to do with saturation. In that sense it is undesirable. From these considerations, a good candidate for a dilute probe is a proton whereas a large nucleus is a nice dense medium. This indeed refines the measurement of the intrinsic nuclear saturation effects and justifies in its own the choice of p-A collisions\footnote{Or at least light-heavy hadrons collisions like the d-Au processes performed at RHIC. For technical reasons, RHIC cannot use protons as projectiles.}. However, the hadronic size is not the only parameter that dertermines to dilute or dense character of a hadron. Saturation also depends on the region of phase space we are looking atn that is, on the kinematics. Let us sketch this in the following elementary example : the exclusive di-hadron production via a single exchange between the projectile and the target represented on figure \ref{DiHadronkinematics}. These considerations are qualitatively unaffected by multiple scatterings and more complicated sub-processes.
By working out kinematic relations, it is not difficult to show that the values of $x_1$ and $x_2$ are completely fixed by the final state kinematics :
\begin{equation}
\label{dihadronkin}
\begin{split}
&x_1=\frac{k_{1,\perp}}{\sqrt{s}}e^{y_1}+\frac{k_{2,\perp}}{\sqrt{s}}e^{y_2}\\
&x_2=\frac{k_{1,\perp}}{\sqrt{s}}e^{-y_1}+\frac{k_{2,\perp}}{\sqrt{s}}e^{-y_2},
\end{split}
\end{equation}
where $y_i=\frac{1}{2}\ln(k_i^+/k_i^-),~i=1,2$ are the respective rapidities of the final partons\footnote{This definition for the rapidity differs from $y=-\ln x$ by an irrelevant constant.} and $\sqrt{s}$ is the total center of mass energy. Beside the number of valence quarks, the saturation scale $Q_s$ also depends on $x$. The dilute-dense kinematics requires that the projectile is probed at values of $x_1\sim 1$ whereas for the target $x_2\ll 1$. Relations \eqref{dihadronkin} shows that in order to be fulfilled one has to consider \emph{forward rapidities}, that is $y_{1,2}>0$. Physically this means that the two measured final partons are right-movers. We come to the conclusion that p-A collisions together with the measurement of forward rapidity regime are the best experimental conditions for studying saturation physics.\\

One can even go further since some processes are dominant depending on the experimental conditions. Nowadays one has essentially two available tools for the exploration of dense hadronic matter. The first one is the RHIC. For technical reasons it does not perform p-A collisions but d-Au ones at energies of 200 GeV/nucleon. The rapidity range goes up to $y\simeq 4$. On the deuteron side one has typically $x_1\sim 10^{-1}$. At this value of $x_1$ the deuteron wave function is dominated by the valence quarks \cite{Gwenlan:2009kr} and the leading di-hadron production process is the radiation of a gluon by this quark as it scatters the gold nucleus off : $q Au\rightarrow qg X$. The explored values of $x_2$ are expected to lie at the boundary of the saturated phase at such energies. Therefore it was a priori difficult to forecast whether saturation will be observed or not on the experimental side and whether theories for saturation will agree with data or not on the theoretical side. The disappearance of back-to-back correlations in di-hadron final states has been indeed experimentally observed in d-Au (and even some kinematic regions of p-p) collisions at RHIC \cite{Adare:2011sc, Braidot:2011zj}. On the contrary, we expect that the effect of the medium will be tiny for gluons with transverse momenta large with respect to $Q_s$, in agreement e.g. with the results \cite{:2012mj,Adler:2003ii}. Even though the validity range of CGC around the transition to saturation is ill defined, CGC-based predictions seem to agree with data \cite{RHIC2012}. One has been able to compute qualitatively \cite{Marquet:2007vb, JalilianMarian:2004da, Baier:2005dv, Nikolaev:2003zf} and even quantitatively \cite{Albacete:2010pg, Stasto:2011ru} these decoherence effects in agreement with experiments. The other accelerator that is expected to reach the fully saturated regime is the LHC. The first p-Pb runs have been performed in winter 2012-2013 and some preliminary data is available yet \cite{:2012xs,:2012mj}. Since the center of mass energy is $\sqrt{s}=5$ TeV/nucleon, the lead nucleus is clearly in the saturated phase. The rapidity range covered is a bit larger than at RHIC : $y<5$ or $6$. The proton is probed in the regime $x_1\sim 10^{-2}-10^{-3}$ where its wave function is dominated by gluons. Then the dominant process is the splitting of a gluon into either a gluon pair or a $q\bar{q}$ pair. We shall focus on the first one : $g Pb\rightarrow gg X$ which has a richer color structure.\\

So far, our problem is motivated by physics. Our task is next two compute the cross-section for the inclusive di-gluon production in order to make predictions for LHC runs in p-Pb collisions. Let us first compute one of the simplest observable in the CGC framework : the single quark production inclusive cross-section. This school case will allow us to introduce the formal devices that will be used for the di-gluon production. Then we will not be lost into peripheral digressions and be able to follow a straight guideline for this more complicated case.

\section[Single quark scattering]{A pedestrian example of practical computation : single quark scattering}

\subsection{Cross-section}
\indent

\begin{figure}[h]
\centering
%(along, up)
\begin{fmfgraph*}(80,60)
    \fmftop{t1,t2,t3}
    \fmfbottom{b1,b2,b3}
    \fmfleft{i}
    \fmfright{o}
      \fmf{dbl_dots}{t2,v,b2}
      \fmf{fermion}{i,v,o}
    \fmffreeze
      \fmflabel{$a, s, p\rightarrow$}{i}
      \fmflabel{$\rightarrow q, r, b$}{o}
\end{fmfgraph*}
\caption{Feynman diagram for the process $qA\rightarrow qA$. The doted line represents the nucleus. One has to keep in mind that the quark has been emitted in the remote past by the proton. \label{singlequarkpA}}
\end{figure}
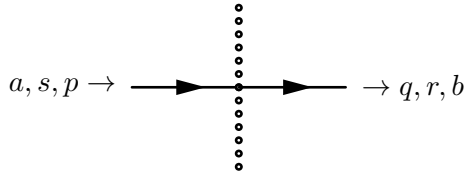

To set the notations and see how calculations work, one considers the simplest example of a single quark scattering off the nucleus. This process is illustrated on figure \ref{singlequarkpA}.
The incoming quark is conveniently chosen as forward, that is its momentum $p$ reads $p=(p^+,0,\mathbf{p})$. The S-matrix element corresponding to this process is very simple since it is nothing but the normalization condition \eqref{BFsinglePSnormalization} detailed in appendix \ref{eikonalprop} for one particle states in presence of a background field :
\begin{equation}
\label{singlequarkSmatrix}
\left\langle\left. \vec{q},r,b\right|\vec{p},s,a\right\rangle_{\mathcal{A}^-}=2p^+\delta^{rs}2\pi\delta(p^+-q^+)\int\rmd^2x_{\perp}\Omega_{ba}(\mathbf{x}_{\perp})e^{-i(\mathbf{q}_{\perp}-\mathbf{p}_{\perp})\cdot\mathbf{x}_{\perp}}.
\end{equation}
$\Omega_{ba}(\mathbf{x}_{\perp})$ is the Wilson line \eqref{WilsonLine} in the fundamental representation. The physical quantity accessible to experimentalists is the cross-section. The way we relate cross-sections to S-matrix elements is explained in appendix \ref{pAcrosssections}. Although the initial state in \eqref{singlequarkSmatrix} seems to be a one-particle state, the meaningful observable is a cross-section rather than a decay rate. The initial nucleus does not appear in the initial quantum state since it is described at the classical level but the physical process is indeed a two-body collision. Let us define as the $\mathcal{M}$-matrix the coefficient of $2\pi\delta(p^+-q^+)$ in formula \eqref{singlequarkSmatrix}. The differential cross-section for the single quark scattering then reads in terms of the $\mathcal{M}$-matrix element as follow :
\begin{equation}
\label{qqCS}
\rmd\sigma(q(p)\rightarrow q(q))=\frac{1}{2p^+}\left|\mathcal{M}(q(p)\rightarrow q(q))\right|^22\pi\delta(p^+-q^+)\frac{\rmd^3q}{(2\pi)^32q^+}.
\end{equation}
This is the cross-section for a process with definite colors and spins in the initial and final states, however, it is easier if one deals with gauge and Lorentz invariant quantities. Then we have to sum over the final state's color and spin and average over the initial state's color and spin. Moreover one has to average also over the background field according to \eqref{fundamentalCGC}. The essential ingredient for computing the cross section is then :
\begin{equation}
\label{qtoqcross}
\begin{split}
\left\langle\overline{\left|\mathcal{M}(q(p)\rightarrow q(q))\right|^2}\right\rangle_Y&=\frac{1}{2N_c}\displaystyle{\sum_{r,s}}\displaystyle{\sum_{a,b}}~4(p^+)^2\delta^{rs}\delta^{rs}\int\rmd^2x_{\perp}\rmd^2y_{\perp}\left\langle\Omega_{ba}(\mathbf{x}_{\perp})\Omega^{\dagger}_{ab}(\mathbf{y}_{\perp})\right\rangle_Ye^{-i\mathbf{q}_{\perp}\cdot(\mathbf{x}_{\perp}-\mathbf{y}_{\perp})}\\
&=\frac{4(p^+)^2}{N_c}\int\rmd^2x_{\perp}\rmd^2y_{\perp}\left\langle\tr\left[\Omega(\mathbf{x}_{\perp})\Omega^{\dagger}(\mathbf{y}_{\perp})\right]\right\rangle_Ye^{-i\mathbf{q}_{\perp}\cdot(\mathbf{x}_{\perp}-\mathbf{y}_{\perp})}.
\end{split}
\end{equation}
In this last expression one sees the emergence of a quantity of special importance known as a \emph{color dipole}, denoted $S$, defined as :
\begin{equation}
\label{dipoledef}
S(\mathbf{x}_{\perp},\mathbf{y}_{\perp})=\frac{1}{N_c}\tr\left[\Omega(\mathbf{x}_{\perp})\Omega^{\dagger}(\mathbf{y}_{\perp})\right].
\end{equation}
In this very simple example we have seen how to perform the calculations and the emergence of trace color operators of the form \eqref{Coloroperator}, the dipole.

\subsection{Dilute regime limit\label{diluteqq}}
\indent

Performing the dilute limit in \eqref{qtoqcross} will enable us to generalize the unintegrated gluon distribution to multiple scatterings. For this purpose, one has to match together two ways for getting the dilute limit. The dilute limit assumes that there is only a single gluon exchanged between the incoming quark and the target.\\

The first way to compute the dilute limit follows from the general collinear factorization, already encountered in \ref{DIS} and justified for high energy. The cross-section for the $qA\rightarrow qX$ process reads in terms of the cross-section corresponding to the $qg\rightarrow q$ process with a gluon carrying a fraction $x_2$ of the target's longitudinal momentum $Q^-$ :
\begin{equation}
\label{dilutesinglequarkcross}
\rmd\sigma(q(p)\rightarrow q(q))=\int_0^1\rmd x_2 G(x_2;Q^2)\rmd\sigma(q(p)g(k=x_2Q)\rightarrow q(q)).
\end{equation}
The frame is chosen so that the incoming quark has a zero transverse momentum and the momentum $k$ transferred by the exchanged gluon is $\mathbf{k}_{\perp}=\mathbf{q}_{\perp}$. In the collinear factorization approximation, the distribution of the gluon is sharply peaked around $\mathbf{k}_{\perp}=\mathbf{0}$ in the transverse plane. Thus equation \eqref{UGPDF} merely reads\footnote{This relation between the integrated and unintegrated gluon distribution is a bit trivial. Actually we shall recover the so-called $k_{\perp}$-factorization result using the less general collinear factorization in which the gluon coming from the target is on-shell. } :
\begin{equation}
x_2 G(x_2;Q^2)\delta^{(2)}(\mathbf{k}_{\perp})=\frac{1}{\pi\mathbf{k}_{\perp}^2}f_Y(\mathbf{k}_{\perp}).
\end{equation}
Furthermore, in the eikonal approximation, the $qg\rightarrow q$ cross-section, averaged over initial spins and colors, reads :
\begin{equation}
\frac{\rmd\sigma(q(p)g(k)\rightarrow q(q))}{\rmd y\rmd^2q_{\perp}}=\frac{g^2p^+k^-}{4N_c\mathbf{q}_{\perp}^2}2\pi\delta^{(4)}(p+k-q).
\end{equation}
$y$ being the rapidity of the final quark, $y=\frac{1}{2}\ln\frac{q^+}{q^-}$. Plugging this last expression into \eqref{dilutesinglequarkcross} and performing the integration over $x_2$ thanks to the delta function which fixes $x_2=q^-/Q^-$ gives :
\begin{equation}
\label{dilutesinglequarkcross1}
\frac{\rmd\sigma(q(p)\rightarrow q(q))}{\rmd y\rmd^2q_{\perp}}= x_2 G(x_2;Q^2)\frac{g^2p^+}{4N_c\mathbf{q}_{\perp}^2}2\pi\delta(p^+-q^+)\delta^{(2)}(\mathbf{q}_{\perp})=\frac{g^2p^+}{2N_c\mathbf{q}_{\perp}^4}f_Y(\mathbf{q}_{\perp})\delta(p^+-q^+).
\end{equation}
Note that the last equality is a device since the unintegrated gluon distribution is a mere delta function in the collinear factorization approximation. Closing our eyes on that point, the last form of \eqref{dilutesinglequarkcross1} is known as $k_{\perp}$-factorization. It is more general than collinear factorization since the gluon exchanged with the target is off-shell in $k_{\perp}$-factorization. This special limit will enable us to write down a generalized unintegrated gluon distribution in presence of non-linear effects read from the structure of the cross-section.\\

On the other hand, one can perform the brute force computation of the $qA\rightarrow qX$ cross-section thanks to \eqref{qtoqcross} together with \eqref{qqCS}. It leads to :
\begin{equation}
\label{dilutesinglequarkcross2}
\frac{\rmd\sigma(q(p)\rightarrow q(q))}{\rmd y\rmd^2q_{\perp}}= \frac{p^+}{(2\pi)^2}\delta(p^+-q^+)\int\rmd^2x_{\perp}\rmd^2y_{\perp}\left\langle S(\mathbf{x}_{\perp},\mathbf{y}_{\perp})\right\rangle_Ye^{-i\mathbf{q}_{\perp}\cdot(\mathbf{x}_{\perp}-\mathbf{y}_{\perp})}.
\end{equation}
The dilute limit \eqref{dilutesinglequarkcross1} allows a natural \emph{definition} of a generalized gluon distribution in presence of non-linear effects. Indeed, we expect that the total cross-section takes the form \eqref{dilutesinglequarkcross1} even in presence of non-linear effects by replacing the unintegrated gluon distribution $f_Y$ by the generalized object under consideration. An obvious identification with \eqref{dilutesinglequarkcross2} leads to the gluon distribution associated to the fundamental dipole $f^{\text{dip}, F}_Y$, defined as :
\begin{equation}
\label{UGDdip}
\frac{f^{\text{dip}, F}_Y(\mathbf{q}_{\perp})}{\mathbf{q}_{\perp}^2}\equiv \frac{2N_c}{g^2(2\pi)^2}\mathbf{q}_{\perp}^2\int\rmd^2x_{\perp}\rmd^2y_{\perp}\left\langle S(\mathbf{x}_{\perp},\mathbf{y}_{\perp})\right\rangle_Ye^{-i\mathbf{q}_{\perp}\cdot(\mathbf{x}_{\perp}-\mathbf{y}_{\perp})},
\end{equation}
where $\mathbf{q}_{\perp}$ is now the total transverse momentum transferred by the target. It is not possible in general to associate generalized gluon distributions to any trace operator, this makes sense only in particular cases. The single quark scattering is one of them, we shall see another example for the $g\rightarrow gg$ process in section \ref{backtoback}.\\

Furthermore, the dilute limit can be also performed directly in \eqref{dilutesinglequarkcross2}. This will lead to a definition of the unintegrated gluon distribution as a background field correlator. By definition, the l.h.s of \eqref{UGDdip} reduces to $f_Y(\mathbf{q}_{\perp})$ in the dilute limit. Concerning the r.h.s the single exchange assumption is an expansion of trace operators up to second order in the background field since the cross-section receives a contribution from both the amplitude and the complex conjugate amplitude. Therefore we will expand the Wilson lines \eqref{WilsonLine} as :
\begin{equation}
\Omega(\mathbf{x}_{\perp})=1+ig\int\rmd x^+\mathcal{A}^-(x^+,\mathbf{x}_{\perp})-\frac{g^2}{2}\int\rmd x^+\rmd y^+\mathcal{P}\left\{\mathcal{A}^-(x^+,\mathbf{x}_{\perp})\mathcal{A}^-(y^+,\mathbf{x}_{\perp})\right\}+\mathcal{O}(\mathcal{A}^3).
\end{equation}
Plugging this expansion into the dipole definition \eqref{dipoledef}, keeping only terms up to second order and then performing the average \eqref{AAtwopfunction}\footnote{Even though we do not make the gaussian approximation, the two-point function \eqref{AAtwopfunction} is the essential ingredient in the dilute limit. Differences between gaussian and non-gaussian source distributions arise only from higher-point correlation functions. Moreover $\langle\mathcal{A}^-\rangle_Y$ is always zero by charge conjugation symmetry.} gives :
\begin{equation}
\left\langle S(\mathbf{x}_{\perp},\mathbf{y}_{\perp})\right\rangle_Y=1+g^2C_F\int\rmd x^+\left(\gamma_Y(x^+,\mathbf{x}_{\perp}-\mathbf{y}_{\perp})-\gamma_Y(x^+,\mathbf{0})\right).
\end{equation}
where $C_F=(N_c^2-1)/2N_c$ is the fundamental representation Casimir. For brevity, let us define
\begin{equation}
\Gamma_Y(\mathbf{x}_{\perp}-\mathbf{y}_{\perp})\equiv\int\rmd x^+\left(\gamma_Y(x^+,\mathbf{0})-\gamma_Y(x^+,\mathbf{x}_{\perp}-\mathbf{y}_{\perp})\right).
\end{equation}
By inserting the dipole expansion into the relation \eqref{UGDdip}, one relate the unintegrated gluon distribution to the Fourier transform of the background field two-point function integrated over $x^+$. Actually, the integrand in \eqref{UGDdip} only depends on the coordinate difference $\mathbf{x}_{\perp}-\mathbf{y}_{\perp}$ leaving the integral of $1$ over the whole transverse plane. Of course the transverse plane has to be cut at the hadron's size. As long as the transverse integrals contain oscillating exponentials, the integration range can be extended to infinity but in this last case, the remaining integral just brings a factor $S_{\perp}$. Hence, the unintegrated gluon distribution reads :
\begin{equation}
\label{UGPDF2p}
\frac{f_Y(\mathbf{q}_{\perp})}{\mathbf{q}_{\perp}^2}= -\frac{(N_c^2-1)S_{\perp}}{(2\pi)^2}\mathbf{q}_{\perp}^2\int\rmd^2x_{\perp}\Gamma_Y(\mathbf{x}_{\perp})e^{-i\mathbf{q}_{\perp}\cdot\mathbf{x}_{\perp}}.s
\end{equation}
This result proves the, so far stated, formula \eqref{MVUGD}. Recalling that $f_Y(\mathbf{q}_{\perp})/\mathbf{q}_{\perp}^2$ is the number of gluons per phase space $\rmd Y\rmd^2p_{\perp}$, it has an alternate definition as a number operator acting on the proton wave function. The classical correspondence is a background field two-point function averaged over the sources in the CGC framework. This justifies the existence of such a relation between the unintegrated gluon distribution and the background field two-point function.

\section{Di-gluon decorrelation at the LHC}
\indent

On the projectile side the relevant part of the proton wave function, which is purely gluonic and dilute, is treated within the collinear factorization. According to appendix \ref{pAcollinear}, this yields the following expression for the p-A cross-section :
\begin{equation}
\label{PgDEF}
\begin{split}
\frac{\rmd\sigma (pA\rightarrow ggX)}{\rmd y_1 \rmd y_2 \rmd^2k_{1,\perp} \rmd^2k_{2,\perp}}&=\frac{1}{256\pi^5(p^+)^2}x_1 G(x_1,Q^2)\left<|\mathcal{M}(g(p)A \rightarrow g(k_1)g(k_2))|^2\right>_Y.
\end{split}
\end{equation}
It is understood that $x_1=p^+/P^+$ with $p^+=k_1^++k_2^+$ (in agreement with plus momentum component conservation). This is a special case of \eqref{pAtogAcrosssection}. The phase space volume element is written, for convenience, in the rapidity representation $y_i=\frac{1}{2}\ln(k^+_i/k^-_i)$ and then one has $\rmd y_i=\rmd k_i^+/k^+_i$.. To the accuracy of interest, the factorization scale $Q$ should be chosen of the order of a typical value of the final transverse momenta, say, of the order of the saturation scale $Q_s(A,Y)$ in the nuclear target. Thus the proton enters trivially into the cross section through prefactors and one can focus only on the sub-process $gA\rightarrow ggX$.

\subsection{The amplitude}
\indent

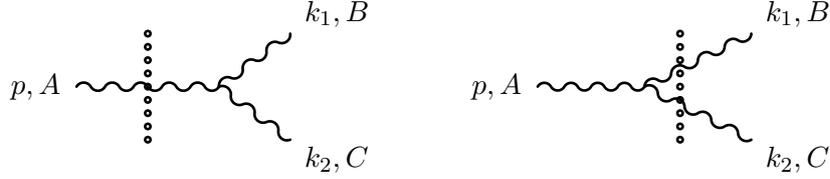
\begin{figure}[h]
\centering
\vspace{1cm}
%(along, up)
\begin{fmfgraph*}(80,40)
 \fmfstraight
    \fmftop{i1,v1,v2,o1}
    \fmfleft{i2}
    \fmfbottom{i3,v3,v4,o2}
      \fmf{dbl_dots}{v1,v3}
      \fmffreeze
      \fmf{photon}{i2,v}
      \fmf{photon}{v,o1}
      \fmf{photon}{v,o2}
      \fmflabel{$p,A$}{i2}
      \fmflabel{$k_1,B$}{o1}
      \fmflabel{$k_2,C$}{o2}
\end{fmfgraph*}
\hspace{3cm}
\begin{fmfgraph*}(80,40)
 \fmfstraight
    \fmftop{i1,v1,v2,o1}
    \fmfleft{i2}
    \fmfbottom{i3,v3,v4,o2}
      \fmf{dbl_dots}{v2,v4}
      \fmffreeze
      \fmf{photon,tension=2}{i2,v}
      \fmf{photon}{v,o1}
      \fmf{photon}{v,o2}
      \fmflabel{$p,A$}{i2}
      \fmflabel{$k_1,B$}{o1}
      \fmflabel{$k_2,C$}{o2}
\end{fmfgraph*}
\vspace{.5cm}
\caption{The two contributions to the gluon's splitting. The dotted line represents the shockwave. One could wonder about a possible contribution arising from the splitting \emph{in} the medium. It turns out that such contributions always involve integrals of regular functions over the $x^+$ source support, which cancel as the source becomes infinitely narrow in this direction.\label{process}}
\end{figure}

The partonic process $gA\to ggX$ involves the two diagrams illustrated in figure \ref{process}.
Using the Feynman rules detailed in appendix \ref{shockwaveFeynrules}, it is straightforward to write down the corresponding contributions to the scattering amplitude :
\begin{equation}
\label{ggsplitsmatrix}
\begin{split}
&i\mathcal{M}\left(g(p,A)A \rightarrow g(k_1,B)g(k_2,C)\right)=-gf^{DBC}\epsilon_{\mu}(p)\epsilon^{\nu*}(k_1)
\epsilon^{\rho*}(k_2)\Gamma_{\sigma\nu\rho}(k_1+k_2,k_1,k_2)\\ 
& ~~~~~~~~~~~~~~~~~~~~~~~~~~~~~~~~~\times
2p^+\beta^{\mu i}(\mathbf{p}_{\perp},p^+)\, \frac{i\beta^{\sigma i}(\mathbf{k}_{1,\perp}+\mathbf{k}_{2,\perp},p^+)}
{(k_1+k_2)^2+i\epsilon}
\int \rmd^2x_{\perp}\tilde{\Omega}_{DA}(\mathbf{x}_{\perp})e^{-i\mathbf{x}_{\perp}\cdot (\mathbf{k}_{1,\perp}+\mathbf{k}_{2,\perp}-\mathbf{p}_{\perp})} \\
&  ~~~~~~~~~~~~~~~~~~~~~~~~~~~~~~~~~
+gf^{AEF}\epsilon^{\mu}(p)\epsilon^*_{\nu}(k_1)\epsilon^*_{\rho}(k_2)\int_{l^+=k^+}\frac{\rmd l^- \rmd^2l_{\perp}}{(2\pi)^3}\Gamma_{\mu\sigma\lambda}(p,l,p-l)\\
&~~ ~~~~~~~~~~~~~~~~~~~~~~~~~~~~~~~\times 2k_1^+\beta^{\nu i}(\mathbf{k}_{1,\perp},k_1^+)\frac{i\beta^{\sigma i}(\mathbf{l}_{\perp},k_1^+)}{l^2+i\epsilon}\int \rmd^2x_{\perp} \tilde{\Omega}_{BE}(\mathbf{x}_{\perp})e^{-i\mathbf{x}_{\perp}\cdot(\mathbf{k}_{1,\perp}-\mathbf{l}_{\perp})} \\
& ~~~~ ~~~~~~~~~~~~~~~~~~~~~~~~~~~~~\times 2k_2^+\beta^{\rho j}(\mathbf{k}_{2,\perp},k_2^+)\frac{i\beta^{\lambda j}(\mathbf{p}_{\perp}-\mathbf{l}_{\perp},k_2^+)}{(p-l)^2+i\epsilon}\int \rmd^2y_{\perp}\tilde{\Omega}_{CF}(\mathbf{y}_{\perp})e^{-i\mathbf{y}_{\perp}\cdot (\mathbf{k}_{2,\perp}-\mathbf{p}_{\perp}+\mathbf{l}_{\perp})}.
\end{split}
\end{equation}
The transverse momentum $\mathbf{p}_{\perp}$of the initial gluon is momentarily kept generic, but it will be eventually set to zero. The polarization indices have not been explicitly written in \eqref{ggsplitsmatrix} to alleviate notations.  The gauge condition $A^+=0$ together with the Ward identity $k\cdot\epsilon(k)=0$ imply the constraints
$\epsilon^+=0$ and $\epsilon^-(k)={k^i\epsilon^i(k)}/{k^+}$. $\Gamma_{\mu\nu\rho}(k,p,q)$ denotes the Lorentz piece of the three-gluon vertex - with the color and $g$ factor omitted: the momentum $k$ is incoming, while $p$ and $q$ are outgoing (see \ref{shockwaveFeynrules} for the explicit expression). The symbol $\beta^{\mu i}(\mathbf{p}_{\perp},k^+)$ 
may be viewed as the `square-root' of the tensorial structure of the gluon propagator in the background field. As  discussed in detail in \ref{shockwaveFeynrules}, this propagator is conveniently written as (in momentum space)\footnote{Since in this chapter we shall only deal with tree-level Green functions, we shall not introduce notations for distinguishing between the exact propagator and the propagator dressed by the background field at tree level. In the next chapter we will need to distinguish them but here we have preferred to alleviate the notations by avoiding proliferation of indices.} :
\begin{equation}
\Delta^{\mu\nu}_{AB}(k^-,\mathbf{k}_{\perp}; q^-,\mathbf{q}_{\perp}; k^+)=\beta^{\mu i}(\mathbf{k}_{\perp},k^+)\beta^{\nu i}(\mathbf{q}_{\perp},k^+)G_{AB}(k^-,\mathbf{k}_{\perp}; q^-,\mathbf{q}_{\perp}; k^+)
\end{equation}
where $G_{AB}(k^-,\mathbf{k}_{\perp}; q^-,\mathbf{k}_{\perp}; k^+)$ is the respective scalar propagator and
\begin{equation}
\label{beta0}
\beta^{\mu i}(\mathbf{q}_{\perp},k^+)=\delta^{\mu-}\frac{q^i}{k^+}+\delta^{\mu i}.
\end{equation}

As a guide to see how \eqref{ggsplitsmatrix} has been derived, let us consider the second diagram shown on figure \ref{process}. Using \eqref{extleg}, the upper final leg attached to the shock wave combined with the propagator running from the branching vertex to the shock wave contributes as 
\begin{equation}
-2k_1^+\epsilon^*_{\nu}(k_1)\beta^{\nu i}(\mathbf{k}_{1,\perp},k_1^+)\frac{i\beta^{\sigma i}(\mathbf{l}_{\perp},k_1^+)}{l^2+i\epsilon}\int \rmd^2x_{\perp}\tilde{\Omega}_{BE}(\mathbf{x}_{\perp})e^{-i\mathbf{x}_{\perp}\cdot(\mathbf{k}_{1,\perp}-\mathbf{l}_{\perp})}.
\end{equation}
$l$ is the momentum running between the vertex and the shock wave through the upper branch. Since the plus component of the momentum remains unaffected by the shock wave, this fixes $l^+=k_1^+$. However, the other components are not fixed and one has to integrate the whole diagram over $l^-$ and $\mathbf{l}_{\perp}$. There is a similar expression for the lower final leg but the final momentum is $k_2$ and the momentum between the vertex and the shock wave is $p-l$ by momentum conservation at the vertex. We have $(p-l)^+=k_2^+$. Finally, the vertex brings a factor $gf^{AEF}\Gamma_{\mu\sigma\lambda}(p,l,p-l)$ and the initial gluon introduces the polarization vector $\epsilon^{\mu}(p)$. The first diagram of figure \ref{process} is obtained in a similar way.\\

In equation \eqref{ggsplitsmatrix}, the integral over $l^-$ is performed using the residue theorem. The result is to set $l$ on-shell (i.e. $l^-=\mathbf{l}_{\perp}^2/2k_1^+$) and to replace  $i/l^2\rightarrow 2\pi/2k_1^+$. (The $i\epsilon$ prescriptions play no role since none of the denominators is vanishing.) As already mentioned, we chose the frame so that $p=(p^+, 0, \mathbf{0})$ and we introduce the $z$ parameter so that $k_1^\mu=(zp^+,k_1^-,\mathbf{k}_{1,\perp})$ and $k_2^\mu=((1-z)p^+,k_2^-,\mathbf{k}_{2,\perp})$, with $k_1$ and $k_2$ on-shell. The value of $z$ is related to the kinematic variables of the produced gluons via
\begin{equation}
\label{zkinem}
z= \frac{k_{1,\perp}e^{y_1}}{k_{1,\perp}e^{y_1}+k_{2,\perp}e^{y_2}}.
\end{equation}
Then the denominators in \eqref{ggsplitsmatrix} can be rewritten as :
\begin{equation}
\begin{split}
&(k_1+k_2)^2=\frac{1}{z(1-z)}((1-z)\mathbf{k}_{1,\perp}-z\mathbf{k}_{2,\perp})^2\\
&(p-l)^2=-\frac{1}{z}\mathbf{l}_{\perp}^2.
\end{split}
\end{equation}
Moreover, we use equations \eqref{propepsilon} and \eqref{propepsilon2} in order to replace the 
polarization 4-vectors by their transverse components alone. (This is possible since, as alluded
to above, the transverse components are the only independent ones in the present set-up.) 
After performing these various manipulations, the amplitude \eqref{ggsplitsmatrix} becomes
 \begin{equation}
\label{gtoggamplitude}
\begin{split}
&i\mathcal{M}\left(g(p,a)A \rightarrow g(k_1,b)g(k_2,c)\right)=g\epsilon^i(p)\epsilon^{j*}(k_1)\epsilon^{k*}(k_2)\times \\
 &~~~~~~~~\times\left[f^{DBC}\frac{2ip^+z(1-z)}{((1-z)\mathbf{k}_{1,\perp}-z\mathbf{k}_{2,\perp})^2}\beta^{\mu i}(\mathbf{k}_{1,\perp}+\mathbf{k}_{2,\perp},p^+)\beta^{\nu j}(\mathbf{k}_{1,\perp},k_1^+)\beta^{\rho k}(\mathbf{k}_{2,\perp},k_2^+)\Gamma_{\mu\nu\rho}(k_1+k_2,k_1,k_2)\right.\\
&~~~~~~~~~~~~~~~~~~~~~~~~\times\int \rmd^2x_{\perp}\tilde{\Omega}_{DA}(\mathbf{x}_{\perp})e^{-i\mathbf{x}_{\perp}\cdot(\mathbf{k}_{1,\perp}+\mathbf{k}_{2,\perp}-\mathbf{p}_{\perp})} \\
 &~~~~~~~~~~~ -f^{AEF}\int\frac{\rmd^2l_{\perp}}{(2\pi)^2}\frac{2ik_2^+z}{\mathbf{l}_{\perp}^2}\beta^{\mu i}(\mathbf{p}_{\perp},p^+)\beta^{\nu j}(\mathbf{l}_{\perp},k_1^+)\beta^{\rho k}(\mathbf{p}_{\perp}-\mathbf{l}_{\perp},k_2^+)\Gamma_{\mu\nu\rho}(p,l,p-l)\\
&~~~~~~~~~~~~~~~~~~~~~~~~ \left.\times\int \rmd^2x_{\perp}\rmd^2y_{\perp}\tilde{\Omega}_{BE}(\mathbf{x}_{\perp})\tilde{\Omega}_{CF}(\mathbf{y}_{\perp})e^{-i\mathbf{x}_{\perp}\cdot(\mathbf{k}_{1,\perp}-\mathbf{l}_{\perp})-i\mathbf{y}_{\perp}\cdot(\mathbf{k}_{2,\perp}-\mathbf{p}_{\perp}+\mathbf{l}_{\perp})}\right].
\end{split}
\end{equation}
Now that we have a rather compact expression for the amplitude, the next goal is to compute the probability.

\subsection{The splitting cross-section}
\indent

Here we come to the main topic, namely, the computation of the cross-section for the partonic process $gA \rightarrow ggX$. This is obtained according to \eqref{PgDEF} where
\begin{equation}
\label{splitProbab}
\left\langle\overline{|\mathcal{M}(g(p)A \rightarrow g(k_1)g(k_2))|^2}\right\rangle_Y\equiv\frac{1}{2(N_c^2-1)}
\displaystyle{\sum_{pol.}}\displaystyle{\sum_{ABC}}
\left\langle \left|\mathcal{M}\big(g(p,A)A \rightarrow g(k_1,B)g(k_2,C)\big)\right|^2\right\rangle_Y.
\end{equation}
For more clarity, the calculation of the r.h.s. of \eqref{splitProbab}
will be split into two stages : first, the sum over polarizations and
next the sum and average over colors (including the CGC average over the target background field).

\subsubsection{Sum over polarizations : Lorentz structure \label{Lorentzpiece}}
\indent

After taking the modulus squared of the amplitude in \eqref{gtoggamplitude}, the sum over polarizations
is readily performed by using
\begin{equation}
\displaystyle{\sum_{pol.}}\epsilon^i(k)\epsilon^{j*}(k)\,=\,\delta^{ij}
\end{equation}
Notice that the r.h.s. of the above equation is independent of the momentum $k$ carried by the
polarization vector. Hence, an expression like $\epsilon^i(p)\beta^{\mu i}(\mathbf{k}_{\perp},k^+)$, after being squared and summed over polarizations, will give a result which depends only upon $k$, and not upon $p$. The computation of the modulus squared of the vertex functions which appear in \eqref{gtoggamplitude}   
- this leads to terms of the form $\sum_{ijk}|\beta^{i\mu}\beta^{j\nu}\beta^{k\rho}\Gamma_{\mu\nu\rho}|^2$- is quite lengthy but straightforward\footnote{
For the reader interested in fully following the calculation, we give the intermediate results. One has to compute essentially three kinds of squares of vertices which correspond to the sum over the polarizations of the square of \eqref{gtoggamplitude}. From the explicit form \eqref{beta0} of the $\beta^{\mu i}$'s one gets :
\begin{equation*}
\begin{split}
&\beta^{\mu i}(\mathbf{k}_{1,\perp}+\mathbf{k}_{2,\perp},p^+)\beta^{\mu' i}(\mathbf{k}_{1,\perp}+\mathbf{k}_{2,\perp},p^+)\beta^{\nu j}(\mathbf{k}_{1,\perp},k_1^+)\beta^{\nu' j}(\mathbf{k}_{1,\perp},k_1^+)\beta^{\rho k}(\mathbf{k}_{2,\perp},k_2^+)\beta^{\rho' k}(\mathbf{k}_{2,\perp},k_2^+)\times\\
&~~~~~~~~~~~~~~~~~~~~~~~~~~~~~~\times\Gamma_{\mu\nu\rho}(k_1+k_2,k_1,k_2)\Gamma_{\mu'\nu'\rho'}(k_1+k_2,k_1,k_2)=\frac{8((1-z)\mathbf{k}_{1,\perp}-z\mathbf{k}_{2,\perp})^2}{z(1-z)}P_{g\leftarrow g}(z).\\
&\beta^{\mu i}(\mathbf{k}_{1,\perp}+\mathbf{k}_{2,\perp},p^+)\beta^{\mu' i}(\mathbf{p}_{\perp},p^+)\beta^{\nu j}(\mathbf{k}_{1,\perp},k_1^+)\beta^{\nu' j}(\mathbf{l}_{\perp},k_1^+)\beta^{\rho k}(\mathbf{k}_{2,\perp},k_2^+)\beta^{\rho' k}(\mathbf{p}_{\perp}-\mathbf{l}_{\perp},k_2^+)\times\\
&~~~~~~~~~~~~~~~~~~~~~~~~~~~~~~\times\Gamma_{\mu\nu\rho}(k_1+k_2,k_1,k_2)\Gamma_{\mu'\nu'\rho'}(p,l,p-l)=\frac{8\mathbf{l}_{\perp}\cdot((1-z)\mathbf{k}_{1,\perp}-z\mathbf{k}_{2,\perp})}{z(1-z)}P_{g\leftarrow g}(z)\\
&\beta^{\mu i}(\mathbf{p}_{\perp},p^+)\beta^{\mu' i}(\mathbf{p}_{\perp},p^+)\beta^{\nu j}(\mathbf{l}_{\perp},k_1^+)\beta^{\nu' j}(\mathbf{l}_{\perp}',k_1^+)\beta^{\rho k}(\mathbf{p}_{\perp}-\mathbf{l}_{\perp},k_2^+)\beta^{\rho' k}(\mathbf{p}_{\perp}-\mathbf{l}_{\perp}',k_2^+)\times\\
&~~~~~~~~~~~~~~~~~~~~~~~~~~~~~~\times\Gamma_{\mu\nu\rho}(p,l,p-l)\Gamma_{\mu'\nu'\rho'}(p,l',p-l')=\frac{8\mathbf{l}_{\perp}\cdot\mathbf{l}_{\perp}'}{z(1-z)}P_{g\leftarrow g}(z).
\end{split}
\end{equation*}}. One eventually obtains (as compared to \eqref{gtoggamplitude}, we shall from now on set $\mathbf{p}_{\perp}=0$) :
\begin{equation}
\label{gtoggeffamplitude0}
\begin{split}
&\overline{|\mathcal{M}(g(p)A \rightarrow g(k_1)g(k_2))|^2}=\frac{16g^2 (p^+)^2z(1-z)}{N_c^2-1}P_{g\leftarrow g}(z)\times \\
&~~~~ \times\displaystyle{\sum_{ABC}}\left|f^{DBC}\frac{(1-z)k_1^i-zk_2^i}{((1-z)\mathbf{k}_{1,\perp}-z\mathbf{k}_{2,\perp})^2}
\int \rmd^2x_{\perp}\tilde{\Omega}_{DA}(\mathbf{x}_{\perp})e^{-i\mathbf{x}_{\perp}\cdot(\mathbf{k}_{1,\perp}+\mathbf{k}_{2,\perp})}\right.- \\
&~~~~~~~~~~ -f^{AEF}\int\frac{\rmd^2l_{\perp}}{(2\pi)^2}\frac{l^i}{\mathbf{l}_{\perp}^2}\left.\int \rmd^2x_{\perp}\rmd^2y_{\perp}\tilde{\Omega}_{BE}(\mathbf{x}_{\perp})\tilde{\Omega}_{CF}(\mathbf{y}_{\perp})e^{-i\mathbf{x}_{\perp}\cdot(\mathbf{k}_{1,\perp}-\mathbf{l}_{\perp})-i\mathbf{y}_{\perp}\cdot(\mathbf{k}_{2,\perp}+\mathbf{l}_{\perp})}\right|^2,
\end{split}
\end{equation}
where  $P_{g\leftarrow g}(z)$ is the DGLAP gluon-to-gluon splitting function :
\begin{equation}
\label{DGLAPgtog}
P_{g\leftarrow g}(z)\equiv \frac{z}{1-z}+\frac{1-z}{z}+z(1-z).
\end{equation}
This result can be rewritten in a more suggestive form by using the following identities,
\begin{equation}
\label{D2Fourier}
\begin{split}
&\int\frac{\rmd^2l_{\perp}}{(2\pi)^2}\frac{l^i}{\mathbf{l}_{\perp}^2}e^{i\mathbf{l}_{\perp}\cdot(\mathbf{x}_{\perp}-\mathbf{y}_{\perp})}=\frac{i}{2\pi}\frac{x^i-y^i}{(\mathbf{x}_{\perp}-\mathbf{y}_{\perp})^2}\\
&\frac{(1-z)k_1^i-zk_2^i}{((1-z)\mathbf{k}_{1,\perp}-z\mathbf{k}_{2,\perp})^2}=\frac{i}{2\pi}\int \rmd^2y_{\perp}\frac{x^i-y^i}{(\mathbf{x}_{\perp}-\mathbf{y}_{\perp})^2}
e^{-i((1-z)\mathbf{k}_{1,\perp}-z\mathbf{k}_{2,\perp})\cdot(\mathbf{x}_{\perp}-\mathbf{y}_{\perp})},
\end{split}
\end{equation}
in which one recognizes the derivative $\partial_x^iG(\mathbf{x}_{\perp}-\mathbf{y}_{\perp})$ of the two-dimensional
Laplace propagator, $G(\mathbf{x}_{\perp})=(1/4\pi)\ln(\mathbf{x}_{\perp}^2)$. In the
present context, this plays the role of the transverse splitting function, as we shall shortly discuss.
Namely, after using \eqref{D2Fourier} and performing some changes in the integration variables,
one can recast \eqref{gtoggeffamplitude0} into the form
\begin{equation}
\label{gtoggeffamplitude}
\begin{split}
&\overline{|\mathcal{M}(g(p)A \rightarrow g(k_1)g(k_2))|^2}=\frac{4g^2(p^+)^2z(1-z)}{\pi^2(N_c^2-1)}P_{g\leftarrow g}(z)\\
&~~~\times \displaystyle{\sum_{ABC}} \left|\int \rmd^2x_{\perp}\rmd^2y_{\perp}\frac{x^i-y^i}{(\mathbf{x}_{\perp}-\mathbf{y}_{\perp})^2} 
e^{-i\mathbf{k}_{1,\perp}\cdot\mathbf{x}_{\perp}-i\mathbf{k}_{2,\perp}\cdot\mathbf{y}_{\perp}}\left[f^{DBC}\tilde{\Omega}_{DA}(\mathbf{b}_{\perp})-f^{AEF}\tilde{\Omega}_{BE}(\mathbf{x}_{\perp})\tilde{\Omega}_{CF}(\mathbf{y}_{\perp})\right]\right|^2
\end{split}
\end{equation}
which admits a transparent physical interpretation: $\mathbf{x}_{\perp}$ and $\mathbf{y}_{\perp}$ are the transverse coordinates
of the two final gluons, whereas $\mathbf{b}_{\perp}\equiv z\mathbf{x}_{\perp} + (1-z)\mathbf{y}_{\perp}$, which is recognized as their
barycenter in the transverse plane, is the respective coordinate of the original gluon. The function
\begin{equation}
 \frac{x^i-y^i}{(\mathbf{x}_{\perp}-\mathbf{y}_{\perp})^2}=(1-z)\frac{x^i-b^i}{(\mathbf{x}_{\perp}-\mathbf{b}_{\perp})^2}=-z\frac{y^i-b^i}{(\mathbf{y}_{\perp}-\mathbf{b}_{\perp})^2} 
\end{equation}
is proportional to the amplitude for splitting a gluon at $\mathbf{x}_{\perp}$ (or at $\mathbf{y}_{\perp}$) from an original gluon
at $\mathbf{b}_{\perp}$. The first terms within the square brackets in \eqref{gtoggeffamplitude} corresponds
to the process where the original gluon interacts with the shockwave prior to splitting. The second
terms describes the other situation, where the splitting occurs before the interaction, so the final
gluons scatter off the shockwave.

\subsubsection{Sum over colors and average over the background field}
\indent

It is now straightforward to explicitly perform the square in \eqref{gtoggeffamplitude} and then
formally average over the background field. This yields
\begin{equation}
\label{prob0}
\begin{split}
&\left\langle\overline{|\mathcal{M}(g(p)A \rightarrow g(k_1)g(k_2))|^2}\right\rangle_Y
=\frac{4g^2 N_c}{\pi^2}(p^+)^2z(1-z)P_{g\leftarrow g}(z) \\
&~~~~~\times \int \rmd^2x_{\perp}\rmd^2y_{\perp}\rmd^2\bar{x}_{\perp} \rmd^2\bar{y}_{\perp}
\frac{(\mathbf{x}_{\perp}-\mathbf{y}_{\perp})\cdot(\bar{\mathbf{x}}_{\perp}-\bar{\mathbf{y}}_{\perp})}{(\mathbf{x}_{\perp}-\mathbf{y}_{\perp})^2(\bar{\mathbf{x}}_{\perp}-\bar{\mathbf{y}}_{\perp})^2}e^{-i\mathbf{k}_{1,\perp}\cdot(\mathbf{x}_{\perp}-\bar{\mathbf{x}}_{\perp})
-i\mathbf{k}_{2,\perp}\cdot(\mathbf{y}_{\perp}-\bar{\mathbf{y}}_{\perp})} \\
& ~~~~~ \times \left\langle\tilde S^{(2)}(\mathbf{b}_{\perp},\bar{\mathbf{b}}_{\perp})-\tilde S^{(3)}(\mathbf{b}_{\perp},\bar{\mathbf{x}}_{\perp},\bar{\mathbf{y}}_{\perp})-\tilde S^{(3)}(\bar{\mathbf{b}}_{\perp},\mathbf{x}_{\perp},\mathbf{y}_{\perp})+\tilde S^{(4)}(\mathbf{x}_{\perp},\mathbf{y}_{\perp},\bar{\mathbf{x}}_{\perp},\bar{\mathbf{y}}_{\perp})\right\rangle_Y,
\end{split}
\end{equation}
which is our main new result in this chapter : the probability for gluon splitting induced by
the interaction with the nucleus (the corresponding cross-section is then easily obtained
according to \eqref{PgDEF}).\\

In \eqref{prob0}, $\mathbf{b}_{\perp}\equiv z\mathbf{x}_{\perp} + (1-z)\mathbf{y}_{\perp}$ and $\bar{\mathbf{b}}_{\perp}\equiv z\bar{\mathbf{x}}_{\perp} + (1-z)\bar{\mathbf{y}}_{\perp}$, where
$\mathbf{x}_{\perp}$ and $\bar{\mathbf{x}}_{\perp}$ denote the transverse coordinates of the first produced gluon (the one with
momentum $\mathbf{k}_{1,\perp}$) in the direct and respectively complex conjugate amplitude, whereas
$\mathbf{y}_{\perp}$ and $\bar{\mathbf{y}}_{\perp}$ similarly refer to the second produced gluon.
The other new notations appearing in \eqref{prob0} are defined as follows:
\begin{equation}
\label{ABCdef}
\begin{split}
&\tilde S^{(2)}(\mathbf{b}_{\perp},\bar{\mathbf{b}}_{\perp})=\frac{1}{N_c(N_c^2-1)}f^{DBC}f^{D'BC}\tilde{\Omega}_{DA}(\mathbf{b}_{\perp})\tilde{\Omega}_{D'A}(\bar{\mathbf{b}}_{\perp})=\frac{1}{N_c^2-1} {\rm Tr}\big[\tilde{\Omega}(\mathbf{b}_{\perp})\tilde{\Omega}^\dagger(\bar{\mathbf{b}}_{\perp})\big]
\\
&\tilde S^{(3)}(\mathbf{b}_{\perp},\bar{\mathbf{x}}_{\perp},\bar{\mathbf{y}}_{\perp})=\frac{1}{N_c(N_c^2-1)}f^{DBC}f^{AEF}\tilde{\Omega}_{DA}(\mathbf{b}_{\perp})\tilde{\Omega}_{BE}(\bar{\mathbf{x}}_{\perp})\tilde{\Omega}_{CF}(\bar{\mathbf{y}}_{\perp})\\
&\tilde S^{(4)}(\mathbf{x}_{\perp},\mathbf{y}_{\perp},\bar{\mathbf{x}}_{\perp},\bar{\mathbf{y}}_{\perp})=\frac{1}{N_c(N_c^2-1)}f^{AEF}f^{AE'F'}\tilde{\Omega}_{BE}(\mathbf{x}_{\perp})\tilde{\Omega}_{CF}(\mathbf{y}_{\perp})\tilde{\Omega}_{BE'}(\bar{\mathbf{x}}_{\perp})\tilde{\Omega}_{CF'}(\bar{\mathbf{y}}_{\perp}).
\end{split}
\end{equation}
The normalization factors in \eqref{ABCdef} are chosen in such a way that the various functions
$\tilde S^{(k)}$, with $k=2,3,4$, approach unity in limit of a vanishing background field. 
Using the identity $f^{AEF}\tilde{\Omega}_{BE}\tilde{\Omega}_{CF}=\tilde{\Omega}_{DA}f^{DBC}$, it is easy to check that
they all can be obtained from  $\tilde S^{(4)}$ :
\begin{equation}
\label{ABCprop}
\begin{split}
\tilde S^{(2)}(\mathbf{b}_{\perp},\bar{\mathbf{b}}_{\perp})&=\tilde S^{(4)}(\mathbf{b}_{\perp},\mathbf{b}_{\perp},\bar{\mathbf{b}}_{\perp},\bar{\mathbf{b}}_{\perp})\\
\tilde S^{(3)}(\mathbf{b}_{\perp},\bar{\mathbf{x}}_{\perp},\bar{\mathbf{y}}_{\perp})&=\tilde S^{(4)}(\mathbf{b}_{\perp},\mathbf{b}_{\perp},\bar{\mathbf{x}}_{\perp},\bar{\mathbf{y}}_{\perp}).
\end{split}
\end{equation}
Physically, the functions $\tilde S^{(k)}$ represent $S$-matrices for the eikonal scattering 
between a system of $k$ gluons in an overall color singlet state and the background field.
For instance, $\tilde S^{(2)}$ corresponds to a gluonic dipole made with the original
gluon in the amplitude times its hermitian conjugate in the complex conjugate amplitude.
Its contribution to \eqref{prob0} represents the probability for the process in which the splitting occurs
after the scattering. Similarly,  $\tilde S^{(4)}$ (a gluonic quadrupole) describes the process
where the splitting occurs prior to the scattering, and the two pieces involving 
$\tilde S^{(3)}$ describe the interference between the two possible time orderings.
The identities \eqref{ABCprop} have a simple physical interpretation: if the two gluons produced
by the splitting are very close to each other, such that one can approximate $\mathbf{x}_{\perp}\simeq\mathbf{y}_{\perp}\simeq\mathbf{b}_{\perp}$,
then there is no difference (in so far as the scattering off the shockwave is concerned) between
this system of two overlapping gluons and their parent gluon prior to its splitting. Not surprisingly, the general structure of the probability for gluon splitting in \eqref{prob0} is very similar to that for the corresponding quark splitting ($qA\to qgX$), as computed in
\cite{Marquet:2007vb}. The main difference refers, as expected, to the replacement of the quark
Wilson lines (in the fundamental representation of the color group) 
by adjoint Wilson lines for gluons. Moreover, \eqref{prob0} generalizes previous results for
gluon splitting \cite{JalilianMarian:2004da, Baier:2005dv,Dominguez:2011wm} obtained in various limits and that
we shall later recover by taking the appropriate limits of  \eqref{prob0}.

\subsubsection{Relation to multipoles in the fundamental representation\label{adjtofund}}
\indent

In order to compute the CGC expectation values of the gluonic multipole operators in  \eqref{ABCdef},
it is convenient to first re-express them in terms of Wilson lines in the fundamental representation (this is particularly useful in view of the large $N_c$ limit, to be discussed next).
This can be done by using several group identities. The first one relates the structure constants - or equivalently, up to a phase, the adjoint representation generators - to a trace of generators in an arbitrary representation $R$ :
\begin{equation}
f^{ABC}=-2i\tr\left([T_R^A;T_R^B]T_R^C\right).
\end{equation}
The $\tr$ symbol with small letters denotes the trace in an arbitrary representation. The second one relates an arbitrary group matrix in some representation $R$ to the adjoint representation matrix $\tilde{U}$ \footnote{Previously, formula \eqref{ABCprop} follows from this identity with $R$ taken to be the adjoint representation as well.} :
\begin{equation}
\label{adjtoR}
UT_R^AU^{\dagger}=T^B_R\tilde{U}_BA=(\tilde{U}^{\dagger})_{AB}T^B_R.
\end{equation}
The last required identity holds only for the fundamental representation $F$ of $SU(N_c)$ (the two previous ones where valid for any semi-simple Lie groups) :
\begin{equation}
(T^A_F)_{ab}(T^A_F)_{cd}=\frac{1}{2}\left(\delta_{ad}\delta_{bc}-\frac{1}{N_c}\delta_{ab}\delta_{cd}\right).
\end{equation}
The net result is that any adjoint Wilson line gets replaced by a pair of Wilson lines in the fundamental 
representation. After straightforward manipulations, the function $\tilde S^{(4)}$ is eventually rewritten as follows
\begin{equation}
\label{Cfunction}
\begin{split}
\tilde S^{(4)}(\mathbf{x}_{\perp},\mathbf{y}_{\perp},\mathbf{u}_{\perp},\mathbf{v}_{\perp})&=\frac{N_c^2}{2(N_c^2-1)}\left[Q(\mathbf{x}_{\perp},\mathbf{y}_{\perp},\mathbf{v}_{\perp},\mathbf{u}_{\perp})S(\mathbf{u}_{\perp},\mathbf{x}_{\perp})S(\mathbf{y}_{\perp},\mathbf{v}_{\perp}) \right.\\
&~~~~~~~ +Q(\mathbf{y}_{\perp},\mathbf{x}_{\perp},\mathbf{u}_{\perp},\mathbf{v}_{\perp})S(\mathbf{x}_{\perp},\mathbf{u}_{\perp})S(\mathbf{v}_{\perp},\mathbf{y}_{\perp}) \\
&-\frac{1}{N_c^2}\left. O(\mathbf{v}_{\perp},\mathbf{x}_{\perp},\mathbf{u}_{\perp},\mathbf{v}_{\perp},\mathbf{y}_{\perp},\mathbf{u}_{\perp},\mathbf{x}_{\perp},\mathbf{y}_{\perp})-\frac{1}{N_c^2}O(\mathbf{v}_{\perp},\mathbf{u}_{\perp},\mathbf{x}_{\perp},\mathbf{v}_{\perp},\mathbf{y}_{\perp},\mathbf{x}_{\perp},\mathbf{u}_{\perp},\mathbf{y}_{\perp})\right], 
\end{split}
\end{equation}
where the various terms in the r.h.s. are multipoles (i.e. single-trace operators) built with Wilson 
lines in the fundamental representation (to be denoted by $\Omega$ and $\Omega^\dagger$). Namely,
we shall need the respective dipole, quadrupole, hexapole, and octupole, defined as
 \begin{equation}
\label{multipoledef}
\begin{split}
S&(\mathbf{x}_{\perp},\mathbf{y}_{\perp})=\frac{1}{N_c}\tr\left[\Omega(\mathbf{x}_{\perp})\Omega^{\dagger}(\mathbf{y}_{\perp})\right]\\
Q&(\mathbf{x}_{\perp},\mathbf{y}_{\perp},\mathbf{u}_{\perp},\mathbf{v}_{\perp})=\frac{1}{N_c}\tr\left[\Omega(\mathbf{x}_{\perp})\Omega^{\dagger}(\mathbf{y}_{\perp})\Omega(\mathbf{u}_{\perp})\Omega^{\dagger}(\mathbf{v}_{\perp})\right]\\
H&(\mathbf{x}_{\perp},\mathbf{y}_{\perp},\mathbf{u}_{\perp},\mathbf{v}_{\perp},\mathbf{w}_{\perp},\mathbf{z}_{\perp})=\frac{1}{N_c}\tr\left[\Omega(\mathbf{x}_{\perp})\Omega^{\dagger}(\mathbf{y}_{\perp})\Omega(\mathbf{u}_{\perp})\Omega^{\dagger}(\mathbf{v}_{\perp})\Omega(\mathbf{w}_{\perp})\Omega^{\dagger}(\mathbf{z}_{\perp})\right]\\
O&(\mathbf{x}_{\perp},\mathbf{y}_{\perp},\mathbf{u}_{\perp},\mathbf{v}_{\perp},\mathbf{w}_{\perp},\mathbf{z}_{\perp},\mathbf{t}_{\perp},\mathbf{s}_{\perp})=\frac{1}{N_c}\tr\left[\Omega(\mathbf{x}_{\perp})\Omega^{\dagger}(\mathbf{y}_{\perp})\Omega(\mathbf{u}_{\perp})\Omega^{\dagger}(\mathbf{v}_{\perp})\Omega(\mathbf{w}_{\perp})\Omega^{\dagger}(\mathbf{z}_{\perp})\Omega(\mathbf{t}_{\perp})\Omega^{\dagger}(\mathbf{s}_{\perp})\right].
\end{split}
\end{equation}
$\tr$ denotes here the trace in the fundamental representation. These are formally the operators which describe the scattering between a quark-antiquark ($q\bar{q}$) color
dipole, a $q\bar{q}q\bar{q}$ color quadrupole, etc, off the background field. We shall generically refer to
such single-trace operators as \emph{multipoles}.
The corresponding expressions for $\tilde S^{(2)}$ and $\tilde S^{(3)}$ follow from \eqref{ABCprop} :
\begin{equation}
\label{ABCasD}
\begin{split}
\tilde S^{(2)}(\mathbf{x}_{\perp},\mathbf{u}_{\perp})&=\frac{N_c^2}{N_c^2-1}\,
\left[S(\mathbf{x}_{\perp},\mathbf{u}_{\perp})S(\mathbf{u}_{\perp},\mathbf{x}_{\perp})-\frac{1}{N_c^2}\right]\\
\tilde S^{(3)}(\mathbf{x}_{\perp},\mathbf{u}_{\perp},\mathbf{v}_{\perp})&=\frac{N_c^2}{2(N_c^2-1)}\,
\left[S(\mathbf{v}_{\perp},\mathbf{u}_{\perp})S(\mathbf{u}_{\perp},\mathbf{x}_{\perp})S(\mathbf{x}_{\perp},\mathbf{v}_{\perp})+S(\mathbf{u}_{\perp},\mathbf{v}_{\perp})S(\mathbf{x}_{\perp},\mathbf{u}_{\perp})S(\mathbf{v}_{\perp},\mathbf{x}_{\perp})-\right.\\
&~~~~~~
\left.-\frac{1}{N_c^2}H(\mathbf{v}_{\perp},\mathbf{x}_{\perp},\mathbf{u}_{\perp},\mathbf{v}_{\perp},\mathbf{x}_{\perp},\mathbf{u}_{\perp})-\frac{1}{N_c^2}H(\mathbf{v}_{\perp},\mathbf{u}_{\perp},\mathbf{x}_{\perp},\mathbf{v}_{\perp},\mathbf{u}_{\perp},\mathbf{x}_{\perp})\right].
\end{split}
\end{equation}
In principle, all such expectation values can be computed by numerically solving the JIMWLK equation
\cite{Rummukainen:2003ns,Lappi:2011ju,Dumitru:2011vk}, with appropriate
initial conditions (say, as provided by the McLerran-Venugopalan model detailed in section \ref{MVmodel}) 
at low energies. Moreover, explicit analytic expressions can be obtained in the Gaussian approximation to the JIMWLK evolution discussed in \ref{gaussianJIMWLK} (the ensuing expressions may be viewed as extrapolations to high-energy of the 
respective formulas in the MV model \cite{Dominguez:2011wm}).
In practice though all these calculations become prohibitively cumbersome with increasing number of Wilson
lines. Important simplifications occurs in the large $N_c$ limit to be discussed next.

\subsection{The large $N_c$ limit \label{largeNc}}
\indent

The limit of a large number of colors ($N_c\gg 1$) is interesting since it preserves the essential physical effects, 
while allowing for important technical simplifications. Indeed, within equations \eqref{Cfunction} and \eqref{ABCasD}, all 
the multipole operators higher than the quadrupole are accompanied by an explicit factor
of $1/N_c^2$ and hence they are suppressed\footnote{Notice that, according to \eqref{multipoledef},
the multipoles are normalized such that they remain of $\mathcal{O}(1)$ as $N_c\rightarrow \infty$.}
as $N_c\to \infty$.  This reduction of the multipole functional space to dipoles and quadrupoles, 
occurring at large $N_c$,  has been recently argued \cite{Dominguez:2012ad} to be a general property, 
which holds for any production process of the dilute-dense type (within the limits of the present, 
CGC-like, factorization).\\

Independently, we shall assume that the source is large in the sense discussed in section \ref{B-JIMWK}, that is the interaction with the target is dominated by independent scatterings. In this case, averages of products of multipoles factorize into products of averages
of individual multipoles. (This is a generic property of multi-trace expectation values.) For instance, the large $N_c$ limit version of the gluonic dipole $S$-matrix, as shown in the first line of  \eqref{ABCasD}, reads
\begin{equation}
 \big\langle\tilde S^{(2)}(\mathbf{x}_{\perp},\mathbf{u}_{\perp})\big\rangle_Y=\left <S(\mathbf{u}_{\perp},\mathbf{x}_{\perp})\right>_Y\left <S(\mathbf{x}_{\perp},\mathbf{u}_{\perp})\right>_Y ~~~~ (N_c\rightarrow\infty).
\end{equation}
In general, the dipole expectation value is \emph{not} symmetric: whenever non-vanishing, the difference\\
$\left <S(\mathbf{u}_{\perp},\mathbf{x}_{\perp})-S(\mathbf{x}_{\perp},\mathbf{u}_{\perp})\right>_Y$ is purely imaginary and $C$-odd and describes the 
amplitude for odderon exchanges in the dipole-target scattering \cite{Kovchegov:2003dm,Hatta:2005as}.
However, if the initial condition for the dipole amplitude at low energy is real,
as is e.g. the case within the context of the MV model,
then this property will be preserved by the JIMWLK evolution up to arbitrarily high energy.
A similar property holds for the quadrupole $S$-matrix: if this is real at $Y=Y_0$ (as is
indeed the case within the MV model), then it remains real for any $Y>Y_0$ ; then
the expectation values of the two quadrupoles which enter \eqref{Cfunction} are
equal to each other : $\left<Q(\mathbf{x}_{\perp},\mathbf{y}_{\perp},\mathbf{v}_{\perp},\mathbf{u}_{\perp})\right>_Y=\left<Q(\mathbf{y}_{\perp},\mathbf{x}_{\perp},\mathbf{u}_{\perp},\mathbf{v}_{\perp})\right>_Y$.\\

To summarize, at large $N_c$ and for initial conditions provided by the MV model,
the target expectation values relevant for the 2-gluon production simplify to
\begin{equation}
\label{largeNcmultipoles}
\begin{split}
&\big<\tilde S^{(2)}(\mathbf{x}_{\perp},\mathbf{u}_{\perp})\big>_Y\simeq\left<S(\mathbf{u}_{\perp},\mathbf{x}_{\perp})\right>_Y^2\\
&\big<\tilde S^{(3)}(\mathbf{x}_{\perp},\mathbf{u}_{\perp},\mathbf{v}_{\perp})\big>_Y\simeq
\left<S(\mathbf{x}_{\perp},\mathbf{u}_{\perp})\right>_Y\left<S(\mathbf{u}_{\perp},\mathbf{v}_{\perp})\right>_Y\left<S(\mathbf{v}_{\perp},\mathbf{x}_{\perp})\right>_Y\\
&\big<\tilde S^{(4)}(\mathbf{x}_{\perp},\mathbf{y}_{\perp},\mathbf{u}_{\perp},\mathbf{v}_{\perp})\big>_Y\simeq
\left<Q(\mathbf{x}_{\perp},\mathbf{y}_{\perp},\mathbf{v}_{\perp},\mathbf{u}_{\perp})\right>_Y\left<S(\mathbf{x}_{\perp},\mathbf{u}_{\perp})\right>_Y\left<S(\mathbf{y}_{\perp},\mathbf{v}_{\perp})\right>_Y.
\end{split}
\end{equation}
This immediately yields the large-$N_c$ version of the squared amplitude in
\eqref{prob0}  :
\begin{equation}
\label{prob1}
\begin{split}
&\left\langle\overline{|\mathcal{M}(g(p)A \rightarrow g(k_1)g(k_2))|^2}\right\rangle_Y=
\frac{4g^2N_c}{\pi^2}(p^+)^2z(1-z)P_{g\leftarrow g}(z) \\
&~~~~~\times \int \rmd^2x_{\perp}\rmd^2y_{\perp}\rmd^2\bar{x}_{\perp}\rmd^2\bar{y}_{\perp} \frac{(\mathbf{x}_{\perp}-\mathbf{y}_{\perp})\cdot(\bar{\mathbf{x}}_{\perp}-\bar{\mathbf{y}}_{\perp})}{(\mathbf{x}_{\perp}-\mathbf{y}_{\perp})^2(\bar{\mathbf{x}}_{\perp}-\bar{\mathbf{y}}_{\perp})^2}e^{-i\mathbf{k}_{1,\perp}\cdot(\mathbf{x}_{\perp}-\bar{\mathbf{x}}_{\perp})-i\mathbf{k}_{2,\perp}\cdot(\mathbf{y}_{\perp}-\bar{\mathbf{y}}_{\perp})} \\
& ~~~~~ \times \left[\left<S(\mathbf{b}_{\perp},\bar{\mathbf{b}}_{\perp})\right>_Y^2\left<S(\mathbf{b}_{\perp},\bar{\mathbf{x}}_{\perp})\right>_Y\left<S(\bar{\mathbf{x}}_{\perp},\bar{\mathbf{y}}_{\perp})\right>_Y\left<S(\bar{\mathbf{y}}_{\perp},\mathbf{b}_{\perp})\right>_Y\right.\\
\\
& ~~~~~\left<S(\bar{\mathbf{b}}_{\perp},\mathbf{x}_{\perp})\right>_Y\left<S(\mathbf{x}_{\perp},\mathbf{y}_{\perp})\right>_Y\left<S(\mathbf{y}_{\perp},\bar{\mathbf{b}}_{\perp})\right>_Y\\
& ~~~~~\left.+\left<Q(\mathbf{x}_{\perp},\mathbf{y}_{\perp},\bar{\mathbf{y}}_{\perp},\bar{\mathbf{x}}_{\perp})\right>_Y\left<S(\mathbf{x}_{\perp},\bar{\mathbf{x}}_{\perp})\right>_Y\left<S(\mathbf{y}_{\perp},\bar{\mathbf{y}}_{\perp})\right>_Y\right],
\end{split}
\end{equation}
where the variables $\mathbf{b}_{\perp}$ and $\bar{\mathbf{b}}_{\perp}$ have been defined after \eqref{prob0}. As a check,
one can easily verify that for a very asymmetric splitting ($z\ll 1$ or $1-z\ll 1$), our \eqref{prob1} reduces, 
as it should, to the respective result in \cite{JalilianMarian:2004da, Baier:2005dv}.\\

Still at large $N_c$, the general Balitsky-JIMWLK hierarchy of coupled evolution equations for the
multipole expectation values boils down to a triangular hierarchy of equations, which can be solved 
one after the other: the dipole $S$-matrix $\left <S\right>_Y$ obeys the closed, 
non-linear, Balitsky-Kovchegov (BK) equation \cite{Balitsky:1995ub,Kovchegov:1998bi}, 
while the quadrupole $S$-matrix $\left <Q\right>_Y$  obeys an inhomogeneous equation in
which the source term and the coefficients of the homogeneous terms depend upon $\left <S\right>_Y$. 
This last equation is still quite complicated, but a good approximation to it - in the form of
an analytic expression relating $\left <Q\right>_Y$ to $\left <S\right>_Y$ - 
can be obtained within the Gaussian approximation to the JIMWLK evolution.
In view of this, \eqref{prob1} is quite explicit  (at least, conceptually) and can be used 
as such for applications to phenomenology. To that aim, one should combine a reasonable approximation
to the dipole $S$-matrix (say, as given by the solution to the BK equation with a running coupling
\cite{Balitsky:2006wa,Kovchegov:2006vj,Balitsky:2008zza}) with the expression for $\left <Q\right>_Y$
valid in the Gaussian approximation and at large $N_c$ 
(as given e.g. in equation(4.26) of reference \cite{Iancu:2011nj}). In practice, the main technical complication 
that we foresee is the calculation of the Fourier transforms in \eqref{prob1}, which may require
numerical techniques (see e.g. \cite{Albacete:2010pg,Albacete:2007yr,Albacete:2010bs,Kuokkanen:2011je} for some
similar calculations).

\subsection{The back-to-back correlation limit\label{backtoback}}
\indent

Let us focus on the hard final gluons phase space region with $k_{1,\perp}, k_{2,\perp} \gg Q_s$. From the phenomenological considerations of section \ref{CGCdihadronpheno}, the effects of multiple scattering may remain important if one is interested in the details of the azimuthal distribution around its peak at $\Delta\Phi=\pi$. Since any gluon exchanged with the target have a typical transverse momentum of order $Q_s$ (see section \ref{QCDPS}), it almost does not affect $k_{1,\perp}$ and $k_{2,\perp}$ separately. However $|\mathbf{k}_{1,\perp}+\mathbf{k}_{2,\perp}|$ is the total transverse momentum transferred by the target and is of order $Q_s$. Although the two final gluons are hard enough to be almost undeviated by multiple scatterings with the target, there are deviations to the exact back-to-back distribution due to these multiple scatterings.
The proper strategy in that sense, as originally proposed in reference \cite{Dominguez:2011wm},
relies on the observation that the relative momentum $\mathbf{P}_{\perp}=(1-z)\mathbf{k}_{1,\perp}-z\mathbf{k}_{2,\perp}$ refers to
the hard splitting which creates the gluon pair, while the total momentum $\mathbf{K}_{\perp}$ 
accounts for the transverse momentum broadening of the two gluons via their 
(comparatively soft) interactions with the target. $\mathbf{P}_{\perp}$ controls the transverse separation $\mathbf{r}_{\perp}=\mathbf{x}_{\perp}-\mathbf{y}_{\perp}$ between
the offspring gluons in the direct amplitude
(and similarly $\bar{\mathbf{r}}_{\perp}=\bar{\mathbf{x}}_{\perp}-\bar{\mathbf{y}}_{\perp}$ in the complex conjugate amplitude), 
whereas $\mathbf{K}_{\perp}$ controls the difference $\mathbf{b}_{\perp}-\bar{\mathbf{b}}_{\perp}$ between the average positions
of the gluons in the direct and the c.c. amplitude, that is, their
transverse fluctuations $\mathbf{x}_{\perp}-\bar{\mathbf{x}}_{\perp}$  
and $\mathbf{y}_{\perp}-\bar{\mathbf{y}}_{\perp}$, which in turn encode the effects of the multiple scattering with the target. 
Accordingly, in this  `back-to-back correlation limit' where $K_\perp\sim Q_s \ll P_\perp$,  the integral in 
\eqref{prob0} is controlled by configurations where the transverse-size variables $\mathbf{r}_{\perp}$ and $\bar{\mathbf{r}}_{\perp}$ 
are small as compared to  the difference $\mathbf{b}_{\perp}-\bar{\mathbf{b}}_{\perp}$ between the center-of-mass variables
(and of course also small as compared to $\mathbf{b}_{\perp}$ and $\bar{\mathbf{b}}_{\perp}$ themselves).  This allows for 
appropriate Taylor expansions of the various multipoles in \eqref{prob0}. Specifically, using the new variables $\mathbf{b}_{\perp}, \mathbf{r}_{\perp}, \bar{\mathbf{b}}_{\perp}, \bar{\mathbf{r}}_{\perp}$ and the
respective conjugate momenta, the r.h.s. of \eqref{prob0} becomes :
\begin{equation}
\label{probBtoB0} 
\begin{split}
&\left\langle\overline{|\mathcal{M}(g(p)A \rightarrow g(k_1)g(k_2))|^2}\right\rangle_Y=\frac{4g^2 N_c}{\pi^2}(p^+)^2z(1-z)P_{g\leftarrow g}(z) \\
&~~~~~~\times \int \rmd^2b_{\perp}\rmd^2r_{\perp}\rmd^2\bar{b}_{\perp}\rmd^2\bar{r}_{\perp}\frac{\mathbf{r}_{\perp}\cdot\bar{\mathbf{r}}_{\perp}}{\mathbf{r}_{\perp}^2\bar{\mathbf{r}}_{\perp}^2}e^{-i\mathbf{K}_{\perp}\cdot(\mathbf{b}_{\perp}-\bar{\mathbf{b}}_{\perp})-i\mathbf{P}_{\perp}\cdot(\mathbf{r}_{\perp}-\bar{\mathbf{r}}_{\perp})} \\
&~~~~~~ \times \left\langle\tilde S^{(2)}(\mathbf{b}_{\perp},\bar{\mathbf{b}}_{\perp})-\tilde S^{(3)}(\mathbf{b}_{\perp},\bar{\mathbf{b}}_{\perp}+(1-z)\bar{\mathbf{r}}_{\perp},\bar{\mathbf{b}}_{\perp}-z\bar{\mathbf{r}}_{\perp}) - \tilde S^{(3)}(\bar{\mathbf{b}}_{\perp},\mathbf{b}_{\perp}+(1-z)\mathbf{r}_{\perp},\mathbf{b}_{\perp}-z\mathbf{r}_{\perp})\right.\\
&~~~~~~ \left.+\tilde S^{(4)}(\mathbf{b}_{\perp}+(1-z)\mathbf{r}_{\perp},\mathbf{b}_{\perp}-z\mathbf{r}_{\perp},\bar{\mathbf{b}}_{\perp}+(1-z)\bar{\mathbf{r}}_{\perp},\bar{\mathbf{b}}_{\perp}-z\bar{\mathbf{r}}_{\perp})\right\rangle_Y.
\end{split}
\end{equation}
We now expand the multipoles inside the integrand around $\mathbf{b}_{\perp}$ and $\bar{\mathbf{b}}_{\perp}$. 
In view of the identities \eqref{ABCprop}, it should be quite clear that the leading non trivial result arises
from expanding $\tilde S^{(4)}$ up to second order in $r^i$ and $\bar{r}^i$ and keeping only the
`off-diagonal' terms which are bilinear in $r^i\bar r^j$. (The `diagonal' terms proportional to either
$r^i r^j$ or $\bar{r}^i\bar{r}^j$ cancel against similar terms arising from the expansion of the
two pieces involving $\tilde S^{(3)}$, and the same happens for the terms which are
linear in $r^i$ or $\bar{r}^i$.) A straightforward calculation gives
\begin{equation}
\label{btobcoloroperator}
\begin{split}
 r^i\bar{r}^j&\left[(1-z)\partial^i_x-z\partial^i_y\right]\left[(1-z)\partial^j_{u}-z\partial^j_{v}
\right] \left\langle\tilde S^{(4)}(\mathbf{x}_{\perp},\mathbf{y}_{\perp},\mathbf{u}_{\perp},\mathbf{v}_{\perp})
\right\rangle_Y\Big |_{\mathbf{b}_{\perp}\mathbf{b}_{\perp}\bar{\mathbf{b}}_{\perp}\bar{\mathbf{b}}_{\perp}}\\
&=\frac{r^i\bar{r}^j}{N_c(N_c^2-1)}\ {\rm Tr}
\Big\langle\left[(1-z)\partial^i\tilde{U}(\mathbf{b}_{\perp})T^A\tilde{U}^{\dagger}(\mathbf{b}_{\perp})-z\tilde{U}(\mathbf{b}_{\perp})T^A\partial^i\tilde{U}^{\dagger}(\mathbf{b}_{\perp})\right]\\
&~~~~~~~~~~~~~~\times
\left[(1-z)\tilde{U}(\bar{\mathbf{b}}_{\perp})T^A\partial^j\tilde{U}^{\dagger}(\bar{\mathbf{b}}_{\perp})-z\partial^j\tilde{U}(\bar{\mathbf{b}}_{\perp})T^A\tilde{U}^{\dagger}(\bar{\mathbf{b}}_{\perp})\right]\Big\rangle_Y \\
&=\frac{r^i\bar{r}^j}{N_c(N_c^2-1)}\ \left[-2z(1-z){\rm Tr}
\Big\langle\partial^i\tilde{U}(\mathbf{b}_{\perp})T^A\tilde{U}^{\dagger}(\mathbf{b}_{\perp})\partial^j\tilde{U}(\bar{\mathbf{b}}_{\perp})T^A\tilde{U}^{\dagger}(\bar{\mathbf{b}}_{\perp})\right.\Big\rangle_Y \\
&~~~~~~~~~~~~~~ +((1-z)^2+z^2){\rm Tr}\left.
\Big\langle\partial^i\tilde{U}(\mathbf{b}_{\perp})T^A\tilde{U}^{\dagger}(\mathbf{b}_{\perp})\tilde{U}(\bar{\mathbf{b}}_{\perp})T^A\partial^j\tilde{U}^{\dagger}(\bar{\mathbf{b}}_{\perp})\Big\rangle_Y \right],
\end{split}
\end{equation}
where the second equality is obtained after using the identity $\tilde{U}T^A\tilde{V}^{\dagger}
= - \big(\tilde{V}T^A\tilde{U}^{\dagger}\big)^{\tau}$, valid for generic color matrices $\tilde{U}$
and $\tilde{V}$ in the adjoint representation. \\

It is convenient to split the final expression in
\eqref{btobcoloroperator} into two pieces, one proportional to $z(1-z)$ 
and another one that is independent of $z$. The $z$-independent 
piece cannot be further simplified (it is proportional to a second derivative of $\tilde S^{(4)}$,
as visible on the first line of \eqref{btobcoloroperator}).
The piece proportional to $z(1-z)$, on the other hand, can be written in a simpler form, 
namely as  a second derivative of $\tilde S^{(2)}$, by
using the same trick as the one used to get the last equality in \eqref{btobcoloroperator}.
After also performing the integrals over $\mathbf{r}_{\perp}$ and $\bar{\mathbf{r}}_{\perp}$ in \eqref{probBtoB0}, according to
\begin{equation}
\label{intrr}
\int \rmd^2r_{\perp}\rmd^2\bar{r}_{\perp}\frac{r^ir^k\bar{r}^k\bar{r}^j}{\mathbf{r}_{\perp}^2\bar{\mathbf{r}}_{\perp}^2}e^{-i\mathbf{P}_{\perp}\cdot(\mathbf{r}_{\perp}-\bar{\mathbf{r}}_{\perp})}
=\pi^2\frac{\partial^2\ln \mathbf{P}_{\perp}^2}{\partial P^i\partial P^k}\frac{\partial^2 \ln \mathbf{P}_{\perp}^2}{\partial P^j\partial P^k}=\frac{4\pi^2}{\mathbf{P}_{\perp}^4}\delta^{ij},
\end{equation}
one finally obtains
\begin{equation}
\label{probBtoB1}
\begin{split} 
&\left\langle\overline{|\mathcal{M}(g(p)A \rightarrow g(k_1)g(k_2))|^2}\right\rangle_Y
=16g^2 N_c\frac{(p^+)^2z(1-z)}{\mathbf{P}_{\perp}^4}P_{g\leftarrow g}(z) \\
&~~~~~~\times \int \rmd^2b_{\perp}\rmd^2\bar{b}_{\perp} e^{-i\mathbf{K}_{\perp}\cdot(\mathbf{b}_{\perp}-\bar{\mathbf{b}}_{\perp})} \Big\langle \partial^i_x\partial^i_u\tilde S^{(4)}(\mathbf{x}_{\perp},\mathbf{b}_{\perp},\mathbf{u}_{\perp},\bar{\mathbf{b}}_{\perp})\Big |_{\mathbf{b}_{\perp}\mathbf{b}_{\perp}\bar{\mathbf{b}}_{\perp}\bar{\mathbf{b}}_{\perp}}-z(1-z)\partial^i_b\partial^i_{\bar{b}}\tilde S^{(2)}(\mathbf{b}_{\perp},\bar{\mathbf{b}}_{\perp})\Big\rangle_Y. 
\end{split}
\end{equation}
Incidentally, \eqref{intrr} confirms that the transverse separations $|\mathbf{r}_{\perp}|$ and $|\bar{\mathbf{r}}_{\perp}|$ 
in both the direct and the complex conjugate amplitude are separately of order $1/P_{\perp}$,
as anticipated.\\

\eqref{probBtoB1} represents the complete result (under the present assumptions) for the production
of a pair of relatively hard gluons, with transverse momenta $k_{1,\perp}, k_{2,\perp} \gg Q_s(A,Y)$.
This generalizes the collinear factorization by including the non-linear
effects accompanying the hard branching process, which describe the multiple scattering
between the gluons involved in the branching and the nuclear target. As manifest on
\eqref{probBtoB1}, these non-linear effects control the magnitude $K_{\perp} \equiv |\mathbf{k}_{1,\perp}+\mathbf{k}_{2,\perp}|$
of the total transverse momentum of the pair: the target expectation values appearing
in the integrand of \eqref{probBtoB1} rapidly decay for transverse separations $|\mathbf{b}_{\perp}-\bar{\mathbf{b}}_{\perp}|
\gg 1/Q_s$, which in turn implies that, typically, $K_\perp \lesssim Q_s$. The bi-local color operators built with the
second derivatives of $\tilde S^{(4)}$ and $\tilde S^{(2)}$ which enter \eqref{probBtoB1} can be
viewed as generalizations of the unintegrated gluon distribution  $f_Y(\mathbf{K}_{\perp})$ in
\eqref{UGPDF} to the non-linear regime. Their definitions are unambiguous since one has to recover the dilute limit \eqref{UGPDF2p}. The first one is associated with $\tilde S^{(2)}$ (the `adjoint dipole gluon distribution'), namely
\begin{equation}
\label{dipUGD}
\begin{split}
\frac{f_Y^{\text{dip}, A}(\mathbf{K}_{\perp})}{\mathbf{K}_{\perp}^2}\equiv\frac{N_c^2-1}{g^2(2\pi)^2N_c}
\int \rmd^2b_{\perp}\rmd^2\bar{b}_{\perp} e^{-i\mathbf{K}_{\perp}\cdot(\mathbf{b}_{\perp}-\bar{\mathbf{b}}_{\perp})}
\Big\langle\partial^i_b\partial^i_{\bar{b}}\tilde S^{(2)}(\mathbf{b}_{\perp},\bar{\mathbf{b}}_{\perp})\Big\rangle_Y,
\end{split}
\end{equation}
is well-known known in the literature, as it enters various inclusive and semi-inclusive
processes involving a dense target, like the total cross-section for deep inelastic scattering
(DIS) and the single-inclusive parton production in DIS and p-A collisions (see 
\cite{Dominguez:2011wm} for a recent overview). The `adjoint quadrupole gluon
distribution' associated with $\tilde S^{(4)}$, that is,
\begin{equation}
\label{quadUGD}
\begin{split}
\frac{f_Y^{\text{quad}, A}(\mathbf{K}_{\perp})}{\mathbf{K}_{\perp}^2}\equiv\frac{N_c^2-1}{g^2(2\pi)^2N_c}
\int \rmd^2b_{\perp}\rmd^2\bar{b}_{\perp}e^{-i\mathbf{K}_{\perp}\cdot(\mathbf{b}_{\perp}-\bar{\mathbf{b}}_{\perp})}
\Big\langle\partial^i_x\partial^i_u\tilde S^{(4)}(\mathbf{x}_{\perp},\mathbf{b}_{\perp},\mathbf{u}_{\perp},\bar{\mathbf{b}}_{\perp})
\Big |_{\mathbf{b}_{\perp}\mathbf{b}_{\perp}\bar{\mathbf{b}}_{\perp}\bar{\mathbf{b}}_{\perp}}\Big\rangle_Y,
\end{split}
\end{equation}
has not been introduced before to our knowledge, 
but its limit at large $N_c$ has been studied
in  \cite{Dominguez:2011wm}. For the physical interpretation
of these objects, $f_Y^{\rm dip}(\mathbf{K}_{\perp})$ and $f_Y^{\rm quad}(\mathbf{K}_{\perp})$, one should
however keep in mind that they involve both `final-state' and `initial-state'
interactions (that is, gluon-target interactions occurring
both before and after the branching process), which cannot be simultaneously 
gauged away by a proper choice of the light-cone gauge for the target ($A^-=0$). Hence,
these quantities do not really measure the gluon occupation number\footnote{Interestingly
though, as pointed out in \cite{Dominguez:2011wm}, the large-$N_c$ decomposition
of $f_Y^{\rm quad}(\mathbf{K}_{\perp})$, cf. \eqref{largeNcmultipoles}, involves a piece (the
last piece in  \eqref{backtobackNc} below) which is
proportional to the Weizs\"acker-Williams gluon distribution and hence represents
the gluon occupation number for a proper choice of the light-cone gauge.}.\\

The large-$N_c$ limit of \eqref{probBtoB1} is also interesting, in particular, 
because it allows us to make contact with the corresponding result in
\cite{Dominguez:2011wm}. Namely, using the approximations \eqref{largeNcmultipoles} for the
color multipoles which appear in \eqref{probBtoB1} one
finds after some algebra 
\begin{equation}
\label{backtobackNc}
\begin{split}
&\left\langle\overline{|\mathcal{M}(g(p)A \rightarrow g(k_1)g(k_2))|^2}\right\rangle_Y=16g^2N_c\frac{(p^+)^2z(1-z)}{\mathbf{P}_{\perp}^4}P_{g\leftarrow g}(z)\int \rmd^2b_{\perp}\rmd^2\bar{b}_{\perp} e^{-i\mathbf{K}_{\perp}\cdot(\mathbf{b}_{\perp}-\bar{\mathbf{b}}_{\perp})} \\
& ~~~~\times\bigg\{[(1-z)^2+z^2]\left\langle S(\mathbf{b}_{\perp},\bar{\mathbf{b}}_{\perp})\right\rangle_Y\partial^i_b\partial^i_{\bar{b}}\left\langle S(\mathbf{b}_{\perp},\bar{\mathbf{b}}_{\perp})\right\rangle_Y-
2z(1-z)\partial^i_b\left\langle S(\mathbf{b}_{\perp},\bar{\mathbf{b}}_{\perp})\right\rangle_Y\partial^i_{b}\left\langle S(\mathbf{b}_{\perp},\bar{\mathbf{b}}_{\perp})
\right\rangle_Y+ \\
& ~~~~+\left\langle S(\mathbf{b}_{\perp},\bar{\mathbf{b}}_{\perp})\right\rangle_Y^2\partial^i_x\partial^i_u\left\langle Q(\mathbf{x}_{\perp},\mathbf{b}_{\perp},\bar{\mathbf{b}}_{\perp},\mathbf{u}_{\perp})\right\rangle_Y\Big |_{\mathbf{b}_{\perp},\mathbf{b}_{\perp},\bar{\mathbf{b}}_{\perp},\bar{\mathbf{b}}_{\perp}}\bigg\},
\end{split}
\end{equation}
which is indeed equivalent to equation (105) in reference \cite{Dominguez:2011wm}, as one can easily check.

\chapter{Initial state factorization in nucleus-nucleus collisions\label{AAcollisions}}
\indent

\setcounter{equation}{0}

In this section we shall study the \emph{factorization} property of inclusive observables in A-A collisions. The 1-loop corrections to inclusive observables contain small $x$, logarithmic divergences as one performs the four-momentum loop integral. These logarithms arise from the CGC cutoff separating the quantum and classical description of fields. In the p-A case we had a single strong classical source for which we know the associated background field. In this case we know that the small $x$ evolution is governed by the JIMWLK equation. However the generalization to the A-A case is not obvious. In A-A collisions, we are in presence of two strong classical sources for which it is impossible to solve the Yang-Mills equations in the whole space-time analytically. Indeed, the non-linear character of the Yang Mills equations breaks the superposition principle. Thus the question is how is the small $x$ evolution for observables governed in nucleus-nucleus collisions ? is the JIMWLK equation still valid ? do we have to generalize it to more complicated background fields ? Fortunately the JIMWLK equation is enough for this purpose. The structure of small $x$ divergences turns out to be inherent to the partonic content of the nucleus and does not depend on any upcoming reaction or measured observables, we say they are \emph{universal}. Since the two colliding nuclei are initially not in causal contact, universality requires that the logarithms corresponding to the two nuclei' wave functions factorize separately. We, hence, expect that the evolution is governed by two copies of the JIMWLK hamiltonian corresponding to the two nuclei. This is what we are going to prove (or at least argue) for particle spectra in A-A collisions.\\

The universal character of small $x$ divergences is nowadays established in dilute dense collisions like DIS. However, proving the factorization property in nucleus-nucleus collisions is much more technical and it has to be done case by case.%%, at least analytically (numerical proof of this property are found in \cite{Krasnitz:2003jw,Krasnitz:2002mn,Krasnitz:2001qu,Krasnitz:2000gz,Krasnitz:1999wc,Krasnitz:1998ns,Lappi:2003bi}). 
There is a powerful formalism developed by Gelis, Lappi and Venugopalan \cite{Gelis:2008rw,Gelis:2008ad,Gelis:2008sz} to deal with this problem of factorization for inclusive observables based on the Schwinger-Keldysh formalism.\\

First we shall write the inclusive quark and gluon spectra as Green functions in the Schwinger-Keldysh formalism. Then we shall see how perturbation theory works and analyze the diagrams that must be taken into account at a given order in perturbation theory. For both gluons and quarks we will work out the leading and next to leading order of perturbation theory. The key point will be the emergence of recursion relations via differential operators perturbing the initial conditions. The new feature is that the proof of this recursion relation for the quark spectrum as well \cite{Gelis:2012ct}. At the end we shall discuss the small $x$ evolution. Although the quantum evolution has been proved to be governed by the JIMWLK equation in the case of gluons, this still has to be proved for quarks.

\section{Inclusive observables}
\indent

Among the possible observables, one distinguishes two classes of them that have different properties :
\begin{itemize}
\item the \emph{exclusive} observables which constrain the final state.
\item the \emph{inclusive} observables which do not constrain the final number of particles.  
\end{itemize}
Exclusive observable may be for instance the cross-section for producing $n$ gluons. As an example of inclusive observable is the \emph{particle spectrum}, i.e. the average number of particle in the final state. Here we shall deal with inclusive observables and especially particle spectra. Let us emphasize a fundamental property of inclusive observables : from the formal point of view the evolution in time of some in-state decomposes onto the basis of out-states in a non trivial way. An exclusive observable only takes into account some particular projection among this set of possible final states while inclusive observables do not. Physically its means that inclusive observables can be expressed as the \emph{causal} propagation of initial conditions.\\

In A-A collisions, observables are very difficult to compute explicitly. As we shall see, even the leading order gluon spectrum obeys the classical Yang-Mills equation of motion in presence of two strong sources in A-A collisions which cannot be solved analytically. Thus the explicit computation of particle spectra is in general not doable analytically. This is not our purpose, we want to find relations among them which do not require their explicit computation.

\subsection{The inclusive spectrum\label{inclusivespectrum}}
\indent

In this section we shall set the framework in order to compute inclusive spectra from Feynman diagrams.

\subsubsection{Primary definition}
\indent

As far as we are concerned with nucleus-nucleus collisions we only consider as in-state the state $|AA;in>$. Indeed we do not expect other incoming particles. Since the two nuclei are described as external classical sources within the CGC framework they actually do not belong to the spectrum and we refer to $|AA>$ as the vacuum denoted $|0>$ in this chapter. The collision is assumed to be observed in approximately the center of mass frame (if the two colliding nuclei are the same, the lab frame and the center of mass frame are the same, otherwise they are not exactly the same but in both of them the longitudinal momentum of the two colliding nuclei are very large). From this symmetric configuration, we do not expect produced particles to be especially forward or backward. For this reason we work here in Minkowski coordinates and not in light-cone ones. According to the above considerations, we shall only deal with amplitudes of the form :
\begin{equation}
\label{AAfundamentalSmatrix}
\left\langle\left.\mathbf{p}_1;...;\mathbf{p}_n; out \right|0, in\right\rangle
\end{equation}
where the $\mathbf{p}_i$'s denote the spatial components of the momentum in Minkowski coordinates. For brevity, we consider only one particle species and the discrete quantum numbers like spin... are dropped. From the Minkowski analog to normalization condition \eqref{statesnormalization} , it is easily seen that the matrix element \eqref{AAfundamentalSmatrix} has a length dimension $n$, the number of particles in the out state. Then the quantity :
\begin{equation}
\rmd P(\mathbf{p}_1;...;\mathbf{p}_n)=\left|\left\langle\left.\mathbf{p}_1;...;\mathbf{p}_n; out \right|0, in\right\rangle\right|^2\frac{\rmd^3p_1}{(2\pi)^32p^0_1}...\frac{\rmd^3p_n}{(2\pi)^32p^0_n}
\end{equation}
is dimensionless and together with the Minkowski analog of the completeness relation \eqref{completenessrelation}\footnote{When performing the integration over the phase space, one has to add a combinatorial factor $1/n!$ to avoid multiple counting.} is interpreted as the probability for vacuum to $|\mathbf{p}_1;...;\mathbf{p}_n; out>$ transition in the corresponding phase space element. The average number of particles in the final state, is therefore given by :
\begin{equation}
\mathcal{N}=\displaystyle{\sum_{n=0}^{\infty}}\frac{n}{n!}\int\rmd P(\mathbf{p}_1;...;\mathbf{p}_n).
\end{equation}
The physical quantity we are interested in is the average number of particles in some phase space volume element. It is given by only a partial integration over the total phase space volume. The simplest one and the one we will consider from now on is the average number of particles in the phase space element $\rmd^3p/(2\pi)^32p^0$, given by :
\begin{equation}
\label{spectrumdefinition}
(2\pi)^32p^0\frac{\rmd \mathcal{N}}{\rmd^3p}=\left|\left\langle\left.\mathbf{p}; out \right|0, in\right\rangle\right|^2+\displaystyle{\sum_{n=1}^{\infty}}\frac{1}{n!}\int \frac{\rmd^3p_1}{(2\pi)^32p^0_1}...\frac{\rmd^3p_n}{(2\pi)^32p^0_n}\left|\left\langle\left.\mathbf{p};\mathbf{p}_1;...;\mathbf{p}_n; out \right|0, in\right\rangle\right|^2.
\end{equation}
Written like this, there seem to be an infinite set of Feynman diagrams to consider for computing the spectrum. Fortunately there is a way out which takes into account the systematic resummation of all the final states. For this purpose we will show that the problem is reduced to the computation of Green functions in the Schwinger-Keldysh formalism. Note that $\mathcal{N}$ is the number of particles in the final state and has nothing to do with the $N$ encountered in the previous chapters which is the number of particles in the hadronic wave function. This is why the spectrum is denoted $\mathcal{N}$ while $N$ is the distribution.

\subsubsection{LSZ reduction formulas and Schwinger-Keldysh Green functions\label{inclusivespectrumandKeldysh}}
\indent

\begin{figure}[h]
\centering
%(along, up)
\begin{fmfgraph*}(300,100)
    \fmftop{t1,t2,t3}
    \fmfbottom{b1,b2,b3}
      \fmf{dashes}{t2,o,b2}
      \fmf{dbl_plain_arrow}{b1,v1,t2}
      \fmf{dbl_plain_arrow}{t1,v1,b2}
      \fmf{dbl_plain_arrow}{b3,v2,t2}
      \fmf{dbl_plain_arrow}{t3,v2,b2}
    \fmffreeze
      \fmf{phantom}{v1,i1,o,i2,v2}
      \fmf{phantom,label=$\rightarrow$}{i1,o}
      \fmf{phantom,label=$\rightarrow$}{o,i2}
      \fmflabel{$p$}{o}
      \fmf{plain, tension=0}{v1,i1}
      \fmf{plain, tension=0}{v2,i2}
      \fmfblob{30.}{v1}
      \fmfblob{30.}{v2}
\end{fmfgraph*}
\\
\vspace{1cm}
\begin{fmfgraph*}(300,100)
    \fmftop{t1,t2,t3}
    \fmfbottom{b1,b2,b3}
    \fmfleft{l1}
    \fmfright{r1}
      \fmf{dashes}{t2,o,b2}
      \fmf{dbl_plain_arrow}{b1,v1,t2}
      \fmf{dbl_plain_arrow}{t1,v1,b2}
      \fmf{dbl_plain_arrow}{b3,v2,t2}
      \fmf{dbl_plain_arrow}{t3,v2,b2}
    \fmffreeze
      \fmf{phantom}{v1,i1,o,i2,v2}
      \fmf{phantom,label=$\rightarrow$}{i1,o}
      \fmf{phantom,label=$\rightarrow$}{o,i2}
      \fmflabel{$p$}{o}
      \fmf{plain}{l1,v1}
      \fmf{plain}{v2,r1}
      \fmf{plain, tension=0}{v1,i1}
      \fmf{plain, tension=0}{v2,i2}
      \fmfv{decor.shape=cross,decor.size=7.}{l1}
      \fmfv{decor.shape=cross,decor.size=7.}{r1}
      \fmfblob{30.}{v1}
      \fmfblob{30.}{v2}
\end{fmfgraph*}
\caption{The two contributions to the spectrum \eqref{spectrumconnectedGF}. Each of them must be understood as two space-time copies, one for the amplitude and another one for the complex conjugate amplitude. The double arrows represent the two colliding nuclei. The upper diagram is the one-point functions contribution where a particle represented by a solid line is created from the sources both in the amplitude and the complex conjugate amplitude. The lower diagram represents the $G_{+-}$ contribution, that is the propagator that runs from the complex conjugate amplitude to the amplitude. The crosses refers to the fact that it is a cut propagator, it carries an on-shell momentum in the infinite past. \label{diagramaticN}}
\end{figure}
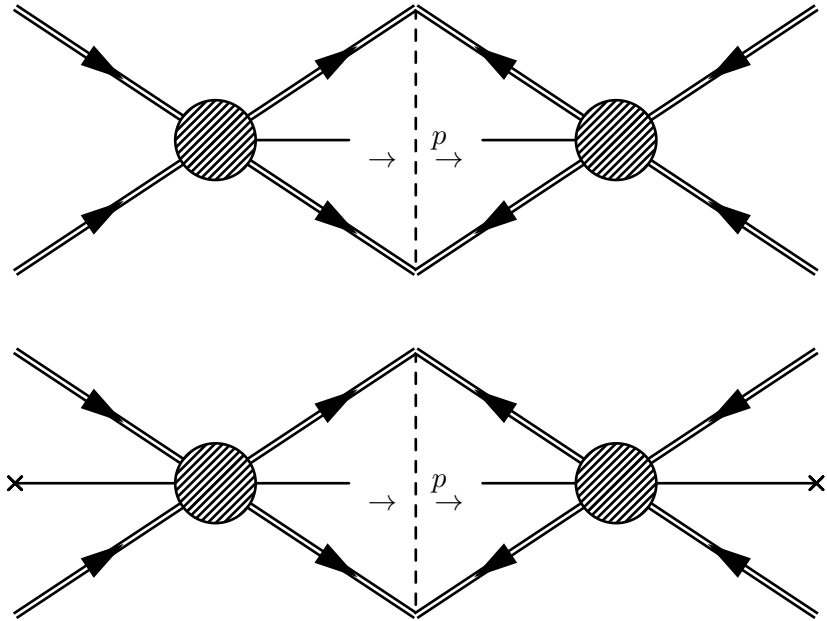

Our aim is to write the matrix elements in \eqref{spectrumdefinition} as Green functions to be precised. To do so, the essential step is the reduction formula. In this section we still not specify the type of field. We work with a generic free field $\phi$ given by the Minkowski analogue to \eqref{genericfield}. The action of the free field takes the generic form :
\begin{equation}
S_{\rm free}=\int\rmd^4x\phi^{\dagger}(x)D_x\phi(x)
\end{equation}
for a complex field, or
\begin{equation}
S_{\rm free}=\frac{1}{2}\int\rmd^4x\phi(x)D_x\phi(x)
\end{equation}
for a real field. $D_x$ denotes the quadratic integral kernel and is a first (half-integral spin) or second (integral spin) order differential operator. For definiteness, we assume that the particles that enter into the spectrum are the particles destroyed by $\phi$ (in other words they are not anti-particles). The reduction formula reads :
\begin{equation}
\label{inclusiveLSZ}
\left\langle\left.\mathbf{p};\mathbf{p}_1;...;\mathbf{p}_n; out \right|0, in\right\rangle=-\frac{i}{\sqrt{Z_{\phi}}}\int\rmd^4xe^{ip.x}u^{\dagger}(p)D_x\left\langle\mathbf{p}_1;...;\mathbf{p}_n; out \left|\phi(x)\right|0, in\right\rangle,
\end{equation}
with $\phi(x)$ the corresponding field operator in Heisenberg picture and $\sqrt{Z_{\phi}}$ the field renormalization factor. Plugging the reduction formula into the spectrum \eqref{spectrumdefinition} gives :
\begin{equation}
\label{spectrumLSZ}
\begin{split}
(2\pi)^32p^0&\frac{\rmd \mathcal{N}}{\rmd^3p}=\frac{1}{Z_{\phi}}\int\rmd^4x\rmd^4ye^{ip\cdot(x-y)}u^{\dagger}(p)D_x\Big[\big\langle 0; in \big|\phi^{\dagger}(y)\big|0, out\big\rangle\big\langle 0; out \big|\phi(x)\big|0, in\big\rangle\\
&+\displaystyle{\sum_{n=1}^{\infty}}\frac{1}{n!}\int \frac{\rmd^3p_1}{(2\pi)^32p^0_1}...\frac{\rmd^3p_n}{(2\pi)^32p^0_n}\big\langle 0, in\big|\phi^{\dagger}(y)\big|\mathbf{p}_1;...;\mathbf{p}_n; out\big\rangle\big\langle\mathbf{p}_1;...;\mathbf{p}_n; out \big|\phi(x)\big|0, in\big\rangle\Big]\overleftarrow{D}^{\dagger}_yu(p).
\end{split}
\end{equation}
In appendix \ref{Keldyshappendix} we detail the methods to write such Green functions summed over out states as Green functions in the Schwinger-Keldysh formalism. The Green functions in \eqref{spectrumLSZ} are generated by adding to the Lagrangian, integrated over the Keldysh contour, a source term, conventionally written :
\begin{equation}
\int\rmd^4x \left(j_+(x)\phi_+(x)-\phi^{\dagger}_-(x)j_-(x)\right)
\end{equation}
where $\phi_+$ denotes the fields living on the forward branch of the Keldysh contour and $\phi_-$, the fields living on the backward one\footnote{Be careful with the $\phi^{\dagger}_-(x)j_-(x)$ term if $\phi$ is a fermionic field. Indeed a functional derivative acting from the left has to go through $\phi^{\dagger}_-$ and is affected by a minus sign if $\phi$ is a Grassmann field :
$$\frac{\delta}{\delta j_-(y)}\int\rmd^4x \phi^{\dagger}_-(x)j_-(x)=\pm \phi^{\dagger}_-(y).$$
The upper sign concerns bosons and the lower one, fermions.
}. Then the spectrum \eqref{spectrumLSZ} reads as a functional derivative of $Z$, the generating partition function \eqref{Keldyshgeneratingpartitionfunction} \footnote{Beware of the order of derivatives if the $\phi$ field is fermionic. The convention given here makes all the derivatives act on their right.} :
\begin{equation}
\begin{split}
(2\pi)^32p^0\frac{\rmd \mathcal{N}}{\rmd^3p}&=\frac{1}{Z_{\phi}}\int\rmd^4x\rmd^4ye^{ip\cdot(x-y)}u^{\dagger}(p)D_x\left[\left.\frac{1}{Z[j_+=j_-=0]}\frac{\delta^2}{\delta j_+(x)\delta j_-(y)}Z[j_+,j_-]\right|_{j_{\pm}=0}\right]\overleftarrow{D}^{\dagger}_yu(p).
\end{split}
\end{equation}
However we prefer to write such expression in terms of connected Green functions. For this purpose we rather use the generating functional of connected Green functions $\mathcal{W}=-i\ln Z$ :
\begin{equation}
\label{spectrumconnectedGFW}
\begin{split}
(2\pi)^32p^0\frac{\rmd \mathcal{N}}{\rmd^3p}&=\frac{1}{Z_{\phi}}\int\rmd^4x\rmd^4ye^{ip\cdot(x-y)}u^{\dagger}(p)D_x\left[\left.i\frac{\delta^2}{\delta j_+(x)\delta j_-(y)}\mathcal{W}[j_+,j_-]\right|_{j_{\pm}=0}-\right.\\
&\left.-\left.\frac{\delta}{\delta j_+(x)}\mathcal{W}[j_+,j_-]\right|_{j_{\pm}=0}\left.\frac{\delta}{\delta j_-(y)}\mathcal{W}[j_+,j_-]\right|_{j_{\pm}=0}\right]\overleftarrow{D}^{\dagger}_yu(p).
\end{split}
\end{equation}
We recognize the one and two-point \emph{connected} Green functions defined so that (see \eqref{Keldyshfreepropagators} for the 2-point function) :
\begin{equation}
\label{Keldysh2and1ptfunctions}
\begin{split}
&i\left.\frac{\delta^2}{\delta j_+(x)\delta j_-(y)}\mathcal{W}[j_+,j_-]\right|_{j_{\pm}=0}=G_{+-}(x,y)=\pm \big\langle 0; in \big|\phi^{\dagger}(y)\phi(x)\big|0, in\big\rangle\\
&\left.\frac{\delta}{\delta j_+(x)}\mathcal{W}[j_+,j_-]\right|_{j_{\pm}=0}=\phi_+(x)=\big\langle 0; in \big|\phi(x)\big|0, in\big\rangle\\
&\left.\frac{\delta}{\delta j_-(y)}\mathcal{W}[j_+,j_-]\right|_{j_{\pm}=0}=\mp\phi_-^{\dagger}(y)=\mp\big\langle 0; in \big|\phi^{\dagger}(y)\big|0, in\big\rangle
\end{split}
\end{equation}
where the upper sign refers to bosons and the lower one to fermions. Plugging these into \eqref{spectrumconnectedGFW} gives our fundamental formula to deal with particle spectra in A-A collision :
\begin{equation}
\label{spectrumconnectedGF}
\begin{split}
(2\pi)^32p^0\frac{\rmd \mathcal{N}}{\rmd^3p}&=\pm\frac{1}{Z_{\phi}}\int\rmd^4x\rmd^4ye^{ip\cdot(x-y)}u^{\dagger}(p)D_x\left[G_{+-}(x,y)+\phi_+(x)\phi^{\dagger}_-(y)\right]\overleftarrow{D}^{\dagger}_yu(p).
\end{split}
\end{equation}
Written in this form it is possible to compute the spectrum with Feynman diagrams order by order in perturbation theory. Indeed, the Green functions \eqref{Keldysh2and1ptfunctions} are the exact Green functions of the interacting theory including all possible radiative corrections. An intuitive picture of this formula is given on figure \ref{diagramaticN}. Note that we made the derivation for the simplest spectrum $\rmd \mathcal{N}/\rmd^3p$ because this the one we will be bothered with. However our derivation is straightforwardly applied to higher correlation spectra. The analog of equation \eqref{spectrumconnectedGF} for the average number of particles per phase space volume $\rmd^3p_1...\rmd^3p_n$ would involve $2n$ integrals, $n$ LSZ operators and $n$ complex conjugate LSZ operators and up to $2n$-point functions. Coming back to spectrum \eqref{spectrumconnectedGF}, our aim is now to precise the perturbative expansion in the A-A problem.

\section{Perturbation theory : power counting and loop expansion\label{powercounting}}
\indent

The natural question that arises is how to handle expression \eqref{spectrumconnectedGF} in the framework of perturbation theory for A-A collisions. As we shall see, the explicit computation of even the lowest order of the spectrum is, in general, a hard task. However it is possible in some cases, to find formal recursion relations between the leading and next to leading order in perturbation theory. This is precisely our goal. For this purpose one has to precise the structure of the theory. From section \ref{Partondistribution}, the hadronic content is dominated by small $x$ partons and mostly gluons that have occupation numbers of order $1/\alpha_s$. The strength of quark sources is suppressed by one power of the coupling constant with respect to the gluon ones. Moreover there is no analogue to the external field approximation for fermions but fortunately the quark distribution depends on the gluon's one and the whole quark content of the nucleus is properly described by quantum fluctuations of gluons\footnote{This is true up to the valence quarks that are spectators and do not enter into consideration.}. We saw in section \ref{nucleardescription} that the nuclei are considered as classical sources $\mathcal{J}$ or equivalently, the associated classical field $\mathcal{A}$. Since the occupation number which is of order $1/\alpha_s\sim 1/g^2$ is given by a background field two-point function, the strength of the background field is $\sim 1/g$. The lagrangian is the Yang-Mills lagrangian $\mathcal{L}_{YM}$ for the gluons corresponding to field $A$. The classical source appears explicitly in the lagrangian through a $\mathcal{J}\cdot A$ term. Moreover one can consider various types of fields interacting through covariant derivatives with gluons. They represent the so called matter content of the theory and the lagrangian, whose explicit form does not matter at this point, is $\mathcal{L}_{\text{mat}}$. Then we write the full lagrangian density as :
\begin{equation}
\mathcal{L}=\mathcal{L}_{\text{mat}}+\mathcal{L}_{YM}+\mathcal{J}\cdot A.
\end{equation}
For the following discussion, we assume that all the non-gauge couplings like Yukawa or scalar self interactions, are turned off. Non abelian gauge invariance requires that all the gauge couplings are proportional to some power of $g$ which is assumed to be small. This is justified on the physical side since high energy collisions lie in the scope of perturbative QCD, that is in the kinematic regime where the coupling constant $g$ is small. If one plugs a source on one of these couplings it will bring a power of $1/g$ that will kill one power of $g$ in the corresponding coupling. This suggests that insertion of sources does not affect the power counting. Therefore at a given order in perturbation theory, one has to resum an infinite set of Feynman diagrams with an arbitrary number of sources inserted. We are going to show this in a more rigorous way also in order to show which topology of diagrams contributes at some given order in perturbation theory. For definiteness we consider some generic diagram with $N_f$ external legs of type $f$, $f$ denoting all possible fields of the theory. In addition, there are $S$ gluon external sources inserted in this diagram. This diagram has $L$ loops, $I_f$ internal lines of type $f$ and $V_i$ vertices of type $i$ that is proportional to some power of $g$. To each vertex of type $i$ there are $n_{if}$ fields of type $f$ attached. It turns out that in all renormalizable theories of Yang-Mills field coupled to arbitrary fields, interaction of type $i$ has a coupling constant dimension $\sum\limits_f n_{if}-2$. To see this, one notices that the gauge field is coupled to only one type of field (that can be the gauge field itself) in some given interaction. Interactions are either cubic and proportional to $g$ (spin $0, 1/2, 1, 3/2$) or quartic and proportional to $g^2$ (spin $0, 1$). The proof of our statement is completed. Obviously the order of the considered diagram, in the coupling constant, is :
\begin{equation}
\label{couplingdimension}
g^{-S+\sum\limits_{i}(\sum\limits_{f}n_{if}-2)V_i}.
\end{equation}
The number of vertices of each type is an unsatisfactory parameter, in the sense it is difficult to control as we draw Feynman diagrams. Fortunately, it is not independent of other parameters that are easy to handle like the number of loops or external legs. One of these relation is the consequence of the following observation : for each type of fields $f$, each internal propagator is attached to two different vertices and each vertex of type $i$ carries $n_{if}$ legs that are either internal or external lines of type $f$. Then for each non gluon field one has : 
\begin{equation}
N_f+2I_f=\sum\limits_i n_{if}V_i.
\end{equation}
Concerning gluons, one has to be careful since gluon internal propagators can be attached either to vertices or to external sources. This is why one has to count sources as vertices with one gluon attached. So for gluons the upper identity is a bit modified according to\footnote{One could have had treated the sources $S$ as vertices as well, each of them being proportional to $1/g$ and with $n_f=\delta_{gf}$.} :
\begin{equation}
N_g+2I_g=\sum\limits_i n_{ig}V_i+S.
\end{equation}
The second identity follows from a topological invariant, the Betti number which is one for a connected graph\footnote{This is the equivalent of the Euler characteristic applied to graphs. It states that the number of loops minus the number of edges plus the number of vertices equals the number of connected components.} :
\begin{equation}
L-\sum\limits_f I_f+S+\sum\limits_iV_i=1.
\end{equation}
Plugging the topological equations above into \eqref{couplingdimension} shows that the $g$ power of the considered diagram is :
\begin{equation}
\label{couplingdimension1}
g^{-S+\sum\limits_f(N_f+2I_f)-S+2(L-\sum\limits_fI_f+S-1)}=g^{2L+\sum\limits_{f}N_f-2}.
\end{equation}
As expected it does not depend on the number of sources, then at a given order the task is to resum all possible insertions of sources. Moreover the contribution of Feynman diagrams becomes smaller as we increase the number of loops and external legs.\\

Let us go back to the spectrum \eqref{spectrumconnectedGF}. According to power counting \eqref{couplingdimension1} one can expand the one and two-point functions as :
\begin{equation}
\begin{split}
\phi_{\pm}(x)&=\sum\limits_L\phi^{(L)}_{\pm}(x)\\
G_{+-}(x,y)&=\sum\limits_LG^{(L)}_{+-}(x,y)~;
\end{split}
\end{equation}
where $\phi^{(L)}_{\pm}$ is the $L$ loop contribution to $\phi_{\pm}$ and is proportional to $g^{2L-1}$ and $G^{(L)}_{+-}$ is the $L$ loop contribution to $G_{+-}$ proportional to $g^{2L}$. thanks to this expansion in powers of the coupling constant, one has a perturbative expansion of the spectrum \eqref{spectrumconnectedGF}. As we shall see the perturbative expansion does not actually start at the same order if one is considering matter or gauge fields. Let us discuss this in detail.

\section[Quark and gluon spectra]{Quark and gluon spectra : computation of leading and next to leading order\label{qgspectra}}
\indent

In this section we shall write the LO and NLO gluon and quark spectra in terms of Green functions. The starting point is formula \eqref{spectrumconnectedGF} but now with specific notations and various indices restored. We are interested in the unpolarized and uncolored spectra : the spins and colors will be summed. For the spectrum \eqref{spectrumconnectedGF} corresponding to gluons we denote $\phi_{\pm}\rightarrow A^{A\mu}_{\pm}$ which is real. The dressed two-point function is denoted $\Delta^{\mu\nu AB}_{+-}(x,y)$. We fix the gauge to be an axial one (the precise choice will be discussed later in \ref{gspectrumgaugefix} and does not matter for present discussion). Then the differential operator $D$ is given by \eqref{axialGF1} to be : $D\rightarrow \delta^{AB}(g^{\mu\nu}\Box-\partial^{\mu}\partial^{\nu})$ which is not exactly the inverse propagator but the extra term in \eqref{axialGF1} is suppressed when the index $\mu$ is contracted with $u^{\dagger}\rightarrow \epsilon^*_{\mu}$ thanks to the gauge condition \eqref{gaugeid}. Moreover the momentum $k$ carried by the final observed gluon is on-shell and then thanks to the on-shell Ward identity \eqref{oswardid} the $\partial^{\mu}\partial^{\nu}$ operator does not contribute as well. So we can make the replacement $D\rightarrow \Box$ in \eqref{spectrumconnectedGF} for the gluon spectrum in axial gauge. Gathering all these considerations leads to the following expression for the gluon spectrum : 
\begin{equation}
\label{spectrumgluons}
\begin{split}
(2\pi)^32k^0\frac{\rmd \mathcal{N}_g}{\rmd^3k}&=\frac{1}{Z_{g}}\int\rmd^4x\rmd^4ye^{ik\cdot(x-y)}\epsilon^*_{\mu}(k)\epsilon_{\nu}(k)\Box_x\Box_y\left[\Delta^{\mu\nu AA}_{+-}(x,y)+A^{\mu A}_+(x)A^{\nu A}_-(y)\right].
\end{split}
\end{equation}
For quarks the one-point function is denoted $\psi^a_{\pm}$ and the two-point function $S^{ab}_{+-}$. The $u$'s in \eqref{spectrumconnectedGF} are the $u$ spinors and the quadratic kernel $D$ is $\delta^{ab}\gamma^0(i\slashed{\partial}-m)$ (in this chapter there is no technicalities arising from keeping a non-zero mass for quarks, so it is kept). Plugging this into \eqref{spectrumconnectedGF} gives the quark spectrum :
\begin{equation}
\label{spectrumquarks}
\begin{split}
(2\pi)^32p^0\frac{\rmd \mathcal{N}_q}{\rmd^3p}&=\frac{1}{Z_{q}}\int\rmd^4x\rmd^4ye^{ip\cdot(x-y)}\bar{u}(p)\left(i\overrightarrow{\slashed{\partial}}_x-m\right)\left[S^{aa}_{+-}(x,y)+\psi^a_+(x)\bar{\psi}^a_-(y)\right]\left(i\overleftarrow{\slashed{\partial}}_y+m\right)u(p).
\end{split}
\end{equation}
We are interested in both of them although the gluon spectrum has already been computed in previous works \cite{Gelis:2008rw,Gelis:2008ad,Gelis:2008sz}, the computation of the Green functions involved in it is an essential ingredient to the quark spectrum as well. Let us see how the perturbative expansion works at the first two orders. First we shall make general observations in order to see how the loop expansion occurs for both the gluon and quark spectrum. Then we shall interpret the meaning of Green functions involved in the two first leading orders of perturbation theory.\\

In formulas \eqref{spectrumgluons} and \eqref{spectrumquarks} we have closed our eyes on the gauge invariance problem. Obviously, the l.h.s of \eqref{spectrumgluons} and \eqref{spectrumquarks} are not gauge invariant. Particle spectra actually make sense in a given gauge in which their proper interpretation is to count the number of \emph{physical} degrees of freedom. This is not true in arbitrary gauge in which the forms \eqref{spectrumgluons} and \eqref{spectrumquarks} may distinguish redundant or unphysical states. We postpone the discussion of gauge fixing to section \ref{gspectrumgaugefix}. We shall see that there is indeed an axial gauge in which \eqref{spectrumgluons} and \eqref{spectrumquarks} take this form. For the moment let us just assume this result.

\subsection{The one-point function : generalities}
\indent

Since the nuclear description only has gluonic external sources, the gauge fields and matter fields one-point functions must be distinguished. We shall emphasize some general properties that tell us more about the loop expansion of spectra \eqref{spectrumgluons} and \eqref{spectrumquarks}.

\subsubsection{Gauge fields\label{LOgluon}}
\indent

Let us consider the leading order $A^{(0)A}_{\mu}(x)$, the one-point function at tree level. By power counting \eqref{couplingdimension1}, it is of order $g^{-1}$. Let us consider the simplest diagram shown on figure \ref{SingleJ}.
\begin{figure}[h]
\centering
%(along, up)
\begin{fmfgraph*}(50,40)
    \fmfleft{i1}
    \fmfright{o1}
      \fmflabel{${\scriptstyle A\mu \pm }$}{i1}
      \fmf{photon}{o1,i1}
      \fmflabel{${\scriptstyle\pm}$}{o1}
      \fmfdot{i1}
      \fmfv{decor.shape=hexagram,decor.filled=full,decor.size=10.}{o1}
\end{fmfgraph*}
\caption{The elementary contribution to one point function. The star represents a classical external source. \label{SingleJ}}
\end{figure}
The difference with usual perturbation theory is that one has to sum over the plus or minus vertices according to the Schwinger-Keldysh formalism Feynman rules detailed in \ref{KeldyshFeynmanrules}. By applying the Schwinger-Keldysh formalism Feynman rules discussed in section \ref{KeldyshFeynmanrules}, diagram \ref{SingleJ} is equal to
\begin{equation}
\begin{split}
i\int \rmd^4y& \left[\Delta^{\mu\nu}_{0\pm+}(x-y)\mathcal{J}^A_{\nu}(y)-\Delta^{\mu\nu}_{0\pm-}(x-y)\mathcal{J}^A_{\nu}(y)\right],
\end{split}
\end{equation}
where $\Delta_0$ denotes the \emph{free} propagator with the trivial color structure dropped out : $\Delta^{AB}_0\rightarrow \delta^{AB}\Delta_0$. The important thing is that the external current $\mathcal{J}$ does not differ on the forward and backward branch of the Keldysh contour and therefore factorizes into the retarded and $\pm$-independent quantity :
\begin{equation}
\begin{split}
i\int \rmd^4y& \Delta^{\mu\nu}_{0R}(x-y)\mathcal{J}^A_{\nu}(y).
\end{split}
\end{equation}
Since any tree diagram contributing to the one-point function is built from such $\pm$-independent elementary bricks, it is obvious that at the end one has $A^{(0)}_-(x)=A^{(0)}_{+}(x)$. Furthermore loop corrections lead to the same conclusion and then $A_-(x)=A_{+}(x)$ at all orders of perturbation theory. From now, one omits the sign index since it does not matter anymore. $A^{(0)}(x)$ satisfies the recursion relation represented on figure \ref{Arec}.
\begin{figure}[h]
\centering
%(along, up)
\begin{fmfgraph*}(50,40)
    \fmfleft{i1}
    \fmfright{o1}
      \fmflabel{$A^{(0)A}_{\mu}={\scriptstyle A\mu \pm }$}{i1}
      \fmf{photon}{o1,i1}
      \fmfdot{i1}
      \fmfblob{10.}{o1}
\end{fmfgraph*}
\hspace{15mm}
\begin{fmfgraph*}(60,40)
    \fmfleft{i1}
    \fmfright{o1}
      \fmf{photon}{o1,i1}
      \fmflabel{${\scriptstyle\pm}~~~+$}{o1}
      \fmflabel{$=~{\scriptstyle A\mu \pm }$}{i1}
      \fmfdot{i1}
      \fmfv{decor.shape=hexagram,decor.filled=full,decor.size=10.}{o1}
\end{fmfgraph*}\\
\hspace{8mm}
\begin{fmfgraph*}(70,40)
    \fmfleft{i1}
    \fmfright{o1,o2}
      \fmf{photon}{o1,v1,o2}
      \fmf{photon}{v1,i1}
      \fmflabel{${\scriptstyle\pm}$}{v1}
      \fmflabel{$+~{\scriptstyle A\mu \pm }$}{i1}
      \fmfdot{i1}
      \fmfblob{10.}{o1,o2}
\end{fmfgraph*}
\hspace{8mm}
\begin{fmfgraph*}(70,40)
    \fmfleft{i1}
    \fmfright{o1,o2,o3}
      \fmf{photon}{o1,v1,o2}
      \fmf{photon,tension=2}{i1,v1}
      \fmf{photon}{v1,o3}
      \fmflabel{\vspace{5mm}\hspace{-5mm}${\scriptstyle\pm}$}{v1}
      \fmflabel{$+~{\scriptstyle A\mu \pm }$}{i1}
      \fmfdot{i1}
      \fmfblob{10.}{o1,o2,o3}
\end{fmfgraph*}
\caption{Recursion relation for $A^{(0)A}_{\mu}$ at leading order.\label{Arec}}
\end{figure}
This recursion relation reads in terms of equations as :
\begin{equation}
\label{classicalA}
A^{(0)\mu A}(x)=i\int \rmd^4y\Delta^{\mu\nu}_{0R}(x-y)\left[\mathcal{J}^A_{\nu}(y)-\frac{\delta V}{\delta A^{A\nu}}[A^{(0)}(y)]\right].
\end{equation}
where $V$ denotes the self gluon interaction :
\begin{equation}
\label{YMinteraction}
V[A]\equiv gf^{ABC}(\partial_{\mu}A^A_{\nu})A^{B\mu}A^{C\nu}+\frac{g^2}{4}f^{ABC}f^{ADE}A^B_{\mu}A^C_{\nu}A^{D\mu}A^{E\nu}
\end{equation}
The notation $\delta V/\delta A$ is a symbolic notation for the vertex with one leg amputated. A problem arises from derivatives in the interaction potential. Thus $\frac{\delta V}{\delta A^{B\nu}}[A]$ is not a mere factor containing the vertex with one leg amputated but a differential operator acting on the surrounding propagators. It should be understood as 
\begin{equation}
\frac{\delta V}{\delta A^{B\nu}}[A]=\frac{\partial V}{\partial A^{B\nu}}[A]+\frac{\partial V}{\partial (\partial_{\rho}A^{C\nu})}[A]\partial_{\rho}.
\end{equation}
At this formal point, this notation is not misleading but its precise meaning must be kept in mind. For definiteness let us suppose we work in generalized axial gauge with a finite gauge fixing parameter $\xi$. In this gauge, the free retarded propagator is a Green function of the differential operator $(g_{\mu\lambda}\Box-\partial_{\mu}\partial_{\lambda}-n_{\mu}n_{\lambda}/\xi)\Delta^{\lambda\nu}_{0R}(x-y)=i\delta^{\mu}_{\nu}\delta^{(4)}(x-y)$. Acting on equation \eqref{classicalA} with the d'Alembert operator gives :
\begin{equation}
\label{EOMA0}
(g_{\mu\lambda}\Box-\partial_{\mu}\partial_{\lambda}+\frac{n_{\mu}n_{\lambda}}{\xi}) A^{(0)A\lambda}(x)=-\mathcal{J}^A_{\mu}(x)+\frac{\partial V}{\partial A^{A\mu}}[A^{(0)}(x)]-\partial_{\nu}\frac{\partial V}{\partial \partial_{\nu}A^{A\mu}}[A^{(0)}(x)]
\end{equation}
which is nothing but the classical Yang-Mills equation in generalized axial gauge. Hence, together with the initial condition $A^{(0)}(x)=0$ for $x^0=-\infty$, $A^{(0)}$ is the classical field.

\subsubsection{Matter fields}
\indent

The situation for the matter fields is very different. Since there are no sources, the analogue of diagram \ref{SingleJ} for matter one-point functions is zero. Actually it is easy to see that this assertion holds for any loop corrections as long as we do not add interactions like $\phi^3$ or Yukawa couplings as it is supposed to be. Therefore the matter one-point function is zero to all order of perturbation theory. The spectrum \eqref{spectrumconnectedGF} only receives contributions from the two-point function in case of matter fields.

\subsection{LO and NLO}
\indent

In the light of these general considerations on the one-point functions one can now identify the starting point of perturbation theory. For gluons the leading (LO) and next to leading orders (NLO) are given by :
\begin{equation}
\label{spectrumgluons1}
\begin{split}
\left.(2\pi)^32k^0\frac{\rmd \mathcal{N}_g}{\rmd^3k}\right|_{LO}&=\int\rmd^4x\rmd^4ye^{ik\cdot(x-y)}\epsilon^*_{\mu}(k)\epsilon_{\nu}(k)\Box_x\Box_y\left[A^{(0)\mu A}(x)A^{(0)\nu A}(y)\right]\sim g^{-2}\\
\left.(2\pi)^32k^0\frac{\rmd \mathcal{N}_g}{\rmd^3k}\right|_{NLO}&=\int\rmd^4x\rmd^4ye^{ik\cdot(x-y)}\epsilon^*_{\mu}(k)\epsilon_{\nu}(k)\Box_x\Box_y\left[\Delta^{(0)\mu\nu AA}_{+-}(x,y)+A^{(1)\mu A}(x)A^{(0)\nu A}(y)\right.\\
&\left.~~~~~~~~~~+A^{(0)\mu A}(x)A^{(1)\nu A}(y)\right]+~\text {counterterm}\sim g^0.
\end{split}
\end{equation}
In the NLO spectrum, the counterterm comes from the expansion of the field renormalization factor $Z_g$ in powers of $g$ : $Z_g=1+g^2C+...$ where $C$ is an infinite constant. Its role is to cancel the UV divergences which are not of concern here. From now it will be droped out also for quarks for the same reasons. The quark spectrum is one order lower since as explained previously the one-point functions are zero. This is easily understood by noticing that the quarks merge from gluons splitting into $q\bar{q}$ pairs and the corresponding amplitude and complex conjugate amplitude are both affected by a factor $g$ with respect to the gluon case in which gluons can directly be produced from the classical sources. Then for quarks the LO and NLO spectrum contributions read :  
\begin{equation}
\label{spectrumquarks1}
\begin{split}
\left.(2\pi)^32p^0\frac{\rmd \mathcal{N}_q}{\rmd^3p}\right|_{LO}&=\int\rmd^4x\rmd^4ye^{ip\cdot(x-y)}\bar{u}(p)\left(i\overrightarrow{\slashed{\partial}}_x-m\right)S^{(0)aa}_{+-}(x,y)\left(i\overleftarrow{\slashed{\partial}}_y+m\right)u(p)\sim g^0\\
\left.(2\pi)^32p^0\frac{\rmd \mathcal{N}_q}{\rmd^3p}\right|_{NLO}&=\int\rmd^4x\rmd^4ye^{ip\cdot(x-y)}\bar{u}(p)\left(i\overrightarrow{\slashed{\partial}}_x-m\right)S^{(1)aa}_{+-}(x,y)\left(i\overleftarrow{\slashed{\partial}}_y+m\right)u(p)\sim g^2.
\end{split}
\end{equation}

\subsubsection{The gluon spectrum}
\indent

Here we will essentially recall known results concerning the gluon's Green functions at LO and NLO. The LO spectrum has already been interpreted in term of the classical field $A^{(0)}$ in section \ref{LOgluon}. The procedure is similar for computing higher order corrections. According to the formula \eqref{spectrumgluons1} the NLO gluon spectrum is given by the tree level two-point function and the one loop correction to the one-point function.\\

\paragraph{The two-point function at tree level}

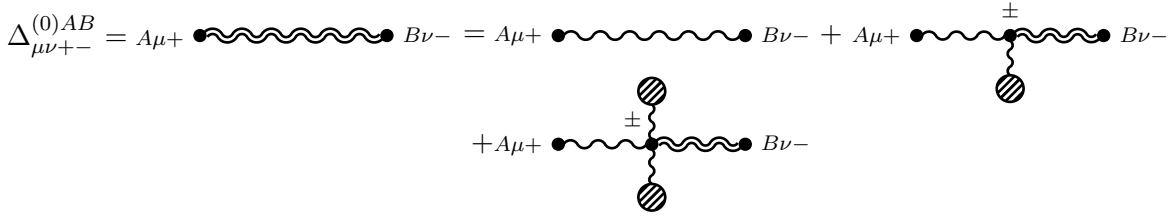
\begin{figure}[h]
\centering
%(along, up)
\begin{fmfgraph*}(70,40)
    \fmfleft{i1}
    \fmfright{o1}
      \fmflabel{$\Delta^{(0)AB}_{\mu\nu+-}={\scriptstyle A\mu + }$}{i1}
      \fmflabel{${\scriptstyle B\nu - }$}{o1}
      \fmf{dbl_wiggly}{i1,o1}
      \fmfdot{i1,o1}
\end{fmfgraph*}
\hspace{20mm}
\begin{fmfgraph*}(70,40)
    \fmfleft{i1}
    \fmfright{o1}
      \fmf{wiggly}{i1,o1}
      \fmflabel{$={\scriptstyle A\mu + }$}{i1}
      \fmflabel{${\scriptstyle B\nu - }$}{o1}
      \fmfdot{i1,o1}
\end{fmfgraph*}
\hspace{20mm}
\begin{fmfgraph*}(70,40)
    \fmfleft{i1}
    \fmfright{o1}
    \fmfbottom{g1}
\fmf{wiggly}{i1,v1}
      \fmf{dbl_wiggly}{v1,o1}
      \fmf{photon,tension=0}{g1,v1}
      \fmflabel{${\scriptstyle\pm}$}{v1}
      \fmflabel{$+~{\scriptstyle A\mu + }$}{i1}
      \fmflabel{${\scriptstyle B\nu - }$}{o1}
      \fmfdot{i1,o1,v1}
      \fmfblob{10.}{g1}
\end{fmfgraph*}\\
\begin{fmfgraph*}(70,40)
    \fmfleft{i1}
    \fmfright{o1}
    \fmfbottom{g1}
    \fmftop{g2}
\fmf{wiggly}{i1,v1}
      \fmf{dbl_wiggly}{v1,o1}
      \fmf{photon,tension=0}{g1,v1}
      \fmf{photon,tension=0}{g2,v1}
      \fmflabel{\hspace{-5mm}${\scriptstyle\pm}$}{v1}
      \fmflabel{$+{\scriptstyle A\mu + }$}{i1}
      \fmflabel{${\scriptstyle B\nu - }$}{o1}
      \fmfdot{i1,o1,v1}
      \fmfblob{10.}{g1,g2}
\end{fmfgraph*}
\caption{Recursion relations for the tree level gluon propagator $\Delta^{(0)AB}_{\mu\nu+-}$. The one point gluon functions correspond to the lowest order $A^{(0)}$. \label{delta0NLO}}
\end{figure}
Let us begin with the two-point function $\Delta^{(0)AB}_{\mu\nu+-}$ that satisfies the diagrammatic recurrence shown on figure \ref{delta0NLO}. The diagrammatic recursion formally reads :
\begin{equation}
\label{recD1}
\begin{split}
\Delta^{(0)AB\mu\nu}_{+-}(x,y)=&\delta^{AB}\Delta^{\mu\nu}_{0+-}(x-y)-i\int \rmd^4z \left\{\Delta^{\mu\rho}_{0++}(x-z)\frac{\delta^2 V}{\delta A^{A\rho}\delta A^{C\sigma}}[A^{(0)}(z)]\Delta^{(0)CB\sigma\nu}_{+-}(z,y)\right.\\
&\left.-\Delta^{\mu\rho}_{0+-}(x-z)\frac{\delta^2 V}{\delta A^{A\rho}\delta A^{C\sigma}}[A^{(0)}(z)]\Delta^{(0)CB\sigma\nu}_{--}(z,y)\right\}.
\end{split}
\end{equation}
Once again we use a rather formal notation which needs to be precised :
\begin{equation}
\label{delta2V}
\frac{\delta^2 V}{\delta A^{A}_{\rho}\delta A^{C}_{\sigma}}=\frac{\partial^2 V}{\partial A^{A}_{\rho}\partial A^{C}_{\sigma}}+\overleftarrow{\partial_{\lambda}}\frac{\partial^2 V}{\partial(\partial_{\lambda}A^{A}_{\rho})\partial A^{C}_{\sigma}}+\frac{\partial^2 V}{\partial A^{A}_{\rho}\partial (\partial_{\lambda}A^{C}_{\sigma})}\overrightarrow{\partial_{\lambda}}.
\end{equation}
There could be in principle second derivatives with respect to $\partial A$ but they are zero in our case since, according to \eqref{YMinteraction}, the interaction has a linear dependence in the first derivatives of the gluon field.
Acting on $x$ in \eqref{recD1} with the free inverse propagator, we get the equation of motion for $\Delta^{(0)}_{+-}$ :
\begin{equation}
\label{elD1}
\begin{split}
(g_{\mu\sigma}\Box_x-\partial_{x\mu}\partial_{x\sigma}+\frac{n_{\mu}n_{\sigma}}{\xi})&\Delta^{(0)AB\sigma\nu}_{+-}(x,y)\\
&=\left(\frac{\partial^2 V}{\partial A^{A\mu}\partial A^{C\sigma}}[A^{(0)}(x)]+\frac{\partial^2 V}{\partial A^{A\mu}\partial (\partial_{\lambda}A^{C\sigma})}[A^{(0)}(x)]\partial_{x\lambda}\right)\Delta^{(0)CB\sigma\nu}_{+-}(x,y)\\
&-\partial_{x\lambda}\left(\frac{\partial^2 V}{\partial (\partial_{\lambda}A^{A\mu})\partial A^{C\sigma}}[A^{(0)}(x)]\Delta^{(0)CB\sigma\nu}_{+-}(x,y)\right).
\end{split}
\end{equation}
There is the same kind of relation, when the derivatives act on $y$. There is a particularly useful representation of this propagator allowed by the fact that the $+-$ propagator is a cut propagator that carries an on-shell momentum in the remote past :
\begin{equation}
\label{D1repr}
\Delta^{(0)AB\mu\nu}_{+-}(x,y)\equiv \displaystyle{\sum_{\lambda,C}}\int \frac{\rmd^3k}{(2\pi)^3 2k^0}\alpha^{(\lambda)A\mu}_{\mathbf{k}C}(x)\alpha^{(\lambda)B\nu*}_{\mathbf{k}C}(y).
\end{equation}
$\mathbf{k}$ is the initial momentum of the gluon, $\lambda$ its polarization and $C$ its color. Since this propagator must reduce to the free one for $x^0$ and $y^0$ approaching $-\infty$, one has the initial condition :
\begin{equation}
\label{freealpha}
\displaystyle{\lim_{x^0\rightarrow -\infty}}\alpha^{(\lambda)A}_{\mathbf{k}C\mu}(x)=\delta^A_C\epsilon^{(\lambda)*}_{\mu}(k)e^{ik.x}.
\end{equation}
The reason it is an $\epsilon^*$ instead of an $\epsilon$ vector and the energy sign in the exponential are obtained by computing $\Delta_{0+-}(x,y)=<0|A(y)A(x)|0>$ in the free theory with free fields operators. $\Delta^{(0)}_{+-}$ is represented by the convolution of two single fields $\alpha_{\mathbf{k}}$ which each satisfies equation \eqref{elD1} (the polarization index is dropped for brevity) :
\begin{equation}
\label{EOMalpha}
\begin{split}
(g_{\mu\nu}\Box-\partial_{\mu}\partial_{\nu}+\frac{n_{\mu}n_{\nu}}{\xi})\alpha^{A\nu}_{\mathbf{k}C}(x)&=\left(\frac{\partial^2 V}{\partial A^{A\mu}\partial A^{B\nu}}[A^{(0)}(x)]+\frac{\partial^2 V}{\partial A^{A\mu}\partial (\partial_{\lambda}A^{B\nu})}[A^{(0)}(x)]\partial_{\lambda}\right)\alpha^{B\nu}_{\mathbf{k}C}(x)\\
&-\partial_{\lambda}\left(\frac{\partial^2 V}{\partial (\partial_{\lambda}A^{A\mu})\partial A^{B\nu}}[A^{(0)}(x)]\alpha^{B\nu}_{\mathbf{k}C}(x)\right).
\end{split}
\end{equation}
This equation is the linearized Yang-Mills equation for a perturbation propagating in the background Yang-Mills field $A^{(0)}$.\\

\paragraph{The one-point function at one loop}

\begin{figure}[h]
\centering
%(along, up)
\begin{fmfgraph*}(50,40)
    \fmfleft{i1}
    \fmfright{o1}
      \fmflabel{$A^{(1)A}_{\mu+}={\scriptstyle A\mu + }$}{i1}
      \fmf{photon}{o1,i1}
      \fmfdot{i1}
      \fmfv{decor.shape=circle,decor.filled=full,decor.size=10.}{o1}
\end{fmfgraph*}
\hspace{15mm}
\begin{fmfgraph*}(70,40)
    \fmfleft{i1}
    \fmfright{o1,o2}
      \fmf{photon}{o1,v1,o2}
      \fmf{photon}{v1,i1}
      \fmflabel{${\scriptstyle\pm}$}{v1}
      \fmflabel{$=~{\scriptstyle A\mu + }$}{i1}
      \fmfdot{i1}
      \fmfblob{10.}{o1}
      \fmfv{decor.shape=circle,decor.filled=full,decor.size=10.}{o2}
\end{fmfgraph*}
\hspace{8mm}
\begin{fmfgraph*}(70,40)
    \fmfleft{i1}
    \fmfright{o1,o2,o3}
      \fmf{photon}{o1,v1,o2}
      \fmf{photon,tension=2}{i1,v1}
      \fmf{photon}{v1,o3}
      \fmflabel{\vspace{5mm}\hspace{-5mm}${\scriptstyle\pm}$}{v1}
      \fmflabel{$~+~{\scriptstyle A\mu +}$}{i1}
      \fmfdot{i1,v1}
      \fmfblob{10.}{o1,o3}
      \fmfv{decor.shape=circle,decor.filled=full,decor.size=10.}{o2}
\end{fmfgraph*}\\
\vspace{15mm}
\begin{fmfgraph*}(60,40)
    \fmfleft{i1}
    \fmfright{o1}
      \fmf{photon}{o1,i1}
      \fmf{dbl_wiggly}{o1,o1}
      \fmflabel{${\scriptstyle\pm}$}{o1}
      \fmflabel{$+~{\scriptstyle A\mu + }$}{i1}
      \fmfdot{i1,o1}
\end{fmfgraph*}
\hspace{27mm}
\begin{fmfgraph*}(60,40)
    \fmfleft{i1}
    \fmfright{o1}
    \fmftop{o2}
      \fmf{photon}{i1,o1,o2}
      \fmf{dbl_wiggly}{o1,o1}
      \fmflabel{${\scriptstyle\pm}$}{o1}
      \fmflabel{$+~{\scriptstyle A\mu + }$}{i1}
      \fmfblob{10.}{o2}
      \fmfdot{i1,o1}
\end{fmfgraph*}
\caption{Recursion relation for $A^{(1)A}_{\mu+}$. The fermion loop has been omitted although it does not vanish, we will not be interested in it. \label{A2rec}}
\end{figure}
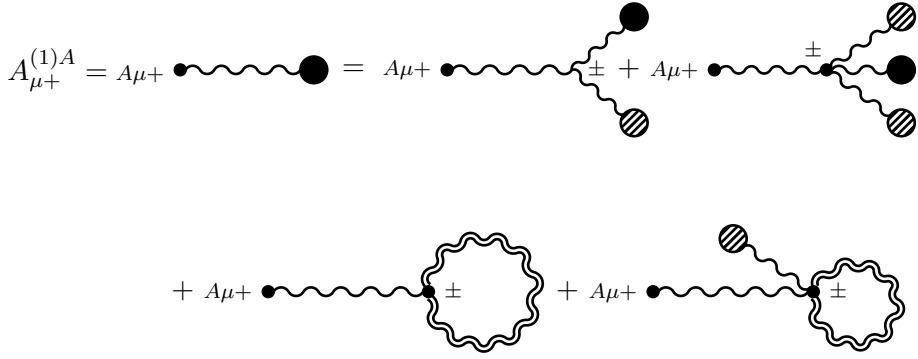

Now let us look at the one loop one-point function $A^{(1)}$. Its diagrammatic recursion relation is shown on figure \ref{A2rec}.
If the Yang-Mills field is coupled to fermion, there is a missing gluon tadpole with a fermion loop in \ref{A2rec}. Such diagram would cause some troubles for writing LO to NLO recursion relations for reasons that will become clear later. However, what we are interested in, is the region of phase space where the logarithms of $1/x$ are large and such fermion loop does not have such divergences and we omit it. The last two diagrams of figure \ref{A2rec} get a $1/2$ symmetry factor. Moreover, the sum over the $\pm$ vertex leads to two terms involving $\Delta^{(0)}_{++}(y,y)$ and $\Delta^{(0)}_{--}(y,y)$ where the two end points are the same. From definitions \eqref{Keldyshfreepropagators} there is no distinction between the four Schwinger-Keldysh propagators if the two end points are taken to be the same. Therefore these two propagators can be chosen to be both $\Delta^{(0)}_{+-}(y,y)$ given by the representation \eqref{D1repr}. Thanks to these, the diagrammatic recurrence represented on \ref{A2rec} formally reads :
\begin{equation}
\label{recA1}
\begin{split}
A^{(1)A}_{\mu}(x)&=-i\int \rmd^4y \Delta_{0R\mu\nu }(x-y)\\
&\times\left[\frac{\delta^2 V}{\delta A^{A}_{\nu}\delta A^{B}_{\sigma}}[A^{(0)}(y)]A^{(1)B}_{\sigma}(y)+\frac{1}{2}\frac{\delta^3 V}{\delta A^{A}_{\nu}\delta A^{B}_{\sigma}\delta A^{C}_{\rho}}[A^{(0)}(y)]\int\frac{\rmd^3k}{(2\pi)^3 2k^0}\alpha^{B}_{\mathbf{k}\sigma}(y)\alpha^{C*}_{\mathbf{k}\rho}(y)\right].
\end{split}
\end{equation}
The second order functional derivative of the interaction is given by \eqref{delta2V}. The third order derivative term is more complicated and the notation has to be understood as follow :
\begin{equation}
\begin{split}
\Delta_{0R\mu\nu }(x-y)\frac{\delta^3 V}{\delta A^{A}_{\nu}\delta A^{B}_{\sigma}\delta A^{C}_{\rho}}\alpha^{B}_{\mathbf{k}\sigma}(y)\alpha^{C*}_{\mathbf{k}\rho}(y)&=\Delta_{0R\mu\nu }(x-y)\frac{\partial^3 V}{\partial A^{A}_{\nu}\partial A^{B}_{\sigma}\partial A^{C}_{\rho}}\alpha^{B}_{\mathbf{k}\sigma}(y)\alpha^{C*}_{\mathbf{k}\rho}(y)\\
&+\partial_{y\lambda}\Delta_{0R\mu\nu }(x-y)\frac{\partial^3 V}{\partial (\partial_{\lambda}A^{A}_{\nu})\partial A^{B}_{\sigma}\partial A^{C}_{\rho}}\alpha^{B}_{\mathbf{k}\sigma}(y)\alpha^{C*}_{\mathbf{k}\rho}(y)\\
&+\Delta_{0R\mu\nu }(x-y)\frac{\partial^3 V}{\partial A^{A}_{\nu}\partial (\partial_{\lambda}A^{B}_{\sigma})\partial A^{C}_{\rho}}\partial_{\lambda}\alpha^{B}_{\mathbf{k}\sigma}(y)\alpha^{C*}_{\mathbf{k}\rho}(y)\\
&+\Delta_{0R\mu\nu }(x-y)\frac{\partial^3 V}{\partial A^{A}_{\nu}\partial A^{B}_{\sigma}\partial (\partial_{\lambda}A^{C}_{\rho})}\alpha^{B}_{\mathbf{k}\sigma}(y)\partial_{\lambda}\alpha^{C*}_{\mathbf{k}\rho}(y).
\end{split}
\end{equation}
In the last term of \eqref{recA1}, the sums over the lower color index and polarization are understood. Acting on the left with the inverse free propagators shows that $A^{(1)}$ satisfies the following equation of motion :
\begin{equation}
\label{EOMA1}
\begin{split}
(g_{\mu\nu}\Box-\partial_{\mu}\partial_{\nu}+\frac{n_{\mu}n_{\nu}}{\xi})A^{(1)A\nu}(x)&=\left(\frac{\partial^2 V}{\partial A^{A\mu}\partial A^{B\nu}}[A^{(0)}(x)]+\frac{\partial^2 V}{\partial A^{A\mu}\partial (\partial_{\lambda}A^{B\nu})}[A^{(0)}(x)]\partial_{\lambda}\right)A^{(1)B\nu}(x)\\
&-\partial_{\lambda}\left(\frac{\partial^2 V}{\partial (\partial_{\lambda}A^{A\mu})\partial A^{B\nu}}[A^{(0)}(x)]A^{(1)B\nu}(x)\right)\\
&+\frac{1}{2}\int\frac{\rmd^3k}{(2\pi)^3 2k^0}\left[\frac{\partial^3 V}{\partial A^{A}_{\nu}\partial A^{B}_{\sigma}\partial A^{C}_{\rho}}[A^{(0)}(x)]\alpha^{B}_{\mathbf{k}\sigma}(x)\alpha^{C*}_{\mathbf{k}\rho}(x)\right. \\
&+\frac{\partial^3 V}{\partial A^{A}_{\nu}\partial (\partial_{\lambda}A^{B}_{\sigma})\partial A^{C}_{\rho}}[A^{(0)}(x)]\partial_{\lambda}\alpha^{B}_{\mathbf{k}\sigma}(x)\alpha^{C*}_{\mathbf{k}\rho}(x)\\
&+\frac{\partial^3 V}{\partial A^{A}_{\nu}\partial A^{B}_{\sigma}\partial (\partial_{\lambda}A^{C}_{\rho})}[A^{(0)}(x)]\alpha^{B}_{\mathbf{k}\sigma}(x)\partial_{\lambda}\alpha^{C*}_{\mathbf{k}\rho}(x)\\
&-\left.\partial_{\lambda}\left(\frac{\partial^3 V}{\partial (\partial_{\lambda}A^{A}_{\nu})\partial A^{B}_{\sigma}\partial A^{C}_{\rho}}[A^{(0)}(x)]\alpha^{B}_{\mathbf{k}\sigma}(x)\alpha^{C*}_{\mathbf{k}\rho}(x)\right)\right].
\end{split}
\end{equation}

\subsubsection{The quark spectrum}
\indent

The LO and NLO quark spectrum \eqref{spectrumquarks1} requires the computation of the two-point function at tree level and at one loop. In the principle, the procedure does not differ from the one for gluons.\\

\paragraph{The leading order two-point function}

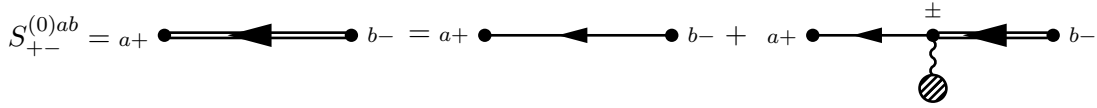
\begin{figure}[h]
\centering
%(along, up)
\begin{fmfgraph*}(70,40)
    \fmfleft{i1}
    \fmfright{o1}
      \fmflabel{$S^{(0)ab}_{+-}={\scriptstyle a+ }$}{i1}
      \fmflabel{${\scriptstyle b- }$}{o1}
      \fmf{dbl_plain_arrow}{o1,i1}
      \fmfdot{i1,o1}
\end{fmfgraph*}
\hspace{15mm}
\begin{fmfgraph*}(70,40)
    \fmfleft{i1}
    \fmfright{o1}
      \fmf{fermion}{o1,i1}
      \fmflabel{$={\scriptstyle a+ }$}{i1}
      \fmflabel{${\scriptstyle b- }$}{o1}
      \fmfdot{i1,o1}
\end{fmfgraph*}
\hspace{16mm}
\begin{fmfgraph*}(90,40)
    \fmfleft{i1}
    \fmfright{o1}
    \fmfbottom{g1}
      \fmf{fermion}{v1,i1}
      \fmf{dbl_plain_arrow}{o1,v1}
      \fmf{photon,tension=0}{g1,v1}
      \fmflabel{${\scriptstyle\pm}$}{v1}
      \fmflabel{$+~~{\scriptstyle a+ }$}{i1}
      \fmflabel{${\scriptstyle b- }$}{o1}
      \fmfdot{i1,v1,o1}
      \fmfblob{10.}{g1}
\end{fmfgraph*}
\caption{Recursion relation for the dressed quark propagator $S^{(0)}_{+-}$ at leading order. \label{Spmrec}}
\end{figure}
Let us begin with the leading order two-point function $S^{(0)}_{+-}$. Diagrammatically it obeys the recursion relation of figure \ref{Spmrec}. This relation formally reads :
\begin{equation}
S^{(0)}_{+-}(x,y)=S_{0+-}(x-y)+ig\int \rmd^4z \left\{S_{0++}(x-z)\slashed{A}^{(0)}(z)S^{(0)}_{+-}(z,y)-S_{0+-}(x-z)\slashed{A}^{(0)}(z)S^{(0)}_{--}(z,y)\right\}
\end{equation}
where the color indices have been dropped for brevity. $S_0$ is the free propagator proportional to $\delta^{ab}$ and it has to be understood that $A_{\mu}$ is a Lie algebra valued quantity, namely $A_{\mu}\rightarrow A_{\mu}^{ab}=A_{\mu}^{A}T_{ab}^{A}$. By applying the free Dirac operator on the left of the previous equation, it is easily seen that the equations of motion are :
\begin{equation}
\label{EOMS1}
(i\slashed{\partial}_x-m)S^{(0)}_{+-}(x,y)=-g\slashed{A}^{(0)}(x)S^{(0)}_{+-}(x,y),
\end{equation}
and the adjoint equation for the hermitian conjugate Dirac operator acting on the right on the $y$ variable. For the same reasons as the tree level gluon two-point function, we can choose the following representation for this propagator (with color indices restored) :
\begin{equation}
\label{S1repr}
S^{(0)ab}_{+-}(x,y)=-\displaystyle{\sum_{s,c}}\int \frac{\rmd^3p}{(2\pi)^3 2p^0}b^{s,a}_{\mathbf{p}c}(x)\bar{b}^{s,b}_{\mathbf{p}c}(y).
\end{equation}
$\mathbf{p}$ is the on-shell momentum of the initial quark in the far past, $s$ its spin and $c$ its color leading to the initial condition :
\begin{equation}
\label{initb1}
\begin{split}
&\displaystyle{\lim_{x^0\rightarrow -\infty}}b^{s,a}_{\mathbf{p}c}(x)=\delta^a_cv^s(p)e^{ip.x}.
\end{split}
\end{equation}
As for gluons, this initial condition is obtained by the eplicit computation of $S_{0+-}(x,y)=-<0|\bar{\psi}(y)\psi(x)|0>$ in the free case. From the equation of motion \eqref{EOMS1}, $b^a_{\mathbf{p}c}$ satisfies (the spin index is dropped out) :
\begin{equation}
\label{EOMb1}
(i\slashed{\partial}-m)b^a_{\mathbf{p}c}(x)=-g\slashed{A}^{(0)ab}(x)b^b_{\mathbf{p}c}(x).
\end{equation}
This classical equation of motion together with \eqref{initb1} enables us to interpret $b_{\mathbf{p}}$ as the spinor describing an anti-quark propagating in the background field $A^{(0)}$.\\

\paragraph{The next to leading order two-point function}

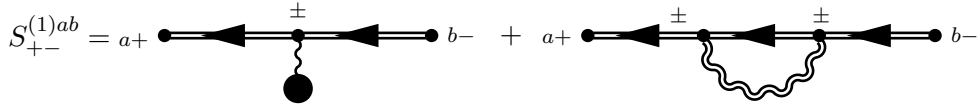
\begin{figure}[h]
\centering
%(along, up)
\begin{fmfgraph*}(100,40)
    \fmfleft{i1}
    \fmfright{o1}
    \fmfbottom{g1}
      \fmf{dbl_plain_arrow}{v1,i1}
      \fmf{dbl_plain_arrow}{o1,v1}
      \fmf{photon,tension=0}{g1,v1}
      \fmflabel{${\scriptstyle\pm}$}{v1}
      \fmflabel{$S^{(1)ab}_{+-}={\scriptstyle a+ }$}{i1}
      \fmflabel{${\scriptstyle b- }$}{o1}
      \fmfdot{i1,v1,o1}
      \fmfv{decor.shape=circle,decor.filled=full,decor.size=10.}{g1}
\end{fmfgraph*}
\hspace{18mm}
\begin{fmfgraph*}(130,40)
    \fmfleft{i1}
    \fmfright{o1}
      \fmf{dbl_plain_arrow}{o1,v2,v1,i1}
      \fmf{dbl_wiggly,tension=0,left}{v2,v1}
      \fmflabel{\vspace{5mm}${\scriptstyle\pm}$}{v1}
      \fmflabel{\vspace{5mm}\hspace{-3mm}${\scriptstyle\pm}$}{v2}
      \fmflabel{$~~+~~{\scriptstyle a+ }$}{i1}
      \fmflabel{${\scriptstyle b- }$}{o1}
      \fmfdot{i1,v1,v2,o1}
\end{fmfgraph*}
\caption{Recursion relation for the dressed quark propagator $S^{(1)ab}_{+-}$ at next to leading order. The double fermion line symbolizes the leading order propagator. \label{recS2}}
\end{figure}
\begin{figure}[ht]
\begin{minipage}[b]{0.5\linewidth}
\centering
\includegraphics[width=.6\textwidth]{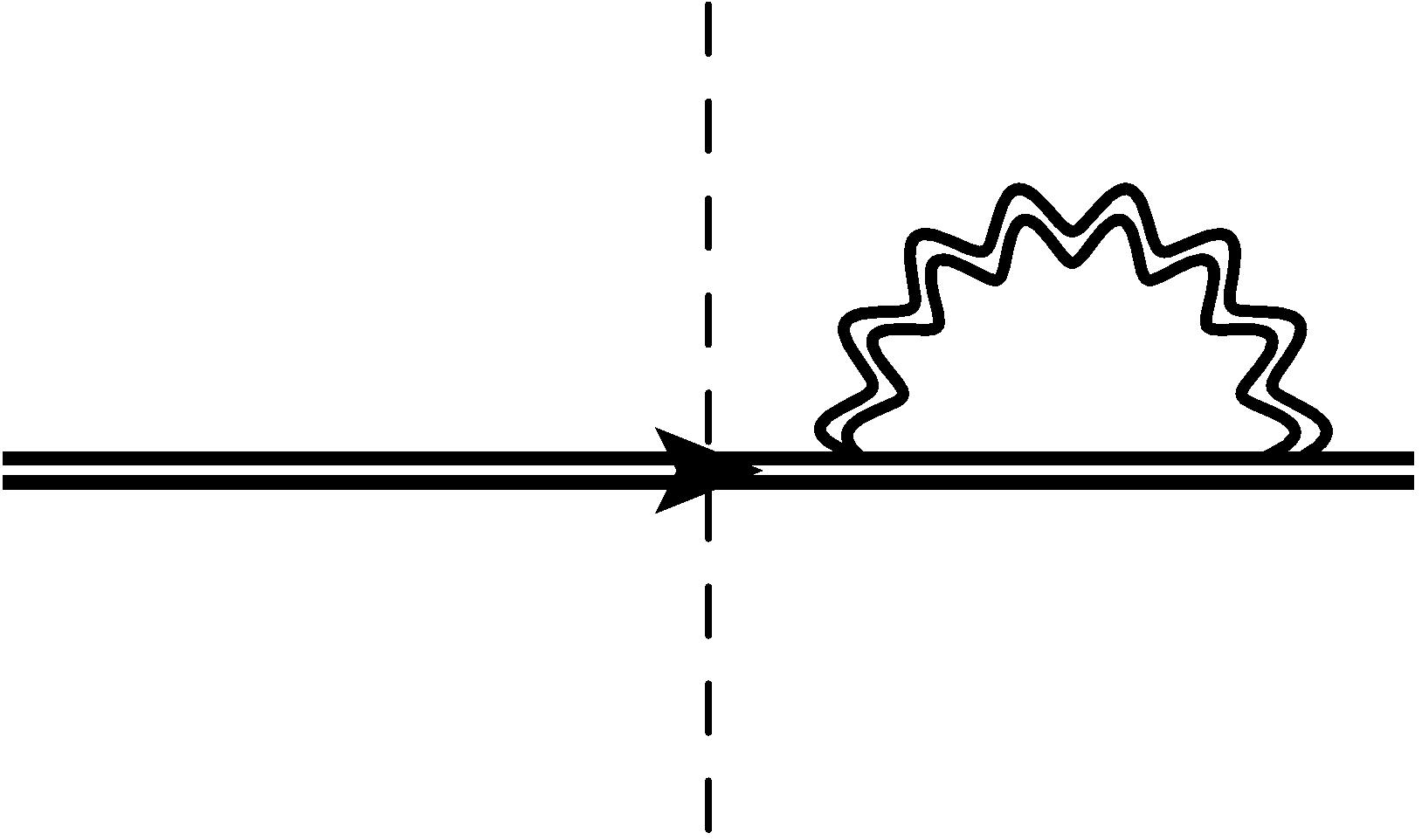}
\end{minipage}
\begin{minipage}[b]{0.5\linewidth}
\centering
\includegraphics[width=.6\textwidth]{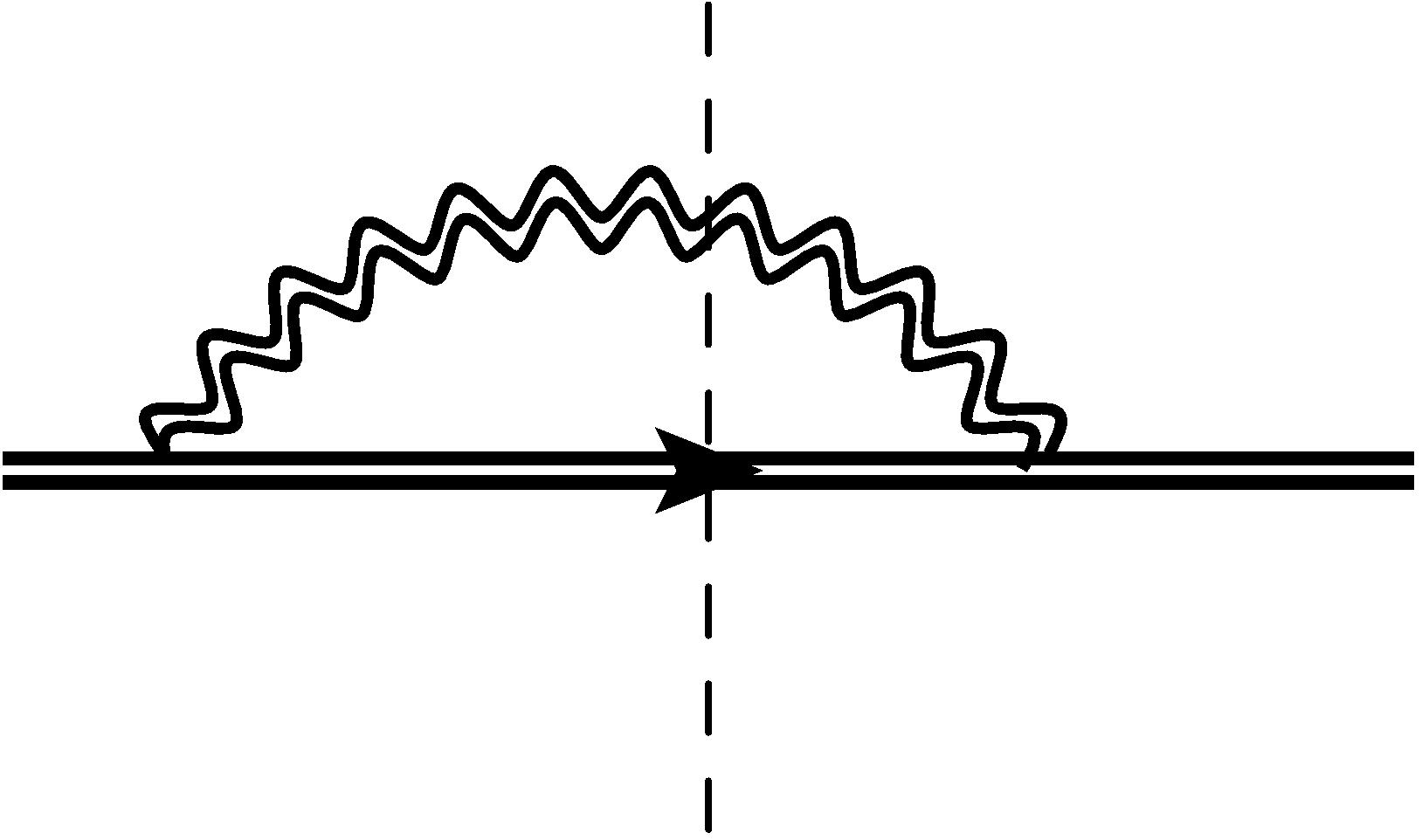}
\end{minipage}
\caption{The two types of contributions to the quark self-energy. The dashed line represent the separation between the amplitude (on its right) and the complex conjugate amplitude (on its left). The nuclei are not represented (they are implicitly contained in the dressing of propagators). The leftmost diagram is a $qg$ virtual fluctuation in the amplitude contributing to $\delta b_{\mathbf{p}}$ (there is the same for $\delta\bar{b}_{\mathbf{p}}$ in the complex conjugate amplitude. The remaining undrawn contributions are the tadpoles. The rightmost diagram is the representation of the connected fluctuation term $\delta_c[b_{\mathbf{p}}\bar{b}_{\mathbf{p}}]$. Both the gluon and the quark within the loop are real, that is, they are on-shell in the infinite past. It is the only contribution to the connected term. \label{fermionfluct}}
\end{figure}

$S^{(0)}_{+-}$ has now a clear interpretation, let us focus on its one-loop correction $S^{(1)}_{+-}$. It receives two kinds of contributions. The first one is a one-loop correction to a gluon one-point function, i.e $A^{(1)}$. At this level it can in principle include the quark tadpole as well. The other contribution is a self-energy correction. Both these contributions are represented on \ref{recS2}. This expansion reads :
\begin{equation}
\label{S2rec}
\begin{split}
S^{(1)}_{+-}(x,y)&=ig\displaystyle{\sum_{\epsilon=\pm1}}\epsilon\int \rmd^4z S^{(0)}_{+\epsilon}(x,z)\slashed{A}^{(1)}(z)S^{(0)}_{\epsilon-}(z,y)\\
&+(ig)^2\displaystyle{\sum_{\epsilon,\eta=\pm1}}\epsilon\eta\int \rmd^4z \rmd^4t S^{(0)}_{+\epsilon}(x,z)\Sigma_{\epsilon\eta}(z,t)S^{(0)}_{\eta-}(t,y).
\end{split}
\end{equation}
$\Sigma$ is a shorthand for the self-energy whose full expression reads :
\begin{equation}
\label{selfenergy}
\Sigma_{\epsilon\eta}(z,t) = T^A\gamma^{\mu}S^{(0)}_{\epsilon\eta}(z,t)T^B\gamma^{\nu}\Delta^{(0)AB}_{\mu\nu~\epsilon\eta}(z,t).
\end{equation}
Performing the sums over the Keldysh contour indices in \eqref{S2rec} and after some algebra thanks to propagator identities \eqref{Keldyshadvret} one can rearrange the terms in the following way :
\begin{equation}
\label{S0split}
\begin{split}
S^{(1)}_{+-}(x,y)&=ig\int \rmd^4z \left[ S^{(0)}_{R}(x,z)\slashed{A}^{(1)}(z)S^{(0)}_{+-}(z,y)+S^{(0)}_{+-}(x,z)\slashed{A}^{(1)}(z)S^{(0)}_{A}(z,y)\right]\\
&+(ig)^2\int \rmd^4z \rmd^4t \left[ \mathcal{S}^{(0)}_{R}(x,z)\Sigma_{R}(z,t)S^{(0)}_{+-}(t,y)\right.\\
&\left.+S^{(0)}_{R}(x,z)\Sigma_{+-}(z,t)S^{(0)}_{A}(t,y)+S^{(0)}_{+-}(x,z)\Sigma_{A}(z,t)S^{(0)}_{A}(t,y)\right],
\end{split}
\end{equation}
where $\Sigma_R\equiv \Sigma_{++}-\Sigma_{+-}$ and $\Sigma_A\equiv \Sigma_{++}-\Sigma_{-+}$. This \emph{definition} is inspired by identities \eqref{Keldyshadvret} and is \emph{not} equivalent to the definition \eqref{selfenergy} where the propagators on the r.h.s are replaced by the retarded or advanced ones respectively. Such definition \eqref{selfenergy} has been stated for the the $++, +-, -+$ and $--$ self-energies only. Rewriting the $S^{(0)}_{+-}$ in the r.h.s of \eqref{S0split} thanks to the representation \eqref{S1repr} suggests the following representation of $S^{(1)}_{+-}$ :
\begin{equation}
\label{S2split}
\mathcal{S}^{(1)ab}_{+-}(x,y)=-\displaystyle{\sum_{s,c}}\int \frac{\rmd^3p}{(2\pi)^3 2p^0}\left[\delta b^{s,a}_{\mathbf{p}c}(x)\bar{b}^{s,b}_{\mathbf{p}c}(y)+b^{s,a}_{\mathbf{p}c}(x)\delta\bar{b}^{s,b}_{\mathbf{p}c}(y)+\delta_c[b^{s,a}_{\mathbf{p}c}(x)\bar{b}^{s,b}_{\mathbf{p}c}(y)]\right].
\end{equation}
with the obvious identifications :
\begin{equation}
\label{S2decomp}
\begin{split}
&\delta b^{s,a}_{\mathbf{p}c}(x)=ig\int \rmd^4y S^{(0)ab}_{R}(x,y)\slashed{A}^{(1)bd}(y)b^{s,d}_{\mathbf{p}c}(y)+(ig)^2\int \rmd^4y \rmd^4z S^{(0)ab}_{R}(x,y)\Sigma^{bd}_{R}(y,z)b^{s,d}_{\mathbf{p}c}(z)\\
&\displaystyle{\sum_{s,c}}\int \frac{\rmd^3p}{(2\pi)^3 2p^0}\delta_c[b^{s,a}_{\mathbf{p}c}(x)\bar{b}^{s,b}_{\mathbf{p}c}(y)]=-(ig)^2\int \rmd^4z \rmd^4t S^{(0)ac}_{R}(x,z)\Sigma^{cd}_{+-}(z,t)S^{(0)db}_{A}(t,y).
\end{split}
\end{equation}
$\delta b_{\mathbf{p}}$ is a fluctuation of $b_{\mathbf{p}}$ occurring in the amplitude (and similarly for $\delta\bar{b}_{\mathbf{p}}$ in the complex conjugate amplitude) and $\delta_c[b_{\mathbf{p}}(x)\bar{b}_{\mathbf{p}}(y)]$ is a connected fluctuation that cannot be split in this way. These terms are represented on figure \ref{fermionfluct} for a physical picture. Our aim is to see the equations of motion satisfied by these quantities.\\

%%\begin{figure}[h]
%%\centering
%(along, up)
%%\begin{fmfgraph*}(120,80)
%%    \fmfleft{i1,ii,i0}
%%    \fmfright{o1,oo,o0}
%%      \fmf{dashes}{i0,i1}
%%      \fmf{dbl_plain_arrow}{ii,a,b,oo}
%%    \fmffreeze
%%      \fmf{dbl_wiggly,tension=0,left}{a,b}
%%      \fmfdot{a,b}
%%\end{fmfgraph*}
%%\hspace{1cm}
%%\begin{fmfgraph*}(150,50)
%%    \fmftop{t1,t2,t3}
%%    \fmfbottom{b1,b2,b3}
%%    \fmfleft{i1}
%%    \fmfright{o1}
%%      \fmf{dashes}{t2,b2}
%%      \fmf{dbl_plain_arrow}{i1,a,b,o1}
%%    \fmffreeze
%%      \fmf{dbl_wiggly,tension=0,left}{a,b}
%%      \fmfdot{a,b}
%%\end{fmfgraph*}
%%\caption{The two types of contributions to the quark self-energy. The dashed line represent the separation between the amplitude (on its right) and the complex conjugate amplitude (on its left). The nuclei are not represented (they are implicitly contained in the dressing of propagators). The leftmost diagram is a $qg$ virtual fluctuation in the amplitude contributing to $\delta b_{\mathbf{p}}$ (there is the same for $\delta\bar{b}_{\mathbf{p}}$ in the complex conjugate amplitude. The remaining undrawn contributions are the tadpoles. The rightmost diagram is the representation of the connected fluctuation term $\delta_c[b_{\mathbf{p}}\bar{b}_{\mathbf{p}}]$. Both the gluon and the quark within the loop are real, that is, they are on-shell in the infinite past. It is the only contribution to the connected term. \label{fermionfluct}}
%%\end{figure}

\begin{figure}[h]
\centering
%(along, up)
\begin{fmfgraph*}(150,25)
    \fmfright{a,b}
    \fmfleft{c,d}
      \fmf{dbl_plain_arrow}{c,o,a}
    \fmffreeze
      \fmf{dbl_wiggly,tension=0}{d,o}
      \fmfdot{o}
      \fmflabel{$p \rightarrow$}{c}
      \fmflabel{$k \rightarrow$}{d}
\end{fmfgraph*}
\caption{Diagrammatic representation of $\xi^{(\mathbf{k}+)}_{\mathbf{p}}$. The time axis runs from left to right. \label{XIvertex}}
\end{figure}
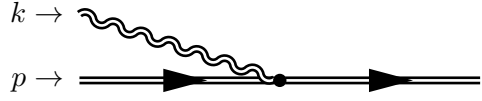

Let us begin with the connected term. Plugging the representations \eqref{D1repr} of $\Delta^{(0)}_{+-}$ and \eqref{S1repr} of $S^{(0)}_{+-}$ into the definition \eqref{selfenergy} of $\Sigma_{+-}$ gives (with sum over initial spins and colors understood) gives :
\begin{equation}
\begin{split}
\Sigma^{cd}_{+-}(z,t) &=-\int \frac{\rmd^3k}{(2\pi)^3 2k^0}\frac{\rmd^3p}{(2\pi)^3 2p^0}\slashed{\alpha}^{ce}_{\mathbf{k}}(z)b^{e}_{\mathbf{p}}(z)\bar{b}^{f}_{\mathbf{p}}(t)\slashed{\alpha}^{fd*}_{\mathbf{k}}(t).
\end{split}
\end{equation}
One has to be careful with the notation $\slashed{\alpha}^{*}_{\mathbf{k}}$. This is a shorthand for $\gamma^{\mu}\alpha^{A*}_{\mathbf{k}\mu}T^A$ which is not the complex conjugate of $\slashed{\alpha}_{\mathbf{k}}$. Plugging this into the full expression \eqref{S2decomp} for the connected fluctuation allows the following representation :
\begin{equation}
\label{crossrepr}
\displaystyle{\sum_{s,c}}\int \frac{\rmd^3p}{(2\pi)^3 2p^0}\delta_c[b^{s,a}_{\mathbf{p}c}(x)\bar{b}^{s,b}_{\mathbf{p}c}(y)]=\displaystyle{\sum_{s,c}}\displaystyle{\sum_{\lambda,C}} \int \frac{\rmd^3k}{(2\pi)^3 2k^0}\frac{\rmd^3p}{(2\pi)^3 2p^0} \xi^{a(\mathbf{k}+)}_{\mathbf{p}s(\lambda),cC}(x)\bar{\xi}^{b(\mathbf{k}-)}_{\mathbf{p}s(\lambda),cC}(y),
\end{equation}
where we have defined :
\begin{equation}
\label{xidef}
\begin{split}
\xi^{a(\mathbf{k}+)}_{\mathbf{p}s(\lambda),cC}(x)&= ig\int \rmd^4y ~S^{(0)ab}_{R}(x,y)(\slashed{\alpha}^{(\lambda)}_{\mathbf{k}C}(y))_{bd}b^{s,d}_{\mathbf{p}c}(y)\\
\bar{\xi}^{a(\mathbf{k}-)}_{\mathbf{p}s(\lambda),cC}(x)&=ig\int \rmd^4y ~\bar{b}^{s,d}_{\mathbf{p}c}(y)(\slashed{\alpha}^{(\lambda)*}_{\mathbf{k}C}(y))_{db}S^{(0)ba}_{A}(y,x).
\end{split}
\end{equation}
The sign $+$ or $-$ refers to the energy sign of the gluon : $+$ represents an outgoing gluon with positive energy or an incoming one with a negative energy, $-$ represents an outgoing gluon with negative energy or an incoming one with positive energy. The notation with a bar is consistent in the sense that $\bar{\xi}^{(\mathbf{k}-)}_{\mathbf{p}}$ is the Dirac conjugate of $\xi^{(\mathbf{k}+)}_{\mathbf{p}}$. $\xi^{(\mathbf{k}+)}_{\mathbf{p}}$ represents the retarded propagation of the initial quark $b_{\mathbf{p}}$ which has absorbed an initial gluon $\alpha_{\mathbf{k}}$. It is represented by the diagram \ref{XIvertex}. For future convenience, it will be also useful to define the two following spinors that are related by Dirac conjugation :
\begin{equation}
\label{xidef2}
\begin{split}
\xi^{a(\mathbf{k}-)}_{\mathbf{p}s(\lambda),cC}(x)&= ig\int \rmd^4y ~S^{(0)ab}_{R}(x,y)(\slashed{\alpha}^{(\lambda)*}_{\mathbf{k}C}(y))_{bd}b^{s,d}_{\mathbf{p}c}(y)\\
\bar{\xi}^{a(\mathbf{k}+)}_{\mathbf{p}s(\lambda),cC}(x)&=ig\int \rmd^4y ~\bar{b}^{s,d}_{\mathbf{p}c}(y)(\slashed{\alpha}^{(\lambda)}_{\mathbf{k}C}(y))_{db}S^{(0)ba}_{A}(y,x).
\end{split}
\end{equation}
From \eqref{xidef}, acting on $\xi^{(\mathbf{k}+)}_{\mathbf{p}}(x)$ with the Dirac operator in presence of the background field $A^{(0)}$ whom $S^{(0)}_{R}(x,y)$ is a Green function one gets the following equation of motion :
\begin{equation}
\label{EOMxi}
\left(\delta_{ab}(i\slashed{\partial}-m)+g\slashed{A}^{(0)}_{ab}(x)\right)\xi^{b(\mathbf{k}+)}_{\mathbf{p}}(x)=-g\slashed{\alpha}^{ab}_{\mathbf{k}}b^b_{\mathbf{p}}(x).
\end{equation}
This equation of motion is the linearized Dirac equation describing the propagation of a fermionic fluctuation in a perturbed background field. \\

The last missing piece for which we want to write an equation of motion is $\delta b_{\mathbf{p}}$. The derivation of the equation of motion satisfied by $\delta b_{\mathbf{p}}$ easily follows by applying the Dirac operator in the background field $A^{(0)}$ on its definition \eqref{S2decomp}. This reads :
\begin{equation}
\label{EOMdeltab}
\begin{split}
\left(i\slashed{\partial}-m+g\slashed{A}^{(0)}(x)\right)\delta b_{\mathbf{p}}(x)=-g\slashed{A}^{(1)}(x)b_{\mathbf{p}}(x)-ig^2\int \rmd^4y~ \Sigma_{R}(x,y)b_{\mathbf{p}}(y).
\end{split}
\end{equation}\\

Our work seems to be finished. However the following step of our formalism to be seen in the next section breaks down with this expression. The trouble arise from $\Sigma_{R}$ that can be split into a more physical way as we shall see. The retarded self-energy can be seen as the sum of two contributions : a virtual quark fluctuation carried by a causal gluon plus a virtual gluon fluctuation carried by a causal quark. The problem is the same as the quark tadpole that has been omitted in the gluon one-point function correction $A^{(1)}$. In order to be consistent one has to also forget about the quark fluctuation of the causal gluon as well. Such contribution does not bring large logarithms of $x$ and can be dropped if we focus on this region of phase space. Let us rewrite $\Sigma_{R}$ in another form to see this. One way to rewrite $\Sigma_{R}$ is :
\begin{equation}
\label{sigmaR1}
\begin{split}
\Sigma_{R}(x,y)&=\Sigma_{++}(x,y)-\Sigma_{+-}(x,y)\\
&=T^A\gamma^{\mu}\left[S^{(0)}_{++}(x,y)\Delta^{(0)AB}_{\mu\nu~++}(x,y)-S^{(0)}_{+-}(x,y)\Delta^{(0)AB}_{\mu\nu~+-}(x,y)\right]\gamma^{\nu}T^B\\
&=T^A\gamma^{\mu}\left[\left(S^{(0)}_{R}(x,y)+S^{(0)}_{+-}(x,y)\right)\Delta^{(0)AB}_{\mu\nu~++}(x,y)+\left(S^{(0)}_{R}(x,y)-S^{(0)}_{++}(x,y)\right)\Delta^{(0)AB}_{\mu\nu~+-}(x,y)\right]\gamma^{\nu}T^B\\
&=T^A\gamma^{\mu}\left[S^{(0)}_{R}(x,y)\Delta^{(0)AB}_{\mu\nu~++}(x,y)+S^{(0)}_{+-}(x,y)\Delta^{(0)AB}_{\mu\nu~R}(x,y)\right]\gamma^{\nu}T^B\\
&=T^A\gamma^{\mu}\left[S^{(0)}_{R}(x,y)\Delta^{(0)AB}_{\mu\nu~-+}(x,y)+S^{(0)}_{+-}(x,y)\Delta^{(0)AB}_{\mu\nu~R}(x,y)\right]\gamma^{\nu}T^B.
\end{split}
\end{equation}
In the previous last step, we used the fact that multiplication by $S^{(0)}_{R}(x,y)$ enforces the two coordinates $x$ and $y$ to be ordered in time. Thus $\Delta^{(0)}_{++}$, which is the Feynman propagator, can be replaced by $\Delta^{(0)}_{-+}$ since only the $<A(x)A(y)>$ ordering survives to the multiplication with $S^{(0)}_{R}(x,y)$. One could also have reversed the roles of gluons and quarks in the previous rewriting exercise which would give :
\begin{equation}
\label{sigmaR2}
\Sigma_{R}(x,y)=T^A\gamma^{\mu}\left[S^{(0)}_{-+}(x,y)\Delta^{(0)AB}_{\mu\nu~R}(x,y)+S^{(0)}_{R}(x,y)\Delta^{(0)AB}_{\mu\nu~+-}(x,y)\right]\gamma^{\nu}T^B.
\end{equation}
By averaging these two forms \eqref{sigmaR1} and \eqref{sigmaR2} for $\Sigma_R$, one gets :
\begin{equation}
\begin{split}
\Sigma_{R}(x,y)&=\frac{1}{2}T^A\gamma^{\mu}\left[\left(S^{(0)}_{-+}(x,y)+S^{(0)}_{+-}(x,y)\right)\Delta^{(0)AB}_{\mu\nu~R}(x,y)\right.\\
\left.+S^{(1)}_{R}(x,y)\left(\Delta^{(0)AB}_{\mu\nu~-+}(x,y)+\Delta^{(0)AB}_{\mu\nu~+-}(x,y)\right)\right]\gamma^{\nu}T^B.
\end{split}
\end{equation}
The first term is a fermion loop which merges onto the retarded gluon. This is the term we will not care about anymore. The second term is the reverse situation : a gluon loop which merges onto the retarded fermion propagator. This term will be the only one of interest. Using the representation \eqref{D1repr} for the gluon propagator it will be convenient for future purposes to write :
\begin{equation}
\label{quarkselfenergy}
\begin{split}
\Sigma_{R}(x,y)&\rightarrow\frac{1}{2}T^A\gamma^{\mu}S^{(0)}_{R}(x,y)\gamma^{\nu}T^B\int\frac{\rmd^3k}{(2\pi)^3 2k^0}\left[\alpha^{A}_{\mathbf{k}\mu}(x)\alpha^{B*}_{\mathbf{k}\nu}(y)+\alpha^{A*}_{\mathbf{k}\mu}(x)\alpha^{B}_{\mathbf{k}\nu}(y)\right].
\end{split}
\end{equation}

All the LO and NLO Green functions have been written in terms of various fields that satisfies identified equations of motion. Solving them can be a hard task but one can always write formal solutions in the general case. This is what we discuss in the next section.

\section{Recursion relations}
\indent

From the various equations of motion of the previous section, one can write formal solutions known as \emph{Green's formulas}. The spirit of a Green's formula is to write the value of a field at some point in space-time in terms of an initial condition, that is the value of the field (and possibly its derivatives) on some suitably chosen boundary surface. We shall see that the NLO gluon and quark spectra are actually given by a perturbation of the LO initial conditions. First we write the Green's formulas for the LO and NLO spectra and then we look at the differences between the LO and NLO structure and try to establish some formal relation among them. The LO to NLO recursion relations will immediately follow.

\subsection{Formal solutions of Green functions\label{Greenfomulae}}
\indent

In this part we work out the general structure of gluons and quarks Green's formulas. Let us start with gluonic equations of motion. We refer to elementary fields, the fields $A^{(0)}$, $\alpha_{\mathbf{k}}$ and $A^{(1)}$ that are respectively solutions of differential equations \eqref{EOMA0}, \eqref{EOMalpha} and \eqref{EOMA1}. These equations of motion can all be summarized into the following form :
\begin{equation}
\label{genericEOMA}
(g_{\mu\nu}\Box-\partial_{\mu}\partial_{\nu}+\frac{n_{\mu}n_{\nu}}{\xi})A^{A\nu}(x)=\mathcal{F}^A_{\mu}[A(x),x].
\end{equation}
$A(x)$ denotes one of these elementary fields and $\mathcal{F}$ is a local functional of the field $A$ but also of possible other fields that bring another space-time position dependence not arising from the field $A$. Although it does not matter at this level, we can notice that $\mathcal{F}$ depends only on gluonic fields and not on quark fields, so that $\mathcal{F}=\mathcal{F}[A(x)]$ only. This would have been the case if we had kept the quark loop in the gluon tadpole $A^{(1)}$. The free Green function $\Delta_0$ corresponding to the above l.h.s operator satisfies :
\begin{equation}
\label{freegluonGF}
(g^{\mu\nu}\Box-\partial^{\mu}\partial^{\nu}+\frac{n^{\mu}n^{\nu}}{\xi})\Delta_{0\nu\rho}(x-y)=i\delta_{\rho}^{\mu}\delta^{(4)}(x-y).
\end{equation}
The color structure of $\Delta_0$ is $\delta^{AB}$ and has been omitted : $\Delta_0^{AB}\rightarrow \delta^{AB}\Delta_0$. For the moment the procedure does not require a given prescription for the Green function. It can be the retarded one but also advanced, Feynman, anti-Feynman or any linear combination of them (with a unit sum of coefficients). To derive the corresponding Green's formula, one multiplies \eqref{genericEOMA} on the left by the free Green function and multiply on the right the adjoint of \eqref{freegluonGF} with $A$. This order is conventional one could have made the converse. This leads to the following system :
\begin{equation}
\left\{
    \begin{array}{ll}
\Delta^{\mu\rho}_{0}(x-y)(g_{\rho\sigma}\overrightarrow{\Box_y}-\overrightarrow{\partial_{y\rho}\partial_{y\sigma}}+\frac{n_{\rho}n_{\sigma}}{\xi})A^{A\sigma}(y)&=\Delta^{\mu\rho}_{0}(x-y)\mathcal{F}^{A\rho}[A(y),y]\\
\Delta^{\mu\rho}_{0}(x-y)(g_{\rho\sigma}\overleftarrow{\Box_y}-\overleftarrow{\partial_{y\rho}\partial_{y\sigma}}+\frac{n_{\rho}n_{\sigma}}{\xi})A^{A\sigma}(y)&=i\delta^{(4)}(x-y)A^{A\mu}(y).
    \end{array}
\right.
\end{equation}
Let us integrate the difference of the second equation minus the first one, times an overall $-i$ factor, over $y$ on some simply connected, four-dimensional, space-time domain $\Omega$. This gives :
\begin{equation}
\begin{split}
A^{A\mu}(x)=-i\int_{\Omega}\rmd^4y\Delta^{\mu\rho}_{0}(x-y)&\left(g_{\rho\sigma}\overleftarrow{\Box}_y-\overleftarrow{\partial_{y\rho}\partial_{y\sigma}}-g_{\rho\sigma}\overrightarrow{\Box}_y+\overrightarrow{\partial_{y\rho}\partial_{y\sigma}}\right)A^{A\sigma}(y)\\
&-i\int_{\Omega}\rmd^4y\Delta^{\mu\rho}_{0}(x-y)\mathcal{F}^{A\rho}[A(y),y].
\end{split}
\end{equation}
The last term in the integrand is a bulk term whose role will be precised later. The first term is a total derivative\footnote{\samepage For arbitrary functions $f$ and $g$, one has $$f(\overrightarrow{\Box}-\overleftarrow{\Box})g=\partial_{\mu}(f\overleftrightarrow{\partial^{\mu}}g)$$ and for two vector fields $A$ and $B$, $$A^{\mu}(\overrightarrow{\partial_{\mu}\partial_{\nu}}-\overleftarrow{\partial_{\mu}\partial_{\nu}})B^{\nu}=\frac{1}{2}\left(\partial_{\mu}(A^{\mu}\overleftrightarrow{\partial^{\nu}}B^{\nu})+\partial_{\nu}(A^{\mu}\overleftrightarrow{\partial^{\mu}}B^{\nu})\right).$$ The symbol $\overleftrightarrow{\partial}$ is the antisymmetric derivative defined as $$\overleftrightarrow{\partial}=\overrightarrow{\partial}-\overleftarrow{\partial}.$$}. Writing the integral of the total derivative as a three-dimensional surface integral over $\partial\Omega$, the boundary of $\Omega$, with a normal vector $l^{\mu}$ and embedded with the measure $\rmd^3\sigma_{y}$ gives :
\begin{equation}
\label{GFgluons}
\begin{split}
A^{A\mu}(x)&=-\frac{i}{2}\oint_{\partial\Omega}\rmd^3\sigma_{y}\left[2\Delta^{\mu\rho}_{0}(x-y)\overleftrightarrow{(l\cdot\partial_y)}A^{A}_{\rho}(y)-\Delta^{\mu\rho}_{0}(x-y)l_{\rho}\overleftrightarrow{\partial_{y\sigma}}A^{A\sigma}(y)-\Delta^{\mu\rho}_{0}(x-y)\overleftrightarrow{\partial_{y\rho}}(l\cdot A^{A}(y))\right]\\
&-i\int_{\Omega}\rmd^4y~\Delta^{\mu\rho}_{0}(x-y)\mathcal{F}^{A\rho}[A(y),y].
\end{split}
\end{equation}
This is the required formula. Note that this expression is no longer singular as $\xi\rightarrow 0$. Then one can enforce the gauge constrain $n\cdot A=0$ at this level without any problem.\\

Let us write the Green's formula for quarks. The general structure of equations \eqref{EOMb1}, \eqref{EOMxi} and \eqref{EOMdeltab} that respectively govern the dynamics of $b_{\mathbf{p}}$, $\xi^{(\mathbf{k}\pm)}_{\mathbf{p}}$ and $\delta b_{\mathbf{p}}$, generically denoted by a spinor $\psi$ which can be any of them, reads as follow :
\begin{equation}
\begin{split}
\left(i\slashed{\partial}-m\right)\psi(x)=\mathcal{G}[\psi(x),x]
\end{split}
\end{equation}
where $\mathcal{G}[\psi(x),x]$ is again a local functional of the $\psi$ field and possibly various other fields. The Green's formula is obtained in the same way as for gluons. The Green function $S_0$ of the free Dirac operators satisfies the equation $S_0(x-y)(i\overleftarrow{\slashed{\partial}}+m)=i\delta^{(4)}(x-y)$ and has a trivial color structure $\delta^{ab}$ which is understood. Following the same procedure as for gluons, one gets the system :
\begin{equation}
\left\{
    \begin{array}{ll}
S_0(x-y)\left(i\overrightarrow{\slashed{\partial}_y}-m\right)\psi(y)&=S_0(x-y)\mathcal{G}[\psi(y),y]\\
S_0(x-y)\left(i\overleftarrow{\slashed{\partial}_y}+m\right)\psi(y)&=i\delta^{(4)}(x-y)\psi(y).
    \end{array}
\right.
\end{equation}
For quarks, one has to integrate the sum of these two equations over $\Omega$ times $-i$ and this gives :
\begin{equation}
\psi(y)=\int_{\Omega}\rmd^4y\left[S_0(x-y)\left(\overrightarrow{\slashed{\partial}_y}+\overleftarrow{\slashed{\partial}_y}\right)\psi(y)+iS_0(x-y)\mathcal{G}[\psi(y),y]\right].
\end{equation}
The total derivative is here obvious in the first term. This last expression is written as the following boundary integral :
\begin{equation}
\label{GFquarks}
\psi(y)=\oint_{\partial\Omega}\rmd^3\sigma_{y}S_0(x-y)\slashed{l}\psi(y)+i\int_{\Omega}\rmd^4y~S_0(x-y)\mathcal{G}[\psi(y),y].
\end{equation}

Solutions for gluonic and quark fields \eqref{GFgluons} and \eqref{GFquarks} are general. To go further one has to find an explicit form for the bulk term. We shall see the emergence of the recursion relations.

\subsection{Boundary operators}
\indent

For all gluon fields $A^{(0)}$, $\alpha_{\mathbf{k}}$ and $A^{(1)}$ equation \eqref{GFgluons} shows us that the field at some point in space-time depends linearly in the value of the field itself and its first derivatives on the surface $\partial\Omega$ that will be referred to as the initial surface. Let us examine the structure of the functional $\mathcal{F}$ defined in \eqref{genericEOMA} for $A^{(0)}$, $\alpha_{\mathbf{k}}$ and $A^{(1)}$ respectively. Its explicit expression can be read directly from \eqref{EOMA0}, \eqref{EOMalpha} and \eqref{EOMA1} respectively. Let us find the corresponding relation between \eqref{EOMA0} and \eqref{EOMalpha} for instance. Equation \eqref{EOMalpha} for $\alpha_{\mathbf{k}}$ is the one obtained by perturbing $A^{(0)}$ in \eqref{EOMA0}. It is natural to expect that a perturbation merges from a perturbed initial condition on the boundary surface. The gluonic Green's formula \eqref{GFgluons} having a dependence on the gluon field's components and first derivatives on the initial surface $\partial\Omega$, specifying the value of the perturbation and of its first derivatives on the initial surface will uniquely determine its "future" evolution\footnote{For the moment considerations are more general as long as we neither specify the prescription of the free Green functions in Green's formulas nor the surface $\partial\Omega$. At this level the terms "future" or "initial" are abusive but this is in view of the upcoming causal interpretation.}. From these considerations we define the shift operator acting on the initial condition :
\begin{equation}
\oint_{\partial\Omega} \rmd^3\sigma_u\left[\delta A\cdot\mathbb{T}^{A}_u\right]=\oint_{\partial\Omega} \rmd^3\sigma_u\left[\delta A^{A}_{\mu}(u)\frac{\partial}{\partial A^{(0)A}_{\mu}(u)}+(\partial_{\nu}\delta A^A_{\mu}(u))\frac{\partial}{\partial (\partial_{\nu}A^{(0)A}_{\mu}(u))}\right].
\end{equation}
To see how this operator acts, let us write, thanks to the explicit form of the functional $\mathcal{F}$ in the r.h.s of \eqref{EOMA0}, the Green's formula \eqref{GFgluons} corresponding to $A^{(0)}$ :
\begin{equation}
\label{GFA0}
\begin{split}
A^{(0)A\mu}(x)&=-\frac{i}{2}\oint_{\partial\Omega}\rmd^3\sigma_{y}\left[2\Delta^{\mu\rho}_{0}(x-y)\overleftrightarrow{(l\cdot\partial_y)}A^{A}_{\rho}(y)-\Delta^{\mu\rho}_{0}(x-y)l_{\rho}\overleftrightarrow{\partial_{y\sigma}}A^{A\sigma}(y)-\Delta^{\mu\rho}_{0}(x-y)\overleftrightarrow{\partial_{y\rho}}(l\cdot A^{A}(y))\right]\\
&-i\int_{\Omega}\rmd^4y~\Delta^{\mu\rho}_{0}(x-y)\left[-\mathcal{J}^A_{\rho}(x)+\frac{\partial V}{\partial A^{A\rho}}[A^{(0)}(x)]-\partial_{\nu}\frac{\partial V}{\partial \partial_{\nu}A^{A\rho}}[A^{(0)}(x)]\right].
\end{split}
\end{equation}
Acting on the initial condition - i.e. the boundary term - the shift operator merely replaces $A^{(0)}$ by $\delta A$ on the surface. Since the shift operator is a linear differential operator it perturbs the bulk term giving exactly the r.h.s of \eqref{EOMalpha}  and by the way the bulk term of the Green's formula corresponding to $\alpha_{\mathbf{k}}$ but with $\delta A$ instead of $\alpha_{\mathbf{k}}$. Its action on $\mathcal{J}$ is zero. Then one can identify $\delta A$ and $\alpha_{\mathbf{k}}$ and one has :
\begin{equation}
\label{A0toalpha}
\alpha^{(\lambda)A}_{\mathbf{k}C\mu}(x)=\oint_{\partial\Omega} \rmd^3\sigma_u\left[\alpha^{(\lambda)}_{\mathbf{k}C}\cdot\mathbb{T}^{A}_u\right]A^{(0)A}_{\mu}(x).
\end{equation}\\

$A^{(1)}$ is obtained from $A^{(0)}$ in a similar way. The bulk term is read from \eqref{EOMA1}. The first piece has the same form as the r.h.s of \eqref{EOMalpha}. This ensures that $A^{(1)}$ will receive a contribution like \eqref{A0toalpha} but with $A^{(1)}$ instead of $\alpha_{\mathbf{k}}$ in the shift operator. To get the remaining piece of the $A^{(1)}$'s bulk term from $A^{(0)}$ one has to perform a double derivative. If one applies the shift operator \eqref{A0toalpha} and its complex conjugate on $A^{(0)}$ and sum over the initial momentum $\mathbf{k}$, polarization $\lambda$ and color $C$ we get the required term. This double derivative contribution does not affect the boundary term in \eqref{GFA0} since this last one is linear in the fields. Gathering these two pieces together gives :
\begin{equation}
\label{A0toA1}
A^{(1)A}_{\mu}(x)=\left[\oint_{\partial\Omega} \rmd^3\sigma_u\left[A^{(1)}\cdot\mathbb{T}^{A}_u\right]+\frac{1}{2} \displaystyle{\sum_{\lambda,C}}\int \frac{\rmd^3k}{(2\pi)^3 2k^0}\oint_{\partial\Omega} \rmd^3\sigma_u \rmd^3\sigma_v\left[\alpha^{(\lambda)}_{\mathbf{k}C}\cdot\mathbb{T}^{A}_u\right]\left[\alpha^{(\lambda)*}_{\mathbf{k}C}\cdot\mathbb{T}^{A}_v\right]\right]A^{(0)A}_{\mu}(x).
\end{equation}
Relations \eqref{A0toalpha} and \eqref{A0toA1} relate the LO and NLO elementary fields and will be essential ingredients for recursion relations.\\

There are analogous relations for fermionic fields $b_{\mathbf{p}}$, $\xi^{(\mathbf{k}+)}_{\mathbf{p}}$ and $\delta b_{\mathbf{p}}$. Let us first look for the relation between $b_{\mathbf{p}}$ and $\xi^{(\mathbf{k}+)}_{\mathbf{p}}$. Plugging \eqref{EOMb1} and \eqref{EOMxi} into the general fermionic Green's formula \eqref{GFquarks} gives :
\begin{equation}
\label{GFbandxi}
\begin{split}
b_{\mathbf{p}}(x)&=\oint_{\partial\Omega}\rmd^3\sigma_{y}S_0(x-y)\slashed{l}b_{\mathbf{p}}(y)-ig\int_{\Omega}\rmd^4y~S_0(x-y)\slashed{A}^{(0)}(y)b_{\mathbf{p}}(y)\\
\xi^{(\mathbf{k}+)}_{\mathbf{p}}(x)&=\oint_{\partial\Omega}\rmd^3\sigma_{y}S_0(x-y)\slashed{l}\xi^{(\mathbf{k}+)}_{\mathbf{p}}(y)-ig\int_{\Omega}\rmd^4y~S_0(x-y)\left[\slashed{A}^{(0)}(y)\xi^{(\mathbf{k}+)}_{\mathbf{p}}(y)+\slashed{\alpha}_{\mathbf{k}}(y)b_{\mathbf{p}}(y)\right].
\end{split}
\end{equation}
These two expressions suggest the definition of a shift operator that acts on the initial value of $\slashed{l}b_{\mathbf{p}}$ and not on its derivatives\footnote{As long as we do not specify the $l$ vector there is no reason for $\slashed{l}$ to be invertible a priori. This is somewhat conventional to derive with respect to $\slashed{l}b_{\mathbf{p}}$ rather than $b_{\mathbf{p}}$. It is more convenient since it explicitly reduces the number of initial degrees of freedom if $\slashed{l}$ is not invertible}. This operator reads :
\begin{equation}
\oint_{\partial\Omega} \rmd^3\sigma_u\left[\delta \psi\cdot\mathbb{T}^{b}_u\right]=\displaystyle{\sum_{s,c}}\int \frac{\rmd^3p}{(2\pi)^3 2p^0}\oint_{\partial\Omega} \rmd^3\sigma_u\left[\slashed{l}\delta \psi^s_{\mathbf{p}c}(u)\frac{\partial}{\partial \slashed{l}b^s_{\mathbf{p}c}(u)}+\delta \bar{\psi}^s_{\mathbf{p}c}(u)\slashed{l}\frac{\partial}{\partial \bar{b}^s_{\mathbf{p}c}(u)\slashed{l}}\right].
\end{equation}
Note that we are not dealing with Grassmann numbers and the side on which the derivative acts on some given functional of quark fields does not matter. The perturbation $\delta \psi$ necessarily carries the same quantum numbers as $b_{\mathbf{p}}$, that is a spatial momentum, a spin and a color. When this operator acts on the first equation of \eqref{GFbandxi}, it replaces $b_{\mathbf{p}}$ by $\delta\psi$. The derivative with respect to $\bar{b}_{\mathbf{p}}$ does not contribute when acting on $b_{\mathbf{p}}$ and has been conventionally inserted in this operator. Thus, the shift operator $\xi^{(\mathbf{k}+)}\cdot\mathbb{T}^{b}_u$ acting on the first line of \eqref{GFbandxi} almost gives the second line, except the last term in the r.h.s. To get the additional bulk term of the second equation \eqref{GFbandxi} one has to add the contribution of the operator \eqref{A0toalpha} that changes $A^{(0)}$ into $\alpha_{\mathbf{k}}$. Of course this operator plays no role on the surface term. The relation between $b_{\mathbf{p}}$ and $\xi^{(\mathbf{k}+)}_{\mathbf{p}}$ then reads :
\begin{equation}
\label{btoxi}
\begin{split}
\xi^{(\mathbf{k}+)}_{\mathbf{p}}(x)&=\oint_{\partial\Omega} \rmd^3\sigma_u\left[\alpha_{\mathbf{k}}\cdot\mathbb{T}^{A}_u+\xi^{(\mathbf{k}+)}\cdot\mathbb{T}^{b}_u\right]b_{\mathbf{p}}(x).
\end{split}
\end{equation}\\

To complete the recursion relations between elementary fields one has to find the one matching $b_{\mathbf{p}}$ to $\delta b_{\mathbf{p}}$. The first interaction term in equation \eqref{EOMdeltab} is similar to the one in \eqref{EOMxi} which ensures a contribution of the same form as \eqref{btoxi} but with $b_{\mathbf{p}}$ replaced by $\delta b_{\mathbf{p}}$ instead of $\xi^{(\mathbf{k}+)}_{\mathbf{p}}$ and $A^{(0)}$ by $A^{(1)}$ instead of $\alpha_{\mathbf{k}}$. But there is also a self-energy term whose final form is given by \eqref{quarkselfenergy}. Let us rewrite the self-energy term in \eqref{EOMdeltab} in the form \eqref{quarkselfenergy} and together with \eqref{xidef} and \eqref{xidef2}, one has :
\begin{equation}
\begin{split}
-ig^2\int\rmd^4y~\Sigma_{R}(x,y)b_{\mathbf{p}}(y)=-\frac{g}{2}\int\frac{\rmd^3k}{(2\pi)^3 2k^0}\left[\slashed{\alpha}_{\mathbf{k}}\xi^{(\mathbf{k}-)}_{\mathbf{p}}(x)+\slashed{\alpha}^*_{\mathbf{k}}\xi^{(\mathbf{k}+)}_{\mathbf{p}}(x)\right].
\end{split}
\end{equation}
It is now straightforward to write this term as shift operators acting on LO fields since such relations are already known thanks to \eqref{A0toalpha} and \eqref{btoxi} :
\begin{equation}
\begin{split}
ig\int&\rmd^4y~\Sigma_{R}(x,y)b_{\mathbf{p}}(y)=\frac{1}{2}\int\frac{\rmd^3k}{(2\pi)^3 2k^0}\oint_{\partial\Omega} \rmd^3\sigma_u\rmd^3\sigma_v\\
&\times\left[\left[\alpha_{\mathbf{k}}\cdot\mathbb{T}^{A}_u\right]\slashed{A}^{(0)}(x)\left[\alpha^*_{\mathbf{k}}\cdot\mathbb{T}^{A}_v+\xi^{(\mathbf{k}-)}\cdot\mathbb{T}^{b}_v\right]b_{\mathbf{p}}(x)+\left[\alpha^*_{\mathbf{k}}\cdot\mathbb{T}^{A}_u\right]\slashed{A}^{(0)}(x)\left[\alpha_{\mathbf{k}}\cdot\mathbb{T}^{A}_v+\xi^{(\mathbf{k}+)}\cdot\mathbb{T}^{b}_v\right]b_{\mathbf{p}}(x)\right]
\end{split}
\end{equation}
but since $A^{(0)}$ does not depend on any spinor field in its initial condition, any shift operator with respect to $b_{\mathbf{p}}$ acting on $A^{(0)}$ gives zero and the above formula can be written as :
\begin{equation}
\begin{split}
ig\int\rmd^4y~\Sigma_{R}(x,y)b_{\mathbf{p}}(y)=\frac{1}{2}&\int\frac{\rmd^3k}{(2\pi)^3 2k^0}\oint_{\partial\Omega} \rmd^3\sigma_u\rmd^3\sigma_v\\
&\times\left[\alpha_{\mathbf{k}}\cdot\mathbb{T}^{A}_u+\xi^{(\mathbf{k}+)}\cdot\mathbb{T}^{b}_u\right]\left[\alpha^*_{\mathbf{k}}\cdot\mathbb{T}^{A}_v+\xi^{(\mathbf{k}-)}\cdot\mathbb{T}^{b}_v\right]\slashed{A}^{(0)}(x)b_{\mathbf{p}}(x).
\end{split}
\end{equation}
$-g\slashed{A}^{(0)}(x)b_{\mathbf{p}}(x)$ is precisely what is in the bulk term of the Green's formula corresponding to $b_{\mathbf{p}}$ according to \eqref{GFbandxi}. Moreover, this double derivative operator gives zero when it acts on the boundary term in the Green's formula for $\delta b_{\mathbf{p}}$. Now we have all the pieces we need to write $\delta b_{\mathbf{p}}$ as a shift operator acting on LO fields. The self-energy term together with the term of the same form as \eqref{btoxi} gives :
\begin{equation}
\label{btodeltab}
\begin{split}
\delta b_{\mathbf{p}}(x)&=\oint_{\partial\Omega} \rmd^3\sigma_u\left[A^{(1)}\cdot\mathbb{T}^{A}_u+\delta b\cdot\mathbb{T}^{b}_u\right]b_{\mathbf{p}}(x)\\
&+\frac{1}{2}\int\frac{\rmd^3k}{(2\pi)^3 2k^0}\oint_{\partial\Omega} \rmd^3\sigma_u\rmd^3\sigma_v\left[\alpha_{\mathbf{k}}\cdot\mathbb{T}^{A}_u+\xi^{(\mathbf{k}+)}\cdot\mathbb{T}^{b}_u\right]\left[\alpha^*_{\mathbf{k}}\cdot\mathbb{T}^{A}_v+\xi^{(\mathbf{k}-)}\cdot\mathbb{T}^{b}_v\right]b_{\mathbf{p}}(x).
\end{split}
\end{equation}
This is the last elementary piece missing.\\

Here is a good place to pause for a comment about fermionic fluctuations. We recall that two types of diagrams have been omitted for obscure reasons up to now. These diagrams are the quark loop dressing the gluon one-point function and the other one was the "half" of the self-energy that has been interpreted as a quark fluctuation. The reason we got rid of them - justified by the fact that they do not lead to large logarithms in the region of phase space we want to look at - is because such contributions cannot be written as shift operators acting on elementary fields. Formally this is due to the vanishing character of fermionic one-point functions or equivalently the absence of quark sources. The construction of a quark tadpole would be similar to the gluonic case provided by the last term of \eqref{A0toA1}. This would have required that the gluon one-point function possibly emerges from two quark's classical sources which is forbidden since they are zero. Indeed, in this case, the fermionic shift operator has to act on the initial condition of $A^{(0)}$. Since $A^{(0)}$ does not depend on any fermionic quantity according to \eqref{EOMA0}, such operator gives zero. The quark fluctuation in the self-energy cannot be treated in this way for analogous reasons.

\subsection{Physical picture of boundary operators}
\indent

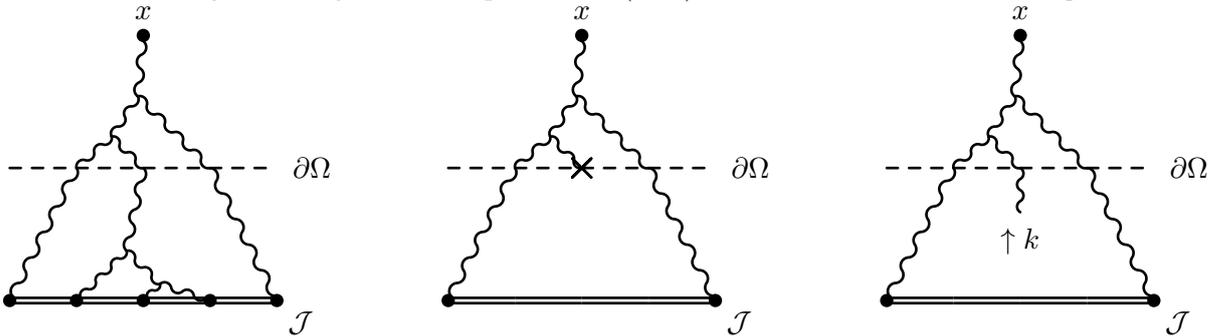
\begin{figure}[h]
\centering
%(along, up)
\begin{fmfgraph*}(100,100)
    \fmfleft{b1,i1,t1}
    \fmfright{b5,o1,t5}
      \fmf{phantom}{t1,t2,t3,t4,t5}
      \fmf{dashes}{i1,a,b,c,o1}
      \fmf{dbl_plain}{b1,b2,b3,b4,b5}
    \fmffreeze
      \fmf{photon,tension=3.5}{t3,v1,v2,a,b1}
      \fmf{photon}{v1,c,b5}
      \fmf{photon}{v2,b,v3,v4,b4}
      \fmf{photon}{v3,b2}
      \fmf{photon}{v4,b3}
      \fmfdot{b1,b2,b3,b4,b5,t3}
      \fmflabel{$\partial\Omega$}{o1}
      \fmflabel{$\mathcal{J}$}{b5}
      \fmflabel{$x$}{t3}
\end{fmfgraph*}
\hspace{2cm}
\begin{fmfgraph*}(100,100)
    \fmfleft{b1,i1,t1}
    \fmfright{b5,o1,t5}
      \fmf{phantom}{t1,t2,t3,t4,t5}
      \fmf{dashes}{i1,a,b,c,o1}
      \fmf{dbl_plain}{b1,b2,b3,b4,b5}
    \fmffreeze
      \fmf{photon,tension=3.5}{t3,v1,v2,a,b1}
      \fmf{photon}{v1,c,b5}
      \fmf{photon}{v2,b}
      \fmfdot{b1,b5,t3}
      \fmfv{decor.shape=cross,decor.size=10.}{b}
      \fmflabel{$\partial\Omega$}{o1}
      \fmflabel{$\mathcal{J}$}{b5}
      \fmflabel{$x$}{t3}
\end{fmfgraph*}
\hspace{2cm}
\begin{fmfgraph*}(100,100)
    \fmfleft{b1,i1,t1}
    \fmfright{b5,o1,t5}
      \fmf{phantom}{t1,t2,t3,t4,t5}
      \fmf{dashes}{i1,a,b,c,o1}
      \fmf{dbl_plain}{b1,b2,b3,b4,b5}
    \fmffreeze
      \fmf{photon,tension=3.5}{t3,v1,v2,a,b1}
      \fmf{photon}{v1,c,b5}
      \fmf{photon}{v2,b}
      \fmf{phantom}{b,v,b3}
      \fmf{photon}{b,v}
      \fmfdot{b1,b5,t3}
      \fmflabel{$\partial\Omega$}{o1}
      \fmflabel{$\mathcal{J}$}{b5}
      \fmflabel{$\uparrow k$}{v}
      \fmflabel{$x$}{t3}
\end{fmfgraph*}
\caption{Illustration of formula \eqref{A0toalpha}. Diagram on the left represents some diagram entering into the tree level one-point function $A^{(0)}(x)$. The central diagram is one of the results of the action of the operator $\mathbb{T}^A$. This operator differentiates $A^{(0)}(x)$ with respect to its initial value on the surface $\partial\Omega$, that is it cuts one leg at the surface in all possible ways. Diagram on the right is the result of the multiplication by $\alpha_{\mathbf{k}}$ that is a dressed external leg coming from the infinite past. To summarize, the shift operator cuts one leg of an LO field in all possible ways on the initial surface and replace it by a new field at this point.\label{A0toalphadiagr}}
\end{figure}

\begin{figure}[h]
\centering
%(along, up)
\begin{fmfgraph*}(100,100)
    \fmfleft{b1,t1}
    \fmfright{b5,t5}
      \fmf{phantom}{t1,t2,t3,t4,t5}
      \fmf{dbl_plain}{b1,b2,b3,b4,b5}
    \fmffreeze
      \fmf{phantom}{t3,a,b,c,d,b3}
    \fmffreeze
      \fmf{fermion}{d,t3}
    \fmffreeze
      \fmf{photon}{a,aa,b1}
      \fmf{photon}{aa,b2}
      \fmf{photon}{b,b5}
      \fmf{photon}{c,b4}
      \fmfdot{b1,b2,b4,b5}
      \fmflabel{$\mathcal{J}$}{b5}
      \fmflabel{$\uparrow p$}{d}
\end{fmfgraph*}
\hspace{3cm}
\begin{fmfgraph*}(100,100)
    \fmfleft{b1,t1}
    \fmfright{b5,t5}
      \fmf{phantom}{t1,t2,t3,t4,t5}
      \fmf{dbl_plain}{b1,b2,b3,b4,b5}
    \fmffreeze
      \fmf{phantom}{t3,a,b,c,d,b3}
    \fmffreeze
      \fmf{fermion}{d,t3}
    \fmffreeze
      \fmf{photon,tension=3.}{a,aa,bb,b1}
      \fmf{photon}{aa,cc,b3}
      \fmf{photon}{bb,cc}
      \fmf{photon}{c,b5}
      \fmfdot{b1,b3,b5}
      \fmflabel{$\mathcal{J}$}{b5}
      \fmflabel{\hspace{.3cm}$\uparrow p$}{d}
\end{fmfgraph*}
\hspace{2cm}
\begin{fmfgraph*}(100,100)
    \fmfleft{b1,t1}
    \fmfright{b5,t5}
      \fmf{phantom}{t1,t2,t3,t4,t5}
      \fmf{dbl_plain}{b1,b2,b3,b4,b5}
    \fmffreeze
      \fmf{phantom}{t3,a,b,c,d,b3}
    \fmffreeze
      \fmf{fermion}{d,t3}
    \fmffreeze
      \fmf{photon,tension=3.}{a,aa,bb,b1}
      \fmf{photon}{aa,c}
      \fmf{photon}{b,b5}
      \fmf{photon}{bb,b2}
      \fmfdot{b1,b2,b5}
      \fmflabel{$\mathcal{J}$}{b5}
      \fmflabel{$\uparrow p$}{d}
\end{fmfgraph*}
\vspace{.5cm}
\caption{First diagram : a typical diagram contributing to $b_{\mathbf{p}}$. Last two diagrams : the two kinds of contributions to $\delta b_{\mathbf{p}}$ corresponding to the tadpole (middle diagram) and the self-energy (rightmost diagram).\label{bdeltab}}
\end{figure}
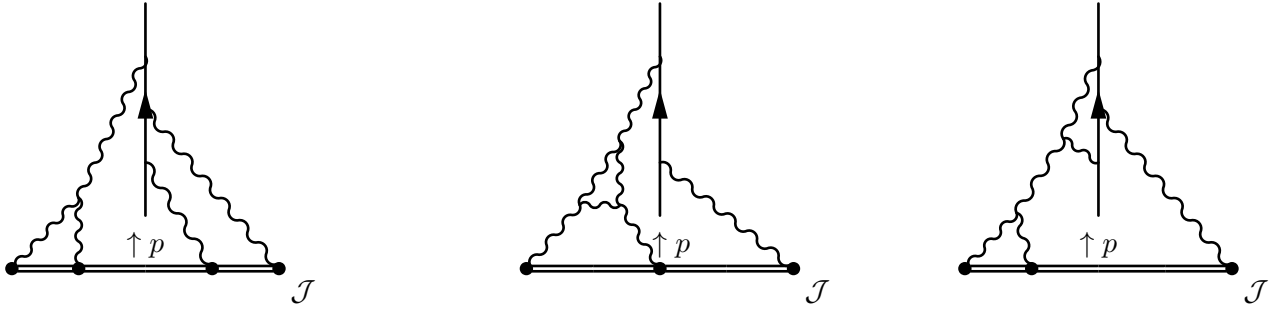

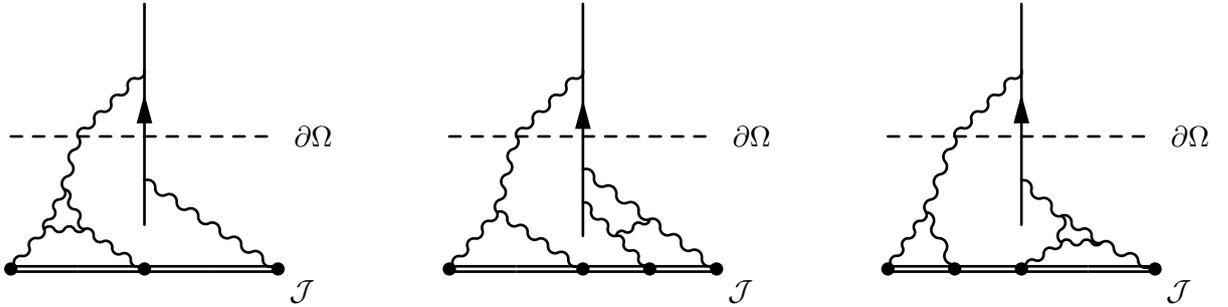
\begin{figure}[h]
\centering
%(along, up)
\begin{fmfgraph*}(100,100)
    \fmfleft{b1,i1,t1}
    \fmfright{b5,o1,t5}
      \fmf{phantom}{t1,t2,t3,t4,t5}
      \fmf{dbl_plain}{b1,b2,b3,b4,b5}
      \fmf{dashes}{i1,v1,v,v2,o1}
    \fmffreeze
      \fmf{phantom}{t3,a,v,b,d,b3}
    \fmffreeze
      \fmf{fermion}{d,t3}
    \fmffreeze
      \fmf{photon,tension=3.}{a,v1,aa,bb,b1}
      \fmf{photon}{aa,cc,b3}
      \fmf{photon}{bb,cc}
      \fmf{photon}{b,b5}
      \fmfdot{b1,b3,b5}
      \fmflabel{$\partial\Omega$}{o1}
      \fmflabel{$\mathcal{J}$}{b5}
\end{fmfgraph*}
\hspace{2cm}
\begin{fmfgraph*}(100,100)
    \fmfleft{b1,i1,t1}
    \fmfright{b5,o1,t5}
      \fmf{phantom}{t1,t2,t3,t4,t5}
      \fmf{dbl_plain}{b1,b2,b3,b4,b5}
      \fmf{dashes}{i1,v1,v,v2,o1}
    \fmffreeze
      \fmf{phantom}{t3,a,v,b,c,d,b3}
    \fmffreeze
      \fmf{fermion}{d,t3}
    \fmffreeze
      \fmf{photon,tension=3.}{a,v1,aa,b1}
      \fmf{photon}{aa,b3}
      \fmf{photon}{b,dd,b5}
      \fmf{photon}{c,ee,b4}
      \fmf{photon,tension=0}{dd,ee}
      \fmfdot{b1,b3,b4,b5}
      \fmflabel{$\partial\Omega$}{o1}
      \fmflabel{$\mathcal{J}$}{b5}
\end{fmfgraph*}
\hspace{2cm}
\begin{fmfgraph*}(100,100)
    \fmfleft{b1,i1,t1}
    \fmfright{b5,o1,t5}
      \fmf{phantom}{t1,t2,t3,t4,t5}
      \fmf{dbl_plain}{b1,b2,b3,b4,b5}
      \fmf{dashes}{i1,v1,v,v2,o1}
    \fmffreeze
      \fmf{phantom}{t3,a,v,b,d,b3}
    \fmffreeze
      \fmf{fermion}{d,t3}
    \fmffreeze
      \fmf{photon,tension=3.}{a,v1,aa,b1}
      \fmf{photon}{aa,b2}
      \fmf{photon,tension=3}{b,cc,bb,b5}
      \fmf{photon}{cc,dd,b3}
      \fmf{photon}{bb,dd}
      \fmfdot{b1,b2,b3,b5}
      \fmflabel{$\partial\Omega$}{o1}
      \fmflabel{$\mathcal{J}$}{b5}
\end{fmfgraph*}
\vspace{.5cm}
\caption{First diagram : $\left[A^{(1)}\cdot\mathbb{T}^{A}_u\right]b_{\mathbf{p}}(x)$. A tree level gluon one-point function is replaced by its one loop correction at the surface. Second and last diagram : $\left[\delta b\cdot\mathbb{T}^{b}_u\right]b_{\mathbf{p}}(x)$. The loop corrections to the tree level spinor $b_{\mathbf{p}}$ is bellow the initial surface. They are of types : self-energy (second diagram) and tadpole (last diagram).\label{lineardeltab}}
\end{figure}

At this point manipulations may seem to be a bit formal. However they do have a very intuitive physical picture that can be represented diagrammatically. Let us say for definiteness that the initial surface is a space-like surface of equal time, lying in the future of the sources and the volume $\Omega$ is the future side of this surface. Moreover we choose the retarded prescription in Green's formulas so that this equal time surface is actually the only one contributing to the boundary integrals. Spatial and future infinity are discarded by causality. The simplest case \eqref{A0toalpha} is illustrated in detail on figure \ref{A0toalphadiagr}.
Let us deal with a more complicated case, formula \eqref{btodeltab} giving $\delta b_{\mathbf{p}}$ as a functional of $b_{\mathbf{p}}$. $b_{\mathbf{p}}$ is an initial free spinor dressed by tree level sources and $\delta b_{\mathbf{p}}$ is its one loop correction, respectively represented on figure \ref{bdeltab}.
Let us see what is the meaning of each term in \eqref{btodeltab}. We begin with the linear term $\left[A^{(1)}\cdot\mathbb{T}^{A}_u+\delta b\cdot\mathbb{T}^{b}_u\right]b_{\mathbf{p}}(x)$. The first one replaces an $A^{(0)}$ by an $A^{(1)}$ on the surface, the second one a tree level spinor $b_{\mathbf{p}}$ by its one loop correction under the surface as represented on figure \ref{lineardeltab}.
The quadratic term in \eqref{btodeltab} $\left[\alpha_{\mathbf{k}}\cdot\mathbb{T}^{A}_u+\xi^{(\mathbf{k}+)}\cdot\mathbb{T}^{b}_u\right]\left[\alpha^*_{\mathbf{k}}\cdot\mathbb{T}^{A}_v+\xi^{(\mathbf{k}-)}\cdot\mathbb{T}^{b}_v\right]b_{\mathbf{p}}(x)$ has essentially two kinds of contributions. $\left[\alpha_{\mathbf{k}}\cdot\mathbb{T}^{A}_u\right]\left[\alpha^*_{\mathbf{k}}\cdot\mathbb{T}^{A}_v\right]b_{\mathbf{p}}(x)$ cuts two gluon one-point functions on the initial surface and match them together to close the loop as represented on figure \ref{quadalpha}. $\left[\alpha_{\mathbf{k}}\cdot\mathbb{T}^{A}_u\right]\left[\xi^{(\mathbf{k}-)}\cdot\mathbb{T}^{b}_v\right]b_{\mathbf{p}}(x)$ and its complex conjugate correspond to a loop correction that crosses the surface as shown on \ref{quadalphaxi}.
$\left[\xi^{(\mathbf{k}+)}\cdot\mathbb{T}^{b}_u\right]\left[\xi^{(\mathbf{k}-)}\cdot\mathbb{T}^{b}_v\right]b_{\mathbf{p}}(x)$ vanishes since the dependence of $b_{\mathbf{p}}$ on its initial condition is linear.

\begin{figure}[h]
\centering
%(along, up)
\begin{fmfgraph*}(100,100)
    \fmfleft{b1,i1,t1}
    \fmfright{b5,o1,t5}
      \fmf{phantom}{t1,t2,t3,t4,t5}
      \fmf{dbl_plain}{b1,b2,b3,b4,b5}
      \fmf{dashes}{i1,v1,v,v2,o1}
    \fmffreeze
      \fmf{phantom}{t3,a,b,v,c,d,e,b3}
    \fmffreeze
      \fmf{fermion}{e,t3}
    \fmffreeze
      \fmf{photon,tension=3.}{a,v1,aa,b1}
      \fmf{photon}{b,dd,b3}
      \fmf{photon}{c,b5}
      \fmf{photon}{d,b4}
      \fmf{photon}{dd,aa}
      \fmfdot{b1,b3,b4,b5}
      \fmflabel{$\partial\Omega$}{o1}
      \fmflabel{$\mathcal{J}$}{b5}
\end{fmfgraph*}
\hspace{2cm}
\begin{fmfgraph*}(100,100)
    \fmfleft{b1,i1,t1}
    \fmfright{b5,o1,t5}
      \fmf{phantom}{t1,t2,t3,t4,t5}
      \fmf{dbl_plain}{b1,b2,b3,b4,b5}
      \fmf{dashes}{i1,v1,v,v2,o1}
    \fmffreeze
      \fmf{phantom}{t3,a,b,v,c,d,e,b3}
    \fmffreeze
      \fmf{fermion}{e,t3}
    \fmffreeze
      \fmf{photon,tension=3.}{a,u,r,v1,aa,bb,b1}
      \fmf{photon}{u,cc,b2}
      \fmf{photon}{aa,cc}
      \fmf{photon}{c,b5}
      \fmf{photon}{d,b4}
      \fmfdot{b1,b2,b4,b5}
      \fmflabel{$\partial\Omega$}{o1}
      \fmflabel{$\mathcal{J}$}{b5}
\end{fmfgraph*}
\vspace{.5cm}
\caption{$\left[\alpha_{\mathbf{k}}\cdot\mathbb{T}^{A}_u\right]\left[\alpha^*_{\mathbf{k}}\cdot\mathbb{T}^{A}_v\right]b_{\mathbf{p}}(x)$ gives two topologies of diagrams depending whether the two cut one-point functions are attached to the quark at different $qg$ vertices (left diagram) or are two different branches of the same gluon one-point function (right diagram).\label{quadalpha}}
\end{figure}
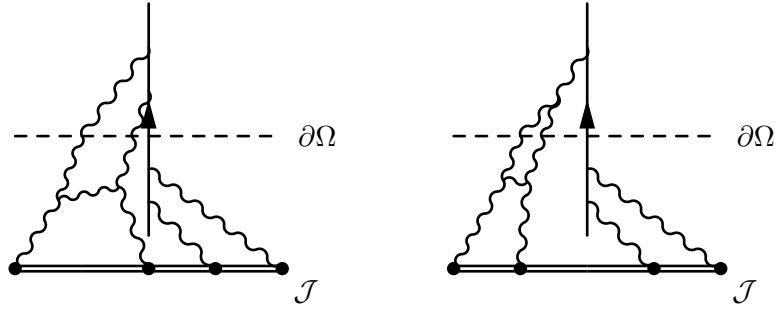
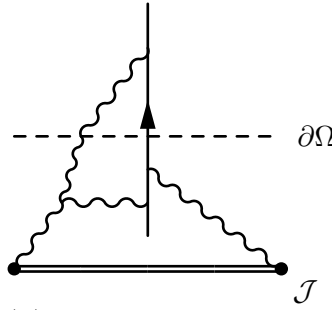
\begin{figure}[h]
\centering
%(along, up)
\begin{fmfgraph*}(100,100)
    \fmfleft{b1,i1,t1}
    \fmfright{b5,o1,t5}
      \fmf{phantom}{t1,t2,t3,t4,t5}
      \fmf{dbl_plain}{b1,b2,b3,b4,b5}
      \fmf{dashes}{i1,v1,v,v2,o1}
    \fmffreeze
      \fmf{phantom}{t3,a,b,v,c,d,e,b3}
    \fmffreeze
      \fmf{fermion}{e,t3}
    \fmffreeze
      \fmf{photon,tension=3.}{a,v1,aa,b1}
      \fmf{photon}{c,b5}
      \fmf{photon}{d,aa}
      \fmfdot{b1,b5}
      \fmflabel{$\partial\Omega$}{o1}
      \fmflabel{$\mathcal{J}$}{b5}
\end{fmfgraph*}
\caption{The $\left[\alpha_{\mathbf{k}}\cdot\mathbb{T}^{A}_u\right]\left[\xi^{(\mathbf{k}-)}\cdot\mathbb{T}^{b}_v\right]b_{\mathbf{p}}(x)$ contribution. The fermion is cut on the initial surface and replaced by a $\xi^{(\mathbf{k}+)}_{\mathbf{p}}$ vertex \ref{XIvertex} whose gluon leg matches the cut gluon one-point function.\label{quadalphaxi}}
\end{figure}

\subsection{Recursion relations}
\indent

In the two previous sections we dealt only with elementary fields, but these are primary ingredients of more complicated combinations of them : Schwinger-Keldysh Green functions. Let us go back to the NLO gluon spectrum \eqref{spectrumgluons1}. Thanks to \eqref{D1repr}, the NLO spectrum reads :
\begin{equation}
\begin{split}
\left.(2\pi)^32k^0\frac{\rmd \mathcal{N}_g}{\rmd^3k}\right|_{NLO}&=\int\rmd^4x\rmd^4ye^{ik\cdot(x-y)}\epsilon^*_{\mu}(k)\epsilon_{\nu}(k)\Box_x\Box_y\left[\int \frac{\rmd^3l}{(2\pi)^3 2l^0}\alpha^{A\mu}_{\mathbf{l}}(x)\alpha^{A\nu*}_{\mathbf{l}}(y)\right.\\
&~~~~~~~~~~\left.+A^{(1)\mu A}(x)A^{(0)\nu A}(y)+A^{(0)\mu A}(x)A^{(1)\nu A}(y)\right].
\end{split}
\end{equation}
Rewriting the functional relations \eqref{A0toalpha} between $\alpha^{A\mu}_{\mathbf{k}}$ and $A^{(0)}$ and \eqref{A0toA1} between $A^{(1)}$ and $A^{(0)}$ respectively. One finds that the NLO spectrum reduces to a shift operator acting on the LO spectrum \eqref{spectrumgluons1}. This is the recursion relation we were looking for and it reads :
\begin{equation}
\label{gluonLOtoNLO}
\begin{split}
\left.(2\pi)^32k^0\frac{\rmd \mathcal{N}_g}{\rmd^3k}\right|_{NLO}&=\left[\oint_{\partial\Omega} \rmd^3\sigma_u\left[A^{(1)}\cdot\mathbb{T}^{A}_u\right]\right.\\
&\left.+\frac{1}{2} \int \frac{\rmd^3l}{(2\pi)^3 2l^0}\oint_{\partial\Omega} \rmd^3\sigma_u \rmd^3\sigma_v\left[\alpha_{\mathbf{l}}\cdot\mathbb{T}^{A}_u\right]\left[\alpha^{*}_{\mathbf{l}}\cdot\mathbb{T}^{A}_v\right]\right]\left.(2\pi)^32k^0\frac{\rmd \mathcal{N}_g}{\rmd^3k}\right|_{LO}.
\end{split}
\end{equation}
This formula is exact in a pure Yang-Mills theory but is up to a fermion loop in QCD.\\

A similar relation for the quark spectrum is straightforward. Let us summarize the role played by the elementary fields $b_{\mathbf{p}}$, $\xi^{(\mathbf{k}+)}_{\mathbf{p}}$ and $\delta b_{\mathbf{p}}$ in the LO and NLO spectrum \eqref{spectrumquarks1}. Let us write the LO spectrum by plugging \eqref{S1repr} into \eqref{spectrumquarks1} :
\begin{equation}
\label{spectrumquarksLO}
\begin{split}
\left.(2\pi)^32p^0\frac{\rmd \mathcal{N}_q}{\rmd^3p}\right|_{LO}&=-\int\rmd^4x\rmd^4ye^{ip\cdot(x-y)}\bar{u}(p)\left(i\overrightarrow{\slashed{\partial}}_x-m\right)\int \frac{\rmd^3q}{(2\pi)^3 2q^0}b_{\mathbf{q}}(x)\bar{b}_{\mathbf{q}}(y)\left(i\overleftarrow{\slashed{\partial}}_y+m\right)u(p).
\end{split}
\end{equation}
The NLO spectrum is obtained thanks to \eqref{S2split} together with \eqref{S2decomp} and \eqref{crossrepr} :
\begin{equation}
\begin{split}
\left.(2\pi)^32p^0\frac{\rmd \mathcal{N}_q}{\rmd^3p}\right|_{NLO}&=-\int\rmd^4x\rmd^4ye^{ip\cdot(x-y)}\bar{u}(p)\left(i\overrightarrow{\slashed{\partial}}_x-m\right)\\
&\times\int \frac{\rmd^3q}{(2\pi)^3 2q^0}\left[\delta b_{\mathbf{q}}(x)\bar{b}_{\mathbf{q}}(y)+b_{\mathbf{q}}(x)\delta\bar{b}_{\mathbf{q}}(y)\right.\\
&+\int \frac{\rmd^3k}{(2\pi)^3 2k^0} \left.\xi^{(\mathbf{k}+)}_{\mathbf{q}}(x)\bar{\xi}^{(\mathbf{k}-)}_{\mathbf{q}}(y)\right]\left(i\overleftarrow{\slashed{\partial}}_y+m\right)u(p).
\end{split}
\end{equation}
Writing the $\xi^{(\mathbf{k}+)}_{\mathbf{p}}$ and $\delta b_{\mathbf{p}}$ fields as shift operators \eqref{btoxi} and \eqref{btodeltab} respectively, everything fits together as a second order differential operator acting on the LO spectrum \eqref{spectrumquarksLO} :
\begin{equation}
\label{quarkLOtoNLO}
\begin{split}
(2\pi)^3&2p^0\left.\frac{\rmd \mathcal{N}_q}{\rmd^3p}\right|_{NLO}=\left[\oint_{\partial\Omega} \rmd^3\sigma_u\left[A^{(1)}\cdot\mathbb{T}^{A}_u+\delta b\cdot\mathbb{T}^{b}_u\right]\right.\\
&\left.+\frac{1}{2}\int\frac{\rmd^3k}{(2\pi)^3 2k^0}\oint_{\partial\Omega} \rmd^3\sigma_u\rmd^3\sigma_v\left[\alpha_{\mathbf{k}}\cdot\mathbb{T}^{A}_u+\xi^{(\mathbf{k}+)}\cdot\mathbb{T}^{b}_u\right]\left[\alpha^*_{\mathbf{k}}\cdot\mathbb{T}^{A}_v+\xi^{(\mathbf{k}-)}\cdot\mathbb{T}^{b}_v\right]\right]\left.(2\pi)^32p^0\frac{\rmd \mathcal{N}_q}{\rmd^3p}\right|_{LO}.
\end{split}
\end{equation}
Recall that this formula does not take into account fermionic fluctuations that do not lead to large logarithms of $x$ and is therefore an approximate formula valid in the small $x$ phase space region. The recursion relations are the fundamental formulas for extracting the small $x$ evolution. Formula \eqref{quarkLOtoNLO} is a new relation that appears for the first time in \cite{Gelis:2012ct}.

\section{Small $x$ logarithms and JIMWLK evolution}
\indent

The small $x$ logarithms are obtained from  LO to NLO recursion relations by an explicit computation of the coefficients in front of the differential operators in \eqref{gluonLOtoNLO} and \eqref{quarkLOtoNLO} for gluons and quarks respectively. In the gluonic case, this computation has been performed in \cite{Gelis:2008rw}. The small $x$ logarithms come from the large $l^3$ integration region over the initial momentum of the cut propagator and the one-loop tadpole $A^{(1)}$ in \eqref{gluonLOtoNLO}. It turns out that the evolution is governed by two JIMWLK hamiltonians corresponding to the two nuclei separately as expected from causality arguments. The full calculation is tricky and the interested reader is sent to the previous references. Let us just mention some important steps of the calculation - the choice of the initial surface and the gauge fixing - before discussing the evolution. 

\subsection{Choice of initial surface}
\indent

\begin{figure}[h]
\begin{center}
\includegraphics[width=0.35\paperwidth]{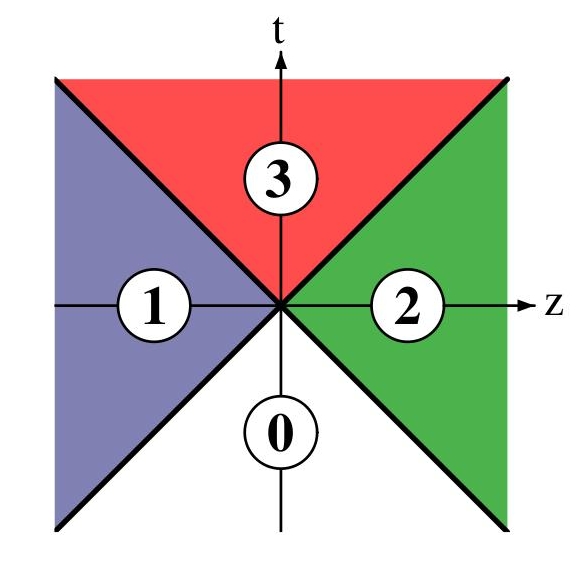}
\caption{Light-cone decomposition for the A-A process : region $(0)$ is the remote past, causally disconnected from the collision process ; region $(1)$ is the causal future of the right-moving nucleus only ; region $(2)$ is the causal future of the left-moving nucleus only and region $(3)$ is the common future of the two nuclei together.\label{spacetimestructure}}
\end{center}
\end{figure}

\begin{figure}[h]
\begin{center}
\includegraphics[width=0.4\paperwidth]{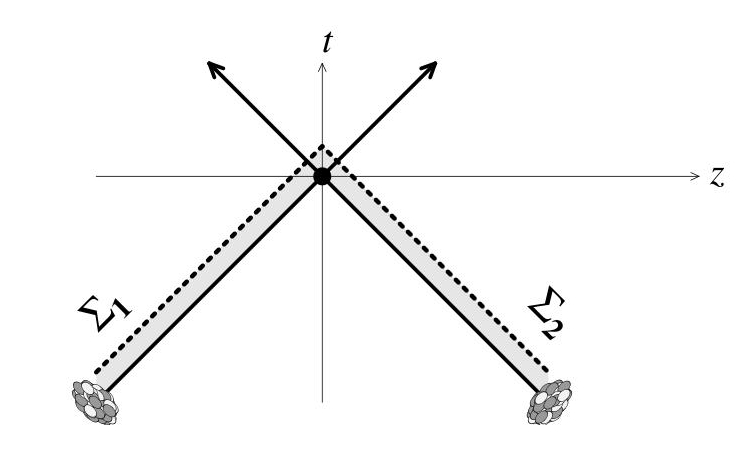}
\caption{Initial surface $\Sigma=\Sigma_1 \cup \Sigma_2$ conveniently chosen at an infinitesimally later time after the collision.\label{initialsurface}}
\end{center}
\end{figure}

In this section we deal with a crucial step for explicitly extracting logarithms from the general expressions \eqref{gluonLOtoNLO} and \eqref{quarkLOtoNLO}. This requires a suitable choice for the boundary surface $\partial\Omega$ that is motivated by the topological distribution of classical sources. The collision between the two nuclei occurring at very high energy in a frame not very different from the center of mass frame, it is convenient to endow space-time with the structure represented on figure \ref{spacetimestructure}.
\begin{itemize}
\item region $(0)$ corresponds to $x^{\pm}$ both negative. This region has no knowledge of the process and the classical field is zero (or, at worth, pure gauge depending on the gauge used).\\
\item regions $(1)$ and $(2)$ are causally disconnected from each other. In each of these regions, the background field is associated to the right-moving and left-moving source respectively. The classical field corresponding to the left-moving source in given by \eqref{classicalA-} and the mirror case (plus and minus indices interchanged) for the right-moving one.\\
\item region $(3)$ the background field is produced by the two sources together and the non-linear character of the Yang-Mills equations allows only numerical solutions. Indeed, due to non-linear terms in the Yang-Mills equation, the superposition principle breaks down : the field produced by several sources does not reduce to the sum of the fields created by these sources separately.
\end{itemize}
In order to get the small $x$ behavior from recursion relations one has, in principle, to deal with all these four regions. Fortunately, the JIMWLK evolution concerns the intrinsic nature of the hadron itself and has nothing to do with some scattering process. Therefore thanks to a suitable gauge fixing, the small $x$ logs can be contained in the initial nucleus wave function and region $(3)$ on figure \ref{spacetimestructure} does not bring such large logarithms. The previous space-time decomposition suggests a natural guess for the initial surface. $\partial\Omega$ is taken to be $\Sigma_1 \cup \Sigma_2$ represented on figure \ref{initialsurface}. The spatial and future infinity does not contribute to the boundary term in Green's formulas \eqref{GFgluons} and \eqref{GFquarks} if one chooses the retarded prescription for the free Green function. This choice is natural since the JIMWLK evolution is encoded in the initial condition. An initial surface embedded in region $(0)$ would not have shown anything interesting : the JIMWLK evolution occurs at a later time and can not be included in boundary operators. Moreover a later time surface would not bring additional information but would lead to tedious or even not doable analytical calculations. The choice $\Sigma_1 \cup \Sigma_2$ of \ref{initialsurface} contains all the JIMWLK behavior - provided an appropriate gauge fixing - and enables simple calculations, since the fields are known analitically in regions $(1)$ and $(2)$.

\subsection{Gauge fixing\label{gspectrumgaugefix}}
\indent

The various space integrals in \eqref{quarkLOtoNLO} are performed on the surface $\Sigma$ represented on \ref{initialsurface}. This implies that all the divergences are indeed contained in the causal past of the surface $\Sigma$. This point is a bit subtle and makes the gauge fixing very important since the time at which the radiative corrections responsible of small $x$ divergences occur differs from one gauge to another one. The condition for all the logarithms to be included "before" $\Sigma$ is ensured if the background field vanishes above the surface. We shall see that one can find such a gauge choice. In region $(0)$, the field is gauge equivalent to the zero field and nothing interesting can happens in any gauge. The logarithms can only arise in the neighborhood of $\Sigma$. In regions $(1)$ and $(2)$ of figure \ref{spacetimestructure}, the background field is known. It is given by an $\mathcal{A}^-_2$ in region $(2)$ whose explicit form is given by \eqref{classicalA-}, in region $(1)$ it is given by the mirror field - with plus and minus components interchanged - $\mathcal{A}^+_1$ (we have introduced $1$ and $2$ indices to distinguish the two fields generated by the two nuclei). In region $(3)$ the background field is analytically unknown but it has been argued \cite{Gelis:2008rw} that possible small $x$ divergences are suppressed by fast oscillating exponentials. Thus the small $x$ logarithms are worked out from regions of space time where the field is known analytically. The covariant gauge that gives the fields $\mathcal{A}^+_1$ and $\mathcal{A}^-_2$ in regions $(1)$ and $(2)$ respectively fulfills the required property that the classical field must lie below the initial surface. Furthermore it has a nice property that makes the explicit computation of quantum fluctuations easier. Let us consider the field $\mathcal{A}^+_1$ associated to the right-moving nucleus - these considerations are obviously transposed to the left-moving one and its respective field $\mathcal{A}^-_2$ by reverting the plus and minus Lorentz components. As already mentioned, at the classical level, the covariant gauge $\partial\cdot A_1$ and the gauge $A^-_1=0$ are equivalent\footnote{As usual $A_1$ denotes the total field, that is the classical field plus the quantum corrections.}. Thus if one instead consider this last gauge $A_1^-=0$, the gluon field does not couple to the classical current $\mathcal{J}^+_1$ associated to the right-moving nucleus. In other words, it means that quantum corrections do not induce current precession. The current is simply conserved to all orders instead of being covariantly conserved. This makes calculations simpler since the first quantum corrections are solutions of the linearized \emph{homogeneous} Yang-Mills equations which can be solved analytically \cite{Gelis:2008rw}.\\

As pointed out at the beginning of section \ref{qgspectra}, the form \eqref{spectrumgluons} of the gluon spectrum is not gauge invariant and we have the stronger constrain that we must fix the gauge so that this formula is interpreted as a number of physical degrees of freedom. To examine this feature let us consider its LO version \eqref{spectrumgluons1} which will be enough to interpret the spectrum as a number of gluons. Let us consider region $(1)$, the causal future of source $(1)$. In covariant gauge, the classical field $A_1^{(0)}=\mathcal{A}_1$ is given by the mirror expression \eqref{classicalA-}, i.e. by a single component field $\mathcal{A}_1^+$. We prefer to work in axial gauge and at the classical level one can equivalently consider instead of the covariant gauge, the $A_1^-=0$ gauge. The contraction of the classical field with the polarization vectors gives zero by gauge condition and thus \eqref{spectrumgluons1} is zero. This situation is unsatisfactory. The gauge $A_1^-=0$ seems to forbid the interpretation of \eqref{spectrumgluons} as a number of gluons.\\

Let us guess the gauge $\tilde{A}_1^+=0$ denoted with a tilde to make the difference with the $A_1^-=0$ gauge. It is an easy exercise to show that, at the classical level, the suitable gauge transformation is provided by :
\begin{equation}
\label{gaugetransformation}
\tilde{\mathcal{A}}_1^{\mu}(x)=\Omega(+\infty,x^-,\mathbf{x}_{\perp})\left(\mathcal{A}_1^{\mu}(x)+\frac{i}{g}\partial^{\mu}\right)\Omega^{\dagger}(+\infty,x^-,\mathbf{x}_{\perp}),
\end{equation}
where $\Omega(+\infty,x^-,\mathbf{x}_{\perp})$ is the Wilson line built from the field $\mathcal{A}_1^{\mu}$ with finite end points :
\begin{equation}
\Omega(y^-,x^-,\mathbf{x}_{\perp})\equiv\mathcal{P}\exp\left[ig\int_{x^-}^{y^-}\rmd z^-\mathcal{A}_1^{+}(z^-,\mathbf{x}_{\perp})\right].
\end{equation}
The $\Omega(\mathbf{x}_{\perp})$ encountered so far is recovered by sending $y^-$ and $x^-$ to $+\infty$ and $-\infty$ respectively. Since $\mathcal{A}_1$ does not depend on the $x^+$ coordinate - recall it is the mirror case of the field considered in previous sections since it is produced by the projectile instead of the target, $\tilde{A}_1$ does not depend on the $x^+$ coordinate as well. Moreover it is vanishing for $x^->0$ and is pure gauge $x^-<0$ :
\begin{equation}
\tilde{\mathcal{A}}_1^{\mu}(x^-,\mathbf{x}_{\perp})=\frac{i}{g}\theta(-x^-)\Omega(\mathbf{x}_{\perp})\partial^{\mu}\Omega^{\dagger}(\mathbf{x}_{\perp})=\frac{i}{g}\theta(-x^-)\delta^{\mu i}\Omega(\mathbf{x}_{\perp})\partial^{i}\Omega^{\dagger}(\mathbf{x}_{\perp}).
\end{equation}
Just note that $\tilde{A}_1^+=0$ does not fix the gauge uniquely and any $x^+$ independent gauge transformation performed on $\tilde{A}_1$ preserves this condition. We have chosen the gauge transformation that cancels the background field above $\Sigma_1$ as required. The situation seems to be much better. Indeed from appendix \ref{BRSTspectrum} we learned that in light-cone gauge the physical degrees of freedom are given by the transverse components. $\tilde{\mathcal{A}}_1^{\mu}$ only has transverse components as expected. Therefore the gauge $\tilde{A}_1^+=0$ seems to provide the proper interpretation of \eqref{spectrumgluons1} as a gluon number.\\

To interpret this from a more rigorous approach one has to see the discussion \ref{gaugespectrum} concerning respectively the $\mathcal{A}_1^{+}$ and $\tilde{\mathcal{A}}_1^{-}$. As shown in \ref{gaugespectrum} these non-transverse degrees of freedom are redundant since they are given in terms of the transverse components by a constrain equation. Moreover we have shown that the longitudinal degrees of freedom are non physical if the momentum $k$ of the gluon is collinear to the gauge fixing vector. Since both $\mathcal{A}_1$ and $\tilde{\mathcal{A}}_1$ do not depend on $x^+$, their Fourier decomposition is proportional to $\delta(k^-)$, that is the $k^-$ component of the momentum of the classical gluons within the nucleus is zero. In the gauge $A_1^-=0$, $k^-=0$ precisely corresponds to this non physical situation where $k$ is aligned along the gauge fixing vector\footnote{Intermediate gluons produced by sources are not exactly on-shell. However we can assume that the momentum distribution essentially contains modes with longitudinal momenta large with respect to the transverse components which are at most of order $Q_s$.} and therefore $\mathcal{A}_1^+$ does not correspond to physical degrees of freedom. This is why we are in trouble in this gauge. On the contrary the condition $k^-=0$ ensures a non vanishing value of $k^+$ which guarantees the physicality of the components of $\tilde{\mathcal{A}}_1$ in the gauge $\tilde{A}_1^+=0$. Moreover at the classical level the constrained component $\tilde{\mathcal{A}}_1^-$ is zero avoiding possible redundant counting. Therefore the leading order gluon spectrum \eqref{spectrumgluons1} makes sense in the $\tilde{A}_1^+=0$ gauge. Furthermore it holds at all order, even though the constrained component $\tilde{A}_1^-$ is non zero, since the contraction of $\tilde{A}_1$ with a polarization vector gives $-\epsilon^i(k)\tilde{A}_1^i$. This ensures a right counting of physical gluons to all order in \eqref{spectrumgluons}.\\

Thus all the calculations performed so far actually hold in the gauge $A^+_1=0$ with the background field vanishing above the $\Sigma_1$ surface. On $\Sigma_1$ the normal vector is given by $l^+=-1$ and $l^-=l^i=0$\footnote{If $x\in \Sigma_1$, then $x=(x^+,0,\mathbf{x}_{\perp})$ and the normal vector is defined so that its scalar product with any vector in $\Sigma_1$ is zero. This is fulfilled if and only if $l=(l^+,0,\mathbf{0}_{\perp})$. The value $l^+=-1$ is required by the Gauss theorem and the orientation of the boundary surface.}. In this gauge and after some partial integrations \eqref{GFgluons} becomes :
\begin{equation}
\begin{split}
\tilde{A}^{A\mu}(x)=i\int_{\Sigma_1}\rmd y^+\rmd^2y_{\perp}\left[2\Delta^{\mu i}_{0}(x-y)\partial_y^+\tilde{A}^{Ai}(y)-\Delta^{\mu+}_{0}(x-y)\right.&\left.\partial_{y\sigma}\tilde{A}^{A\sigma}(y)-(\partial_{y\rho}\Delta^{\mu\rho}_{0}(x-y))\tilde{A}^{A+}(y))\right]\\
&+\text{~~bulk term}.
\end{split}
\end{equation}
the value of the field at some point depends only on four initial degrees of freedom $\tilde{A}_1^-, \partial^-\tilde{A}_1^i$ and $\partial\cdot\tilde{A}_1$.\\

The proper interpretation of \eqref{spectrumquarks} as a number of quarks immediately follows. The counted gluons being physical, so are the quarks created by them in the gauge $A^+_1=0$.

\subsection{Evolution}
\indent

So far we have all the tools to perform the explicit computation of the recursion relations \eqref{gluonLOtoNLO} and \eqref{quarkLOtoNLO}. If one consider the $\Sigma_1$ (resp. $\Sigma_2$) piece of the initial surface, the coefficients of differential operators in \eqref{gluonLOtoNLO} and \eqref{quarkLOtoNLO} are the ones in the gauge $\tilde{A}_1^+=0$ (resp. $\tilde{A}_2^-=0$). The LO quark and gluon spectra are also functionals of the $\tilde{\mathcal{A}}_1$ and $\tilde{\mathcal{A}}_2$ field\footnote{Even the LO spectra are very complicated and one can not compute them explicitely analytically.}. However, from the gauge transformation \eqref{gaugetransformation} all these fields and coefficient functions are functional of the classical field $\mathcal{A}_1^+$ in the gauge $A_1^-=0$ and $\mathcal{A}_2^-$ in the gauge $A_2^+=0$ arising through Wilson lines. Thus one can equivalently see the differential operators in \eqref{gluonLOtoNLO} and \eqref{quarkLOtoNLO} as operators differentiating with respect to $\mathcal{A}_1^+$ for the right-moving source and to $\mathcal{A}_2^-$ for the left-moving one. The recursion relation for gluons \eqref{gluonLOtoNLO} turns out to be a sum of two JIMWLK hamitonians, \eqref{JIMWLKham} and its mirror version for the right-moving nucleus, in the leading log approximation (see \cite{Gelis:2008rw}) :
\begin{equation}
\label{gluonJIMWLK}
\begin{split}
\left.\frac{\rmd N_g}{\rmd^3k}\right|_{NLO}&=\left[\ln(x_1)\mathcal{H}\left[\mathcal{A}^{+}_1,\frac{\delta}{\delta\mathcal{A}_1^{+}}\right]+\ln(x_2)\mathcal{H}\left[\mathcal{A}^{-}_2,\frac{\delta}{\delta\mathcal{A}_2^{-}}\right]\right]\left.\frac{\rmd N_g}{\rmd^3k}\right|_{LO}.
\end{split}
\end{equation}
$x_1$ and $x_2$ are the longitudinal momentum fraction cutoff for the right-moving and left-moving source respectively that separate the classical and quantum partons of the nuclear wave function in the CGC framework. Equation \eqref{gluonJIMWLK} shows the required factorization property of small $x$ logs imposed by causality. The full resummation of one-loop corrections exponentiates and since the two JIMWLK hamiltonians do not talk to each other, it indeed factorizes.\\

The physical meaning of \eqref{gluonJIMWLK} is that the small $x$ evolution is an initial state effect and can be put into the hadron wave function. Since this evolution is an intrinsic property of the nucleus it has nothing to do with the observed particles in the final state. We thus expect that relation \eqref{gluonJIMWLK} holds for the quark spectrum as well. However this has not been worked out explicitly from \eqref{quarkLOtoNLO} yet. However, from the LO to NLO relation \eqref{quarkLOtoNLO}, we can guess how the calculation has to work. The current problem is only computational but the way to do is more or less understood. The LO gluon spectrum is a functional of the background field on the initial surface only whereas the LO quark spectrum is, in addition, a functional of the dressed spinor $b_{\mathbf{p}}$ and its complex conjugate. In \eqref{quarkLOtoNLO} the terms which contain only $\mathbb{T}^A$ operators automatically give the sum of two JIMWLK hamiltonians \eqref{gluonJIMWLK} since this piece is directly inherited from the gluon case. However these operators act only on the explicit dependence of the quark spectrum upon the background field. The missing piece might arise from the terms containing $\mathbb{T}^b$ operators in \eqref{quarkLOtoNLO}. Actually the dependence of the quark spectrum on $b_{\mathbf{p}}$ is an \emph{implicit} dependence on the background field. Indeed, the solution of the equation of motion \eqref{EOMb1} is of the form\footnote{The full solution is more subtle but the fine details do not matter for the present discussion.} $b_{\mathbf{p}}(x)\sim \Omega(\mathbf{x}_{\perp})v(p)e^{ip\cdot x}$. Then one can write $\mathbb{T}^b$ as a derivative with respect to the Wilson lines and therefore with respect to the background field. To compute the JIMWLK evolution \eqref{gluonJIMWLK} from \eqref{quarkLOtoNLO} for the quark spectrum one has to take into account the explicit and implicit dependences of the spectrum on the background field in the sense given above. If the dependences of $\mathbb{T}^b$ on the background field is made explicit, then the quark spectrum is a functional of the background field only and the leading log relation \eqref{gluonJIMWLK} holds with the JIMWLK hamiltonians acting on all the background field dependences : the initially explicit one and the one arising from the implicit dependence via $b_{\mathbf{p}}$.

\chapter{Traces computation\label{invtensors}}
\indent

This chapter is somewhat out of the main scope. Cross-sections computed in p-A collisions often require the computation of traces involving an arbitrary number of gauge group generators. In order to go further in the calculation of some special limits of section \ref{pAchapter}, I tried find methods for explicitly working these traces out. As an example for motivating this work, let us consider the quadrupole distribution \eqref{quadUGD} and let us focus on the color operator $\partial^i_x\partial^i_u\tilde S^{(4)}(\mathbf{x}_{\perp},\mathbf{b}_{\perp},\mathbf{u}_{\perp},\bar{\mathbf{b}}_{\perp})\Big |_{\mathbf{b}_{\perp}\mathbf{b}_{\perp}\bar{\mathbf{b}}_{\perp}\bar{\mathbf{b}}_{\perp}}$. For some purposes - like making the comparison with the Weizs\"acker-Williams distribution which is commonly written in a similar form - it may be useful to rewrite it, together with $\partial^i\mathcal{A}^-=\mathcal{F}^{i-}$, as the following adjoint representation trace :
\begin{equation}
\label{trace0}
\begin{split}
\partial^i_x\partial^i_u&\tilde S^{(4)}(\mathbf{x}_{\perp},\mathbf{b}_{\perp},\mathbf{u}_{\perp},\bar{\mathbf{b}}_{\perp})\Big |_{\mathbf{b}_{\perp}\mathbf{b}_{\perp}\bar{\mathbf{b}}_{\perp}\bar{\mathbf{b}}_{\perp}}=\frac{(ig)^2}{N_c(N_c^2-1)}\int\rmd x^+\rmd y^+\Tr\left[\tilde{\Omega}(+\infty,x^+,\mathbf{b}_{\perp})\mathcal{F}^{i-}(x^+,\mathbf{b}_{\perp})\right.\\
&\times\left.\tilde{\Omega}(x^+,-\infty,\mathbf{b}_{\perp})T^A\tilde{\Omega}^{\dagger}(\mathbf{b}_{\perp})\tilde{\Omega}(+\infty,y^+,\bar{\mathbf{b}}_{\perp})\mathcal{F}^{i-}(y^+,\bar{\mathbf{b}}_{\perp})\tilde{\Omega}(y^+,-\infty,\bar{\mathbf{b}}_{\perp})T^A\tilde{\Omega}^{\dagger}(\bar{\mathbf{b}}_{\perp})\right]
\end{split}
\end{equation}
where $\Omega(x^+,y^+,\mathbf{x}_{\perp})$ is the Wilson line built from the field $\mathcal{A}^{-}$ with finite end points :
\begin{equation}
\Omega(x^+,y^+,\mathbf{x}_{\perp})\equiv\mathcal{P}\exp\left[ig\int_{x^+}^{y^+}\rmd z^+\mathcal{A}^{-}(z^+,\mathbf{x}_{\perp})\right]
\end{equation}
(the $\Omega(\mathbf{x}_{\perp})$ encountered so far is recovered by sending $y^+$ and $x^+$ to $+\infty$ and $-\infty$ respectively). We define $\mathcal{F}^{i-}_{(+)}(x^+,\mathbf{b}_{\perp})$, the parallel transported field strength at $x^+=+\infty$ on the gauge bundle\footnote{From the geometrical point of view, gauge theories are interpreted as the principal bundle that locally looks like $M\times G$ where $M$ is a differential manifold referring to the space-time and $G$ is the gauge group. This is easily understood since gauge invariance is local and associate to each point in space-time a group matrix lying on the fiber $G$. In this context the gauge field $A$ is interpreted as the connection one-form and the Wilson lines encode the parallel transport on the principal bundle along a curve drawn in the base space manifold. This is why Wilson lines naturally arise when one deals with gauge covariant quantities. For more details about the geometrical meaning of gauge theories and how Wilson lines appears naturally in this context, see for instance \cite{Nakahara}.}, as
\begin{equation}
\mathcal{F}^{i-}_{(+)}(x^+,\mathbf{b}_{\perp})\equiv \tilde{\Omega}(+\infty,x^+,\mathbf{b}_{\perp})\mathcal{F}^{i-}(x^+,\mathbf{b}_{\perp})\tilde{\Omega}^{\dagger}(+\infty,x^+,\mathbf{b}_{\perp}).
\end{equation}
Thus after some algebra and together with \eqref{adjtoR}, \eqref{trace0} can be written
\begin{equation}
\begin{split}
\partial^i_x\partial^i_u&\tilde S^{(4)}(\mathbf{x}_{\perp},\mathbf{b}_{\perp},\mathbf{u}_{\perp},\bar{\mathbf{b}}_{\perp})\Big |_{\mathbf{b}_{\perp}\mathbf{b}_{\perp}\bar{\mathbf{b}}_{\perp}\bar{\mathbf{b}}_{\perp}}=\frac{(ig)^2}{N_c(N_c^2-1)}\int\rmd x^+\rmd y^+\mathcal{F}^{Ai-}_{(+)}(x^+,\mathbf{b}_{\perp})\mathcal{F}^{Bi-}_{(+)}(y^+,\bar{\mathbf{b}}_{\perp})\\
&\tilde{\Omega}_{CE}(\mathbf{b}_{\perp})\tilde{\Omega}^{\dagger}_{ED}(\bar{\mathbf{b}}_{\perp})\Tr\left[T^AT^CT^BT^D\right].
\end{split}
\end{equation}
To be completed, this calculation requires the computation of a trace of four adjoint representation generators $\Tr[T^AT^CT^BT^D]$. Through this example we have seen how traces of multiple generators arise.\\

This chapter is more formal, focusing on the mathematical aspect and trying to be as rigorous as possible. These notes are self-taught in the sense that literature for this specific purpose is rather poor or too formal. Then I tried to develop methods lying at the frontier between the mathematical and physical languages, keeping in mind that the main goal is an explicit result for traces involving three or four generators. These are explicitly computed in this chapter. I also show the general method that enables us to go, in principle, further. For the back-to-back limit of the di-gluon decoherence problem one has to compute such traces up to four generators, this is why I do not go further and this is, I think, the limit for making everything by hand without help of some calculus software. The main idea is to decompose products of generators into irreducible representations of the permutation group. This can be performed in an easy systematic way by use of Young tableaux. In the first part we briefly set the notations and conventions. Then we define the traces as invariant tensors and emphasize some basic properties. In the third part we deal with the permutation group and match it to invariant tensors. We also introduce the Young tableaux. On this basis, we next begin the computation of invariant tensors up to rank four with the general methods inherited from the correspondence with the permutation group. In the following section we take benefits from the cyclic symmetry of traces to simplify the procedure. At the end we compute traces up to four generators thanks to this last powerful method. The used references are the following textbooks : for generalities on Lie algebras and their representations \cite{WeinbergII,Zuber}, the permutation group \cite{Jones} and an intuitive approach (for physicists) to Young tableaux \cite{LandauLifshitzQM}.

\section{Introduction and conventions}
\indent

We shall deal only with simple Lie groups, denoted $G$ for which the generators are chosen hermitian and do obey the commutation relations :
\begin{equation}
\left[T^A;T^B\right]=if^{ABC}T^C.
\end{equation}
Simplicity implies that the Killing form is $\delta^{AB}$ and the structure constants are real and completely anti-symmetric. $A, B, C...$ are the indices of the adjoint representation running from $1$ to $d_G$ the group dimension. To set the notations we introduce the symmetrized and antisymmetrized products respectively defined as :
\begin{equation}
\begin{split}
&T^{\{A_1}T^{A_2}...T^{A_n\}}=\frac{1}{n!}(T^{A_1}T^{A_2}...T^{A_n}+perm.)\\
&T^{[A_1}T^{A_2}...T^{A_n]}=\frac{1}{n!}(T^{A_1}T^{A_2}...T^{A_n}\pm perm.).
\end{split}
\end{equation}
In the last line, the sign of the permutation depends on whether the permutation is even $(+)$ or odd $(-)$. 

\section{General properties of traces}

\subsection{Invariant tensors}
\indent

Among the various invariant tensors it is possible to construct, we consider :
\begin{equation}
\tr\left[T^{A_1}T^{A_2}...T^{A_n}\right].
\end{equation}
The $T$'s are the group generators in some arbitrary representation $R$. The number $n$ is called its \emph{rank}. Such tensor is invariant in the sense that it is unchanged under the action of an $R$ representation matrix of the group on the generators. That is for any matrix $U$ in the representation $R$ of $G$, one has :
\begin{equation}
\label{traceinv}
\tr\left[UT^{A_1}U^{-1}UT^{A_2}U^{-1}...UT^{A_n}U^{-1}\right]=\tr\left[T^{A_1}T^{A_2}...T^{A_n}\right].
\end{equation}
But this property involves another one which is non trivial : invariant tensors are also invariant under the action of the adjoint representation. Indeed, let $U$ be in representation $R$ and $\tilde{U}$ the corresponding adjoint representation matrix, we have the relation :
\begin{equation}
UT^AU^{-1}=T^B\tilde{U}^{BA}.
\end{equation}
Plugging this into \eqref{traceinv} gives :
\begin{equation}
\label{traceadjinv}
\tr\left[T^{B_1}T^{B_2}...T^{B_n}\right]\tilde{U}^{B_1A_1}\tilde{U}^{B_2A_2}...\tilde{U}^{B_nA_n}=\tr\left[T^{A_1}T^{A_2}...T^{A_n}\right].
\end{equation}
This last formula shows the invariance of trace tensors under the action of the adjoint representation.

\subsection{Trace algebra}
\indent

The arising question is in term of what we can write these traces ? Available tensors are $\delta^{AB}$ and $f^{ABC}$ but they are not sufficient. There is another class of tensors which cannot be expressed, in general, in terms of $\delta^{AB}$ and $f^{ABC}$, these are the so called \emph{D-symbols} defined as :
\begin{equation}
D^{A_1A_2...A_n}=\tr\left[T^{\{A_1}T^{A_2}...T^{A_n\}}\right].
\end{equation}
From their definition, the D-symbols are totally symmetric. As we shall see, any invariant tensor of rank $n$ is in general a polynomial in $\delta^{AB}$, $f^{ABC}$ and the D-symbols with up to $n$ indices. The numerical value of the D-symbols depends on the representation. 

\section{The permutation group and its irreducible representations}
\indent

In this section we first introduce the permutation group and its notations. Then we make the correspondence with products of generators. Finally we introduce a very efficient tool for finding all the irreducible representations of the permutation group : the Young tableaux.

\subsection{The permutation group, properties and notations}
\indent

An element $\sigma$ of the permutation group of $n$ elements, denoted $\mathfrak{S}_n$, assigns to the $n$ elements of an alphabet, say $(A_1, A_2, ..., A_n)$, the corresponding set $(\sigma(A_1),\sigma(A_2), ..., \sigma(A_n))$ with the following constrains :
\begin{itemize}
\item $\sigma(A_k)$ must be an element $A_l$ of the alphabet.
\item $\sigma(A_k)\neq\sigma(A_l)$ for all $k\neq l$.
\end{itemize}
In other words $\sigma$ is a one to one application from the alphabet onto itself. $\sigma$ realizes a permutation of the elements of the alphabet.
The group structure is therefore obvious : (i) the identity, denoted $e$, satisfies $e(A_k)=A_k$ for all $k$ ; (ii) since it is bijective, $\sigma$ has an inverse denoted $\sigma^{-1}$ ; (iii) it is associative. The dimension of $\mathfrak{S}_n$ is $n!$ . Indeed it corresponds to the number of ways of arranging the members of the alphabet. For $n>2$, $\mathfrak{S}_n$ is non abelian. $\mathfrak{S}_1$ and $\mathfrak{S}_2$ are abelian. It is convenient to introduce a matrix notation for $\sigma\in\mathfrak{S}_n$ which is the following :
\begin{equation}
\sigma=\left( \begin{array}{cccc}
A_1 & A_2 & ... & A_n \\
\sigma(A_1) & \sigma(A_2) & ... & \sigma(A_n)
\end{array} \right).
\end{equation}
Note that the order of the columns does not matter. The important thing is that in the same column, we must have the assigned value by $\sigma$ of the first row in the second one. The composition of two permutations in matrix form is illustrated as follow. Let $\sigma_1, \sigma_2 \in \mathfrak{S}_4$ defined by :
\begin{equation}
\sigma_1=\left( \begin{array}{cccc}
A & B & C & D \\
B & D & C & A
\end{array} \right)~~~
\sigma_2=\left( \begin{array}{cccc}
A & B & C & D \\
A & C & B & D
\end{array} \right).
\end{equation}
Then :
\begin{equation}
\sigma_2\sigma_1=\left( \begin{array}{cccc}
A & B & C & D \\
A & C & B & D
\end{array} \right)\left( \begin{array}{cccc}
A & B & C & D \\
B & D & C & A
\end{array} \right)=\left( \begin{array}{cccc}
A & B & C & D \\
C & D & B & A
\end{array} \right).
\end{equation}
To evaluate the composition, we start from a member of the alphabet in the upper row of the rightmost matrix, we read the value it takes after the first permutation in the second row, we start from this read value in the first row of the leftmost matrix and see the final value in the second row. For instance the first column is obtained by $A\rightarrow B\rightarrow C$. A more compact and really useful notation is the \emph{cyclic notation}. A cycle is obtained by starting from a member of the first row, we read the value given by the action of the permutation in the second row. Let's take this former value in the first row and so on, until we come back to the initial value. Ce cycle decomposition of $\sigma_1$ and $\sigma_2$ are respectively :
\begin{equation}
\sigma_1=\left( \begin{array}{ccc}
A & B & D \\
B & D & A
\end{array} \right)\left( \begin{array}{c}
C\\
C
\end{array} \right)~~~
\sigma_2=\left( \begin{array}{cc}
B & C\\
C & B
\end{array} \right)\left( \begin{array}{c}
A\\
A
\end{array} \right)\left( \begin{array}{c}
D\\
D
\end{array} \right).
\end{equation}
We can shrink the notation further. The identical pieces can be omitted. We only keep the non trivial cycles. These cycles can be denoted without any ambiguity on one line as :
\begin{equation}
\sigma_1=\left(A B D\right)~~~
\sigma_2=\left( B C\right)
\end{equation}
It means that for $\sigma_1$ $A$ becomes $B$, which becomes $D$, which becomes $A$. And similarly for $\sigma_2$. The order of letters in the cycle matters but it is defined up to a cyclic permutation. For instance :
\begin{equation}
\sigma_1=\left(A B D\right)=\left(B D A\right)=\left(D A B\right).
\end{equation}
As an other example, the cycle decomposition of $\sigma_2\sigma_1$ is :
\begin{equation}
\sigma_2\sigma_1=\left(A C B D\right).
\end{equation}

\subsection{Relation between invariant tensors and permutation group}
\indent

In order to write down $\Tr\left[T^{A_1}T^{A_2}...T^{A_n}\right]$ in terms of the various D-symbols, the structure constants and the Killing form, one has to isolate the various symmetries of the trace. For this purpose we need to decompose the product $T^{A_1}T^{A_2}...T^{A_n}$ into a totally symmetric piece, a totally anti-symmetric piece and other hybrid pieces which are symmetric and/or antisymmetric under the permutation of a subset of indices\footnote{For $n>2$ one can not decompose such a tensor only into a totally symmetric and a totally antisymmetric piece.}. One has to decompose the product $T^{A_1}T^{A_2}...T^{A_n}$ into specific linear combinations of its permutations. All the permutations span a $n!$ dimensional space. But there are some linear combinations of these permutations that have specific properties. For instance the totally symmetric piece in invariant under the action of the whole permutation group. It furnishes a trivial representation of the permutation group. In the same way, the totally antisymmetric piece furnishes a projective representation of the permutation group : it is unchanged up to an overall sign depending on whether the permutation is even or odd. As we shall see, there are some subsets of specific linear combinations of permutations that are stable under the action of the permutation group. From a formal point of view, we say that $T^{A_1}T^{A_2}...T^{A_n}$ and its permutations furnish a reducible representation of the permutation group and what we are looking for is the decomposition of $T^{A_1}T^{A_2}...T^{A_n}$ into irreducible representations. We shall now build these irreducible representations.

\subsection{Decomposition into irreducible representations, the Young tableaux technique}
\indent

As already mentioned the totally symmetrized and antisymmetrized tensors both furnish one dimensional irreducible representations but for $n>2$ there are other components that are more complicated and with higher dimensions\footnote{From the formal point of view of $n=2$, the reason of the simplicity is because $\mathfrak{S}_2={e, (AB)}$ obviously commute and can be diagonalized simultaneously in the basis $T^AT^B$ and $T^BT^A$. This is not possible to diagonalize simultaneously all the elements of $\mathfrak{S}_n$ for $n>2$ since it is no longer abelian. This is the reason for the higher dimensional representations in the case $n>2$.}. To find the irreducible representations of the permutation group, there is a very efficient tool known as the Young tableaux. It has been shown that the irreducible representations of the permutation group are in one to one correspondence with the Young tableau. Here is the recipe : let us consider the alphabet $(A_1, A_2, ..., A_n)$. We take $n_1$ elements of this alphabet, $n_2$ elements of the remaining members and so on until we have taken all the members. From this procedure, we get a partition of $n$ : $n_1+n_2+...=n$. Assuming we have chosen $n_1 \geq n_2 \geq ...$, the corresponding Young tableau has a first row of size $n_1$, a second row of size $n_2$ and so on. All the rows are conventionally aligned on the left and are of decreasing length as we go down in the tableau. For example, the Young tableau corresponding to $n=21$, decomposed into $21=6+5+5+3+1+1$, is :
\begin{equation}
\yng(6,5,5,3,1,1)
\end{equation}
The first row contains the arbitrarily chosen $n_1$ members of the alphabet in each box, the second row, the $n_2$ other members and so on... Now the procedure is the following : draw all Young tableaux corresponding to all the partitions of $n$. Symmetrize the tensor among the variables in each line. We therefore get a tensor symmetric with respect to blocks of indices.  Then, we have to antisymmetrize. Antisymmetrization of two variables in the same line gives obviously zero because they have been symmetrized. Thus we can antisymmetrize only with respect to variables in different rows. Once we have chosen one variable in each row, we can put them all in the first column because the tableau is symmetric under the interchange of variables in the same row. We antisymmetrize with respect to the first column. And repeat the operation on the Young tableau amputated of the first column. We do this in all possible ways, that is by choosing any combination of variables in different rows. A given Young tableau furnishes an irreducible representation of the permutation groups. At this point we must mention several points :
\begin{itemize}
\item After antisymmetrization with respect to variables in different rows the resulting tensor is, in general, no longer symmetric with respect to the variables in the same row. The only exception concerns variables present in a one dimensional columns that have therefore not been antisymmetrized.
\item All the tensors we get at the end are, in general, not independent and some of them can be written as linear combinations of other tensors within the same representation.
\item We could as well antisymmetrize with respect to the columns at first and then symmetrize with respect to the rows. It would just have given another (but equivalent) basis of the representation.
\end{itemize}
We will see explicit examples and uses in the next sections.

\section{Explicit computation of traces}
\indent

Here we come to a first attempt for traces computations using the correspondence with the permutation group and Young tableaux.

\subsection{$\tr\left[T^AT^B\right]$}
\indent

Let us begin by the simplest and well know case of two generators. It is obvious that $T^AT^B$ can be written as :
\begin{equation}
T^AT^B=\frac{1}{2}\{T^A;T^B\}+\frac{1}{2}[T^A;T^B].
\end{equation}
As already mentioned the decomposition of $\mathfrak{S}_2$ into irreducible representations is simple. But we shall see how it works with Young tableaux. $T^AT^B$ is written as.
\begin{equation}
\young(A)\otimes\young(B)=\young(AB)\oplus\young(A,B)
\end{equation}
$\young(AB)$ correspond to the one dimensional symmetric representation and $\young(A,B)$ to the one dimensional antisymmetric representation. Taking the trace, only the symmetric piece survives and we get :
\begin{equation}
\tr\left[T^AT^B\right]=\tr\left[T^{\{A}T^{B\}}\right]=D^{AB}.
\end{equation}
In the special case of $D^{AB}$, we can go further and show that $D^{AB}=\text{const.}\delta^{AB}$. Here is the proof : $D^{AB}$ is some group matrix in the adjoint representation. The generators in the adjoint representation are :
\begin{equation}
\left(\tilde{T}^{B}\right)_{AC}=if^{ABC}.
\end{equation}
We consider the commutator of $D^{AB}$ with $\tilde{T}^{A}$. This reads :
\begin{equation}
\begin{split}
\frac{i}{2}f^{BAC}&\tr\left[T^CT^D\right]-\frac{i}{2}\tr\left[T^BT^C\right]f^{CAD}\\
&=\frac{1}{2}\tr\left[\left[T^B;T^A\right]T^D-T^B\left[T^A;T^D\right]\right]=0.
\end{split}
\end{equation}
In the last line we used the cyclic symmetry of the trace. We have shown that $D^{AB}$ commutes with all the generators and therefore with all matrices of the group in the adjoint representation (this statement corresponds to the special case $n=2$ of \eqref{traceinv}). Then, by the Schur lemma, $D^{AB}$ must be a multiple of the identity that we conveniently write :
\begin{equation}
D^{AB}=D\delta^{AB}.
\end{equation}

\subsection{$\tr\left[T^AT^BT^C\right]$\label{tr3T}}
\indent

Here we come to non trivial decompositions. According to the general procedure, we write the symbolics for $T^AT^BT^C$ :
\begin{equation}
\begin{split}
\young(A)\otimes\young(B)\otimes\young(C)&=\left(~\young(AB)\oplus\young(A,B)~\right)\otimes\young(C)\\
&=\young(ABC)\oplus\young(AB,C)\oplus\young(AC,B)\oplus\young(A,B,C).
\end{split}
\end{equation}
Before going further, we can wonder why the tableau $\young(BC,A)$ does not appear in the Clebsch-Gordan decomposition because the $2+1$ partition of $3$ must involve all the ways to take two indices to put in the first line. As we mentioned the set of linear combinations of permutations obtained in the \yng(2,1) decomposition does not necessarily form a basis of the irreducible representations. Actually the $\young(BC,A)$ representation is redundant since it belongs to $\young(AB,C)\oplus\young(AC,B)$. This can be seen by a dimensionality argument. The dimension of $\young(A)\otimes\young(B)\otimes\young(C)$ is $3!=6$. The totally symmetric and antisymmetric representations (single row and single column respectively) are one-dimensional. The dimension of $\young(AB,C)$ is two because we can antisymmetrize $A$ and $C$ or $B$ and $C$ and similarly for $\young(AC,B)$. Hence the dimension counting of the Clebsch-Gordan decomposition is $6=1+2+2+1$ forbidding the $\young(BC,A)$ representation. Now let us look for basis of each irreducible representations. The basis of $\young(ABC)$ and $\young(A,B,C)$ are respectively $T^{\{A}T^BT^{C\}}$ and $T^{[A}T^BT^{C]}$. Then we can write :
\begin{equation}
\label{3Tdecomp}
T^AT^BT^C=T^{\{A}T^BT^{C\}}+T^{[A}T^BT^{C]}+\frac{1}{3}\left(2T^AT^BT^C-T^BT^CT^A-T^CT^AT^B\right).
\end{equation}
The basis of $\young(AB,C)$ is given by the two combinations (the permutations act on the indices) :
\begin{equation}
\frac{1}{4}\left( \begin{array}{c}
e-(AB)\\
e-(AC)
\end{array} \right)(e+AB)T^AT^BT^C
\end{equation}
and the basis of $\young(AC,B)$, by :
\begin{equation}
\frac{1}{4}\left( \begin{array}{c}
e-(BC)\\
e-(BA)
\end{array} \right)(e+AC)T^AT^BT^C.
\end{equation}
We could write them more explicitly and find a linear combination of them to rewrite the rightmost term of \eqref{3Tdecomp} because it lies in the $\young(AB,C)\oplus\young(AC,B)$ representation according to the Clebsch-Gordan decomposition. But there is a way out since it is easily seen that the rightmost term in \eqref{3Tdecomp} vanishes under the trace because of cyclic symmetry. The required decomposition is then :
\begin{equation}
\label{trace3T}
\begin{split}
\tr\left[T^AT^BT^C\right]&=\tr\left[T^{\{A}T^BT^{C\}}\right]+\tr\left[T^{[A}T^BT^{C]}\right]\\
&=D^{ABC}+\frac{1}{2}\tr\left[\left[T^A;T^B\right]T^C\right] ~~~~\text{by cyclicity}\\
&=D^{ABC}+\frac{iD}{2}f^{ABC}.
\end{split}
\end{equation}

\subsection{$\tr\left[T^AT^BT^CT^D\right]$, a naive computation\label{tr4T}}
\indent

As we will see soon, the previous techniques become really tedious at this point. The Clebsch-Gordan decomposition of $T^AT^BT^CT^D$ reads :
\begin{equation}
\label{CG4Ttrace}
\begin{split}
\young(A)\otimes\young(B)\otimes\young(C)\otimes\young(D)&=\left(~\young(ABC)\oplus\young(AB,C)\oplus\young(AC,B)\oplus\young(A,B,C)~\right)\otimes\young(D)\\
&=\young(ABCD)\oplus\young(ABC,D)\oplus\young(ABD,C)\oplus\young(ACD,B)\oplus\young(AB,CD)\oplus\young(AC,BD)\oplus\\
&~~~~~~~\oplus\young(AB,C,D)\oplus\young(AC,B,D)\oplus\young(AD,B,C)\oplus\young(A,B,C,D)
\end{split}
\end{equation}
This decomposition reads in dimensions of the representations :
\begin{equation}
4!=24=1+3+3+3+2+2+3+3+3+1.
\end{equation}
There are two difficulties :
\begin{itemize}
\item the explicit expression for basis in each representation is a tedious work without a computer because they involve a large number of terms.
\item it is not really obvious to see the terms that will be killed by the trace. The totally antisymmetric representation vanishes under the trace because for a given index configuration, there are three others obtained by cyclic permutation. One of them is even, the other two are odd and everything cancels out. But there will be other pieces that will be killed by the trace. Some of them are subspaces of irreducible representations that are difficult to read directly from Young tableaux.
\end{itemize}
Actually, the computation of a trace involving $n+1$ generators does not require much more energy than the computation of a trace involving $n$ generators with the techniques we used so far, provided we take benefits from the cyclic symmetry of traces. This is treated in the next section.

\section{Cyclic symmetry, coset space and isomorphism}
\indent

The method developed above applies to general tensor with no particular symmetries : it decomposes a product of generators into irreducible representations of the permutation group and not necessarily its trace. However, in the case of interest, the trace is invariant under cyclic permutations. This symmetry lowers the dimension of the (a)symmetry under permutations.

\subsection{The cyclic group}
\indent

The cyclic group denoted $C_n$ is the symmetry group of the $n$-sided oriented polygon. Its dimension is $n$. A \emph{cyclic permutation}, is a permutation which takes the alphabet $(A_1, ..., A_k, A_{k+1}, ..., A_n)$ and gives $(A_{k+1}, ..., A_n, A_1,..., A_k)$ for some $k=1,...,n$. As easily seen the composition of two cyclic permutations is again a cyclic permutation and its group structure is therefore inherited from $\mathfrak{S}_n$. Cyclic permutations form a subgroup of $\mathfrak{S}_n$. The dimension of the cyclic permutation group is $n$ which corresponds to the number of ways of choosing the number $k$. This group is isomorphic to $C_n$. Note this subgroup is not a normal subgroup : a cyclic permutation does not commute, in general, with an arbitrary permutation of $\mathfrak{S}_n$. This can be summarized as follow : there are $\sigma\in\mathfrak{S}_n$ so that
\begin{equation}
\sigma^{-1}C_n\sigma \neq C_n.
\end{equation}
This has important consequences to be seen in the following.

\subsection{Trace symmetry}
\indent

Since the trace involving $n$ matrices is invariant under cyclic permutations. It is natural to identify permutations that differ only by a cyclic permutation. For $\sigma, \sigma' \in \mathfrak{S}_n$, we introduce the equivalence relation\footnote{It is obviously reflexive, symmetric and transitive from the group structure of $C_n$.} : $\sigma\sim\sigma'$ if there is $c\in C_n$ so that $\sigma=c\sigma'$. For a given $\sigma\in\mathfrak{S}_n$ we define the \emph{right coset} $C_n\sigma$. It is the set of all the elements of $C_n$ multiplied on the right by $\sigma$. $\sigma\sim\sigma'$ obviously give the same coset : $C_n\sigma=C_n\sigma'$. According to the general properties of cosets, this equivalence relation gives a partition of the group and each coset contain the same number of elements. Since the cyclic permutations do not affect the trace, all members of the same coset lead to equivalent permutations of indices within the trace. The permutation of indices within the trace is defined up to a cyclic permutation.\\

It would be tempting to identify the symmetry group of traces as the coset space $\mathfrak{S}_n/C_n$. However, since $C_n$ is not a normal subgroup, $\mathfrak{S}_n/C_n$ cannot be endowed with a canonical group structure. Since arbitrary permutations do not commute with cyclic ones, we have in general 
\begin{equation}
C_n\sigma_1C_n\sigma_2 \neq C_n\sigma_1\sigma_2~,
\end{equation}
for arbitrary $\sigma_1$ and $\sigma_2$ in $\mathfrak{S}_n$. In other words the composition of any two element of the cosets $C_n\sigma_1$ and $C_n\sigma_2$ does not generally lies in the coset $C_n\sigma_1\sigma_2$. This is a problem since one would like to take benefit of a group structure for trace (a)symmetry. Fortunately there is a way to face this problem. Since all the elements of the same coset are equivalent we acting on a trace, one can reduce the permutation group $\mathfrak{S}_n$ by picking up an element in each different coset. This element is called a \emph{representative} of the coset. All elements within the same coset can be identified to this representative. Which representative can we choose ? because of the cyclic symmetry, given a permutation $\sigma$ in $\mathfrak{S}_n$, there is a unique element in $C_n\sigma$ called $\sigma_r$($=c\sigma$ for some $c$ in $C_n$) so that $\sigma_r(A_n)=A_n$. The (arbitrary) choice we make is to represent the coset $C_n\sigma$ by the element that leaves the last index unchanged. There exist such an element in each coset and it is unique\footnote{{\bf Proof :}
Given $\sigma\in \mathfrak{S}_n$ represented by :
$$
\sigma=\left( \begin{array}{cccc}
A_1 & A_2 & ... & A_n \\
\sigma(A_1) & \sigma(A_2) & ... & \sigma(A_n)
\end{array} \right)\equiv\left( \begin{array}{cccc}
A_1 & A_2 & ... & A_n \\
A_{\sigma(1)} & A_{\sigma(2)} & ... & A_{\sigma(n)}
\end{array} \right).
$$
A cyclic permutation transforms $A_{\sigma(k)}$ into $A_{\sigma(k)+l}$ for all $k$ and fixed $l=0, ..., n-1$ (we conventionally set $A_{n+m}=A_m$). The equivalence class of $\sigma$ is given by all the cyclic permutations of $\sigma$, i.e. all possible $l$ for which $A_{\sigma(k)}\rightarrow A_{\sigma(k)+l}$ for all $k$. But the condition $A_{\sigma(n)+l}=A_n$ fixes $l$ to a \emph{unique} value $l=n-\sigma(n)$. Since $\sigma(n)$ is between 1 and $n$, then $l$ is between $0$ and $n-1$ as expected and guarantees the existence. The proof is completed.}. In matrix form the chosen representative reads :
\begin{equation}
\label{cosetrepr}
\sigma_r=\left( \begin{array}{cccc}
A_1 & ... & A_{n-1} & A_n \\
\sigma_r(A_1) & ... & \sigma_r(A_{n-1}) & A_n
\end{array} \right).
\end{equation}
Then we construct the coset algebra as follow : all elements of the coset $C_n\sigma$ are identified to the representative $\sigma_r$ in the sense given above. The composition of any two elements in different cosets, say $\sigma_1$ and $\sigma_2$, is \emph{defined} as follow :
\begin{equation}
\label{cosetalgebra}
\sigma_1\sigma_2 \rightarrow \sigma_{r1}\sigma_{r2}.
\end{equation}
The composition of two representatives $\sigma_{r1}\sigma_{r2}$ is obviously of the form \eqref{cosetrepr}. And is therefore candidate to be a representative. A point still needs to be clarified : are all the possible permutations in the form \eqref{cosetrepr} in a coset of $\mathfrak{S}_n$ ? We already know that there is a \emph{unique} permutation of the form \eqref{cosetrepr} in each coset. Two different permutations of the form \eqref{cosetrepr} necessarily lie in two different cosets because of their uniqueness. The order of each coset is the order of $C_n$ (the number of different cyclic permutations), that is $n$. Then there are $n!/n=(n-1)!$ different cosets in $\mathfrak{S}_n$ which precisely corresponds to the number of permutations in the form \eqref{cosetrepr}.\\

Moreover, the associativity, the existence of the inverse and the identity are then automatically checked from \eqref{cosetrepr}. Therefore the coset algebra \eqref{cosetalgebra} is isomorphic to $\mathfrak{S}_{n-1}$. To precise the terminology, the isomorphism is between the representatives (in the sense given above) of $\mathfrak{S}_n/C_n$ and $\mathfrak{S}_{n-1}$. There is no canonical isomorphism between $\mathfrak{S}_n/C_n$ and $\mathfrak{S}_{n-1}$ due to the fact that $C_n$ is not a normal subgroup.\\

A more intuitive way to see all these things is the following : since the trace is invariant under a cyclic permutation, for a given configuration of the indices, one can always, by a cyclic permutation, carry the $A_n$ index at the rightmost place. Then two inequivalent permutations in the sense given earlier correspond to two different ways of arranging the $n-1$ indices $(A_1, ..., A_{n-1})$. The cyclic symmetry of the trace enables us to consider the irreducible representations of $\mathfrak{S}_{n-1}$ instead of $\mathfrak{S}_{n}$ for a trace involving $n$ generators.

\subsection{Totally symmetric and antisymmetric representations}
\indent

Here we emphasize general properties of these representations. Because of the cyclic symmetry, we have :
\begin{equation}
\label{totsym}
D^{A_1...A_n}=\Tr\left[T^{\{A_1}T^{A_2}...T^{A_{n-1}}T^{A_n\}}\right]=\Tr\left[T^{\{A_1}T^{A_2}...T^{A_{n-1}\}}T^{A_n}\right]
\end{equation}
and
\begin{equation}
\label{totantisym}
\Tr\left[T^{[A_1}T^{A_2}...T^{A_{n-1}}T^{A_n]}\right]=\left\{
    \begin{array}{ll}
         &\Tr\left[T^{[A_1}T^{A_2}...T^{A_{n-1}]}T^{A_n}\right]~~~\text{if n odd}\\
         & 0 ~~~~~~~\text{if n even.}
    \end{array}
\right.
\end{equation}

\section{Explicit computation of traces (continued)}
\indent

Now we take advantage of the cyclic symmetry to compute the rank three and four tensors of sections \ref{tr3T} and \ref{tr4T}, respectively, in an easier way as previously. The way to proceed automatically gets rid of the irreducible representations of $\mathfrak{S}_{n}$ that are killed by the trace.

\subsection{$\tr\left[T^AT^BT^C\right]$ almost for free}
\indent

Using the cyclic symmetry, the previous techniques enable us to decompose only the $A$ and $B$ indices into irreducible representations which is merely :
\begin{equation}
\begin{split}
\tr\left[T^AT^BT^C\right]&=\frac{1}{2}\tr\left[\{T^A;T^B\}T^C\right]+\frac{1}{2}\tr\left[[T^A;T^B]T^C\right]\\
&=D^{ABC}+\frac{iD}{2}f^{ABC}.
\end{split}
\end{equation}
In the last line, we have used \eqref{totsym} to get the third rank D-symbol. By this method, result \eqref{trace3T} has been recovered very easily. The advantage is that we do not have to worry about the components vanishing under the trace. They are automatically canceled from the beginning.\\

As an example, we can show that the third rank, adjoint representation, D-symbols vanish for any simple Lie group :
\begin{equation}
\begin{split}
\tilde{D}^{ABC}&=\frac{1}{2}\Tr\left[\left\{\tilde{T}^{A};\tilde{T}^{B}\right\}\tilde{T}^{C}\right]\\
&=\frac{-i}{2}\left(f^{DAE}f^{EBF}+f^{DBE}f^{EAF}\right)f^{FCD}\\
&=\frac{-i}{2}\left(f^{DAE}f^{EBF}f^{FCD}+f^{FBE}f^{EAD}f^{DCF}\right)\\
&=\frac{-i}{2}\left(f^{DAE}f^{EBF}f^{FCD}-f^{EBF}f^{DAE}f^{FCD}\right)=0.
\end{split}
\end{equation}
This can be easily seen from the fact that $\Tr\left[\tilde{T}^{A}\tilde{T}^{B}\tilde{T}^{C}\right]$ is completely antisymmetric.

\subsection{Jacobi-type identities}
\indent

For higher rank invariant tensor computation, we will need to use simplifying formulas. One of them is the ordinary Jacobi identity :
\begin{equation}
\label{jacobi1}
\begin{split}
&\left[\tilde{T}^{A};\tilde{T}^{B}\right]_{CD}=if^{ABE}\tilde{T}^{E}_{CD}\\
&\Leftrightarrow~~~~ f^{ABE}f^{CDE}+f^{CAE}f^{BDE}+f^{BCE}f^{ADE}=0.
\end{split}
\end{equation}
We can get another identity which is specific to trace algebra :
\begin{equation}
\label{jacobi2}
\begin{split}
&\Tr\left[T^A\left\{T^B;\left[T^C;T^D\right]\right\}\right]+\Tr\left[T^C\left\{T^A;\left[T^B;T^D\right]\right\}\right]+\Tr\left[T^B\left\{T^C;\left[T^A;T^D\right]\right\}\right]=0\\
\Leftrightarrow~~~~ & D^{ABE}f^{CDE}+D^{CAE}f^{BDE}+D^{BCE}f^{ADE}=0.
\end{split}
\end{equation}
This will be useful to simplify mixed $f-D$-symbols terms.

\subsection{$\tr\left[T^AT^BT^CT^D\right]$ made easier}
\indent

To compute this trace, we start from decomposition \eqref{3Tdecomp} :
\begin{equation}
\Tr\left[T^AT^BT^CT^D\right]=\Tr\left[\left(T^{\{A}T^BT^{C\}}+T^{[A}T^BT^{C]}+\frac{1}{3}\left(2T^AT^BT^C-T^BT^CT^A-T^CT^AT^B\right)\right)T^D\right].
\end{equation}
Let us look at the different terms one by one.

\paragraph{$\tr\left[T^{\{A}T^BT^{C\}}T^D\right]$ :}

This term corresponds to the totally symmetric component $\yng(3)$ and equals $D^{ABCD}$ by \eqref{totsym}.

\paragraph{$\tr\left[T^{[A}T^BT^{C]}T^D\right]$ :}

This term corresponds to the totally antisymmetric component of $\mathfrak{S}_3$, $\yng(1,1,1)$. This is equal to :
\begin{equation}
\begin{split}
\tr\left[T^{[A}T^BT^{C]}T^D\right]&=\frac{1}{6}\left(\tr\left[T^{A}\left[T^B;T^{C}\right]T^D\right]+\tr\left[T^{B}\left[T^C;T^{A}\right]T^D\right]+\tr\left[T^{C}\left[T^A;T^{B}\right]T^D\right]\right)\\
&=\frac{1}{6}\tr\left[T^{A}\left[T^B;T^{C}\right]T^D\right]+\text{cycl.}(ABC).
\end{split}
\end{equation}
We note that :
\begin{equation}
\tr\left[T^{A}\left[T^B;T^{C}\right]T^D\right]+\text{cycl.}(ABC)=\frac{1}{2}\tr\left[\left\{T^{A};\left[T^B;T^{C}\right]\right\}T^D\right]+\frac{1}{2}\tr\left[\left[T^{A};\left[T^B;T^{C}\right]\right]T^D\right]+\text{cycl.}(ABC).
\end{equation}
The second term in the r.h.s, together with its cyclic permutations of $(ABC)$, vanishes because it is the Jacobi identity \eqref{jacobi1}, leaving only the first one. Then $\tr\left[T^{[A}T^BT^{C]}T^D\right]$ equals :
\begin{equation}
\begin{split}
\tr\left[T^{[A}T^BT^{C]}T^D\right]&=\frac{1}{6}\left(if^{BCE}D^{ADE}+if^{CAE}D^{BDE}+if^{ABE}D^{CDE}\right)\\
&=\frac{i}{6}\left(D^{ADE}f^{BCE}+D^{BDE}f^{CAE}+D^{CDE}f^{ABE}\right).
\end{split}
\end{equation}

\paragraph{$\frac{1}{3}\tr\left[\left(2T^AT^BT^C-T^BT^CT^A-T^CT^AT^B\right)T^D\right]$ :}

This contribution is a linear superposition of the two inequivalent $\yng(2,1)$ representations. Instead of looking for an explicit decomposition into these two different representations, this term can easily be computed directly thanks to \eqref{trace3T} :
\begin{equation}
\begin{split}
\frac{1}{3}\tr&\left[\left(2T^AT^BT^C-T^BT^CT^A-T^CT^AT^B\right)T^D\right]=\frac{1}{3}\left(\tr\left[T^AT^B\left[T^C;T^D\right]\right]+\tr\left[T^BT^C\left[T^D;T^A\right]\right]\right)\\
&=\frac{i}{3}\left(f^{CDE}\left(D^{ABE}+\frac{iD}{2}f^{ABE}\right)+f^{DAE}\left(D^{BCE}+\frac{iD}{2}f^{BCE}\right)\right)\\
&=\frac{2i}{6}\left(f^{CDE}D^{ABE}+f^{DAE}D^{BCE}\right)-\frac{D}{6}\left(f^{CDE}f^{ABE}+f^{DAE}f^{BCE}\right)\\
&=\frac{2i}{6}\left(D^{ABE}f^{CDE}-D^{BCE}f^{ADE}\right)-\frac{D}{6}\left(f^{ABE}f^{CDE}-f^{BCE}f^{ADE}\right)
\end{split}
\end{equation}\\

Let us stick all the pieces together. This gives :
\begin{equation}
\label{trace3T}
\begin{split}
\tr\left[T^AT^BT^CT^D\right]&=D^{ABCD}+\frac{i}{6}\left(D^{ADE}f^{BCE}+D^{BDE}f^{CAE}+D^{CDE}f^{ABE}+\right.\\
&\left. +2D^{ABE}f^{CDE}-2D^{BCE}f^{ADE}\right)+\frac{D}{6}\left(f^{BCE}f^{ADE}-f^{ABE}f^{CDE}\right)\\
&=D^{ABCD}+\frac{i}{2}\left(D^{ADE}f^{BCE}-D^{BCE}f^{ADE}\right)+\frac{D}{6}\left(f^{BCE}f^{ADE}-f^{ABE}f^{CDE}\right).
\end{split}
\end{equation}
To get the last line, we have used the second Jacobi identity \eqref{jacobi2}. Formula \eqref{trace3T} is the full formula for the trace involving four generators. Taking advantage of the trace symmetry has made the calculation much simpler than in our first attempt above.\\

One can in principle work out higher rank tensors but this becomes cumbersome analytically. We have been lucky to be able compute the $\yng(2,1)$ term directly for the trace of four generators, however the systematic procedure requires to find the two explicit representations of $\yng(2,1)$. With the simplification thanks to the trace symmetry trick, the trace of four generators has the same complexity as decomposing $T^AT^BT^C$ into irreducible representations of the permutation group. Similarly for computing the trace involving five generators, one has to find the explicit representations of the Clebsch-Gordan decomposition \eqref{CG4Ttrace} for a product of four generators which is greatly simplified with a computer.

\appendix

\chapter{The external field approximation in non abelian gauge theories\label{nonabelianBF}}
\indent

\setcounter{equation}{0}

Here we shall deal with the external field approximation in ultra-relativistic collisions. We shall emphasize the conditions for fast particles that radiates soft gauge bosons to be considered as a classical source or equivalently, an external classical gauge field $\mathcal{A}$ added to the action. Especially we will point out the difference between the abelian and the non abelian case for the domain of validity of this approximation. Because of the color algebra of non abelian gauge theories, the external field approximation really differs from the abelian case : in the abelian case, one can approximate any single fast particle radiating soft photons by a classical source whereas in the non abelian case one cannot and one must consider a very large number of hard sources in order to justify the external field approximation as we shall see.

\section{Soft gauge bosons radiation by an ultra-relativistic charged particle}
\indent

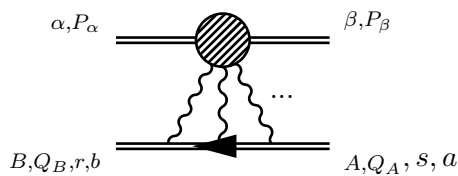
\begin{figure}[h]
\centering
\vspace{1cm}
%(along, up)
\begin{fmfgraph*}(80,40)
 \fmfstraight
    \fmftop{i1,v1,VU,v2,o1}
    \fmfbottom{i2,v3,v4,v5,o2}
      \fmf{dbl_plain}{i1,o1}
      \fmf{dbl_plain_arrow}{o2,i2}
      \fmfblob{20.}{VU}
      \fmffreeze
      \fmf{photon,label=$...$}{v5,VU}
      \fmf{photon}{v4,VU}
      \fmf{photon}{v3,VU}
      \fmflabel{$_{\alpha,P_{\alpha}}$}{i1}
      \fmflabel{$_{\beta,P_{\beta}}$}{o1}
      \fmflabel{$_{B,Q_{B},r,b}$}{i2}
      \fmflabel{$_{A,Q_A},s,a$}{o2}
\end{fmfgraph*}
\caption{Typical process considered where a heavy charged particle represented with an arrow interacts with another system through soft gauge bosons exchanges.\label{alphaAbetaB}}
\end{figure}

To fix the conventions, the reaction we are going to consider is $\alpha A \rightarrow \beta B$ represented on diagram \ref{alphaAbetaB}.
$\alpha$ and $\beta$ are any asymptotic multi-particle states with total momenta $P_{\alpha}=(P_{\alpha}^+,0,\mathbf{0})$, and $P_{\beta}=(P_{\beta}^+,P_{\beta}^-,\mathbf{P}_{\beta ,\perp})$ respectively. $A$ and $B$ are respectively the in and out state of a fast one-particle state considered as a fermion for definiteness, but the procedure can be mimicked for arbitrary spin (we shall use the eikonal vertex approximation, detailed in section \ref{eikonalvertex} that has the same structure for any kind of field). The frame is chosen so that the $A$ state carries a momentum pointing for definiteness in the minus direction of light cone coordinates : $Q_A=n^{\mu}Q^-$ with $n^{\mu}=(0,1,\mathbf{0})$. We shall focus on the piece of phase space where the gauge bosons exchanged between the fast fermion and the other particles involved in the process are soft. The soft gauge boson radiation condition reads $Q_B\simeq Q_A$, that is the radiated bosons have components of their momenta that are very small compared to the energy of the fast fermion. Since we consider ultra-relativistic reactions, masses are neglected. It will be convenient to introduce the Green function $\mathcal{G}^{A_1...A_n}_{\beta\alpha,\mu_1...\mu_n}(l_1,...,l_n)$ which is represented on figure \ref{GreenMultigluon}.
\begin{figure}[h]
\centering
\vspace{1cm}
%(along, up)
\begin{fmfgraph*}(150,40)
 \fmfstraight
    \fmftop{i1,a,b,c,o1}
    \fmfbottom{i2,d,e,f,o2}
      \fmf{dbl_plain}{i1,b}
      \fmf{dots}{b,c}
      \fmf{dbl_plain}{c,o1}
      \fmfblob{10.}{a}
      \fmfblob{10.}{b}
      \fmfblob{10.}{c}
      \fmffreeze
      \fmf{photon,label=$\uparrow~l_1$}{d,a}
      \fmf{photon,label=$\uparrow~l_2$}{e,b}
      \fmf{photon,label=$\uparrow~l_n$}{f,c}
      \fmflabel{$_{\alpha,P_{\alpha}}$}{i1}
      \fmflabel{$_{\beta,P_{\beta}}$}{o1}
      \fmflabel{$_{A_1, \mu_1}$}{d}
      \fmflabel{$_{A_2, \mu_2}$}{e}
      \fmflabel{$_{A_n, \mu_n}$}{f}
\end{fmfgraph*}
\caption{Diagrammatic representation of $\mathcal{G}^{A_1...A_n}_{\beta\alpha,\mu_1...\mu_n}(l_1,...,l_n)$\label{GreenMultigluon}}
\end{figure}
With this notation, the S-matrix element\footnote{The total four-momentum delta function will play a crucial role since in the derivation of external field approximation, this is why we do not consider the $\mathcal{M}$-matrix.} corresponding to diagram \ref{alphaAbetaB} reads :
\begin{equation}
\begin{split}
(ig)^n&(2\pi)^4\delta^{(4)}(Q_A+P_{\alpha}-Q_B-P_{\beta})\int \frac{\rmd^4l_1}{(2\pi)^4}...\frac{\rmd^4l_n}{(2\pi)^4}(2\pi)^4\delta^{(4)}(Q_A-l_1-...-l_n-Q_B)\\
&\times\mathcal{G}^{A_1...A_n}_{\beta\alpha,\mu_1...\mu_n}(l_1,...,l_n)\frac{i\Pi^{\mu_1\nu_1}(l_1)}{l_1^2+i\epsilon}...\frac{i\Pi^{\mu_n\nu_n}(l_n)}{l_n^2+i\epsilon}\displaystyle{\sum_{\sigma\in\mathfrak{S}_n}}(T^{A_{\sigma(1)}}...T^{A_{\sigma(n)}})_{ba}\\
&\times\overline{u}^r(Q_B)\gamma_{\nu_{\sigma(1)}}\frac{i(\slashed{Q_A}-\slashed{l}_{\sigma(2)}-...-\slashed{l}_{\sigma(n)})}{(Q_A-l_{\sigma(2)}-...-l_{\sigma(n)})^2+i\epsilon}...\frac{i(\slashed{Q_A}-\slashed{l}_{\sigma(n)})}{(Q_A-l_{\sigma(n)})^2+i\epsilon}\gamma_{\nu_{\sigma(n)}}u^s(Q_A).
\end{split}
\end{equation}
Now we consider the piece of the integral over the $l_i$'s so that they are soft : $l_i\ll Q_A$. In the numerators of the propagators one can neglect the $l_i$'s with respect to $Q_A$ and since $Q_B\simeq Q_A$, the numerator sandwiched between the $u$'s spinors gives a factor $2Q_{A\nu_{\sigma(1)}}...2Q_{A\nu_{\sigma(n)}}$ which is fully symmetric in its Lorentz indices. Moreover, one can expand the denominators in powers of $l_i$'s recalling that $Q_A^2=0$. Therefore the upper expression becomes :
\begin{equation}
\label{softSmatrix0}
\begin{split}
2ig^n&\delta^{rs}(2\pi)^4\delta^{(4)}(Q_A+P_{\alpha}-Q_B-P_{\beta})Q_{A\nu_1}...Q_{A\nu_n}\int \frac{\rmd^4l_1}{(2\pi)^4}...\frac{\rmd^4l_n}{(2\pi)^4}(2\pi)^4\delta^{(4)}(Q_A-l_1-...-l_n-Q_B)\\
&\times\mathcal{G}^{A_1...A_n}_{\beta\alpha,\mu_1...\mu_n}(l_1,...,l_n)\frac{i\Pi^{\mu_1\nu_1}(l_1)}{l_1^2+i\epsilon}...\frac{i\Pi^{\mu_n\nu_n}(l_n)}{l_n^2+i\epsilon}\\
&\times\displaystyle{\sum_{\sigma\in\mathfrak{S}_n}}(T^{A_{\sigma(1)}}...T^{A_{\sigma(n)}})_{ba}\frac{1}{Q_A\cdot(l_{\sigma(2)}+...+l_{\sigma(n)})-i\epsilon}...\frac{1}{Q_A\cdot l_{\sigma(n)}-i\epsilon}+~~~\text{hard emissions}\\
=2ig^n&\delta^{rs}(2\pi)^4\delta^{(4)}(Q_A+P_{\alpha}-Q_B-P_{\beta})Q_A^-n_{\nu_1}...n_{\nu_n}\int \frac{\rmd^4l_1}{(2\pi)^4}...\frac{\rmd^4l_n}{(2\pi)^4}(2\pi)^4\delta^{(4)}(Q_A-l_1-...-l_n-Q_B)\\
&\times\mathcal{G}^{A_1...A_n}_{\beta\alpha,\mu_1...\mu_n}(l_1,...,l_n)\frac{i\Pi^{\mu_1\nu_1}(l_1)}{l_1^2+i\epsilon}...\frac{i\Pi^{\mu_n\nu_n}(l_n)}{l_n^2+i\epsilon}\\
&\times\displaystyle{\sum_{\sigma\in\mathfrak{S}_n}}(T^{A_{\sigma(1)}}...T^{A_{\sigma(n)}})_{ba}\frac{1}{(l_{\sigma(2)}+...+l_{\sigma(n)})^+-i\epsilon}...\frac{1}{l^+_{\sigma(n)}-i\epsilon}+~~~\text{hard emissions}.
\end{split}
\end{equation}
Practically instead of considering the fermion with a definite momentum, one rather consider a wave function :
\begin{equation}
\label{fermionwavefunction}
\left|\right.Q_A,s,a\left.\right\rangle \rightarrow \left|\right.\Phi_Q\left.\right\rangle\equiv\displaystyle{\sum_{s,a}}\int\frac{\rmd Q_A^-\rmd^2Q_{A,\perp}}{(2\pi)^3\sqrt{2Q_A^-}}\chi^{sa}_Q(Q_A)\left|\right.Q_A,s,a\left.\right\rangle~;
\end{equation}
where $\chi_Q$ is a sharply peaked function around $Q^{\mu}=n^{\mu}Q^-$ and the normalization condition reads :
\begin{equation}
\label{chinormalization}
\left\langle\Phi_Q\left|\right.\Phi_Q\right\rangle=1 =\displaystyle{\sum_{s,a}}\int\frac{\rmd Q_A^-\rmd^2Q_{A,\perp}}{(2\pi)^3}\left|\chi^{sa}_Q(Q_A)\right|^2.
\end{equation}
Therefore, one has to integrate \eqref{softSmatrix0} over $Q_A^-, \mathbf{Q}_{A,\perp}$ and $Q_B^-, \mathbf{Q}_{B,\perp}$ and to sum over initial and final fermion's spins and colors with respective weights $\chi^{sa}_Q(Q_A)/(2\pi)^3\sqrt{2Q_A^-}$ and $\chi^{r*}_Q(Q_B)/(2\pi)^3\sqrt{2Q_B^-}$. Although \eqref{softSmatrix0} has been derived for a fermion that strictly points in the minus direction, it remains valid even though the fermion state is integrated over some narrow transverse momenta range because the sharpness condition of $\chi_Q$ around $Q$ enables us to replace $Q_A$ and $Q_B$ by $Q$ in the integrand except within the delta functions. The soft emission requirement means that $l_i^+\ll P_{\alpha,\beta}^+$ and $l_i^-\ll Q^-_{A,B}$ and then the delta functions can be approximated by :
\begin{equation}
\begin{split}
&\delta(Q_A^--Q_B^--P_{\beta}^-)\delta(Q_A^--l_1^--...-l_n^--Q_B^-)\simeq \delta(P_{\beta}^--l_1^--...-l_n^-)\delta(Q_A^--Q_B^-)\\
&\delta(P_{\alpha}^+-Q_B^+-P_{\beta}^+)\delta(Q_A^+-l_1^+-...-l_n^+-Q_B^+)\simeq \delta(P_{\alpha}^+-P_{\beta}^+)\delta(l_1^++...+l_n^+)\\
&\delta^{(2)}(\mathbf{Q}_{A,\perp}-\mathbf{Q}_{B,\perp}-\mathbf{P}_{\beta,\perp})\delta^{(2)}(\mathbf{Q}_{A,\perp}-\mathbf{l}_{1,\perp}-...-\mathbf{l}_{n,\perp}-\mathbf{Q}_{B,\perp})=   \delta^{(2)}(\mathbf{Q}_{A,\perp}-\mathbf{Q}_{B,\perp}-\mathbf{P}_{\beta,\perp})\\
&~~~~~~~~~~~~~~~~~~~~~~~~~~~~~~~~~~~~~~~~~~~~~~~~~~~~~~~~~~~~~~~~~~~~~~~~~~~~~~~~~~~~~~\times\delta^{(2)}(\mathbf{P}_{\beta,\perp}-\mathbf{l}_{1,\perp}-...-\mathbf{l}_{n,\perp}).
\end{split}
\end{equation}
Plugging this together with \eqref{fermionwavefunction} into \eqref{softSmatrix0} gives :
\begin{equation}
\label{softSmatrix}
\begin{split}
ig^n&n_{\nu_1}...n_{\nu_n}\int \frac{\rmd^4l_1}{(2\pi)^4}...\frac{\rmd^4l_n}{(2\pi)^4}(2\pi)^4\delta^{(4)}(P_{\alpha}+l_1+...+l_n-P_{\beta})2\pi\delta(l_1^++...+l_n^+)\mathcal{G}^{A_1...A_n}_{\beta\alpha,\mu_1...\mu_n}(l_1,...,l_n)\\
&\times\frac{i\Pi^{\mu_1\nu_1}(l_1)}{l_1^2+i\epsilon}...\frac{i\Pi^{\mu_n\nu_n}(l_n)}{l_n^2+i\epsilon}\displaystyle{\sum_{s,a,b}}\int\frac{\rmd Q_A^-\rmd^2Q_{A,\perp}}{(2\pi)^3}\chi^{sa}_Q(Q_A)\chi^{sb*}_Q(Q_A^-,\mathbf{Q}_{A,\perp}-\mathbf{P}_{\beta,\perp})\\
&\times\displaystyle{\sum_{\sigma\in\mathfrak{S}_n}}(T^{A_{\sigma(1)}}...T^{A_{\sigma(n)}})_{ba}\frac{1}{(l_{\sigma(2)}+...+l_{\sigma(n)})^+-i\epsilon}...\frac{1}{l^+_{\sigma(n)}-i\epsilon}+~~~\text{hard emissions}.
\end{split}
\end{equation}
Before going further, let us pause on the abelian case in which no technicalities arise from the $T^A$'s. We shall see that actually this color structure makes the straightforward generalization to the non abelian case impossible.

\section{The abelian case}
\indent

\begin{figure}[h]
\centering
\vspace{1cm}
%(along, up)
\begin{fmfgraph*}(50,40)
 \fmfstraight
    \fmftop{a,aa,VU,b,bb}
    \fmfbottom{c,cc,v4,d,dd}
    \fmf{dbl_plain_arrow}{cc,d}
      \fmf{dots}{c,cc}
      \fmf{dots}{d,dd}
      \fmffreeze
      \fmf{photon,label=$\uparrow l$}{v4,VU}
      \fmflabel{$_{\mu}$}{VU}
\end{fmfgraph*}
\hspace{.5cm}
$\displaystyle{\rightarrow}$
\hspace{.5cm}
\begin{fmfgraph*}(20,40)
 \fmfstraight
    \fmftop{VU}
    \fmfbottom{v4}
      \fmffreeze
      \fmf{photon,label=$\uparrow l$}{v4,VU}
      \fmfv{decor.shape=hexagram,decor.filled=full,decor.size=10.}{v4}
      \fmflabel{$_{\mu}$}{VU}
\end{fmfgraph*}
\hspace{.5cm}
$=2i\pi e\delta(l^+)\frac{i\Pi^{\mu +}(l)}{-\mathbf{l}_{\perp}^2}$
\caption{Feynman rule for soft photons.\label{newFeynrule}}
\end{figure}
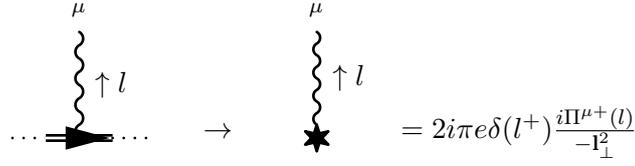

In the abelian case, we perform, in formula \eqref{softSmatrix}, the substitution : $gT^A\rightarrow e$ where $e$ is the electric charge of the fast fermion. The Green function $\mathcal{G}$ no longer carries color indices. Furthermore we make the additional assumption that the transverse momentum transferred by the fast fermion via the gauge bosons is arbitrary small : $\mathbf{P}_{\beta,\perp}$ is neglected with respect to $\mathbf{Q}_{A,\perp}$. In this case one recognizes in \eqref{softSmatrix} the normalization condition \eqref{chinormalization}. Moreover we use the following key identity showing the importance of the delta function $\delta(l_1^++...+l_n^+)$ :
\begin{equation}
\label{identitywithdeltas}
\begin{split}
2i\pi\delta(l_1^++...+l_n^+)&\displaystyle{\sum_{\sigma\in\mathfrak{S}_n}}\frac{1}{(l_{\sigma(2)}+...+l_{\sigma(n)})^+-i\epsilon}...\frac{1}{l^+_{\sigma(n)}-i\epsilon}=(2i\pi)^n\delta(l_1^+)...\delta(l_n^+).
\end{split}
\end{equation}
Plugging this into the abelian analog to \eqref{softSmatrix} gives :
\begin{equation}
\label{softabelianSmatrix}
\begin{split}
(ie)^nn_{\nu_1}&...n_{\nu_n}\int \frac{\rmd^4l_1}{(2\pi)^4}...\frac{\rmd^4l_n}{(2\pi)^4}(2\pi)^4\delta^{(4)}(P_{\alpha}+l_1+...+l_n-P_{\beta})(2\pi)^n\delta(l_1^+)...\delta(l_n^+)\\
&\times\mathcal{G}_{\beta\alpha,\mu_1...\mu_n}(l_1,...,l_n)\frac{i\Pi^{\mu_1\nu_1}(l_1)}{l_1^2+i\epsilon}...\frac{i\Pi^{\mu_n\nu_n}(l_n)}{l_n^2+i\epsilon}+~~~\text{hard emissions}.
\end{split}
\end{equation}
The inspection of \eqref{softabelianSmatrix} shows us that only the plus component of the momentum is conserved in the $\alpha\rightarrow\beta$ sub-process. This last formula is the same as if one has introduced a new Feynman rule which corresponds to each boson exchanged between the fast fermion and the $\alpha\rightarrow\beta$ sub-process : $-2i\pi e\delta(l^+)\Pi^{\mu +}(l)/\mathbf{l}_{\perp}^2$, represented on figure \ref{newFeynrule}. 
In Feynman rules, one has to integrate over all undetermined momenta transferred by the external source keeping in mind that it does not transfer plus components. The best way to interpret this new Feynman rule is to work in Lorenz (also called Landau or covariant) gauge : $\partial_{\mu}A^{\mu}=0$ in which the gauge field propagator numerator reads :
\begin{equation}
\Pi^{\mu\nu}(l)=-g_{\mu\nu}+\frac{l^{\mu}l^{\nu}}{l^2}.
\end{equation}
Since the momentum exchanged has $l^+=0$, $\Pi^{\mu +}(l)=-n^{\mu}$. If we denote $\mathcal{A}^{\mu}(x)$ the Fourier transform of the Feynman rule \ref{newFeynrule}, its only non vanishing component is :
\begin{equation}
\mathcal{A}^{-}(x)=-2\pi e\int \frac{\rmd^4l}{(2\pi)^4}\delta(l^+)\frac{e^{-il.x}}{\mathbf{l}_{\perp}^2}=e\delta(x^+)\Delta_{\perp}^{-1}(|\mathbf{x}_{\perp}|).
\end{equation}
In other words, $\mathcal{A}^{\mu}(x)$ is a classical field in the sense that it satisfies the classical equations of motion :
\begin{equation}
\Box\mathcal{A}^{\mu}(x)=-\mathcal{J}^{\mu}(x)~;
\end{equation}
where $\mathcal{J}^{\mu}(x)=en^{\mu}\delta(x^+)\delta^{(2)}(\mathbf{x}_{\perp})$ is a point like source localized in the $x^+=0$ plane traveling in the backward direction. The fast fermion that radiates soft bosons has the same effect as if one had added a classical source to the lagrangian or equivalently the corresponding classical field in all the vertices of the theory in all possible fashions.

\section{The non abelian case for a single emission with extended sources}
\indent

The trick \eqref{identitywithdeltas} used in the abelian case breaks down in the non abelian case in formula \eqref{softSmatrix} since each permutation on the momenta within the propagators is accompanied by a specific permutation of the generators which does not factorize. One expects that the external field approximation is no longer valid in the non abelian case, however such an approximation can be made in the special case of a single boson exchanged as represented on figure \ref{alphaAbetaBsingleboson}.
\begin{figure}[h]
\centering
\vspace{1cm}
%(along, up)
\begin{fmfgraph*}(80,40)
 \fmfstraight
    \fmftop{i1,VU,o1}
    \fmfbottom{i2,v4,o2}
      \fmf{dbl_plain}{i1,o1}
      \fmf{dbl_plain_arrow}{o2,i2}
      \fmfblob{20.}{VU}
      \fmffreeze
      \fmf{photon}{v4,VU}
      \fmflabel{$_{\alpha,P_{\alpha}}$}{i1}
      \fmflabel{$_{\beta,P_{\beta}}$}{o1}
      \fmflabel{$_{B,Q_{B},r,b}$}{i2}
      \fmflabel{$_{A,Q_A},s,a$}{o2}
\end{fmfgraph*}
\caption{\label{alphaAbetaBsingleboson}}
\end{figure}
Despite this restriction which represents the leading order in perturbation theory, one can remove the assumption made in the abelian case that $\mathbf{P}_{\beta,\perp}$ is neglected with respect to $\mathbf{Q}_{A,\perp}$ and work with more general wave functions. We consider formula \eqref{softSmatrix} for $n=1$ :
\begin{equation}
\label{softsingleSmatrix0}
\begin{split}
ig&n_{\nu}\int \frac{\rmd^4l}{(2\pi)^4}(2\pi)^4\delta^{(4)}(P_{\alpha}+l-P_{\beta})2\pi\delta(l^+)\mathcal{G}^{A}_{\beta\alpha,\mu}(l)\frac{i\Pi^{\mu\nu}(l)}{l^2+i\epsilon}\\
&\times\displaystyle{\sum_{s,a,b}}\int\frac{\rmd Q_A^-\rmd^2Q_{A,\perp}}{(2\pi)^3}\chi^{sa}_Q(Q_A)\chi^{sb*}_Q(Q_A^-,\mathbf{Q}_{A,\perp}-\mathbf{P}_{\beta,\perp})(T^{A})_{ba}+~~~\text{hard emission}.
\end{split}
\end{equation}
It is convenient to introduce the Fourier transform of $\chi_Q$ over the transverse components :
\begin{equation}
\chi^{sa}_Q(Q_A)\equiv\int \rmd^2x_{\perp}\Psi^{sa}(Q^-_A,\mathbf{x}_{\perp})e^{-i\mathbf{Q}_{A\perp}.\mathbf{x}_{\perp}}.
\end{equation}
With this transformation, the second line of \eqref{softsingleSmatrix0} becomes
\begin{equation}
\begin{split}
\displaystyle{\sum_{s,a,b}}&\int \rmd^2x_{\perp}e^{-i\mathbf{P}_{\beta,\perp}.\mathbf{x}_{\perp}}\int\frac{\rmd Q_A^-}{2\pi}\Psi^{sb*}(Q^-_A,\mathbf{x}_{\perp}))(T^{A})_{ba}\Psi^{sa}(Q^-_A,\mathbf{x}_{\perp}))\equiv \int \rmd^2x_{\perp}e^{-i\mathbf{P}_{\beta,\perp}.\mathbf{x}_{\perp}}\rho^A(\mathbf{x}_{\perp}).
\end{split}
\end{equation}
$\rho^A(\mathbf{x}_{\perp})$ is a \emph{classical} color charge distribution in the transverse plane. Plugging this into \eqref{softsingleSmatrix0} and noticing that $l$ is fixed by kinematics, the hard emission piece can be omitted provided $P_{\alpha}$ and $P_{\beta}$ are "almost the sames". This leads to :
\begin{equation}
\begin{split}
-g&n^{\mu}2\pi\delta(P_{\alpha}^+-P_{\beta}^+)\mathcal{G}^{A}_{\beta\alpha,\mu}(P_{\beta}-P_{\alpha})\frac{1}{P_{\beta,\perp}^2}\int \rmd^2x_{\perp}e^{-i\mathbf{P}_{\beta,\perp}.\mathbf{x}_{\perp}}\rho^A(\mathbf{x}_{\perp})\\
&=\mathcal{G}^{A}_{\beta\alpha,\mu}(P_{\beta}-P_{\alpha})\int \rmd^4xgn^{\mu}\delta(x^+)\left[D*\rho^A\right](\mathbf{x}_{\perp})e^{i(P_{\beta}-P_{\alpha}).x}.\\
\end{split}
\end{equation}
Where $D$ is the two-dimensional Laplace operator's Green function :
\begin{equation}
D(\mathbf{x}_{\perp})=-\int\frac{\rmd^2k_{\perp}}{(2\pi)^2}\frac{e^{i\mathbf{k}_{\perp}}\cdot\mathbf{x}_{\perp}}{\mathbf{x}_{\perp}^2}=\frac{1}{2\pi}\ln|\mathbf{x}_{\perp}|.
\end{equation}
Similarly to the abelian case, we define $\mathcal{A}^{A\mu}(x)$ to be :
\begin{equation}
\label{classicalA}
\mathcal{A}^{A\mu}(x)\equiv gn^{\mu}\delta(x^+)\left[D*\rho^A\right](\mathbf{x}_{\perp}).
\end{equation}
Which is a solution of the classical Yang-Mills equations with an extended (in the transverse plane) classical source :
\begin{equation}
\mathcal{J}^{A\mu}(x)= gn^{\mu}\delta(x^+)\rho^A(\mathbf{x}_{\perp}).
\end{equation}
In Lorenz gauge $\mathcal{A}^{A\mu}(x)$ indeed satisfies $\Delta_{\perp}\mathcal{A}^{A\mu}=\mathcal{J}^{A\mu}$. This is consistent with the classical picture of an ultra-relativistic nucleus \cite{Gelis:2005pt}. Unlike the abelian case, the insertion of a classical field in the Feynman rules breaks down. The above calculation cannot be generalized as we consider two or more bosons exchanged with the fast fermion because of intermediate propagators. In the non abelian case the external field approximation cannot be performed for a single source. The way out is to consider a very large number of them.

\section[The external field approximation in non abelian gauge theories]{Dense media : the external field approximation and non abelian gauge theories reconciled}
\indent

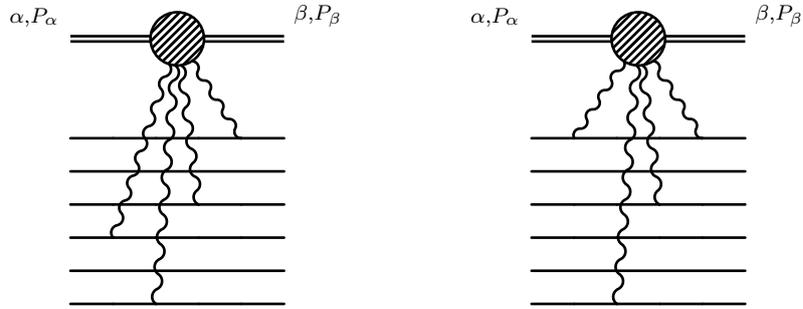
\begin{figure}[h]
\centering
\vspace{1cm}
%(along, up)
\begin{fmfgraph*}(80,100)
 \fmfstraight
    \fmfleft{i9,i8,i7,i6,i5,i4,i3,i2,i1}
    \fmfright{o9,o8,o7,o6,o5,o4,o3,o2,o1}
      \fmf{dbl_plain}{i1,A,o1}
      \fmf{plain}{o4,a4,b4,c4,d4,i4}
      \fmf{plain}{o5,i5}
      \fmf{plain}{o6,a6,b6,c6,d6,i6}
      \fmf{plain}{o7,a7,b7,c7,d7,i7}
      \fmf{plain}{o8,i8}
      \fmf{plain}{o9,a9,b9,c9,d9,i9}
      \fmfblob{20.}{A}
      \fmffreeze
      \fmf{photon}{a4,A}
      \fmf{photon}{b6,A}
      \fmf{photon}{d7,A}
      \fmf{photon}{c9,A}
      \fmflabel{$_{\alpha,P_{\alpha}}$}{i1}
      \fmflabel{$_{\beta,P_{\beta}}$}{o1}
\end{fmfgraph*}
\hspace{3.cm}
\begin{fmfgraph*}(80,100)
 \fmfstraight
    \fmfleft{i9,i8,i7,i6,i5,i4,i3,i2,i1}
    \fmfright{o9,o8,o7,o6,o5,o4,o3,o2,o1}
      \fmf{dbl_plain}{i1,A,o1}
      \fmf{plain}{o4,a4,b4,c4,d4,i4}
      \fmf{plain}{o5,i5}
      \fmf{plain}{o6,a6,b6,c6,d6,i6}
      \fmf{plain}{o7,a7,b7,c7,d7,i7}
      \fmf{plain}{o8,i8}
      \fmf{plain}{o9,a9,b9,c9,d9,i9}
      \fmfblob{20.}{A}
      \fmffreeze
      \fmf{photon}{a4,A}
      \fmf{photon}{b6,A}
      \fmf{photon}{d4,A}
      \fmf{photon}{c9,A}
      \fmflabel{$_{\alpha,P_{\alpha}}$}{i1}
      \fmflabel{$_{\beta,P_{\beta}}$}{o1}
\end{fmfgraph*}
\caption{On the left, a diagram contributing to the non abelian external field approximation and on the right a diagram \emph{not} contributing.\label{nonabelianmultigluons}}
\end{figure}

The derivation of the external field approximation can be extended in the non abelian case if instead of a single fermion one consider a very large number of them :
\begin{equation}
\left|\right.\Phi\left.\right\rangle\rightarrow \displaystyle{\sum_{s_1...s_n}}\displaystyle{\sum_{a_1...a_n}}\int\frac{\rmd Q_{A,1}^-\rmd^2Q_{A,1,\perp}}{(2\pi)^3\sqrt{2Q_{A,1}^-}}...\frac{\rmd Q_{A,n}^-\rmd^2Q_{A,n,\perp}}{(2\pi)^3\sqrt{2Q_{A,n}^-}}\chi^{s_1...s_n}_{a_1,...,a_n}(Q_A)\left|\right.Q_{A,1},s_1,a_1;...;Q_{A,n},s_n,a_n\left.\right\rangle~~~\text{with}~~n\rightarrow\infty.
\end{equation}
$\chi$ is still sharply peaked around large values of momenta pointing in the minus direction and is completely antisymmetric under the interchange of pairs of $(Q_{A,i},s_i,a_i)$ by Pauli's principle. When the number of particles is very large the most probable double exchange comes from two different sources just by combinatorial arguments\footnote{The number of ways to attach two bosons to the same fermion goes like $n$ while to two different fermions, it goes like $n(n-1)/2 \sim n^2$ as $n$ is large.}. These interactions with different fermions are independent single exchanges with color structure not talking to each others.  One can therefore add the classical field \eqref{classicalA} in all the vertices of the lagrangian. Doing this one omits double, triple... boson exchanges with the same source and just consider diagrams with several bosons coming from different fermions as represented on \ref{nonabelianmultigluons}. This is why in QCD a large number of particles radiating soft gluons can be described by a classical source. The large number condition is necessary and is the fundamental difference with the abelian case.

\chapter{Eikonal propagation of particles in a background field\label{eikonalprop}}
\indent

\setcounter{equation}{0}

\begin{figure}[h]
\centering
\vspace{1cm}
%(along, up)
\begin{fmfgraph*}(150,40)
 \fmfstraight
    \fmftop{i1,a,b,c,o1}
    \fmfbottom{i2,d,e,f,o2}
      \fmf{fermion}{i1,a}
      \fmf{fermion}{a,b}
      \fmf{dots}{b,c}
      \fmf{fermion}{c,o1}
      \fmfblob{10.}{a}
      \fmfblob{10.}{b}
      \fmfblob{10.}{c}
      \fmffreeze
      \fmf{photon,label=$\uparrow~l_1$}{d,a}
      \fmf{photon,label=$\uparrow~l_2$}{e,b}
      \fmf{photon,label=$\uparrow~l_n$}{f,c}
      \fmflabel{$_{\alpha,p}$}{i1}
      \fmflabel{$_{\beta,q}$}{o1}
      \fmfv{decor.shape=hexagram,decor.filled=full,decor.size=10.}{d}
      \fmfv{decor.shape=hexagram,decor.filled=full,decor.size=10.}{e}
      \fmfv{decor.shape=hexagram,decor.filled=full,decor.size=10.}{f}
\end{fmfgraph*}
\caption{Multiple scattering between a one-particle state and a background field.\label{eikonalscattering}}
\end{figure}
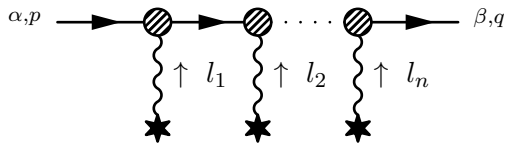

One would like to derive the Feynman rule for a hard particle propagating "freely" in a classical background field whose deep existence has been justified in the previous section \ref{nonabelianBF}. The "free" assumptions means that the particles do not interact with a field other than the background field. This is justified since the background field, in covariant or $A^+=0$ gauge, is very narrow in the direction of the propagation of the particle and thus contributions to the amplitude from processes occurring inside the background field vanish in the limit where the background field becomes infinitely narrow. The motivation follows from the observation that a field generated by a dense medium has a strength of order $1/g$ which imposes a ressummation to all orders. Such evolution of a state is represented on figure \ref{eikonalscattering}. Here we assume $\alpha$ and $\beta$ to be one-particle states of arbitrary spin and furnishing arbitrary, non trivial, representations of the gauge group. We shall work in the eikonal approximation, that is, the particle is undeviated as it travels through the background field. The eikonal approximation is in accordance with the softness of the gauge bosons exchanged : the momentum of the incoming particle is almost unchanged by the interaction with the background field. We take the initial momentum of the incoming particle to be $p=(p^+,0,\mathbf{0})$ with $p^+$ very large with respect to all the exchanged momenta - actually $p$ may even be off-shell without changing the following calculations.
Before deriving the corresponding evolution operator, one has to examine the structure of the eikonal vertex for different types of fields. We will actually find that this structure is general and depends only on the gauge representation of the corresponding field but not on its spin.

\section{Structure of the eikonal vertex\label{eikonalvertex}}
\indent

In this section we study the structure of the vertices in the eikonal approximation for every usual kind of fields.

\subsection{Dirac fermions}
\indent

Let us start with the Dirac fermions whose interaction with the background field is described by the interaction lagrangian $g\bar{\psi}\slashed{\mathcal{A}}\psi$. The vertex reads $igT^A_R\gamma^{\mu}$. The eikonal approximation is equivalent to the statement that the incoming and outgoing momenta of the fermion are almost the same. Thus for $q\simeq p$ :
\begin{equation}
\overline{u}^r(q)igT^A_R\gamma^{\mu}u^s(p)\simeq ig\delta^{rs}2p^{\mu}T^A_R.
\end{equation}
$\mu=+$ is the only relevant component since it is plugged on a $\mathcal{A}^{-}$. This identity has been used in \ref{nonabelianBF} where we used fermions because their Feynman rule is very simple. However we shall see that one has the same rule for scalars and vector fields with more complicated interactions.

\subsection{Gauge fields}
\indent

A more subtle case concerns the gauge fields which have cubic and quartic interaction vertices. In the gauge $A^+=0$, it is easily seen that the quartic vertex is zero when any two fields are replaced by the classical field $\mathcal{A}^{\mu}=\delta^{\mu-}\mathcal{A}^{-}$. Thus we are only left with the cubic vertex. For definiteness the background field $\mathcal{A}^{-}$ caries the momentum $l$ and the color index $A$. The Feynman rule for the 3-gluon vertex then reads :
\begin{equation}
\epsilon_{(\lambda)\mu}(p)\epsilon^{*}_{(\lambda')\nu}(q)ig\tilde{T}^A\left[g^{\mu+}(2l-q)^{\nu}+g^{\mu\nu}(p+q)^{+}-g^{\nu+}(2l+p)^{\mu}\right].
\end{equation}
The background field being $x^-$ independent, one has $p^+=q^+$ by momentum conservation. Moreover $\epsilon^{+}=0$ by gauge condition. Thus only the central term survives : 
\begin{equation}
\delta_{\lambda\lambda'}2ig\tilde{T}^Ap^{+}.
\end{equation}
Eikonal assumption makes no distinction between fermions and gauge fields concerning the structure of the vertex.

\subsection{Vectors and scalars}
\indent

Similarly, massive vectors and scalars - in some representation $R$ of the gauge group, may have interactions quadratic in the background field proportional to $\mathcal{A}^{\mu}\mathcal{A}_{\mu}=0$. Thus, like in the case of gauge fields one has to consider only the cubic vertex\footnote{For massive vector fields, $V^{\mu}$, there is an interaction of the form $\sim V_{\mu}^{\dagger}\mathcal{A}^{\mu}\mathcal{A}^{\nu}V_{\nu}$ which is non zero. Actually the vector case is not very interesting since the only charged massive vectors in the standard model follows from gauge fields in a spontaneously broken symmetry and do not have such interactions with other gauge fields. Since it will not be considered, we will close our eyes on this term.}, linear in the background field. The interaction of two vectors with a gauge field reads :
\begin{equation}
\epsilon_{(\lambda)\nu}(p)\epsilon^{*}_{(\lambda')\rho}(q)igT^A_R(p+q)^{+}\simeq \delta_{\lambda\lambda'}2igT^A_Rp^+.
\end{equation}
The epsilon vectors are replaced by one in the case of scalars.\\

\section{Eikonal evolution}
\indent

The eikonal vertex having always the same structure one can write a general form for the S-matrix corresponding to process \ref{eikonalscattering} that do not take the spin of the particle under consideration into account. The diagram \ref{eikonalscattering} with $n$ insertions of the background field reads :
\begin{equation}
\begin{split}
(ig)^n\delta^{rs}&2p_{\mu_1}...2p_{\mu_n}\int \frac{d^4l_1}{(2\pi)^4}...\frac{d^4l_n}{(2\pi)^4}(2\pi)^4\delta^{(4)}(p+l_1+...+l_n-q)
\frac{i}{(p+l_1+...+l_{n-1})^2+i\epsilon}...\frac{i}{(p+l_1)^2+i\epsilon}\\
&\times\int d^4x_1...d^4x_ne^{il_1\cdot x_1+...+il_n\cdot x_n}\theta(x_n^+-x^+_{n-1})...\theta(x_2^+-x_1^+)T^{A_n}_R...T^{A_n}_R\mathcal{A}^{A_n\mu_n}(x_n)...\mathcal{A}^{A_1\mu_1}(x_1).
\end{split}
\end{equation}
From now we denote $\mathcal{A}^{A}T^{A}_R\equiv \mathcal{A}$, the representation label will be implicit for brevity. Moreover $p_{\mu}\mathcal{A}^{\mu}=p^+\mathcal{A}^{-}$ and the explicit solution \eqref{classicalA} for $\mathcal{A}^{-}$ shows us that it does not depend on $x^-$ and the corresponding integrals can be easily performed. The integrals over the $x_i^-$ bring $n$ delta functions $2\pi\delta(l_i^+)$. On the other components of $l$, one performs the change of variables $k_m=l_1+...+l_m$ for all $m=1, ..., n$. The denominators are expanded as $(p+k_m)^2=2p^+k^-_m-\mathbf{k}_{m,\perp}^2$. The S-matrix then becomes :
\begin{equation}
\label{eikonalSmatrix00}
\begin{split}
&(ig)^n\delta^{rs}2\pi\delta(p^+-q^+)2p^+\int \frac{dk_1^-d^2k_{1,\perp}}{(2\pi)^3}...\frac{dk_n^-d^2k_{n,\perp}}{(2\pi)^3}2\pi\delta(k_n^--q^-)(2\pi)^2\delta^{(2)}(\mathbf{k}_{n,\perp}-\mathbf{q}_{\perp})\\
&~\times\frac{i}{k^+_{n-1}-\mathbf{k}_{n-1,\perp}^2/2p^++i\epsilon}...\frac{i}{k^+_1-\mathbf{k}_{1,\perp}^2/2p^++i\epsilon}\int dx_1^+d^2x_{1,\perp}...dx_n^+d^2x_{n,\perp}\theta(x_n^+-x^+_{n-1})...\theta(x_2^+-x_1^+)\\
&~\times e^{ik^-_1(x_1-x_2)^++...+ik^-_nx_n^+}e^{-i\mathbf{k}_{1,\perp}\cdot(\mathbf{x}_{1,\perp}-\mathbf{x}_{2,\perp})-...-i\mathbf{k}_{n,\perp}\cdot\mathbf{x}_{n,\perp}}\mathcal{A}^{-}(x_n^+,\mathbf{x}_{n,\perp})...\mathcal{A}^{-}(x_1^+,\mathbf{x}_{1,\perp}).
\end{split}
\end{equation}
The integral over $k_n$ is trivial because of the delta function. The integrals over the others $k^-_m$ are performed thanks to the residue theorem. This gives the $\theta$ constraint we already have and replaces all the $k^-_m$ by $\mathbf{k}_{m,\perp}^2/2p^+$ in the exponentials. After these manipulations, one is left only with gaussian integrals over $\mathbf{k}_{m,\perp}^2$ :
\begin{equation}
\begin{split}
\int \frac{d^2k_{m,\perp}}{(2\pi)^2}&\exp\left[i\frac{\mathbf{k}_{m,\perp}^2}{2p^+}(x_m-x_{m+1})^+-i\mathbf{k}_{m,\perp}\cdot(\mathbf{x}_{m,\perp}-\mathbf{x}_{m+1,\perp})\right]\\
&=\frac{p^+}{2i\pi(x_{m+1}-x_{m})^+}\exp\left[i\frac{p^+}{2}\frac{(\mathbf{x}_{m,\perp}-\mathbf{x}_{m+1,\perp})^2}{(x_{m+1}-x_{m})^+)}\right]\longrightarrow\delta^{(2)}(\mathbf{x}_{m,\perp}-\mathbf{x}_{m+1,\perp})~~~\text{for}~~p^+\rightarrow\infty.
\end{split}
\end{equation}
Plugging this into \eqref{eikonalSmatrix00} gives :
\begin{equation}
\begin{split}
(ig)^n&\delta^{rs}2\pi\delta(p^+-q^+)2p^+\int d^2x_{\perp}\int dx_1^+...dx_n^+\theta(x_n^+-x^+_{n-1})...\theta(x_2^+-x_1^+)\\
&\times e^{iq^-x_n^+-i\mathbf{q}_{\perp}\cdot\mathbf{x}_{n,\perp}}\mathcal{A}^{-}(x_n^+,\mathbf{x}_{\perp})...\mathcal{A}^{-}(x_1^+,\mathbf{x}_{,\perp}).
\end{split}
\end{equation}
Since the support of $\mathcal{A}^{-}$ is very narrow in the plus direction and $q^-$ is assumed to be very small, one can set $ e^{iq^-x_n^+}\simeq 1$. The $\theta$ constraints can be summarized by introducing the path ordering operator along the plus coordinate direction (beware of the combinatorial factor $1/n!$ that includes all the ways to relabel the coordinates) :
\begin{equation}
\frac{(ig)^n}{n!}\delta^{rs}2\pi\delta(p^+-q^+)2p^+\int d^2x_{\perp}\int dx_1^+...dx_n^+e^{-i\mathbf{q}_{\perp}\cdot\mathbf{x}_{\perp}}\mathcal{P}\left\{\mathcal{A}^{-}(x_n^+,\mathbf{x}_{\perp})...\mathcal{A}^{-}(x_1^+,\mathbf{x}_{,\perp})\right\}.
\end{equation}
thanks to this result, the sum of all the diagrams of the form \ref{eikonalscattering} with an arbitrary number of insertions of external fields gives us the S-operator for the eikonal propagation of a single-particle state in a background field (with explicit representation restored) :
\begin{equation}
\label{BFsinglePSnormalization}
\left<\vec{q},s,b|\vec{p},r,a\right>_{\mathcal{A}^-}=\delta^{rs}2\pi\delta(p^+-q^+)2p^+\int d^2x_{\perp}e^{-i\mathbf{q}_{\perp}\cdot\mathbf{x}_{\perp}}\mathcal{P}\exp\left\{\int dx^+\mathcal{A}^{A-}(x^+,\mathbf{x}_{\perp})T^A_R\right\}_{ba}.
\end{equation}
The notation with an index $\mathcal{A}^-$ on Heisenberg state denotes that the evolution operator is given only by the interaction with the background field which is the leading one in perturbation theory. This last formula reduces to the ordinary normalization for Heisenberg states in the vacuum\footnote{Recalling that $\mathbf{p}_{\perp}$ has been chosen to be zero.} : $\left<q,s,b|p,r,a\right>=\delta^{rs}\delta_{ab}2p^+2\pi\delta(p^+-q^+)(2\pi)^2\delta^{(2)}(\mathbf{p}_{\perp}-\mathbf{q}_{\perp})$.

\chapter{p-A collisions cross-sections\label{pAcrosssections}}
\indent

\setcounter{equation}{0}

The aim of this technical section is to set the conventions used for S-matrix and cross-section in p-A collisions, used in chapter \ref{pAchapter}. The nuclear target being described by a CGC, one faces to the problem of explicit space-time dependence of Feynman rules that violates momentum conservation. So in a first part, we shall see how it is consistent to build a cross-section from the S-matrix for a system that is no longer spatially homogeneous. Furthermore, in chapter \ref{pAchapter}, partons emitted in the remote past by the proton were considered as initial states. Radiated partons promoted as initial states enter into the collinear factorization assumption discussed in section \ref{DIS}. In a second part we shall detail how the proton can be added in a very simple way in the total p-A cross section from the cross-section corresponding to the sub-process involving an initial gluon. The deep justification of collinear factorization has already been discussed in chapter \ref{QCDsat} and will be considered as an assumption. The aim of this section is just to get properly normalized expressions for cross-sections.

\section{The target : S-matrix and cross-section in presence of a background field}
\indent

The presence of the background field makes the system explicitly position-dependent which forbids momentum conservation laws since the system is no more invariant under space-time translations. Thus the S-matrix cannot be proportional to a four momentum delta function as in the usual case of homogeneous systems considered in textbooks on quantum field theory. Formulas given in the literature for decay rates, cross-sections... makes this momentum conservation explicit. The question is then how such formulas are changed when we consider inhomogeneous systems. We shall only look for the formula that relates the cross-section to the S-matrix but generalizations to decay rates and more than two-particle initial states is straightforward.\\

We consider the process in which an arbitrary (possibly multi-particle) state $\alpha$ composed of right-moving particles\footnote{In the case of a multi-particle state, the right-moving condition means that all the particles have a positive momentum component along the $z$ axis. This does not mean anyway that all their transverse momenta vanish which is not possible, in general, for a system of particles, by any choice of frame.} scatters off the left-moving nucleus $A$ to produce a right-moving final state $\beta$ and the nucleus with possibly other left-moving particles together denoted $X$.  The final unobserved state together with the nucleus, $X$, and is eventually summed over. This is why one deals with \emph{inclusive} cross-sections. The nucleus is described by a classical field $\mathcal{A}^-$ whose explicit form is given by \eqref{classicalA}. Then we adopt the following notation for the S-matrix, labeled with a background field index :
\begin{equation}
S(A\alpha\rightarrow X\beta)\rightarrow S_{\mathcal{A}^-}(\alpha\rightarrow \beta)=\left< \beta; out|\alpha; in\right>_{\mathcal{A}^-}.
\end{equation}
The "in" and "out" labels refer to interactions between quantum fields themselves and not with the background field\footnote{Comparing to the notation in \eqref{BFsinglePSnormalization}, all the interactions are turned on. The ones with the background field are resummed to all orders while the ones between quantum fields only are treated perturbatively.}. They are dropped if one considers "free" fields interacting just with the background field (all other types of interactions are turned off). First let us see the structure of the S-matrix. The point is to take advantage of the non-dependence of $\mathcal{A}^-$ in the $x^-$ light-cone coordinate. This remnant of translational invariance leads to the conservation of the plus component of the total four-momentum. Then on can parametrize the S-matrix as follow :
\begin{equation}
\label{Mmatrixdef}
S_{\mathcal{A}^-}(\alpha\rightarrow \beta)=2i\pi\delta(P_{\alpha}^+-P^+_{\beta})\mathcal{M}(\alpha\rightarrow \beta),
\end{equation}
where $P_{\alpha}$ and $P_{\beta}$ are the total four-momenta of the states $\alpha$ and $\beta$ respectively. One wants to interpret the S-matrix as a probability. For this purpose we shall focus on the case of a two-body collision, that is a single particle scattering off the nucleus. It is fulfilled if $\alpha$ is a one-particle state, labeled by the quantum number $\vec{p}=(p^+,\mathbf{p}_{\perp})$\footnote{For brevity we consider that the state just carries the spatial momentum quantum number, generalization with various other discrete quantum numbers is trivial.}. Instead of considering an initial state of definite momentum, one rather considers a superposition of states $|\phi_{\vec{p}}>$ sharply peaked around a given value of $\vec{p}$, normalized to unity, as already encountered in \eqref{fermionwavefunction}. The final state $\beta$ is conveniently decomposed as the $n$-particle state $|\vec{q}_1,...,\vec{q}_n>$. It is a rather easy exercise to show that
\begin{equation}
\rmd P(\vec{p}\rightarrow \vec{q}_1,...,\vec{q}_n)=\left|\left< \vec{q}_1,...,\vec{q}_n; out|\phi_{\vec{p}}; in\right>_{\mathcal{A}^-}\right|^2\frac{\rmd^3q_1}{(2\pi)^3 2q_1^+}...\frac{\rmd^3q_n}{(2\pi)^3 2q_n^+}
\end{equation}
is dimensionless and together with the completeness relation \eqref{completenessrelation} is a nice candidate for being interpreted as a differential probability. How to go from probability to cross-section is found in the literature (see for instance \cite{PeskinSchroeder}). Actually the procedure does not differ from the one for homogeneous systems (the trick used by introducing a wave function in the initial state regulates one of the two delta functions arising from the square modulus when one integrates over the impact parameter) and the result is very similar. The cross-section reads in term of the $\mathcal{M}$-matrix defined in \eqref{Mmatrixdef} as :
\begin{equation}
\label{BFcrosssection}
\rmd\sigma(\vec{p}\rightarrow \vec{q}_1,...,\vec{q}_n)=\frac{1}{2p^+}2\pi\delta(p^+-q_1^+-...-q_n^+)\left|\mathcal{M}(\vec{p}\rightarrow \vec{q}_1,...,\vec{q}_n)\right|^2\frac{\rmd^3q_1}{(2\pi)^3 2q_1^+}...\frac{\rmd^3q_n}{(2\pi)^3 2q_n^+}.
\end{equation}
Although the form of formula \eqref{BFcrosssection} is rather obvious from what we already know in the homogeneous case, the absence or presence of four-momentum conservations delta functions changes the dimension of what we defined to be the amplitude $\mathcal{M}$. Here the cross-section \eqref{BFcrosssection} has the dimension of a surface, as expected.

\section{The projectile : collinear factorization\label{pAcollinear}}
\indent

While we are dealing with cross-sections in p-A collisions, here is a good place to talk about collinear factorization. Collinear factorization is an approximation that consist in neglecting the - small - virtuality of a parton emitted by the incoming proton. This is justified in the infinite momentum frame where the parton can be considered as on-shell with a very good approximation. Let us denote by $k$ the four-momentum of the emitted parton. Under the assumption of a soft emission one can roughly neglect its transverse components and its virtuality is therefore zero. Denoting $P=(P^+,0,\mathbf{0})$ the four-momentum of the proton, the radiated parton has a single non vanishing component $k^+=x_1P^+$. Since it is on-shell, it makes sense to consider the former as an asymptotic state and therefore to deal with a cross-section corresponding to the collision between the parton and the nucleus. The situation is then the same as the one encountered in \ref{DIS} : the total cross-section for the pA collision is the sum of cross-sections for the parton-nucleus sub-process with an initial parton of momentum $k$ weighted by the number of partons carrying the corresponding momentum fraction $x_1$. This is precisely in this context that we introduced the integrated parton distributions. Let us say for definiteness that the parton of interest is a gluon, the weight function is the integrated gluon distribution $G$ defined in \eqref{UGPDF}. The relation alluded to above reads in this case :
\begin{equation}
\rmd\sigma(p\rightarrow \beta)=\int\rmd x_1 G(x_1,Q^2)\rmd\sigma(g(k=x_1P)\rightarrow \beta).
\end{equation}
One can even go a step further since the cross-section is proportional to a plus-component momentum conservation delta function as shown in \eqref{BFcrosssection}. Thus the integral over $x_1$ can be performed and this gives :
\begin{equation}
\label{pAtogAcrosssection}
\rmd\sigma(p\rightarrow \vec{q}_1,...,\vec{q}_n)=\frac{\pi}{(p^+)^2} x_1G(x_1,Q^2)\left|\mathcal{M}(g(k)\rightarrow \vec{q}_1,...,\vec{q}_n)\right|^2\frac{\rmd^3q_1}{(2\pi)^3 2q_1^+}...\frac{\rmd^3q_n}{(2\pi)^3 2q_n^+}
\end{equation}
with $x_1$ fixed at the value $x_1=(q_1^++...+q_n^+)/P^+$ and as in the previous section $\beta$ has been detailed as $|\vec{q}_1,...,\vec{q}_n>$.\\

So far we got the framework for dealing both with the nucleus and the proton however there is a flaw in the justification of collinear factorization. Considering the parton emitted by the proton as an initial state implicitly assumes it has been emitted in the remote past, a non obvious situation. There is a very simple way out as one deals with di-hadron correlations : an on-shell massless parton cannot split into two (or more) on-shell massless hadrons unless it receives kicks from the target. By this simple kinematic argument the existence of the process is ensured if and only if hadrons collide the nucleus and have therefore been emitted \emph{before} the collision between to proton and the nucleus. Conversely, a parton emitted by the proton \emph{after} the collision cannot split.

\chapter[Yang-Mills theory in a background field]{Yang-Mills action and propagator in a background field and Feynman rules\label{shockwaveFeynrules}}
\indent

\setcounter{equation}{0}

In this appendix, we detail the framework and derive Feynman rules for a Yang-Mills theory in presence of a shockwave widely used in section \ref{pAchapter}. The total field $A$ is the sum of the background field $\mathcal{A}$ and the quantum field $\alpha$. In a first part, we write the action that concerns the physical degrees of freedom $\alpha$ propagating in the background field $\mathcal{A}$ in the general case. Then we find an explicit form for the corresponding propagator and emphasize some of its properties for our specific case, that is a single $\mathcal{A}^-$-component background field generated by a source left-moving close to the $x^+=0$ plane. At the end we derive the momentum space Feynman rules for the shockwave. The background field is a non trivial function of coordinates and it would be, in principle, not really easy to work in momentum space representation. However, we shall see that the coordinate dependence of the shockwave is mere enough to enable the use of momentum space Feynman rules.
 
\section{The Yang Mills action in a background field\label{BFaction}}
\indent

This section concerns the general case of a Yang-Mills field coupled to classical sources. Although it has naturally emerged from section \ref{nonabelianBF} that all possible insertions of a background field in all the interactions is related to the existence of a classical current, it is actually a general property that external classical sources can always be reabsorbed is this way in the theory. We are going to show that one can get rid of the explicit dependence of the action $S_{YM}[A]+\int\mathcal{J}\cdot A$ in the current if one sees the theory as a quantum gauge field $\alpha$ interacting with the classical field $\mathcal{A}$ produced by the sources. The additional term $\int\mathcal{J}\cdot A$ in the action can be rewritten thanks to the equations of motion :
\begin{equation}
\label{eulerlagrange}
\mathcal{J}=-\left.\frac{\delta S_{YM}[A]}{\delta A}\right|_{\mathcal{A}}.
\end{equation}
$\mathcal{A}$ is by definition a classical field, i.e. it satisfies the Euler-Lagrange equations. For convenience we write the total gauge field $A$ as $\mathcal{A}+\alpha$, $\alpha$ being the quantum field. Plugging, instead of $\mathcal{J}$, the right hand side of the equations of motion \eqref{eulerlagrange} into the action $S[A=\mathcal{A}+\alpha]=S_{YM}[A]+\int\mathcal{J}\cdot A$, denoted from now $S_{\mathcal{A}}[\alpha]$, gives\footnote{The notation $\int a\cdot\frac{\delta S_{YM}[A]}{\delta A}$ is a shorthand for functional differentiation with respect to $A$ and all its derivatives. The action does not contain derivatives of higher rank than one so the notation means that one $A$ (resp. its first derivative) is replaced by $a$ (resp. the first derivative of $a$) in each monomial in all possible fashions.}
\begin{equation}
\label{newaction}
\begin{split}
S_{\mathcal{A}}[\alpha]&=S_{YM}[\mathcal{A}+\alpha]+\int (\mathcal{A}+\alpha)\cdot\mathcal{J}\\
&=S_{YM}[\mathcal{A}]-\int\left.\mathcal{A}\cdot\frac{\delta S_{YM}[A]}{\delta A}\right|_{\mathcal{A}}+\displaystyle{\sum_{n \geq 2}}\int\left.\frac{\alpha^n}{n!}\cdot\left(\frac{\delta }{\delta A}\right)^nS_{YM}[A]\right|_{\mathcal{A}}
\end{split}
\end{equation}
This action depends only on $\mathcal{A}$ and $\alpha$, the external source disappeared. It describes the interaction of the quantum field $\alpha$ with the background field $\mathcal{A}$ as expected. The first two terms in the last line of \eqref{newaction} depend only on $\mathcal{A}$. They correspond to vacuum bubbles that do not contribute to connected amplitudes. For this reason, they can be omitted. Up to these two terms $S_{\mathcal{A}}[\alpha]$ reads :
\begin{equation}
\begin{split}
S_{\mathcal{A}}[\alpha]&=\frac{1}{2}\int \rmd^4x \rmd^4y \alpha^{A\mu}(x)(\Delta^{-1})^{AB}_{\mu\nu}(x,y)\alpha^{B\nu}(y)+\int \rmd^4 x \mathcal{L}_{int}(x),
\end{split}
\end{equation}
where $\Delta^{-1}$ is the background field inverse propagator :
\begin{equation}
\label{dressedprop}
(\Delta^{-1})^{AB}_{\mu\nu}(x,y)=\delta^{(4)}(x-y)\left[g_{\mu\nu}\mathcal{D}^2_x-\mathcal{D}_{x,\mu}\mathcal{D}_{x,\nu}-2ig\mathcal{F}_{\mu\nu}(x)\right]_{AB},
\end{equation}
with $\mathcal{D}$ the covariant derivative with respect to the background field and $\mathcal{F}$ is the background field strength. $\mathcal{L}_{int}$ is the piece that contains the interactions between three or four $\alpha$ fields. We do not write them explicitly, we just say that it contains the ordinary three and four gluon vertices between $\alpha$'s but also a three $\alpha$'s vertex which arises from the four gluons vertex with the insertion of one $\mathcal{A}$.

\section{Gluon propagator in a background field}
\indent

Here we shall write the explicit propagator in the particular case of interest where the only non vanishing component of the background field is $\mathcal{A}^-$.

\subsection{Equations of motion and relation to the scalar propagator}
\indent

The dressed propagator is the Green function of the kernel \eqref{dressedprop} that satisfies the classical\footnote{Here, the term "classical" should be understood as the configuration of $\alpha$ which extremizes the action $S_{\mathcal{A}}[\alpha]$ in presence of a given background field $\mathcal{A}$.} equations of motion :
\begin{equation}
\label{EOMdressed}
\left[\delta^{\mu}_{\lambda}\mathcal{D}^2_x-\mathcal{D}_{x}^{\mu}\mathcal{D}_{x,\lambda}-2ig\mathcal{F}^{\mu}_{~\lambda}(x)\right]_{AC}\Delta^{\lambda\nu}_{CB}(x,y)=i\delta^{AB}g^{\mu\nu}\delta^{(4)}(x-y).
\end{equation}
As in chapter \ref{pAchapter}, $\Delta^{\mu\nu}$ denotes the tree-level propagator, dressed with the background field. Since the only non vanishing component of the background field is the $-$ one, then the covariant derivative is merely $\mathcal{D}^{\mu}=(\partial^+,\partial^--ig\mathcal{A}^-,-\nabla_{\perp})$ and the only non vanishing components of the field strength are $\mathcal{F}^{i-}=-\mathcal{F}^{-i}=\partial^i\mathcal{A}^-$. Using the fact that the background field does not depend on $x^-$, one can perform a Fourier transform over the minus coordinates : 
\begin{equation}
\label{FTdressed}
\Delta^{\mu\nu}_{AB}(x,y)=\int\frac{\rmd k^+}{2\pi}\Delta^{\mu\nu}_{AB}(\vec{x},\vec{y};k^+)e^{-ik^+(x^--y^-)}.
\end{equation}
Here $\vec{x}$ denotes $(x^+,\mathbf{x}_{\perp})$, the "spatial" components.\\

Taking $\mu=-$ and $\nu=i$ in \eqref{EOMdressed} and the corresponding adjoint equation gives :
\begin{equation}
\label{propcomprelation}
\begin{split}
&\Delta^{-i}_{AB}(\vec{x},\vec{y};k^+)=\frac{i}{k^+}\partial^j_x\Delta^{ji}_{AB}(\vec{x},\vec{y};k^+)\\
&\Delta^{i-}_{AB}(\vec{x},\vec{y};k^+)=-\frac{i}{k^+}\partial^j_y\Delta^{ij}_{AB}(\vec{x},\vec{y};k^+)~;
\end{split}
\end{equation}
and taking $\mu=\nu=-$ gives in \eqref{EOMdressed} :
\begin{equation}
\begin{split}
\Delta^{--}_{AB}(\vec{x},\vec{y};k^+)&=\frac{i}{k^+}\partial^i_x\Delta^{i-}_{AB}(\vec{x},\vec{y};k^+)\\
&=\frac{1}{(k^+)^2}\partial^i_x\partial^j_y\Delta^{ij}_{AB}(\vec{x},\vec{y};k^+).
\end{split}
\end{equation}
All the components of the dressed propagator are written as differential operators acting on $\Delta^{ij}$, to be determined.\\

To say more about $\Delta^{ij}$, one has to take $\mu\nu=ij$ in \eqref{EOMdressed} and use \eqref{propcomprelation}. We find that :
\begin{equation}
\label{deltaijeq}
\left[-2ik^+\mathcal{D}^--\Delta_{\perp}\right]_{AC}\Delta^{ij}_{CB}(\vec{x},\vec{y};k^+)=-i\delta^{AB}\delta^{ij}\delta^{(3)}(\vec{x}-\vec{y}).
\end{equation}
This equation must be compared with the equation of motion of a charged massless scalar field ($G$ denotes its propagator) in the adjoint representation of the gauge group\footnote{Such a field is necessarily complex since it is charged.} :
\begin{equation}
\label{EOMscalar}
\mathcal{D}^2_{AC}G^{CB}(\vec{x},\vec{y};k^+)=\left[-2ik^+(\partial^--ig\mathcal{A}^-)-\Delta_{\perp}\right]_{AC}G^{CB}(\vec{x},\vec{y};k^+)=-i\delta^{AB}\delta^{(3)}(\vec{x}-\vec{y}).
\end{equation}
Equations \eqref{deltaijeq} and \eqref{EOMscalar} are the same, up to a trivial $\delta^{ij}$ factor, with the same boundary conditions. Hence one necessarily has $\Delta^{ij}=\delta^{ij}G$.\\

At the end, all components of the gluon's propagator are expressed in terms of a single scalar function as :
\begin{equation}
\label{gluonpropagator}
\begin{split}
&\Delta^{ij}_{AB}(\vec{x},\vec{y};k^+)=\delta^{ij}G^{AB}(\vec{x},\vec{y};k^+)\\
&\Delta^{-i}_{AB}(\vec{x},\vec{y};k^+)=\frac{i}{k^+}\partial^i_xG^{AB}(\vec{x},\vec{y};k^+)\\
&\Delta^{i-}_{AB}(\vec{x},\vec{y};k^+)=-\frac{i}{k^+}\partial^i_yG^{AB}(\vec{x},\vec{y};k^+)\\
&\Delta^{--}_{AB}(\vec{x},\vec{y};k^+)=\frac{1}{(k^+)^2}\partial^i_x\partial^i_yG^{AB}(\vec{x},\vec{y};k^+).
\end{split}
\end{equation}
The task reduces to find the scalar propagator, the gluon's one will be trivially given thanks to \eqref{gluonpropagator}.

\subsection{Prescription and explicit form}
\indent

Here shall work out the solution of equation \eqref{EOMscalar}. For this purpose, it is helpful to consider the following feature : in $k^+$ space, the Feynman prescription $i\epsilon$ and the retarded prescription - strictly speaking, the $x^+$ ordered one - $ik^+\epsilon$ are the same as long as $k^+$ is positive as it is going to be the case in all our calculations\footnote{The nuclei being a left-mover, we consider only processes in which the particles that interact with the background field are right-movers.}. Thus the Feynman propagator is non vanishing only if the endpoint lies in the forward light cone of the starting point. In other words its support is causal.\\

To solve equation \eqref{EOMscalar} one has to take the following considerations into account. We can show that the $x^+$ dependence of the background field $\mathcal{A}^-$ is the same as the $x^+$ dependence of the source, whatever it is \cite{Gelis:2005pt}. The left-moving nucleus being localized close to the $x^+=0$ plane by Lorentz contraction the $\mathcal{A}^-$'s $x^+$-dependence is something that looks like a delta function. In order to have well defined calculations, we rather use a representation of the delta function $\delta_{\eta}(x^+)$ which is a narrow function whose support is between $x^+=0$ and $x^+=\eta$. Actually, no contributions to physical processes arise from the region $0<x^+<\eta$ as $\eta$ becomes infinitesimally small\footnote{The contribution of this region to amplitudes always involves integrations of regular functions over this range which vanish as $\eta$ goes to zero.}. Therefore, we are interested in three cases concerning the scalar propagator : $x^+$ and $y^+$ both negative, both greater than $\eta$ and $x^+>\eta$ and $y^+<0$ - the converse being zero by the causality argument discussed above.\\

The first two cases are rather trivial, since equation \eqref{EOMscalar} for the scalar propagator reduces to the free one. The same argument holds for the gluon's propagator, which reduces to the free propagator $\Delta_0$. Then for $x^+$ and $y^+$ both negative or both greater than $\eta$ :
\begin{equation}
\label{freegluon}
\Delta_{AB}^{\mu\nu}(\vec{x},\vec{y};k^+)=\delta^{AB}\Delta^{\mu\nu}_0(\vec{x}-\vec{y};k^+)=i\delta^{AB}\int\frac{\rmd k^-\rmd^2k_{\perp}}{(2\pi)^3}\frac{\Pi^{\mu\nu}(k)}{k^2+i\epsilon}e^{-ik^-(x^+-y^+)+i\mathbf{k}_{\perp}\cdot(\mathbf{x}_{\perp}-\mathbf{y}_{\perp})},
\end{equation}
where
\begin{equation}
\Pi^{\mu\nu}(k)=-g^{\mu\nu}+\frac{k^{\mu}n^{\nu}+k^{\nu}n^{\mu}}{k^+},
\end{equation}
and $n^{\mu}=\delta^{\mu-}$ is the gauge fixing vector. It is easy to check that \eqref{gluonpropagator} is fulfilled in this trivial case.\\

The last case, $x^+>\eta$ and $y^+<0$, is more interesting. The first step requires a convolution formula among propagators. Let us start from the equations of motion (in real space and with a general background field) :
\begin{equation}
\begin{split}
&G(x,z)\overrightarrow{\mathcal{D}^2}_zG(z,y)=-iG(x,z)\delta^{(4)}(z-y)\\
&G(x,z)\overleftarrow{\mathcal{D}^{\dagger}}^2_zG(z,y)=-iG(z,y)\delta^{(4)}(x-z).
\end{split}
\end{equation}
Integrating over $z$ $i$ times the second equation minus the first one over a volume $\Omega$ that contains $x$ but not $y$ gives :
\begin{equation*}
\begin{split}
G(x,y)&=-i\int_{\Omega} \rmd^4zG(x,z)\left[\overleftrightarrow{\Box}_z-2ig(\partial.\mathcal{A}(z))-2ig(\overleftarrow{\partial}^{\mu}_z\mathcal{A}_{\mu}(z)+\mathcal{A}_{\mu}(z)\overrightarrow{\partial}^{\mu}_z)\right]G(z,y)\\
&=-i\int_{\Omega} \rmd^4z\partial_{\mu}\left[G(x,z)\left[\overleftrightarrow{\partial^{\mu}_z}-2ig\mathcal{A}^{\mu}(z)\right]G(z,y)\right]\\
&=i\oint_{\partial\Omega} \rmd^3\sigma^{\mu}_zG(x,z)\left[\overleftarrow{\mathcal{D}^{\dagger}_{\mu}}_z-\overrightarrow{\mathcal{D}_{\mu}}_z\right]G(z,y),
\end{split}
\end{equation*}
where the last line has been obtained thanks to the Gauss theorem. We take the volume $\Omega$ to be the half-space $z^+>\eta$. Since the propagators are retarded, only the surface $z^+=\eta$ will contribute to the boundary term. The corresponding oriented surface vector is $\rmd^3\sigma^{\mu}_z=-\delta^{\mu -}\rmd z^-\rmd^2z_{\perp}$. Moreover for the considered, $x^-$-independent background field, we can plug the Fourier representation \eqref{FTdressed} for the scalar propagator in the previous equation. This leads to :
\begin{equation}
\label{GFdressedscalar}
G^{AB}(\vec{x},\vec{y};k^+)=2k^+\int_{z^+=\eta} \rmd^2z_{\perp}G^{AC}(\vec{x},\vec{z};k^+)G^{CB}(\vec{z},\vec{y};k^+).
\end{equation}
In this formula the propagator running from $\vec{z}$ to $\vec{x}$ can be replaced by the free one $G_0$ since there is no background field in this region. The problem reduces to find $G^{CB}(\vec{z},\vec{y};k^+)$. This is done by writing the identity :
\begin{equation}
\label{Gfirstrec}
G^{CB}(\vec{z},\vec{y};k^+)=\int_0^{\eta}\rmd u^+\partial^-G^{CB}(u^+,\mathbf{z}_{\perp},\vec{y};k^+)+\delta^{CB}G_0(0,\mathbf{z}_{\perp},\vec{y};k^+).
\end{equation}
The second term in the r.h.s. has been replaced by the free propagator since the endpoints lie in the region in which the background field is zero. The integrand in the r.h.s is computed thanks to the equation of motion \eqref{EOMscalar}. In the region $u^+\in [0;\eta]$, the variation of the background field is very sharp and thus the $\partial^-$ derivative is large and is of the order of the - strong - background field. Therefore one can neglect the transverse derivatives which are negligible with respect to what happens in the $u^+$ direction. Moreover $u^+>0$ and $y^+<0$ by assumption, so the delta function does not contribute. One can replace $\partial^-$ by $ig\mathcal{A}^-$ in the integral :
\begin{equation}
G^{CB}(\vec{z},\vec{y};k^+)=ig\int_0^{\eta}\rmd u^+\mathcal{A}^-_{CD}(u^+,\mathbf{z}_{\perp})G^{DB}(u^+,\mathbf{z}_{\perp},\vec{y};k^+)+\delta^{CB}G_0(0,\mathbf{z}_{\perp},\vec{y};k^+).
\end{equation}
Then we recursively plug the full r.h.s. in place of $G^{DB}(u^+,\mathbf{z}_{\perp},\vec{y};k^+)$. This will bring a path ordered product of $\mathcal{A}^-$ along the plus direction which exponentiates at the end giving the Wilson line \eqref{WilsonLine} times the free propagator $G_0$ :
\begin{equation}
G^{CB}(\vec{z},\vec{y};k^+)=\Omega_{CB}(\mathbf{z}_{\perp})G_0(0,\mathbf{z}_{\perp},\vec{y};k^+)=\Omega_{CB}(\mathbf{z}_{\perp})G_0(\vec{z}-\vec{y};k^+)+\mathcal{O}(\eta).
\end{equation}
Plugging this expression into the convolution formula \eqref{GFdressedscalar} leads to :
\begin{equation}
\label{dresseddomain}
\begin{split}
G^{AB}(\vec{x},\vec{y};k^+)&=2k^+\int_{z^+=0} \rmd^2z_{\perp}G_0(\vec{x}-\vec{z};k^+)\tilde{\Omega}_{AB}(\mathbf{z}_{\perp})G_0(\vec{z}-\vec{y};k^+).
\end{split}
\end{equation}
Note that the initial surface has been moved to $z^+=0$. This generates irrelevant $\eta$ order terms which cancel as $\eta\rightarrow 0$. This expression is intuitive in the sense that the scalar freely propagates until the surface $z^+=0$, then the effect of the shockwave is a convolution in the transverse plane with a Wilson line which encodes the color precession due to the shockwave and at the end the scalar propagates freely again above the surface $z^+=0$. Formula \eqref{dresseddomain} is the required expression for the dressed propagator in the least trivial case $x^+>\eta$ and $y^+<0$.\\

To write down the gluon's propagator, one can summarize relations \eqref{gluonpropagator} into the following shorthand notation :
\begin{equation}
\label{gluonscalar}
\begin{split}
&\Delta^{\mu\nu}_{AB}(\vec{x},\vec{y};k^+)=\mathcal{O}^{\mu\nu}_{\mathbf{x}_{\perp},\mathbf{y}_{\perp},k^+}G^{AB}(\vec{x},\vec{y};k^+)~;
\end{split}
\end{equation}
where $\mathcal{O}^{\mu\nu}_{\mathbf{x}_{\perp},\mathbf{y}_{\perp},k^+}$ is the differential operator : 
\begin{equation}
\begin{split}
&\mathcal{O}^{\mu\nu}_{\mathbf{x}_{\perp},\mathbf{y}_{\perp},k^+}=\left(\delta^{\mu-}\frac{i}{k^+}\partial^i_x+\delta^{\mu i}\right)\left(-\delta^{\nu-}\frac{i}{k^+}\partial^i_{y}+\delta^{\nu i}\right).
\end{split}
\end{equation}
The best way is to write the free scalar propagators in \eqref{gluonscalar} in Fourier space\footnote{We denote $\vec{p}=(p^-,\mathbf{p}_{\perp})$, the conjugate variable to $\vec{x}$.} :
\begin{equation}
\label{dressedpropfourier}
\begin{split}
\Delta^{\mu\nu}_{AB}(\vec{x},\vec{y};k^+)&=2k^+\mathcal{O}^{\mu\nu}_{\mathbf{x}_{\perp},\mathbf{y}_{\perp},k^+}\int_{p^+=q^+=k^+}\frac{\rmd p^-\rmd^2p_{\perp}}{(2\pi)^3}\frac{\rmd q^-\rmd^2q_{\perp}}{(2\pi)^3}G_0(p)G_0(q)e^{-i\vec{p}\cdot\vec{x}+i\vec{q}\cdot\vec{y}}\\
&~~~~~~~~~~~~~~~~~\times\int \rmd^2z_{\perp}\tilde{\Omega}_{AB}(\mathbf{z}_{\perp})e^{-i\mathbf{z}_{\perp}\cdot(\mathbf{p}_{\perp}-\mathbf{q}_{\perp})}\\
&=2k^+\int_{p^+=q^+=k^+}\frac{\rmd p^-\rmd^2p_{\perp}}{(2\pi)^3}\frac{\rmd q^-\rmd^2q_{\perp}}{(2\pi)^3}\frac{i\beta^{\mu i}(\mathbf{p}_{\perp},k^+)}{p^2+i\epsilon}
\frac{i\beta^{\nu i}(\mathbf{q}_{\perp},k^+)}{q^2+i\epsilon}e^{-i\vec{p}\cdot\vec{x}+i\vec{q}\cdot\vec{y}}\\
&~~~~~~~~~~~~~~~~~\times\int \rmd^2z_{\perp}\tilde{\Omega}_{AB}(\mathbf{z}_{\perp})e^{-i\mathbf{z}_{\perp}\cdot(\mathbf{p}_{\perp}-\mathbf{q}_{\perp})}~;
\end{split}
\end{equation}
where we have introduced the following parametrization for the Fourier space representation of the operator $\mathcal{O}^{\mu\nu}_{\mathbf{x}_{\perp},\mathbf{y}_{\perp},k^+}$ : 
\begin{equation}
\label{betadef}
\begin{split}
&\mathcal{O}^{\mu\nu}_{\mathbf{p}_{\perp},\mathbf{q}_{\perp},k^+}=\left(\delta^{\mu-}\frac{p^i}{k^+}+\delta^{\mu i}\right)\left(\delta^{\nu-}\frac{q^i}{k^+}+\delta^{\nu i}\right)\equiv\beta^{\mu i}(\mathbf{p}_{\perp},k^+)\beta^{\nu i}(\mathbf{q}_{\perp},k^+),
\end{split}
\end{equation}
with
\begin{equation}
\label{betadef}
\beta^{\mu i}(\mathbf{p}_{\perp},k^+)=\delta^{\mu-}\frac{p^i}{k^+}+\delta^{\mu i}.
\end{equation}
In some sense $\beta$ play the role of generalized polarization vectors in the case where the incoming and outgoing momenta are no the same.

\subsection{Identities involving the $\beta^{\mu i}$'s}
\indent

As a consistency condition, when the momentum is conserved through the shockwave, we get :
\begin{equation}
\beta^{\mu i}(\mathbf{k}_{\perp},k^+)\beta^{\nu i}(\mathbf{k}_{\perp},k^+)=\Pi^{\mu\nu}(k).
\end{equation}
That is the expression of the fact that we recover the free case \eqref{freegluon} when the Wilson line reduces to the identity in \eqref{dressedpropfourier} - up to a momentum conservation delta function.\\
Moreover, by the gauge condition $\epsilon_-=\epsilon^+=0$ and the Ward identity $k.\epsilon(k)=0$, one has :
\begin{equation}
\label{propepsilon}
\epsilon_{\mu}(k)\beta^{\mu i}(\mathbf{k}_{\perp},k^+)=-\epsilon^i(k)
\end{equation}
and
\begin{equation}
\label{propepsilon2}
\beta^{\mu i}(\mathbf{k}_{\perp},k^+)\epsilon^i(k)=\epsilon^{\mu}(k).
\end{equation}
These identities are useful since they enable us to deal with Lorentz or transverse indices indifferently.

\section{Momentum space Feynman rules\label{momentumfeynrules}}
\indent

Although we are considering inhomogeneous systems - the shockwave being space-time dependent - it is possible and even easier to use momentum space Feynman rules. Among the Feynman rules, one recovers the usual ones for non abelian gauge theories and as usual, integrations are performed over all undetermined momenta.\\

To set the notation, we will denote as follows the Lorentz piece of the 3-gluons vertex\footnote{The 4-gluons vertex is not used here, this is why it is omitted but the standard Feynman rule holds.} :
\begin{equation}
\Gamma_{\mu\nu\rho}(k,p,q)\equiv g_{\mu\nu}(k+p)_{\rho}+g_{\nu\rho}(q-p)_{\mu}-g_{\mu\rho}(k+q)_{\nu}.
\end{equation}
For convenience $k$ has been chosen to be incoming and $p$ and $q$ outgoing.\\

The only technicalities arise from the shockwave that lead to specific rules to be detailed in the following.

\subsection{The dressed gluon's propagator}
\indent

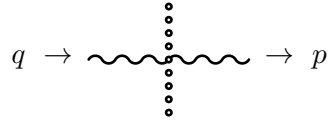
\begin{figure}[h]
\centering
%(along, up)
\begin{fmfgraph*}(60,40)
    \fmfleft{i1}
    \fmfright{o1}
    \fmftop{a,b,c}
    \fmfbottom{d,e,f}
      \fmf{dbl_dots}{e,b}
      \fmffreeze
      \fmf{photon}{i1,o1}
      \fmflabel{$\rightarrow~p$}{o1}
      \fmflabel{$q~\rightarrow$}{i1}
\end{fmfgraph*}
\caption{Gluon coming into the shockwave with momentum $q$ and exiting with momentum $p$. \label{propSW}}
\end{figure}

According to \eqref{dressedpropfourier} the Feynman rule corresponding to figure \ref{propSW}, that is the propagator of a gluon which comes into the shockwave with momentum $q$ and which exits with momentum $p$, reads :
\begin{equation}
\begin{split}
2k^+\frac{i\beta^{\mu i}(\mathbf{p}_{\perp},k^+)}{p^2+i\epsilon}
\frac{i\beta^{\nu i}(\mathbf{q}_{\perp},k^+)}{q^2+i\epsilon} \int \rmd^2z_{\perp}\tilde{\Omega}_{AB}(\mathbf{z}_{\perp})e^{-i\mathbf{z}_{\perp}\cdot(\mathbf{p}_{\perp}-\mathbf{q}_{\perp})}.
\end{split}
\end{equation}
$k^+$ is the $+$ component of both $p$ and $q$ that is conserved through the diagram.

\subsection{External legs attached to the shockwave}
\indent

\begin{figure}[h]
\centering
%(along, up)
\begin{fmfgraph*}(60,40)
    \fmfleft{i1}
    \fmfright{o1}
    \fmftop{a,b,c}
    \fmfbottom{d,e,f}
      \fmf{dbl_dots}{e,b}
      \fmffreeze
      \fmf{photon,label=$k\rightarrow~~~~\rightarrow p$}{i1,o1}
      \fmfblob{20.}{o1}
      \fmfdot{i1}
\end{fmfgraph*}
\caption{Arbitrary process involving an initial gluon that goes through the shockwave at first. \label{initgluon}}
\end{figure}
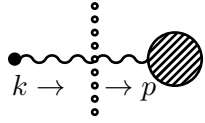

An interesting - and subtle - question is : what is the rule corresponding to an external leg directly attached to the shockwave ? In words, reduction formulas tell us that the most external free propagators in green functions are removed and replaced by polarization vectors $\epsilon^{\mu}$ or $\epsilon^{\mu~*}$ depending on whether the gluons are initial ones or final ones respectively. It would be tempting to say that, for instance, an initial gluon involved in some arbitrary process shown on figure \ref{initgluon}, will bring a contribution :
\begin{equation}
\label{extlegguess}
\begin{split}
2k^+\epsilon_{\mu}(k)\beta^{\mu i}(\mathbf{k}_{\perp},k^+)\int_{p^+\equiv k^+} \frac{\rmd p^-\rmd^2p_{\perp}}{(2\pi)^3}\frac{i\beta^{\nu i}(\mathbf{p}_{\perp},k^+)}{p^2+i\epsilon}\int \rmd^2z_{\perp}\tilde{\Omega}_{AB}(\mathbf{z}_{\perp})e^{-i\mathbf{z}_{\perp}\cdot(\mathbf{p}_{\perp}-\mathbf{k}_{\perp})}\times M^B_{\nu}(p)~;
\end{split}
\end{equation}
where $M^B_{\nu}(p)$ is the Green function corresponding to the rest of the process represented by the bubble on figure \ref{initgluon}, whatever it is\footnote{It may even contain other external legs.}. But this formula has a wrong sign as we shall see.

To derive the Feynman rule corresponding to the process shown on figure \ref{initgluon}, one has to consider the reduction formula\footnote{To check the prefactor $-i$ one can consider the free case where it reduces to the ordinary rule $\epsilon_{\mu}(k)M^{A\mu}(k)$. For sure it has to be taken as a mnemonic rather than a rigorous proof.} :
\begin{equation}
\label{reduc}
-i\int \rmd^4x\rmd^4ye^{-ik.x}\epsilon^{\mu}(k)\Box_x\Delta^{BA}_{\nu\mu}(y,x)M^{B\nu}(y),
\end{equation}
where $M^{A\nu}(y)$ is the Fourier transform of $M^{A\nu}(k)$ with an incoming momentum $k$, that is\footnote{Beware of the unconventional sign in the exponential due to the fact that the momentum is incoming rather than outgoing.} :
\begin{equation}
M^{A\nu}(y)=\int \frac{d^4k}{(2\pi)^4}M^{A\nu}(k)e^{ik.y}.
\end{equation}

Using \eqref {dressedpropfourier} it is straightforward to write the reduction formula \eqref{reduc} :
\begin{equation}
\label{extleg0}
\begin{split}
-i\int &\rmd^4x\rmd^4ye^{-ik.x}\epsilon^{\mu}(k)\Box_x\Delta^{BA}_{\nu\mu}(y,x)M^{B\nu}(y)\\
&=-2k^+\epsilon_{\mu}(k)\beta^{\mu i}(\mathbf{k}_{\perp},k^+)\int_{p^+\equiv k^+} \frac{\rmd p^-\rmd^2p_{\perp}}{(2\pi)^3}
\frac{i\beta^{\nu i}(\mathbf{p}_{\perp},k^+)}{p^2+i\epsilon}\int \rmd^2z_{\perp}\tilde{\Omega}_{BA}(\mathbf{z}_{\perp})e^{-i\mathbf{z}_{\perp}\cdot(\mathbf{p}_{\perp}-\mathbf{k}_{\perp})}M^{A}_{\nu}(p).
\end{split}
\end{equation}
Indeed, this formula differs from \eqref{extlegguess} by a sign.\\

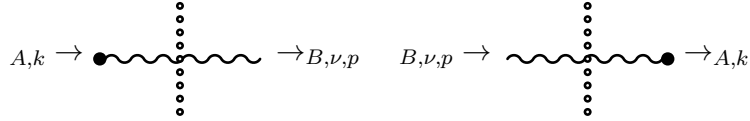
\begin{figure}[h]
\centering
%(along, up)
\begin{fmfgraph*}(60,40)
    \fmfleft{i1}
    \fmfright{o1}
    \fmftop{a,b,c}
    \fmfbottom{d,e,f}
      \fmf{dbl_dots}{e,b}
      \fmffreeze
      \fmf{photon}{i1,o1}
      \fmfdot{i1}
      \fmflabel{$_{A, k}\rightarrow$}{i1}
      \fmflabel{$\rightarrow _{B, \nu, p}$}{o1}
\end{fmfgraph*}
\hspace{3cm}
\begin{fmfgraph*}(60,40)
    \fmfleft{i1}
    \fmfright{o1}
    \fmftop{a,b,c}
    \fmfbottom{d,e,f}
      \fmf{dbl_dots}{e,b}
      \fmffreeze
      \fmf{photon}{i1,o1}
      \fmfdot{o1}
      \fmflabel{$\rightarrow _{A, k}$}{o1}
      \fmflabel{$_{B, \nu, p}\rightarrow$}{i1}
\end{fmfgraph*}
\caption{Feynman diagrams corresponding respectively to the two formulas \eqref{extleg}. \label{schockwavefeynrules}}
\end{figure}

We now have the Feynman rules for external legs attached directly to the shockwave, which correspond respectively to the two figures represented on \ref{schockwavefeynrules} :
\begin{equation}
\label{extleg}
\begin{split}
&-2k^+\epsilon_{\mu}(k)\beta^{\mu i}(\mathbf{k}_{\perp},k^+)\frac{i\beta^{\nu i}(\mathbf{p}_{\perp},k^+)}{p^2+i\epsilon}\int \rmd^2z_{\perp}\tilde{\Omega}_{BA}(\mathbf{z}_{\perp})e^{-i\mathbf{z}_{\perp}\cdot(\mathbf{p}_{\perp}-\mathbf{k}_{\perp})}\\
&-2k^+\epsilon^*_{\mu}(k)\beta^{\mu i}(\mathbf{k}_{\perp},k^+)\frac{i\beta^{\nu i}(\mathbf{p}_{\perp},k^+)}{p^2+i\epsilon}\int \rmd^2z_{\perp}\tilde{\Omega}_{AB}(\mathbf{z}_{\perp})e^{-i\mathbf{z}_{\perp}\cdot(\mathbf{k}_{\perp}-\mathbf{p}_{\perp})},
\end{split}
\end{equation}
where $p^+=k^+$.

\chapter{Schwinger-Keldysh formalism \label{Keldyshappendix}}
\indent

In this section we shall see how inclusive observables can be reinterpreted in the Schwinger-Keldysh formalism. This formalism is generally encountered in finite temperature quantum field theory but here it finds another application at zero temperature. Here, it is actually a way to compute in - in Green functions, the sum over final states beeing interpreted as disconnected vacuum bubbles in this formalism. In the first part we study how general Green functions with final state summed can be written as in-in amplitudes with a generalized $S$-operator, the Schwinger-Keldysh evolution operator. Inclusive spectra encountered in section \ref{inclusivespectrum} are special cases of such Green functions. Then we shall deal with the path integral formulation. At the end we detail Feynman rules since propagators and vertices have more general definitions in this formalism than in the context of "ordinary" quantum field theory.

\section{Schwinger-Keldysh $S$-operator}
\indent

\begin{figure}[h]
\begin{center}
\includegraphics[width=0.4\paperwidth]{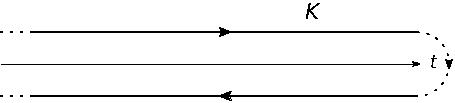}
\caption{Graphic representation of the Keldysh contour. The two branches are shifted with respect to the time axis for clarity but the contour lies on the real axis.\label{keldcontour}}
\end{center}
\end{figure}

As seen in section \ref{inclusivespectrumandKeldysh} the inclusive spectrum is rewritten thanks to a reduction formula as a correlator  between two field operators, one in the conjugate and another one in complex conjugate amplitude with a sum over final states. It is rather straightforward, to show that more general spectra $\rmd \mathcal{N}$ per multi-particle phase space element $\rmd^3p_1...\rmd^3p_k$ can be written in a similar form as equation \eqref{inclusiveLSZ} with a time-ordered product of $k$ fields in the amplitude and an antitime-ordered product of $k$ hermitian conjugate fields in the complex conjugate amplitude. We shall see that it can be rewritten in the Schwinger-Keldysh formalism. Particle spectra are specific cases of the more general Green function :
\begin{equation}
\label{inclusiveGF}
\displaystyle{\sum_{\gamma}}\left\langle \beta ; in\left|\bar{\mathcal{T}}\left\{A_1(x_1)...A_p(x_p)\right\}\right|\gamma;out\right\rangle\left\langle \gamma ; out\left|\mathcal{T}\left\{B_1(y_1)...B_q(y_q)\right\}\right|\alpha;in\right\rangle.
\end{equation}
$\mathcal{T}$ and $\bar{\mathcal{T}}$ respectively denote the time and antitime ordered product, the $A_i$'s and $B_j$'s are operators built from fields in the Heisenberg picture and the sum over $\gamma$ is a generic notation for summing over all the possible multi-particle states properly normalized so that $\sum_{\gamma}|\gamma><\gamma|=1$. If we consider inclusive spectra, $p=q$ and the operators are single field operators in \eqref{inclusiveGF}. After some algebra (the detailed calculation can be found in the general references \cite{WeinbergI,PeskinSchroeder}), the Green function \eqref{inclusiveGF} can be written in terms of interaction picture operators $a_i$'s and $b_j$'s between Heisenberg states :
\begin{equation}
\label{inclusiveGFinteraction}
\begin{split}
\displaystyle{\sum_{\gamma}}&\left\langle \beta \left|\bar{\mathcal{T}}\left\{a_1(x_1)...a_p(x_p)\exp\left(+i\int^{+\infty}_{-\infty}\rmd tV(t)\right)\right\}\right|\gamma\right\rangle\\
&\times\left\langle \gamma\left|\mathcal{T}\left\{b_1(y_1)...b_q(y_q)\exp\left(-i\int^{+\infty}_{-\infty}\rmd tV(t)\right)\right\}\right|\alpha\right\rangle~;
\end{split}
\end{equation}
where $V$ is the interaction Hamiltonian written in terms of interaction picture field operators. The sum over $\gamma$ is trivial because of the completeness relations. Then the operators are ordered in the following way. The ones coming from the amplitude are time ordered on the right and the ones from the complex conjugate amplitude are on the left and antitime-ordered. Then everything works like if the integration contour of the interaction starts from $-\infty$ to $+\infty$ and then comes back to $-\infty$. This integration contour is the so called \emph{Keldysh} contour denoted $K$, represented on figure \ref{keldcontour}.
The $a_i$'s and $b_j$'s are ordered along this contour. We say they are \emph{path-ordered} and we denote $\mathcal{K}$ the corresponding operator that orders operators along the Keldysh contour. With these new notations, \eqref{inclusiveGFinteraction} becomes :
\begin{equation}
\left\langle \beta \left|\mathcal{K}\left\{a_1(x_1)...a_p(x_p)b_1(y_1)...b_q(y_q)e^{-i\int_{K}\rmd tV(t)}\right\}\right|\alpha\right\rangle,
\end{equation}
or between states in the interaction picture :
\begin{equation}
\label{inclusiveGFkeldysh}
\left\langle \beta ; in\left|\mathcal{K}\left\{A_1(x_1)...A_p(x_p)B_1(y_1)...B_q(y_q)\right\}\right|\alpha;in\right\rangle.
\end{equation}
We now have a precise meaning for the in - in Green functions. The S-operator for such amplitude is a non trivial way to rewrite $1$. Indeed if all the operators $A_i$'s and $B_j$'s are set to unity, we find the scalar product between two in-states which is trivial since they are orthonormal. It is equivalent to the operator equation :
\begin{equation}
\label{unitKeldyshSoperator}
\mathcal{K}\left\{e^{-i\int_{K}\rmd tV(t)}\right\}=1
\end{equation}
But then how can we generate Green functions ? This is given by a generalized Schwinger action principle. The Green functions are generated by adding a term to the interaction :
\begin{equation}
V(t)\rightarrow V_{\epsilon}(t) = V(t)+\int\rmd^3x~\epsilon_i(t,\mathbf{x})a_i(t,\mathbf{x})
\end{equation}
and then by taking functional derivatives with respect to $\epsilon$ and sending it to zero. As long as $\epsilon$ is non zero, the S operator with $V$ replaced by $V_{\epsilon}$ does not satisfy \eqref{unitKeldyshSoperator}. This is the right way to deal with this S operator that seems somewhat trivial at first sight. We now have an operator formulation of the Schwinger-Keldysh evolution, we will now look at the corresponding path integral formulation.

\section{Path integral formulation and connected amplitudes}
\indent

We start from formula \eqref{inclusiveGFkeldysh} in the special case where $\alpha$ and $\beta$ are taken to be the vacuum. There is no difference with the usual path integral formulation of quantum field theory whose rederivation would be useless since it can be found in any general references like \cite{WeinbergI,PeskinSchroeder}. The path integral corresponding to an evolution along the Keldysh contour, requires two sets of fields $\phi_+$ and $\phi_-$, $\phi$ denoting generically the whole field content of the theory and being respectively the fields living on the forward and backward time branch of the Keldysh contour. In some sense the Schwinger-Keldysh path integral is the squared modulus of the ordinary path integral. In the path integral formulation, the vacuum in - vacuum in Green function \eqref{inclusiveGF} reads :
\begin{equation}
\label{KeldyshpathintegralGreenfunction}
\begin{split}
&\left\langle 0 ; in\left|\mathcal{K}\left\{A_1(x_1)...A_p(x_p)B_1(y_1)...B_p(y_q)\right\}\right|0;in\right\rangle=\frac{1}{Z}\int\mathcal{D}\phi_-\mathcal{D}\phi_+ A_1[\phi_-(x_1)]...A_p[\phi_-(x_p)]\times\\
&~~~~~~~~~~~~~~~~~~~~~~~~~~~~~\times B_1[\phi_+(y_1)]...B_p[\phi_+(y_q)]e^{i\int^{+\infty}_{-\infty} \rmd t L[\phi_+,t]-i\int^{+\infty}_{-\infty} \rmd t L[\phi_-,t]}
\end{split}
\end{equation}
where $L$ is the lagrangian and $Z$ is the partition function corresponding to the Schwinger-Keldysh path integral :
\begin{equation}
Z=\int\mathcal{D}\phi_-\mathcal{D}\phi_+ e^{i\int^{+\infty}_{-\infty} \rmd t L[\phi_+,t]-i\int^{+\infty}_{-\infty} \rmd t L[\phi_-,t]}.
\end{equation}
The factor $1/Z$ in \eqref{KeldyshpathintegralGreenfunction} ensures the proper normalization of the Green functions. Following the general functional methods, the Green function \eqref{KeldyshpathintegralGreenfunction} can be obtained from the generating functional $Z[j_+,j_-]$ defined as :
\begin{equation}
\label{Keldyshgeneratingpartitionfunction}
\begin{split}
Z[j_+,j_-]=&\int\mathcal{D}\phi_-\mathcal{D}\phi_+ \exp\left(i\int^{+\infty}_{-\infty} \rmd t \left[L[\phi_+,t]+ \int\rmd^3xj^i_+(t,\mathbf{x})B_i[\phi_+(t,\mathbf{x})]\right]\right.\\
&-\left.i\int^{+\infty}_{-\infty} \rmd t \left[L[\phi_-,t]+ \int\rmd^3xj^j_-(t,\mathbf{x})A_j[\phi_-(t,\mathbf{x})]\right]\right).
\end{split}
\end{equation}
The partition function is given by $Z=Z[j_+=j_-=0]$ and the Green function \eqref{KeldyshpathintegralGreenfunction}, by the functional derivative :
\begin{equation}
\begin{split}
\left\langle 0 ; in\left|\mathcal{K}\left\{A_1(x_1)...A_p(x_p)B_1(y_1)...B_q(y_q)\right\}\right|0;in\right\rangle=\frac{i^{p-q}}{Z}\left.\frac{\delta^{p+q}}{\delta j^1_-(x_1)...\delta j^p_-(x_p)\delta j^1_+(y_1)...\delta j^q_+(y_q)}Z[j_+,j_-]\right|_{j_{\pm}=0}.
\end{split}
\end{equation}
Beware of the order of derivatives when one deals with Grassmann fields. The convention given here makes the derivatives act from the left to the right. The Green function \eqref{KeldyshpathintegralGreenfunction} is a vacuum-vacuum amplitude which is not necessarily connected. Since any amplitude always factorizes into product of connected Green functions, it is more convenient to work with the generating functional for connected amplitudes only, denoted $\mathcal{W}[j_+,j_-]$. Its relation to $Z[j_+,j_-]$ is the following :
\begin{equation}
\label{Keldyshconnectedgenerating}
e^{i\mathcal{W}[j_+,j_-]}=Z[j_+,j_-].
\end{equation}
So far, we do have an expression for general inclusive Green functions - that is with the final state summed - as a path integral with time integrated over the Keldysh contour. Naively it can be seen as an ordinary path integral corresponding to the amplitude and a complex conjugate one corresponding to the complex conjugate amplitude but it does \emph{not} factorize into two independent path integrals. The product of two path integrals would be interpreted as the modulus squared of a vacuum in - vacuum out amplitude which is different from summing over all possible final states which leads to non trivial boundary conditions at $t=+\infty$, $\phi_+(+\infty,\mathbf{x})=\phi_-(+\infty,\mathbf{x})$, that forbids the factorization of the Schwinger-Keldysh path integral into two ordinary path integrals. The Schwinger-Keldysh path integral form of particle spectra encountered in section \ref{inclusivespectrumandKeldysh} is obtained from \eqref{KeldyshpathintegralGreenfunction} by taking the $A_i$'s to be single field operators and the $B_i$'s their complex conjugate.

\section{Feynman rules\label{KeldyshFeynmanrules}}
\indent

In this section we detail the new features brought by the Schwinger-Keldysh formalism which is a generalization of the well known Feynman rules. When one draws Feynman diagrams in coordinate space, the integration in time is performed along the whole Keldysh contour instead of the ordinary time axis. This leads to more complicated Feynman rules depending on whether the coordinates under consideration lie on the forward or backward branch. Obviously ordinary Feynman rules are recovered if we consider only field operators on the forward time branch.

\subsection{Propagator}
\indent

In the Schwinger-Keldysh formalism, the exact propagator is defined to be :
\begin{equation}
G^{kl}(x,y)=\left\langle 0;in\left|\mathcal{K}\left\{\phi^k(x)\phi^{l\dagger}(y)\right\}\right|0;in\right\rangle.
\end{equation}
The ordering depends on which branch of the Keldysh contour the points $x$ and $y$ are evaluated or equivalently if the fields are plus or minus fields. There are four configurations :
\begin{equation}
G^{kl}(x,y)\equiv \left\{
    \begin{array}{llll}
G^{kl}_{++}(x,y)&~~~~~\text{if }x^0 \text{ and } y^0 \text{ are on the forward branch} \\
G^{kl}_{-+}(x,y)&~~~~~\text{if }x^0 \text{ is on the backward branch and } y^0 \text{ on the forward one} \\
G^{kl}_{+-}(x,y)&~~~~~\text{if }x^0 \text{ is on the forward branch and } y^0 \text{ on the backward one} \\
G^{kl}_{--}(x,y)&~~~~~\text{if }x^0 \text{ and } y^0 \text{ are on the backward branch} 
    \end{array}
\right.
\end{equation}
where
\begin{equation}
\label{Keldyshfreepropagators}
\begin{split}
G^{kl}_{++}(x,y)&=\left\langle 0;in\left|\mathcal{T}\left\{\phi^k(x)\phi^{l\dagger}(y)\right\}\right|0;in\right\rangle\\
&=\theta(x^0-y^0)\left\langle 0;in\left|\phi^k(x)\phi^{l\dagger}(y)\right|0;in\right\rangle\pm\theta(y^0-x^0)\left\langle 0;in\left|\phi^{l\dagger}(y)\phi^k(x)\right|0;in\right\rangle\\
G^{kl}_{-+}(x,y)&=\left\langle 0;in\left|\phi^k(x)\phi^{l\dagger}(y)\right|0;in\right\rangle\\
G^{kl}_{+-}(x,y)&=\pm\left\langle 0;in\left|\phi^{l\dagger}(y)\phi^k(x)\right|0;in\right\rangle\\
G^{kl}_{--}(x,y)&=\pm\left\langle 0;in\left|\bar{\mathcal{T}}\left\{\phi^k(x)\phi^{l\dagger}(y)\right\}\right|0;in\right\rangle\\
&=\pm\theta(x^0-y^0)\left\langle 0;in\left|\phi^{l\dagger}(y)\phi^k(x)\right|0;in\right\rangle+\theta(y^0-x^0)\left\langle 0;in\left|\phi^k(x)\phi^{l\dagger}(y)\right|0;in\right\rangle.
\end{split}
\end{equation}
The upper and lower signs are for bosons and fermions respectively. The four kinds of propagators are not all independent. As it can be checked directly from \eqref{Keldyshfreepropagators}, they are related by the identity :
\begin{equation}
G^{kl}_{++}(x-y)+G^{kl}_{--}(x-y)=G^{kl}_{-+}(x-y)+G^{kl}_{+-}(x-y).
\end{equation}
The retarded and advanced propagators are given by :
\begin{equation}
\label{Keldyshadvret}
\begin{split}
G^{kl}_{R}(x-y)&=G^{kl}_{++}(x-y)-G^{kl}_{+-}(x-y)=G^{kl}_{-+}(x-y)-G^{kl}_{--}(x-y)\\
&=\theta(x^0-y^0)\left\langle 0;in\left|\left[\phi^k(x);\phi^{l\dagger}(y)\right]_{\mp}\right|0;in\right\rangle\\
G^{kl}_{A}(x-y)&=G^{kl}_{++}(x-y)-G^{kl}_{-+}(x-y)=G^{kl}_{+-}(x-y)-G^{kl}_{--}(x-y)\\
&=-\theta(y^0-x^0)\left\langle 0;in\left|\left[\phi^k(x);\phi^{l\dagger}(y)\right]_{\mp}\right|0;in\right\rangle.
\end{split}
\end{equation}
$[~]_{-}$ denotes the commutator and $[~]_+$ the anti-commutator depending whether the field $\phi$ is bosonic or fermionic respectively. These are generic notations for the exact propagators. Analogous straightforward definitions hold for free, 1-loop, background field dressed and so on... propagators just by choosing the corresponding in vacuum and evolution operator.

\subsection{Vertices\label{keldyshvertices}}
\indent

From the very first physical requirements, vertex operators must be local and real in coordinate space. Since one integrates in coordinate space over the whole Keldysh contour, one has to sum over the minus and plus fields if one attaches a line to a vertex. From \eqref{KeldyshpathintegralGreenfunction}, the vertices evaluated on the forward branch are given by ordinary Feynman rules whereas they take a minus sign when the are evaluated on the backward branch. Obviously this holds in momentum space as well. To illustrate, let us consider the tree level 4-point vertex of the neutral scalar $\phi^4$ theory. One has to sum over the plus and minus branch of the Keldysh contour which gives two diagrams with opposite Feynman rules shown on figure \ref{Keldyshphi4vertex}.
\begin{figure}[h]
\centering
%(along, up)
\begin{fmfgraph*}(40,40)
    \fmfleft{i1,i2}
    \fmfright{o1,o2}
      \fmf{plain}{i1,v,o2}
      \fmf{plain}{i2,v,o1}
      \fmflabel{$_{\pm}$}{v}
\end{fmfgraph*}
\hspace{1cm}$=\mp i\lambda$
\caption{Feynman rule for the $\phi^4$ vertex in the Schwinger-Keldysh formalism. \label{Keldyshphi4vertex}}
\end{figure}

\chapter[Physical spectrum in light-cone gauge]{Physical spectrum of non abelian gauge theories in light-cone gauge\label{BRSTspectrum}}
\indent

In this appendix we shall look at the physical spectrum of light-cone quantized non abelian gauge theories. We recover the well known result that ghost and anti-ghosts are absent from the physical spectrum but the main result is that the physical degrees of freedom of the gauge field are its transverse components. We will see that the light-cone gauge condition $n\cdot A=0$ with $n^{\mu}=\delta^{\mu\pm}$ is physical, in the sense that physical amplitudes only contain the independent, unconstrained degrees of freedom, that is the transverse components of the gauge field. The key point for this proof is the Becchi-Rouet-Stora-Tyutin (BRST) symmetry \cite{Becchi:1974xu,Becchi:1974md,Becchi:1975nq,Tyutin:1975qk}. In a first part we introduce this symmetry and use it to find a physical criterion for the existence of states in the spectrum or not. This point is one of the most beautiful correspondence between physics and mathematics, since the BRST symmetry is closely related to the cohomology theory. It can be used for showing very general (and crucial) properties of non abelian gauge theories like unitarity and renormalizability. Here we will not look at these but just, in a second part, see which states are present or absent from the spectrum.

\section{BRST symmetry}
\indent

The BRST symmetry is a residual symmetry of the lagrangian once the gauge is fixed. Considerations of this section hold for any gauge fixing but for definiteness, we shall work in axial gauge. We consider $\mathcal{L}$ the Yang-Mills lagrangian coupled to matter fields, generically denoted $\psi$, with the gauge fixing term \eqref{Lgf} and the ghost term \eqref{Lghost}. For the moment we shall work with a finite gauge parameter $\xi$. The trick is to rewrite the gauge fixing term thanks to a gaussian transformation in the path integral :
\begin{equation}
\exp\left[\frac{i}{2\xi}\int\rmd^4x(n\cdot A(x))^2\right]=N\int\mathcal{D}B\exp\left[i\int\rmd^4x\left(\frac{\xi}{2}B^2(x)+n\cdot A^A(x)B^A(x)\right)\right].
\end{equation}
The $B$ field is an adjoint representation auxiliary field - i.e. it satisfies algebraic equations of motions - known as the Nakanishi-Lautrup field. The constant $N$ is an inconsequential infinite normalization constant that disappears from connected amplitudes by a proper regularization scheme. After the gaussian transformation, the full lagrangian reads :
\begin{equation}
\label{Ltotaux}
\mathcal{L}_{\rm tot}=\mathcal{L}[A,\psi]+\frac{\xi}{2}B^2+B^An\cdot A^A+\bar{\omega} n\cdot D\omega.
\end{equation}
Note that written in this form, there is no problem in taking the $\xi=0$ limit. In this case, the integration over $B$ will give the constraint $\delta(n.A)$ as expected. Let us now look at the symmetry of the lagrangian \eqref{Ltotaux}. It turns out that this symmetry, known as the BRST symmetry, enables us to do some formal developments with a strong physical meaning.

\subsection{BRST invariance and physicality condition}
\indent

The lagrangian \eqref{Ltotaux} is invariant under the continuous, hermitian transformation of infinitesimal Grassmann parameter $\theta$ (i.e. $\theta$ anti-commutes with fermions, ghosts and anti-ghosts but commutes with bosons and auxiliary fields) :
\begin{equation}
\label{BRSTtrans}
\begin{split}
\delta_{\theta}\psi^a&=ig\theta\omega^A(T^A)_{ab}\psi^b\\
\delta_{\theta}A_{\mu}^A&=\theta D_{\mu}^{AB}\omega^B\\
\delta_{\theta}\bar{\omega}^A&=-\theta B^A\\
\delta_{\theta}\omega^A&=-\theta\frac{g}{2}f^{ABC}\omega^B\omega^C\\
\delta_{\theta}B^A&=0.
\end{split}
\end{equation}
These transformations are known as the BRST transformations. It is an easy exercise to check the invariance directly. However we can note that the invariance of $\mathcal{L}$ is obvious since the transformation of $\psi$ and $A$ is nothing but an infinitesimal gauge transformation of parameter $g\theta\omega$ under which $\mathcal{L}$ is invariant by construction. The remarkable property of the BRST transformation is that it is \emph{nilpotent}, that is $\delta_{\theta}\delta_{\theta'}=0$. To see the consequences of the nilpotency property, we introduce the BRST generator $Q$ whose action on the fields is defined so that $\delta_{\theta}\phi\equiv i\theta[Q;\phi]_{\mp}$, where $\phi$ denotes indifferently the matter, gauge, auxiliary, ghost and anti-ghost fields and the $\mp$ sign denotes a commutation relation if $\phi$ is a c-number field and an anti-commutation relation if $\phi$ is Grassmann. From \eqref{BRSTtrans}, $\phi$ and $\delta_{\theta}\phi$ have the same statistics and since the $\theta$ parameter and ghosts are Grassmann and therefore $\phi$ and $[Q;\phi]_{\mp}$ have opposite statistics. The nilpotency condition reads in terms of $Q$ as :
\begin{equation}
\delta_{\theta}\delta_{\theta'}\phi=0=i\theta[Q;i\theta'[Q;\phi]_{\mp}]_{\mp}=-\theta\theta'[Q;[Q;\phi]_{\mp}]_{\pm}=-\theta\theta'[Q^2;\phi].
\end{equation}
This is fulfilled in operator form, either if $Q^2$ is proportional to unity or if it is zero. If we define the \emph{ghost number} to be $+1$ for a ghost and $-1$ for an anti-ghost, the generator $Q$ carries a $+1$ ghost number by \eqref{BRSTtrans} and cannot be proportional to the unit operator. In other words the action of $Q^2$, if non zero, "creates" two additional ghosts and cannot be proportional to the identity. Thus in operator form we have :
\begin{equation}
\label{Qnilpotence}
Q^2=0.
\end{equation}\\

We shall see in a moment the implication of the nilpotency condition. But first one has to look at the action of $Q$ on states. The point is that, thanks to \eqref{BRSTtrans} the lagrangian \eqref{Ltotaux} can be written in the form\footnote{The action of the BRST operator on a functional composed of several fields is made by a commutator if the functional is bosonic and an anti-commutator if the functional is fermionic. Here is the proof : the starting point is to consider the product of two fields $\phi_1$ and $\phi_2$ and then the recursion is obvious for functionals of higher order in the fields. We associate them respective phases $\epsilon_1$ and $\epsilon_2$ with $\epsilon_i=+ 1$ if $\phi_i$ is a boson and $\epsilon_i=- 1$ if $\phi_i$ is a fermion, $i=1,2$. The BRST transformation of $\phi_1\phi_2$ reads :
\begin{equation*}
\begin{split}
\delta_{\theta}(\phi_1\phi_2)&=\delta_{\theta}(\phi_1)\phi_2+\phi_1\delta_{\theta}(\phi_2)\\
&=i\theta\left[\left(Q\phi_1-\epsilon_1\phi_1Q\right)\phi_2+\epsilon_1\phi_1\left(Q\phi_2-\epsilon_2\phi_2Q\right)\right]\\
&=i\theta\left[Q\phi_1\phi_2-\epsilon_1\epsilon_2\phi_1\phi_2Q\right].
\end{split}
\end{equation*}
The sign of the bracket depends on $\epsilon_1\epsilon_2$, that is, of the \emph{total} statistics of the product $\phi_1\phi_2$ and the proof is completed.
} :
\begin{equation}
\mathcal{L}_{\rm tot}=\mathcal{L}[A,\psi]-i\left\{Q;\frac{\xi}{2}\bar{\omega}^AB^A+\bar{\omega}^A n\cdot A^A\right\}\equiv\mathcal{L}[A,\psi]+i\left\{Q;\mathcal{F}\right\},
\end{equation}
where the brackets denote the anti-commutator. It turns out that whatever is the gauge fixing, the lagrangian can always be written in this form with a different function $\mathcal{F}$ - precisely $n\cdot A$ is replaced by the gauge fixing function in general and the BRST action on the gauge fixing function gives the same ghost lagrangian as the one obtained by the Fadeev-Popov method in the corresponding gauge. Therefore an infinitesimal gauge transformation will induce a variation $i[Q;\delta\mathcal{F}]$ in the gauge fixed lagrangian. Let us consider two arbitrary states $|\alpha>$ and $|\beta>$. Of course, they satisfy the gauge invariant normalization condition $<\alpha|\beta>=\delta_{\alpha\beta}$. Thus the variation $\delta<\alpha|\beta>$ under a gauge transformation must be zero but this implies :
\begin{equation}
0=\delta\left\langle\alpha\left.\right|\beta\right\rangle=-\int\rmd^4x\left\langle\alpha\left|\left\{Q;\delta\mathcal{F}(x)\right\}\right|\beta\right\rangle.
\end{equation}
In order for this condition to be fulfilled for an arbitrary infinitesimal gauge transformation, one must have :
\begin{equation}
\label{physcond}
Q\left|\alpha\right\rangle=Q\left|\beta\right\rangle=0.
\end{equation}
Therefore the action of the BRST generator on a state must be zero. This condition is known as the \emph{physicality condition}. Such states which are in the kernel of $Q$ are called \emph{BRST-closed}.

Let $\mathcal{H}$ be the total Hilbert space spanned with states containing arbitrary numbers of particles associated with matter, gauge, auxiliary, ghost and anti-ghost fields. The physicality condition \eqref{physcond} together with the nilpotency condition \eqref{Qnilpotence} shares the Hilbert space into three pieces :
\begin{itemize}
\item some states may not satisfy the physicality condition, i.e. $Q|\alpha>\neq 0$. Such states $|\alpha>$ do not enter in the spectrum of the theory.\\
\item among the physical states that are BRST-closed, some of them can be \emph{BRST-exact}, that is, there is $|\beta>\in\mathcal{H}$ so that $|\alpha>=Q|\beta>$.\\
\item the last class of states are those that are BRST-closed but not exact. That is they satisfy the physicality condition \eqref{physcond} but cannot be written as $|\alpha>=Q|\beta>$.
\end{itemize}
$Q$ being an application of $\mathcal{H}$ into itself, we have just seen that the physical Hilbert space is made of states in the kernel of $Q$. Let us consider an arbitrary physical state $|\alpha>$ not necessarily BRST-exact and a state $|\beta>$ which is BRST-exact, that is it can be written $|\beta>=Q|\gamma>$. By the physicality condition, the scalar product between $|\alpha>$ and $|\beta>$ is zero : $<\alpha|\beta>=<\alpha|Q|\gamma>=0$. Therefore distinguishing, in the spectrum, two states that only differ by a BRST-exact state is redundant. This introduces an equivalence relation $\sim$ among physical states :
\begin{equation}
\left|\alpha\right\rangle\sim\left|\beta\right\rangle~~~\text{if there is}~~~\left|\gamma\right\rangle\in\mathcal{H}~~~\text{so that}~~~\left|\alpha\right\rangle=\left|\beta\right\rangle+Q\left|\gamma\right\rangle.
\end{equation}
Two physical states are said equivalent if they differ by a BRST-exact state. As argued the BRST-exact contribution will not contribute to matrix element and equivalent states lead to the same matrix element. Such structure is known in mathematics as a \emph{cohomology} and equivalent states are said \emph{cohomologous}. This allows some formal developments to see the structure of the physical Hilbert space.

\subsection{BRST cohomology}
\indent

From previous considerations, the physical Hilbert space is composed of the elements of $Ker~Q$ modulo elements of the image $Q\mathcal{H}$. This quotient space is known as the BRST \emph{cohomology ring} $H^*$ :
\begin{equation}
Ker~Q/Q\mathcal{H}\equiv H^*.
\end{equation}
It will be convenient for the following to decompose the total Hilbert space $\mathcal{H}$ into Hilbert spaces at fixed ghost number $\mathcal{H}_p$ with $p\in\mathbb{Z}$. $\mathcal{H}_p$ is the vector space spanned by the states containing $n$ ghosts and $m$ anti-ghosts - and all other possible particles - so that $n-m=p$. Thus the total Hilbert space reads :
\begin{equation}
\mathcal{H}=\displaystyle{\bigoplus_{p\in\mathbb{Z}}}~\mathcal{H}_p.
\end{equation}
The BRST generator $Q$ increases the ghost number of one unit since its action on any field or functional of fields generates an additional ghost field. Thus the BRST operator makes the following hierarchy between the $\mathcal{H}_p$'s known as the \emph{cochain complex} :
\begin{equation}
...\xrightarrow{Q}\mathcal{H}_{-1}\xrightarrow{Q}\mathcal{H}_{0}\xrightarrow{Q}\mathcal{H}_{1}\xrightarrow{Q}...
\end{equation}
The $p$-th BRST \emph{homology group}, denoted $H^p$, is the analog of the BRST cohomology ring but at fixed ghost number : it is the kernel of $Q$ in $\mathcal{H}_p$, denoted $Ker_p~Q$, modulo an element of the image of $\mathcal{H}_{p-1}$ by $Q$
\begin{equation}
H^p=Ker_p~Q/Q\mathcal{H}_{p-1}. 
\end{equation}
The physical Hilbert space, given by the BRST cohomology ring $H^*$, thus reads in terms of the BRST cohomology groups as :
\begin{equation}
H^*=\displaystyle{\bigoplus_{p\in\mathbb{Z}}}~H^p.
\end{equation}\\

The physical spectrum will be obtained by explicitly writing a basis for $H^*$. The decomposition into cohomology groups at fixed ghost numbers is convenient since we shall see that the physical Hilbert space is given by $H^0$ only.

\section{Physical spectrum}
\indent

After these formal developments around the BRST cohomology let us compute the cohomology groups explicitly. For this purpose one has to see the action of the BRST operator $Q$ on various states. Since we are interested in this action on asymptotic states it is sufficient to consider the zero coupling limit of the BRST transformations \eqref{BRSTtrans} and the corresponding fields are free fields. The zero coupling limit of \eqref{BRSTtrans} reads :
\begin{equation}
\label{zerocouplingBRST}
\begin{split}
\left[Q;\psi\right]_{\mp}&=\left\{Q;\omega\right\}=\left[Q;B\right]=0\\
\left[Q;A_{\mu}\right]&=-i\partial_{\mu}\omega\\
\left\{Q;\bar{\omega}\right\}&=iB.
\end{split}
\end{equation}
Moreover, denoting generically by $\phi_i$ any of these field, with $i$ denoting possible Lorentz, color... indices, we expand these fields into normal modes :
\begin{equation}
\phi_i(x)=\int\frac{\rmd^3k}{(2\pi)^32k^0}\left[\varphi_{i\mathbf{k}}e^{-ik\cdot x}+\varphi^*_{i\mathbf{k}}e^{ik\cdot x}\right].
\end{equation}
To maintain Lorentz invariance, $\varphi_{i\mathbf{k}}$ and $\varphi^*_{i\mathbf{k}}$ must be interpreted as annihilation and creation operators respectively\footnote{For complex matter fields $\varphi^*_{i\mathbf{k}}$ is actually $\varphi^*_{c,i\mathbf{k}}$, the creation operator for an anti-particle. We shall see that there is no trouble with this subtlety for the discussion to come for these kinds of fields.} :
\begin{equation}
\begin{split}
&\varphi_{i\mathbf{k}}\left|0\right\rangle=0\\
&\varphi^*_{i\mathbf{k}}\left|0\right\rangle=\left|\varphi^*_{i\mathbf{k}}\right\rangle.
\end{split}
\end{equation}
The way to find the basis for the cohomology groups is to take a physical state $|\alpha>$ which is BRST-closed but not BRST-exact\footnote{There exists at least one such state, the BRST-invariant vacuum. Thus the existence does not cause trouble.} and to add a quanta $\varphi^*_{i\mathbf{k}}$. The resulting state may be not physical, that is not BRST-closed. It it is BRST-closed, it may be BRST-exact and thus homologous to zero. In both of these cases it does not enter into the physical spectrum.

\subsection{The matter content}
\indent

Let us begin by the simplest case : states with particles associated to matter fields. The normal mode expansion of a matter field is written :
\begin{equation}
\psi_i(x)=\int\frac{\rmd^3k}{(2\pi)^32k^0}\left[\psi_{i\mathbf{k}}e^{-ik\cdot x}+\psi^*_{c,i\mathbf{k}}e^{ik\cdot x}\right].
\end{equation}
From the BRST transformations \eqref{zerocouplingBRST}, we have, by equating the exponential coefficients :
\begin{equation}
\left[Q;\psi_{i\mathbf{k}}\right]_{\mp}=\left[Q;\psi^*_{c,i\mathbf{k}}\right]_{\mp}=0.
\end{equation}
Obviously if we had considered $\psi^*_i(x)$ or, just took the complex conjugate of the previous (anti-)commutation relations, would have led to the (anti-)commuting character of 
$\psi^*_{i\mathbf{k}}$ with $Q$. Let us take a physical (i.e. BRST-closed) state $|\alpha>$ and add a matter quanta $\psi^*_{i\mathbf{k}}$. The BRST action on this state is :
\begin{equation}
Q\psi^*_{i\mathbf{k}}\left|\alpha\right\rangle=\left[Q;\psi^*_{i\mathbf{k}}\right]_{\mp}\left|\alpha\right\rangle=0.
\end{equation}
Thus the state with an additional matter particle satisfies the physicality condition. From the transformations \eqref{zerocouplingBRST} one cannot get such state by the action of $Q$ on another state since there is no BRST transformation that create a matter field. Therefore $\psi^*_{i\mathbf{k}}|\alpha>$ is not closed and appears in the spectrum. States containing only matter particles are all inequivalent and therefore the whole matter Hilbert space $\mathcal{H}_{\rm mat}$ spanned by all the matter states is physical :
\begin{equation}
\mathcal{H}_{\rm mat}\subset H^*.
\end{equation}
This is a well known result that gauge invariance does not constrain the matter spectrum.

\subsection{Anti-ghosts and auxiliary fields}
\indent

Let us look at a more interesting case by considering the closed system of equations in \eqref{zerocouplingBRST} involving the auxiliary fields and the anti-ghosts. Expanding these to fields in normal modes
\begin{equation}
\begin{split}
B(x)&=\int\frac{\rmd^3k}{(2\pi)^32k^0}\left[b_{\mathbf{k}}e^{-ik\cdot x}+b^*_{\mathbf{k}}e^{ik\cdot x}\right]\\
\bar{\omega}(x)&=\int\frac{\rmd^3k}{(2\pi)^32k^0}\left[\bar{\omega}_{\mathbf{k}}e^{-ik\cdot x}+\bar{\omega}^*_{\mathbf{k}}e^{ik\cdot x}\right]
\end{split}
\end{equation}
together with \eqref{zerocouplingBRST} leads to\footnote{Relations among the annihilation operators are not interesting for our purpose and are not written.} :
\begin{equation}
\begin{split}
&\left[Q;b^*_{\mathbf{k}}\right]=0\\
&\left\{Q;\bar{\omega}^*_{\mathbf{k}}\right\}=ib^*_{\mathbf{k}}.
\end{split}
\end{equation}
It follows that a physical state $|\alpha>$ with an additional anti-ghost quanta $\bar{\omega}^*_{\mathbf{k}}$ satisfies :
\begin{equation}
Q\bar{\omega}^*_{\mathbf{k}}\left|\alpha\right\rangle=\left\{Q;\bar{\omega}^*_{\mathbf{k}}\right\}\left|\alpha\right\rangle=ib^*_{\mathbf{k}}\left|\alpha\right\rangle.
\end{equation}
On the one hand $b^*_{\mathbf{k}}|\alpha>$ is non zero and thus $\bar{\omega}^*_{\mathbf{k}}|\alpha>$ does not satisfies the physicality condition. This means that anti-ghost are absent from the physical spectrum. On the other hand $b^*_{\mathbf{k}}|\alpha>$ is BRST-exact and therefore cohomologous to zero. Auxiliary fields are also absent from the physical spectrum.\\

The fact that there is no anti-ghost in the physical spectrum has an important consequence on the cohomology groups of negative ghosts numbers. For negative $p$, all the states in $\mathcal{H}_p$ contain at least $-p$ anti-ghosts (possibly more for states containing ghosts as well). Since $Q\bar{\omega}^*_{\mathbf{k}}|\alpha>\neq 0$, it means that the kernel of $\mathcal{H}_p$ is empty and so is the corresponding cohomology group :
\begin{equation}
H^p=\{\emptyset\}~~~\forall p<0.
\end{equation}

\subsection{Ghosts and gauge fields\label{gaugespectrum}}
\indent

We now come to the main aim : the determination of physical gauge field's degrees of freedom in axial gauge. For general vector fields, the time component leads to negative norm states. Actually it is not really a problem since it is not a physical degree of freedom but it is given by a constrain in terms of the physical ones. For massless vector field, gauge invariance enables to constrain one of the three remaining degrees of freedom. There are two degrees of freedom only and we shall see that they are given by the transverse components in light-cone gauge. From now on, the gauge fixing parameter $\xi$ is sent to zero so that the gauge condition $n\cdot A=0$ is strict. Let us expand the ghost and gauge field in normal modes as previously :
\begin{equation}
\begin{split}
\omega(x)&=\int\frac{\rmd^3k}{(2\pi)^32k^0}\left[\omega_{\mathbf{k}}e^{-ik\cdot x}+\omega^*_{\mathbf{k}}e^{ik\cdot x}\right]\\
A_{\mu}(x)&=\int\frac{\rmd^3k}{(2\pi)^32k^0}\left[a_{\mathbf{k}\mu}e^{-ik\cdot x}+a^*_{\mathbf{k}\mu}e^{ik\cdot x}\right].
\end{split}
\end{equation}
\eqref{zerocouplingBRST} leads to the following relations among creation operators :
\begin{equation}
\label{BRSTaomega}
\begin{split}
&\left[Q;a^{\mu*}_{\mathbf{k}}\right]=k^{\mu}\omega^*_{\mathbf{k}}\\
&\left\{Q;\omega^*_{\mathbf{k}}\right\}=0.
\end{split}
\end{equation}
By the second equation a ghost quanta added to a physical state $|\alpha>$ satisfies the physicality condition $Q\omega^*_{\mathbf{k}}|\alpha>= 0$. The question is whether this state is BRST-exact or not. Let us first contract the first of these equations with $n^{\mu}$ projected on $|\alpha>$. By the gauge condition $n\cdot a^*_{\mathbf{k}}=0$ we have :
\begin{equation}
0=\left(n\cdot k\right)\omega^*_{\mathbf{k}}\left|\alpha\right\rangle.
\end{equation}
Therefore, as long as $n\cdot k\neq 0$, the state $\omega^*_{\mathbf{k}}|\alpha>$ is zero. Let us now contract the first equation \eqref{BRSTaomega} with $k^{\mu}$ and project it on $|\alpha>$. Since the states are labeled by the spatial components of an on-shell momentum one has $k^2=0$ and thus :
\begin{equation}
\label{BRSTWard}
Q\left(k\cdot a^*_{\mathbf{k}}\right)\left|\alpha\right\rangle=k^2\omega^*_{\mathbf{k}}\left|\alpha\right\rangle=0.
\end{equation}
$k\cdot a^*_{\mathbf{k}}|\alpha>$ satisfies the physicality condition. However recall that $a^*_{\mathbf{k}}$ is proportional to the polarization vector $\epsilon_{\mu}^*(k)$. From, for instance, \eqref{oswardid}, the polarization vector satisfies the on-shell Ward identity $k\cdot \epsilon^*(k)$ as the light-cone gauge is strictly enforced ($\xi=0$). Hence $k\cdot a^*_{\mathbf{k}}|\alpha>$ satisfies the physicality condition because it is zero and equation \eqref{BRSTWard} brings no useful information for the cohomologies. The last contraction we can perform on the first equation \eqref{BRSTaomega} is a contraction with $\epsilon^{\mu}(k)$. This gives :
\begin{equation}
Q\left(\epsilon(k)\cdot a^*_{\mathbf{k}}\right)\left|\alpha\right\rangle=\left(k\cdot\epsilon(k)\right)\omega^*_{\mathbf{k}}\left|\alpha\right\rangle=0.
\end{equation}
Moreover, for a gauge fixing vector $n^{\mu}=\delta^{\mu\pm}$, we have (with explicit polarization $\lambda=\pm 1$ written) :
\begin{equation}
Q\left(\epsilon_{(\lambda)}(k)\cdot a^*_{\mathbf{k}}\right)\left|\alpha\right\rangle=-Q\left(\epsilon^i_{(\lambda)}(k) a^{i*}_{\mathbf{k}}\right)\left|\alpha\right\rangle=0.
\end{equation}
Since the $\epsilon^i_{(\lambda)}$'s with $\lambda=\pm 1$ form a basis of the transverse plane, this is fulfilled if and only if 
\begin{equation}
Qa^{i*}_{\mathbf{k}}\left|\alpha\right\rangle=0.
\end{equation}
The $a^{i*}_{\mathbf{k}}|\alpha>$ are the two independent states satisfying the physicality condition. $a^{\mp*}_{\mathbf{k}}|\alpha>$ is zero by gauge condition and $a^{\pm*}_{\mathbf{k}}|\alpha>$ is given by a constrain and actually disappears from physical amplitudes due to contractions with polarization vectors.\\

We have seen that the physical gluon degrees of freedom are given by their transverse components and that the ghosts states are zero for $n\cdot k \neq 0$. We are a little bit in trouble with the states having $n\cdot k = 0$ since there is no incompatibility with this condition and the Ward identity. Indeed, both these conditions are fulfilled if $k^{\mu}$ is proportional to $n^{\mu}$ for light-like $n^{\mu}$. There is a way out by noticing that the form $n^{\mu}=\delta^{\mu\pm}$ is \emph{not} Lorentz invariant. This implicitly means that the Lorentz frame is fixed when one imposes the gauge condition $A^{\mp}=0$. One can always choose a frame in which all the free particles composing the system are seen with a non vanishing $k^{\mp}$ component. Thus making such choice of frame and \emph{then} imposing the gauge condition ensures that $n\cdot k$ never vanishes. These arguments will be useful when we deal with light-cone quantized systems where the particles all move forward or backward. The choice $n^{\mu}=\delta^{\mu\pm}$ is then canonical for determining a physical gauge. However, there is a more rigorous derivation : $n\cdot k = 0$ means that $k^{\mp}=0$ and since $k$ is on-shell, it has just a single non vanishing component $k^{\pm}$. Taking $\mu=\pm$ in the first equation \eqref{BRSTaomega} leads to $Qa^{\pm*}_{\mathbf{k}}|\alpha>=k^{\pm}\omega^*_{\mathbf{k}}|\alpha>$ which is a priori non-zero\footnote{Even if it is zero this component of the gauge field disappears from physical amplitudes when contracted with a polarization vector.}. Thus $a^{\pm*}_{\mathbf{k}}|\alpha>$ is non physical\footnote{Moreover the Ward identity is trivially verified for such $k$ and we have, by the way, the proof that $a^{\pm*}_{\mathbf{k}}|\alpha>$ is non physical in this case which could not have been proved from \eqref{BRSTWard} since $k^{\mp}a^{\pm*}_{\mathbf{k}}=0$.} and $\omega^*_{\mathbf{k}}|\alpha>$ is BRST-exact and cohomologous to zero and the proof is completed. The physical spectrum contains no ghosts. By the same considerations made above when we have proved the absence of anti-ghosts in the spectrum, we conclude that :
\begin{equation}
H^p=\{\emptyset\}~~~\forall p>0.
\end{equation}
Hence the physical spectrum is given by the cohomology group at ghost number 0 :
\begin{equation}
H^*=H^0.
\end{equation}\\

If we summarize everything, we have seen that there are neither ghosts nor anti-ghosts nor auxiliary fields in the physical spectrum. The whole matter spectrum $\mathcal{H}_{\rm mat}$ is physical. The physical spectrum of gauge fields is the subset $\mathcal{H}^{\perp}_{\rm gf}$ of the total gauge field Hilbert space spanned only by the physical degrees of freedom, that are, the transverse components. The BRST cohomology groups have therefore been explicitly endowed with a basis :
\begin{equation}
H^p = \left\{
    \begin{array}{ll}
         \mathcal{H}_{\rm mat}\otimes \mathcal{H}^{\perp}_{\rm gf}~~~&~~~ \mbox{if}~p=0 \\
        \{\emptyset\} ~~~&~~~ \mbox{if}~p\neq 0.
    \end{array}
\right.
\end{equation}

\end{fmffile}

\bibliographystyle{utcaps}
\bibliography{bibliography}

\end{document}